%% file: Main.tex
\documentclass[11pt,letterpaper]{article}

\usepackage{graphicx}
\usepackage{mathrsfs}
\usepackage{amsthm}
\usepackage{latexsym}
\usepackage{amssymb}
\usepackage{amsmath}

\newtheorem{theorem}{Theorem}[section]
\newtheorem{lemma}{Lemma}[section]
\newtheorem{property}{Property}[section]
\newtheorem{corollary}{Corollary}[section]
\newtheorem{claim}{Claim}[section]
\newtheorem{definition}{Definition}[section]

\def\II{\mathbf{I}}  
\newcommand{\nc}{\newcommand} \def\twt{\text{wt}} \def\fD{{\frak D}}
\nc{\Pf}{{\rm Pf}} \nc{\PfS}{{\rm PfS}} \def\uu{\mathbf{u}}
\nc{\GP}{Grassmann-Pl{\"u}cker } \nc{\adj}{{\rm adj}}
\def\AA{\mathbf{A}} \def\eval{\text{\sf EVAL}} \def\CC{\mathbf{C}}
\def\DD{\mathbf{D}}  \def\FF{\mathbf{F}}
 \def\calQ{\mathcal{Q}} \def\00{\mathbf{0}}
\def\HH{\mathbf{H}} \def\qq{\mathbf{q}} \def\calU{\mathcal{U}}
\def\calR{\mathcal{R}} \def\xx{\mathbf{x}} \def\lcm{\text{lcm}}
\def\calL{\mathcal{L}} \def\calD{\mathcal{D}} \def\yy{\mathbf{y}}
\def\kk{\mathbf{k}} \def\pp{\mathbf{p}} \def\hh{\mathbf{h}}
\def\PP{\mathbf{P}} \def\QQ{\mathbf{Q}} 
\def\GC{$\mathcal{GC}$} \def\11{\mathbf{1}} \def\oo{\omega}
\def\EE{\mathbf{E}} \def\RR{\mathbf{R}} \def\vv{\mathbf{v}}
\def\ww{\mathbf{w}} \def\rr{\mathbf{r}} \def\GG{\mathbf{G}}
\def\dd{\mathbf{d}} \def\aa{\mathbf{a}} \def\bb{\mathbf{b}}
\def\calF{\mathcal{F}}  \def\bft{\mathbf{t}}
\def\zqt{\mathbb{Z}_{\fqq}} \def\cc{\mathbf{c}} \def\SS{\mathbf{S}}
\def\ww{\mathbf{w}} \def\zz{\mathbf{z}} 
\def\lcm{\text{{lcm}}}  \def\TT{\mathbf{T}}
\def\ext{\textbf{ext}} \def\extt{\emph{\textbf{ext}}}
\def\wt{\widetilde} \def\BB{\mathbf{B}} \def\gg{\mathbf{g}}
\def\ee{\mathbf{e}} \def\fqq{{\mathcal{Q}}} \def\lin#1{#1^{\text{lin}}}
\def\fq{\mathcal{Q}}  
 \def\fa{\boldsymbol{\frak a}}
\def\fb{\boldsymbol{\frak b}} 
\def\bcF{\boldsymbol{\mathcal{F}}} \def\cS{\mathcal{S}} \def\cT{\mathcal{T}}
\def\hq{\widehat{\pi}}  \def\COUNT{\text{\sf COUNT}}
\def\cc{\mathbf{c}} \def\dd{\mathbf{d}} 
 \def\mm{\mathbf{m}} \def\nn{\mathbf{n}}
\def\calT{\mathcal{T}} \def\calS{\mathcal{S}} \def\KK{\mathbf{K}}
\def\LL{\mathbf{L}} \def\calK{\mathcal{K}} \def\calL{\mathcal{L}}
\def\XX{\mathbf{X}} \def\YY{\mathbf{Y}} 
\def\evalp{\text{\sf EVALP}} \def\calA{\mathcal{A}} \def\eqk{=_{\mu}}
\def\lek{\le_{\mu}} \def\tmatrix{\left(\begin{matrix}b&c\\b'&c'\end{matrix}\right)}
\def\tone{\left(\begin{matrix}b_1&c_1\\b_1'&c_1'\end{matrix}\right)}
\def\ttwo{\left(\begin{matrix}b_2&c_2\\b_2'&c_2'\end{matrix}\right)}
\def\calHH{\boldsymbol{\mathcal{H}}} \def\calH{\mathcal{H}} 
\def\calW{\mathcal{W}} \def\WW{\mathbf{W}} \def\hpi{\widehat{\pi}}

\begin{document}

\title{{\bf 
Graph Homomorphisms with Complex Values:\vspace{0.03in}\\
  A Dichotomy Theorem\vspace{0.32in}}}

\vspace{0.2in}
\author{Jin-Yi Cai\thanks{University of Wisconsin-Madison:
{\tt jyc@cs.wisc.edu}}
\and Xi Chen\thanks{Columbia University:
{\tt csxichen@gmail.com}}
\and Pinyan Lu\thanks{Microsoft Research Asia:
{\tt pinyanl@microsoft.com}}
}

\date{}
\maketitle
\bibliographystyle{plain}

\begin{abstract}
Graph homomorphism has been studied intensively. Given an $m
\times m$ symmetric matrix $\AA$, the graph homomorphism function is
defined as
\[
Z_\AA (G) = \sum_{\xi:V\rightarrow [m]}\hspace{0.07cm}
\prod_{(u,v)\in E} A_{\xi(u),\xi(v)},
\]
where $G = (V, E)$ is any undirected graph. The function $Z_\AA (\cdot)$
can encode many interesting graph properties, including counting
vertex covers and $k$-colorings. We study the computational
complexity of $Z_\AA (\cdot)$ for arbitrary
  symmetric matrices $\AA$ with algebraic complex values.  Building on work
  by Dyer and Greenhill \cite{DyerGreenhill},
  Bulatov and Grohe~\cite{BulatovGrohe}, and especially the recent beautiful work by
Goldberg, Grohe, Jerrum and Thurley~\cite{GoldbergGJT}, we prove a
complete dichotomy theorem for this problem.
We show that $Z_\AA (\cdot)$ is either computable in polynomial-time
or \#P-hard, depending explicitly on the matrix $\AA$.
We further prove that the tractability criterion on $\AA$
  is polynomial-time decidable.
\end{abstract}
\newpage

\tableofcontents\newpage

\input{Introduction}

\section{Pinning Lemmas and Preliminary Reductions}\label{pinninglemma}

In this section, we prove two pinning lemmas, one for $\eval(\AA)$ and one for
  $\eval(\CC,\fD)$, where $(\CC,\fD)$ satisfies certain conditions.
The proof of the first lemma~is very similar to \cite{GoldbergGJT},
  but the second one has some complications.\vspace{-0.1cm}

\subsection{A Pinning Lemma for $\eval(\AA)$}


Let $\AA$ be an $m\times m$ symmetric complex matrix.
We define $\evalp(\AA)$ as follows:
The input is a triple $(G,w,i)$, where $G=(V,E)$ is an undirected graph,
  $w\in V$ is a vertex, and $i\in [m]$;
The output is
$$
Z_{\AA}(G,w,i)=\sum_{\xi:V\rightarrow [m],\hspace{0.06cm}\xi(w)=i}
  \twt_{\AA}(\xi).
$$
It is easy to see that $\eval(\AA)\le \evalp(\AA)$.
The following lemma shows that the other direction also holds:

\begin{lemma}[First Pinning Lemma]\label{pinning1}
$\evalp(\AA)\equiv \eval(\AA)$.
\end{lemma}

We define the following equivalence relation over $[m]$ (note that
  we do not know, given $\AA$, how to compute this
  relation efficiently, but we know it exists.
The lemma only proves, non-constructively, the existence of a
  polynomial-time reduction. Also see \cite{Lovasz2006}):
$$
i\sim j\ \ \text{if for any undirected graph $G=(V,E)$ and $w\in V$,
  $Z_{\AA}(G,w,i)=Z_{\AA}(G,w,j)$.}
$$
This relation divides the set $[m]$ into $s$
  equivalence classes $\calA_1,\ldots,\calA_s$,
  for some positive integer $s$.
For any $t\ne t'\in [s]$, there exists a pair $P_{t,t'}=(G,w)$,
  where $G$ is an undirected graph and $w$ is a vertex of $G$,
  such that (again,
  we do not know how to compute such a pair efficiently, but it always
  exists by the definition of the equivalence relation $\sim$.)
$$
Z_{\AA}(G,w,i)=Z_{\AA}(G,w,j)
  \ne Z_{\AA}(G,w,i')=Z_{\AA}(G,w,j'),\ \ \
\text{for all $i,j\in \calA_t$ and $i',j'\in \calA_{t'}$.}$$

Now for any subset $S\subseteq [s]$, we define a problem $\eval(\AA,S)$
  as follows:
The input is a pair $(G,w)$, where $G=(V,E)$ is an undirected graph
  and $w$ is a vertex in $G$;
The output is
$$
Z_{\AA}(G,w,S)=\sum_{\xi:V\rightarrow [m],\hspace{0.06cm}\xi(w)\in \bigcup_{t\in S} \calA_t}
  \twt_{\AA}(\xi).\label{ZAGWS}
$$
Clearly, if $S=[s]$, then $\eval(\AA,S)$ is exactly $\eval(\AA)$.
We prove the following claim:

\begin{claim}\label{againclaim}
If $S\subseteq [s]$ and $|S|\ge 2$, then there exists a partition
  $\{S_1,\ldots,S_k\}$ of $S$ for some $k>1$ and
$$
\eval(\AA,S_d)\le \eval(\AA,S),\ \ \ \text{for all $d\in [k]$}.$$
\end{claim}

Before proving this claim, we use it to prove the First Pinning Lemma.

\begin{proof}[Proof of Lemma \ref{pinning1}]
Let $(G,w,i)$ be an input of $\evalp(\AA)$, and $i\in \calA_t$ for
  some $t\in [s]$.
We will use Claim \ref{againclaim} to prove that $\eval(\AA,\{t\})\le
  \eval(\AA)$.
If this is true, then we are done because
$$
Z_{\AA}(G,w,i)=\frac{1}{|\calA_t|}\cdot
Z_{\AA}(G,w,\{t\}). 
$$

To prove $\eval(\AA,\{t\})\le \eval(\AA)$,
  we apply Claim \ref{againclaim} above to $S=[s]$; when $s=1$, Lemma \ref{pinning1}~is
  trivially true.
By Claim \ref{againclaim}, there exists a partition $\{S_1,\ldots,S_k\}$
  of $S$, for some $k>1$, such that
$$
\eval(\AA,S_d)\le \eval(\AA,S)\equiv\eval(\AA),\ \ \ \text{for all $d\in [k]$.}
$$
Without loss of generality, assume $t\in S_1$.
If $S_1=\{t\}$, then we are done;
otherwise we have $t\in S_1$ and $|S_1|\ge 2$.
In this case, we just rename $S_1$ to be $S$ and repeat the process above.
Because $|S|$ is strictly monotonically decreasing after each iteration,
  this procedure\vspace{0.003cm} will stop at some time, and we conclude that
  $\eval(\AA,\{t\})\le \eval(\AA)$.
\end{proof}

\begin{proof}[Proof of Claim \ref{againclaim}]
Let $t\ne t'$ be two integers in $S$ (as $|S|\ge 2$, such $t\ne t'$ exist).
We let $P_{t,t'}=(G^*,w^*)$ where $G^*=(V^*,E^*)$.
It defines the following equivalence relation $\sim^*$ over $S$: For $a,b\in S$,
$$
a \sim^*  b\ \ \text{if $Z_{\AA}(G^*,w^*,i)= Z_{\AA}(G^*,w^*,j)$,\ \  where
  $i\in \calA_a$ and $j\in \calA_b$.}
$$
This equivalence relation $\sim^*$ is clearly well-defined, being independent of
  our choices of $i\in \calA_a,j\in \calA_b$.
It gives us equivalence classes $\{S_1,\ldots,S_k\}$, a partition of $S$.
Because $(G^*,w^*)=P_{t,t'}$, by the definition of $\sim^*$,
  $t$ and $t'$ belong to different classes and thus, $k\ge 2$.
For each $d\in [k]$, we let $X_d$ denote
$$
X_d=Z_{\AA}(G^*,w^*,i),\ \ \ \text{where $i\in \calA_a$ and $a\in S_d$.}
$$
This number $X_d$ is well-defined, and is
  independent of the choices of $a\in S_d$ and $i\in \calA_a$.
Moreover, the definition of the equivalence relation $\sim^*$ implies that
$$X_{d}\ne X_{d'},\ \ \ \ \text{for all $d\ne d'\in [k]$.}$$

Next, let $G$ be an undirected graph and $w$ be a vertex.
We show that, by querying $\eval(\AA,S)$ as an oracle,
  one can compute $Z_{\AA}(G,w,S_d)$ efficiently for all $d$.\vspace{0.006cm}

To this end, for each $p:0\le p\le k-1$ we construct a graph $G^{[p]}=(V^{[p]},E^{[p]})$ as follows.
$G^{[p]}$ is the disjoint union of $G$ and $p$ independent copies of $G^*$, except that
  the $w$ in $G$ and the $w^*$'s in all copies of $G^*$ are
  identified as one single vertex $w'\in V^{[p]}$ and thus,
$$
\big|V^{[p]}\big|= |V|+p\cdot |V^*|-p.
$$
In particular, $G^{[0]}=G$.
We have the following collection of equations: \vspace{0.001cm}
$$
Z_{\AA}(G^{[p]},w',S)=\sum_{d\in [k]} (X_d)^p\cdot Z_{\AA}(G,w,S_d),
\ \ \ \ \text{for every $p\in [0:k-1]$.}\vspace{0.001cm}
$$
Because $X_d\ne X_{d'}$ for all $d\ne d' $, this is a Vandermonde
  system and we can solve it to get $Z_{\AA}(G,w,S_d)$ for~all $d\in [k]$.
As both $k$ and the size of the graph $G^*$ are constants
  that are independent of $G$, this gives us a polynomial-time reduction from
  $\eval(\AA,S_d)$ to $\eval(\AA,S)$, for every $d\in [k]$.
\end{proof}

\subsection{A Pinning Lemma for $\eval(\CC,\fD)$}

Let $\CC$ be the bipartisation of $\FF\in \mathbb{C}^{m\times m}$
  (so $\CC$ is $2m\times 2m$).
Let $\fD=\{\DD^{[0]},\ldots,\DD^{[N-1]}\}$ be a sequence
  of $N$ $2m\times 2m$ diagonal matrices.
We use $\evalp(\CC,\fD)$ to denote the following problem:
The input is a triple $(G,w,i)$, where $G=(V,E)$ is an undirected graph,
  $w\in V$, and $i\in [2m]$;
The output is
$$
Z_{\CC,\fD}(G,w,i)=\sum_{\xi:V\rightarrow [2m],\hspace{0.06cm}\xi(w)=i}
  \twt_{\CC,\fD}(\xi).\label{ZCDI}
$$
It is easy to see that $\eval(\CC,\fD)\le \evalp(\CC,\fD)$. 
However, unlike problems $\evalp(\AA)$ and $\eval(\AA)$ we can only
  prove the other direction when the pair $(\CC,\fD)$ satisfies the following condition:
\begin{enumerate}
\item[] \hspace{-0.5cm}({\sl Pinning})\ \ Every entry\vspace{-0.09cm} of $\FF$
  is a power of $\oo_N$, for some positive integer $N$; $\frac{1}{\sqrt{m}}\cdot
  \FF$ is a unitary \\ matrix; and $\DD^{[0]}$ is the $2m\times 2m$
  identity matrix.\vspace{0.05cm}\label{PINNINGCONDITION}
\end{enumerate}

\begin{lemma}[Second Pinning Lemma]\label{pinning2}
If $(\CC,\fD)$ satisfies \emph{({\sl Pinning})}, then
  $\evalp(\CC,\fD)\equiv \eval(\CC,\fD).$
\end{lemma}

\begin{corollary}
If $(\CC,\fD)$ satisfies the condition \emph{({\sl Pinning})},
  then the problem of computing $Z_{\CC,\fD}^{\rightarrow}$ as well as
  $Z_{\CC,\fD}^{\leftarrow}$ is polynomial time reducible to $\eval(\CC,\fD)$.
\end{corollary}

We define the following equivalence relation over $[2m]$:
$$
i\sim j\ \ \text{if for any undirected graph $G=(V,E)$ and $w\in V$,
  $Z_{\CC,\fD}(G,w,i)=Z_{\CC,\fD}(G,w,j)$.}
$$
This relation divides $[2m]$ into $s$
  equivalence classes $\calA_1,\calA_2,\ldots,\calA_s$, for some positive integer $s$.
For any $t\ne t'\in [s]$ there exists a $P_{t,t'}=(G,w)$,
  where $G$ is an undirected graph and $w$ is a vertex,
  such that
$$
Z_{\CC,\fD}(G ,w,i)=Z_{\CC,\fD}(G,w,j)
  \ne Z_{\CC,\fD}(G ,w,i')=Z_{\CC,\fD}(G ,w,j'),\ \ \ \text{
for all $i,j\in \calA_t$ and $i',j'\in \calA_{t'}$.}$$

Now for any subset $S\subseteq [s]$, we define $\eval(\CC,\fD,S)$
  as follows:
The input is a pair $(G,w)$, where $G=(V,E)$ is an undirected graph
  and $w$ is a vertex in $G$;
The output is
$$
Z_{\CC,\fD}(G,w,S)=\sum_{\xi:V\rightarrow [2m],\hspace{0.06cm}\xi(w)\in
  \bigcup_{t\in S} \calA_t} \twt_{\CC,\fD}(\xi).\label{ZCDS}
$$

Clearly, when $S=[s]$, $\eval(\CC,\fD,S)$ is exactly $\eval(\CC,\fD)$.
We prove the following claim:

\begin{claim}\label{againclaim2}
If $S\subseteq [s]$ and $|S|\ge 2$, there exists a partition
  $\{S_1, \ldots,S_k\}$ of $S$ for some $k>1$, such that\vspace{-0.1cm}
$$\eval(\CC,\fD,S_d)\le \eval(\CC,\fD,S),\ \ \ \text{for all $d\in [k]$.}
$$
\end{claim}


Lemma \ref{pinning2} then follows from Claim \ref{againclaim2}.
Its proof is exactly the same as the one of Lemma \ref{pinning1}
  using Claim \ref{againclaim}, so we omit it here.



\begin{proof}[Proof of Claim \ref{againclaim2}]
Let $t\ne t'$ be two integers in $S$ (as $|S|\ge 2$, such $t\ne t'$ exist).
We let $P_{t,t'}=(G^*,w^*)$ where $G^*=(V^*,E^*)$.
It defines the following equivalence relation over $S$: For $a,b\in S$,
$$
a\sim^* b\ \ \text{if $Z_{\CC,\fD}(G^*,w^*,i)=Z_{\CC,\fD}(G^*,w^*,j)$,\ \ where
  $i\in \calA_a$ and $j\in \calA_b$.}
$$
This gives us equivalence classes $\{S_1,\ldots,S_k\}$, a partition of $S$.
Since $(G^*,w^*)=P_{t,t'}$, $t$ and $t'$ belong to different classes
  and thus, $k\ge 2$.
For each $d\in [k]$, we let $Y_d$ denote
$$
Y_d=Z_{\CC,\fD}(G^*,w^*,i),\ \ \ \text{where $i\in \calA_a$ and $a\in S_d$.}
$$
The definition of the equivalence relation implies that $Y_d\ne Y_{d'}$
  for all $d\ne d'\in [k]$.

Now let $G$ be an undirected graph and $w$ be a vertex.
We show that by querying $\eval(\CC,\fD,S)$ as an oracle,
  one can compute $Z_{\CC,\fD}(G,w,S_d)$ efficiently for all $d\in [k]$.\vspace{0.005cm}

\begin{figure}
\center
\includegraphics[height=5cm]{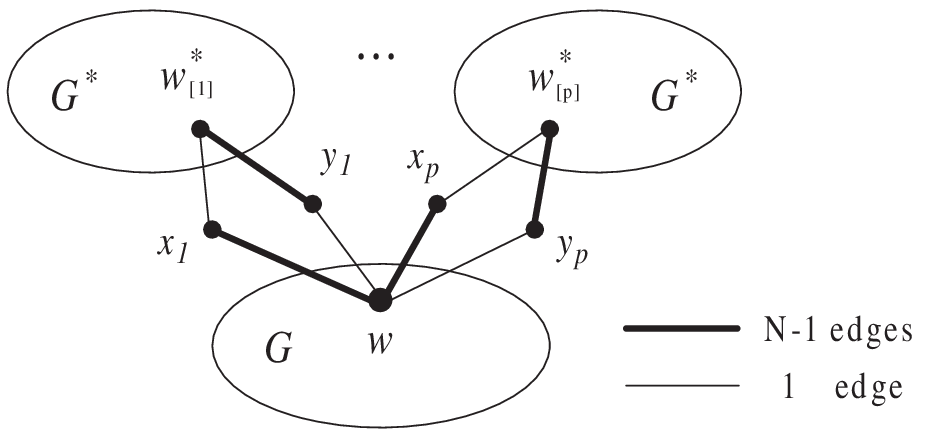}
\caption{Graph $G^{[p]}$, $p\in [0:k-1]$.}\label{figure_1}
\end{figure}

To this end, for each integer $p\in [0:k-1]$, we construct a graph
$G^{[p]}=(V^{[p]},E^{[p]})$ as follows: $G^{[p]}$ contains $G$ and
$p$ independent copies of $G^*$. The vertex $w$ in $G$ is then
\emph{connected appropriately} to the $w^*$ of each $G^*$ (see
Figure \ref{figure_1}). More precisely, we have $V^{[p]}$ as a
disjoint union:\vspace{0cm}
$$
V^{[p]}=V\cup \Big(\bigcup_{i=1}^p\hspace{0.05cm}
  \{v_{[i]}\hspace{0.06cm}\big|\hspace{0.06cm}v\in V^*\}\Big)
  \cup\{x_1,\ldots,x_p,y_1,\ldots,y_p\},
$$
where $x_1,\ldots,x_p,y_1,\ldots,y_p$ are new vertices,
and $E^{[p]}$ contains precisely the following edges:\vspace{0.06cm}
\begin{enumerate}
\item If $uv\in E$, then $uv\in E^{[p]}$; If $uv\in E^*$, then $u_{[i]}v_{[i]}
  \in E^{[p]}$ for all $i\in [p]$;\vspace{-0.1cm}
\item One edge between $(w_{[i]}^*,x_i)$ and $(y_i,w)$ for each $i\in [p]$; and\vspace{-0.1cm}

\item $N-1$ edges between $(x_i,w)$ and $(w_{[i]}^*,y_i)$ for each $i\in [p]$.\vspace{-0.09cm}
\end{enumerate}
In particular, we have $G^{[0]}=G$.

We have the following collection of equations: For $p\in [0:k-1]$,
  $Z_{\CC,\fD}(G^{[p]},w,S)$ is equal to\vspace{0.05cm}
$$
\sum_{\substack{i\in \cup_{a\in S}\calA_a\vspace{0.06cm}\\i_1,...,i_p\in [2m]}}
  Z_{\CC,\fD}(G,w,i) \left(\prod_{j=1}^p Z_{\CC,\fD}(G^*,w^*,i_j)\right)
  \prod_{j=1}^p \left(\sum_{x\in [2m]} C_{i_j,x}\overline{C_{i,x}}
  \sum_{y\in [2m]} \overline{C_{i_j,y}}C_{i,y}\right).\vspace{-0.06cm}
$$
Note that $$\deg(x_i) = \deg(y_i) = N$$ and the changes to the degrees
  of $w$ and $w_{[i]}^*$ are all multiples of $N$.
So by ({\sl Pinning}), there are
  no new vertex weight contributions from $\fD$.

Also by ({\sl Pinning}), we have $$\sum_{x\in [2m]} C_{i_j,x}\overline{C_{i,x}}
  =\langle \FF_{i_j,*},\FF_{i,*}\rangle=0$$ unless $i=i_j$.
Therefore, we have
$$
Z_{\CC,\fD}(G^{[p]},w,S)=m^{2p}\cdot\hspace{-0.2cm} \sum_{i\in \cup_{a\in S}\calA_a} Z_{\CC,
  \fD}(G,w,i)\big(Z_{\CC,\fD} (G^*,w^*,i)\big)^p=
m^{2p}\cdot\sum_{d\in [k]}\left(Y_d \right)^p\cdot Z_{\CC,\fD}(G,w,S_d).
$$
Because $Y_d\ne Y_{d'}$ for all $d\ne d'$, this is a Vandermonde system
  and we can solve it to get $Z_{\CC,\fD}(G,w,S_d)$ for all $d$.
As both $k$ and the size of the graph $G^*$ are constants
  that are independent of $G$, this gives us a polynomial-time reduction from
  $\eval(\CC,\fD,S_d)$ to $\eval(\CC,\fD,S)$ for every $d\in [k]$.
\end{proof}

\subsection{Reduction to Connected Matrices}

The following lemma allows us to focus on the connected components of $\AA$:

\begin{lemma}\label{connected}
Let $\AA\in \mathbb{C}^{m\times m}$ be a symmetric
  matrix with components $\AA_1,\AA_2,\ldots,\AA_s$. Then
\begin{itemize}
\item[--] If $\eval(\AA_i)$ is $\#$P-hard for some
  $i\in [s]$, then $\eval(\AA)$ is $\#$P-hard;\vspace{-0.1cm}
\item[--] If $\eval(\AA_i)$ is polynomial-time computable for every $i \in [s]$,
then so is $\eval(\AA)$.
\end{itemize}
\end{lemma}
\begin{proof}
Lemma \ref{connected} follows directly from the First Pinning Lemma (Lemma \ref{pinning1}).
\end{proof}
The main dichotomy Theorem~\ref{main-in-intro}
will be proved by showing that for every connected
$\AA\in \mathbb{C}^{m\times m}$, the problem
 $\eval(\AA)$ is either solvable in polynomial-time,
or $\#$P-hard.

\section{Proof Outline of the Case: $\AA$ is Bipartite}\label{outlinebipartite}

We now give an overview of the proof of Theorem \ref{main-in-intro} for the
  case when $\AA$ is connected and bipartite.
The proof consists of two parts: a hardness
  part and a tractability part.
The hardness part is further divided into three major steps
  in which we gradually ``simplify'' the problem being considered.
In each of the three steps, we consider an $\eval$ problem
  passed down by the previous step (Step 1 starts with $\eval(\AA)$ itself)
  and show that
\begin{itemize}
\item[--] either the problem is $\#$P-hard; or\vspace{-0.16cm}
\item[--] the matrix that defines the problem
  satisfies certain structural properties; or\vspace{-0.16cm}
\item[--] the problem is polynomial-time equivalent to a new
  $\eval$ problem, and the matrix that \\defines the new problem
  satisfies certain structural properties.
\end{itemize}
One can view the three steps as three filters which remove \#P-hard
  $\eval(\AA)$ problems using different arguments.
Finally, in the tractability part, we show that all the $\eval$ problems
  that survive the three filters are indeed polynomial-time solvable.

\subsection{Step 1: Purification of Matrix $\AA$}

We start with $\eval(\AA)$, where $\AA\in \mathbb{C}^{m\times m}$
  is a fixed symmetric, connected, and bipartite matrix with \emph{algebraic} entries.
It is easy to see that if $m=1$, then $\eval(\AA)$ is tractable.
So in the discussion below, we always assume $m>1$.
In this step, we show that $\eval(\AA)$ is either {$\#$P-hard} or {polynomial-time
  equivalent} to $\eval(\AA')$, in which $\AA'$ is also an
  $m\times m$ matrix but has a very nice structure.

\begin{definition}\label{purified}
Let $\AA\in \mathbb{C}^{m\times m}$ be a symmetric, connected and bipartite matrix.
We say it is a \emph{purified bipartite matrix} if
there exist positive rational numbers $\mu_1, \ldots,\mu_m$
  and an integer $1\le k<m$ such that
\begin{enumerate}
\item $A_{i,j}=0$ for all $i,j\in [k]$;
  $A_{i,j}=0$ for all $i,j\in [k+1:m]$; and\vspace{-0.1cm}
\item $A_{i,j}/(\mu_i\mu_j)=A_{j,i}/(\mu_i\mu_j)$ is a root of unity for all
  $i\in [k]$ and $j\in [k+1:m]$.
\end{enumerate}
In other words, there exists a $k \times (m-k)$ matrix
$\BB$ of the form
\[\BB = \left(\begin{matrix}
\mu_1 \\
& \mu_2 \\
& & \ddots\\
& & & \mu_k
\end{matrix}\right)
\left( \begin{matrix}
\zeta_{1,1} & \zeta_{1,2} & \ldots & \zeta_{1,m-k} \\
\zeta_{2,1} & \zeta_{2,2} & \ldots & \zeta_{2,m-k} \\
\vdots      & \vdots      & \ddots & \vdots \\
\zeta_{k,1} & \zeta_{k,2} & \ldots & \zeta_{k,m-k}
\end{matrix}\right)
\left(\begin{matrix}
\mu_{k+1} \\
& \mu_{k+2} \\
& & \ddots\\
& & & \mu_{m}
\end{matrix}\right),\]
where every $\mu_i$ is a positive rational number
  and every $\zeta_{i,j}$ is a root of unity,
and 
\[\AA = \left( \begin{matrix} {\bf 0} & \BB \\
                \BB^{T} & {\bf 0}
       \end{matrix} \right).\]
\end{definition}

\begin{theorem}\label{bi-step-1}
Let $\AA\in \mathbb{C}^{m\times m}$
  be a symmetric, connected, and bipartite matrix with algebraic entries,\\ 
  then either $\eval(\AA)$ is $\#$P-hard or
  there exists an $m\times m$ purified bipartite matrix
  $\AA'$\vspace{-0.06cm} such that $$\eval(\AA)\equiv\eval(\AA').\vspace{-0.06cm}$$ \emph{(}By
  Definition \ref{purified},  $\AA'$ is symmetric and thus, $\eval(\AA')$
  is well defined.\emph{)}
\end{theorem}

\subsection{Step 2: Reduction to Discrete Unitary Matrix}

Now let $\AA\in \mathbb{C}^{m\times m}$ denote a
  purified bipartite matrix.
We prove that $\eval(\AA)$ is either $\#$P-hard or
  polynomial-time equivalent to $\eval(\CC,{\frak D})$
for some $\CC$ and ${\frak D}$,  where the matrix
  $\CC$ is the bipartisation of a discrete unitary matrix, which
is to be defined in the next definition.

\begin{definition}
Let $\FF\in \mathbb{C}^{m\times m}$ be a \emph{(}not necessarily symmetric\emph{)}
  matrix with entries $(F_{i,j})$.
We say $\FF$ is an \emph{$M$-discrete unitary matrix}, for some positive
  integer $M$, if it satisfies the following conditions:
\begin{enumerate}
\item[1.] Every entry $F_{i,j}$ of $\FF$ is a root of unity,
  and $M = \text{\rm lcm}\hspace{0.07cm}\big\{\text{the order of $F_{i,j}:i,j\in [m]$}\big\}$;
\item[2.] $F_{1,i}=F_{i,1}=1$ for all $i\in [m]$; and
\item [3.] For all $i\ne j\in [m]$, $\langle\FF_{i,*},\FF_{j,*}\rangle=0$
  and $\langle \FF_{*,i},\FF_{*,j}\rangle=0$.\vspace{0.1cm}
\end{enumerate}
\end{definition}

Some simplest examples of discrete unitary matrices can be found in Section \ref{highlevel}.
Also note that the tensor product of any two discrete unitary matrices
is also a discrete unitary matrix.

\begin{theorem}\label{bi-step-2}
Let $\AA\in \mathbb{C}^{m\times m}$ be a purified bipartite matrix. Then
  either \emph{1).} $\eval(\AA)$ is tractable; or
 \emph{2).} $\eval(\AA)$ is $\#$P-hard; or \emph{3).} there exists a triple
  $((M,N),\CC,\fD)$ such that $\eval(\AA)\equiv \eval(\CC,\fD)$,
  and $((M,N),\CC,\fD)$ satisfies the following four
  conditions $(\calU_1)$--$(\calU_4)$:\label{CONDITIONU}
\begin{itemize}
\item[]\hspace{-0.5cm}$(\calU_1)$
$M$ and $N$ are positive integers that satisfy $2\hspace{0.05cm}|
  \hspace{0.05cm}N$ and $M\hspace{0.05cm}|\hspace{0.05cm} N$.
$\CC\in \mathbb{C}^{2n\times 2n}$ for some $n\ge 1$, and
  $$\fD=\{\DD^{[0]}, \DD^{[1]},\ldots,\DD^{[N-1]}\}$$ is a
  sequence of $N$ $2n\times 2n$ diagonal matrices over $\mathbb{C}$;
\item[]\hspace{-0.5cm}$(\calU_2)$
$\CC$ is the bipartisation of an $M$-\emph{discrete unitary
  matrix} $\FF\in \mathbb{C}^{n\times n}$. \emph{(}Note that matrices
  $\CC$  and\\  $\FF$
uniquely determine each other\emph{)}; 
\item[]\hspace{-0.5cm}$(\calU_3)$
For all $i\in [2n]$, $D_{i}^{[0]}=1$.
For all $r\in [N-1]$, we have
\begin{eqnarray*}
&\exists\hspace{0.04cm} i\in [n],\hspace{0.06cm} D_{i}^{[r]}
  \ne 0\ \ \Longrightarrow\ \ \exists\hspace{0.04cm} i'\in [n],
  \hspace{0.06cm}D_{i'}^{[r]}=1,&\text{and}\\[0.2ex]
&\exists\hspace{0.04cm} i\in [n+1:2n],\hspace{0.06cm}
  D_{i}^{[r]}\ne 0\ \ \Longrightarrow\ \
  \exists\hspace{0.04cm} i'\in [n+1:2n],\hspace{0.06cm}D_{i'}^{[r]}=1;&
\end{eqnarray*}

\item[]\hspace{-0.5cm}$(\calU_4)$ For all $r \in [N-1]$ and all $i\in [2n]$,
  $D^{[r]}_i\in \mathbb{Q}(\oo_N)$ and $\big|D_i^{[r]}\big|\in \{0,1\}$.\vspace{0.1cm}
\end{itemize}
\end{theorem}

\subsection{Step 3: Canonical Form of $\CC$, $\FF$ and $\fD$}

After the first two steps, the original problem $\eval(\AA)$ is
  shown to be either tractable; or \#P-hard; or
polynomial-time equivalent to a new problem
  $\eval(\CC,\fD)$.
We also know there exist positive integers $M$ and $N$ such
  that $((M,N),\CC,\fD)$ satisfies conditions $(\calU_1)$--$(\calU_4)$.

For convenience, we still use $2m$ to denote the number of
  rows of $\CC$ and $\DD^{[r]}$, though it should be noted
  that this new $m$ is indeed the $n$ in Theorem \ref{bi-step-2},
  which is different from the $m$ used in the first two steps.
We also denote the upper-right $m\times m$ block of $\CC$ by $\FF$.

In this step, we adopt the following convention: Given an
  $n\times n$ matrix, we use $[0:n-1]$, instead of $[n]$,
  to index its rows and columns.
For example, we index the rows of $\FF$ using $[0:m-1]$
  and index the rows of $\CC$ using $[0:2m-1]$.

We start with the special case when $M=1$.
Because $\FF$ is $M$-discrete unitary, we must have $m=1$.
In this case, it is easy to check that $\eval(\CC,\fD)$
  is tractable:\vspace{-0.16cm} $\CC$ is a 2 by 2 matrix 
$$
\left( \begin{matrix} 0 & 1 \\ 1 & 0 \end{matrix} \right);\vspace{-0.16cm}
$$
$Z_{\CC,{\frak D}}(G)$ is $0$ unless $G$ is bipartite;
for connected and bipartite $G$, there are at most two assignments
$\xi : V \rightarrow \{0,1\}$ which could yield non-zero values;
finally, for a graph $G$ with connected components $G_i$ $Z_{\CC,\fD}(G)$
  is the product of $Z_{\CC,\fD}(G_i)$'s.\vspace{0.005cm}


For the general case when the parameter $M>1$ we further investigate the structure of
  $\FF$ as well as the diagonal matrices in
 $\fD$, and derive three necessary conditions on them for
  the problem $\eval(\CC,\fD)$
  to be \emph{not $\#$P-hard}.
In the tractability part, we prove that these conditions
  are actually sufficient for it to be polynomial-time computable.

\subsubsection{Step 3.1: Entries of $\DD^{[r]}$ are either $0$ or Powers of $\oo_N$}

Suppose $((M,N),\CC,\fD)$ satisfies conditions $(\calU_1)$--$(\calU_4)$ and $M>1$.
In the first step, we show that either $\eval(\CC,\fD)$ is \#P-hard
  or every entry of $\DD^{[r]}$ (in $\fD$), $r\in [N-1]$, is either $0$
  or a power of $\oo_N$.

\begin{theorem}\label{step30}
Suppose $((M,N), \CC,\fD)$ satisfies $(\calU_1)$--$(\calU_4)$ and integer $M>1$,
  then either the problem $\eval(\CC,\fD)$~is \#P-hard~or $((M,N),\CC,\fD)$
  satisfies the following additional condition $(\calU_5)$:\vspace{-0.02cm}
\begin{enumerate}
\item[] \hspace{-0.5cm} $(\calU_5)$\ \ For all $r\in [N-1]$ and $i\in [0:2m-1]$,
   $D^{[r]}_i$ is either $0$ or a power of $\oo_N$.\label{CONDITIONU5}
\end{enumerate}
\end{theorem}

\subsubsection{Step 3.2: Fourier Decomposition}

Second, we show that either problem $\eval(\CC,\fD)$ is \#P-hard or we can permute
  the rows and columns of $\FF$, so that the new $\FF$ is the tensor
  product of a collection of \emph{Fourier matrices}, to
be defined in the next definition.

\begin{definition}\label{FourierMatrix}
Let $q>1$ be a prime power,
  and $k\ge 1$ be an integer such that $\gcd\hspace{0.04cm}(k,q)=1$.
We call the following $q\times q$
  matrix $\bcF_{q,k}$ a $(q,k)$-\emph{Fourier matrix}:
  The $(x,y)^{th}$ entry, where $x,y\in [0:q-1]$, is
$$
\omega_{q}^{kxy}=e^{2\pi i\big({kxy}/{q}\big)}.
$$
In particular, when $k=1$, we use $\bcF_{q}$ to denote $\bcF_{q,1}$ for short.
\end{definition}

\begin{theorem}\label{bi-step-3}
Suppose $((M,N),\CC,\fD)$ satisfies conditions $(\calU_1)$--$(\calU_5)$, and integer $M>1$.
Then either $\eval(\CC,\fD)$~is $\#$P-hard or there exist
\begin{enumerate}
\item two permutations $\Sigma$ and $\Pi$ from $[0:m-1]$ to $[0:m-1]$; and\vspace{-0.15cm}
\item a sequence $q_1,q_2,\ldots,q_d$ of $d$ prime powers, for some $d\ge 1$,
\end{enumerate}
such that\vspace{-0.2cm}
\begin{equation}\label{normalform}
\FF_{\Sigma,\Pi}=
\bigotimes_{i\in [d]}\hspace{0.06cm} \bcF_{q_i}.
\end{equation}\end{theorem}

Suppose there do exist permutations $\Sigma$, $\Pi$ and prime powers
  $ q_1,\ldots,q_d$  such that
  $\FF_{\Sigma,\Pi}$ satisfies (\ref{normalform}), then we
  let $\CC_{\Sigma,\Pi}$ denote the bipartisation of $\FF_{\Sigma,\Pi}$
  and $\fD_{\Sigma,\Pi}$ denote a sequence of $N$ $2m\times 2m$
  diagonal matrices in~which the $r^{th}$ matrix is
$$
\left(\begin{matrix}
D^{[r]}_{\Sigma(0)}\\
& \ddots\\
& & D^{[r]}_{\Sigma(m-1)}\\
& & & D^{[r]}_{\Pi(0)+m}\\
& & & & \ddots\\
& & & & & D^{[r]}_{\Pi(m-1)+m}
\end{matrix}\right),\ \ \ \text{$r\in [0:N-1]$.}
$$
It is clear that permuting the rows and
columns of matrices $\CC$ and every $\DD^{[r]}$
by the same permutation pair $(\Sigma,\Pi)$
  does not affect the complexity of $\eval(\CC,\fD)$,
  so $\eval(\CC_{\Sigma,\Pi},\fD_{\Sigma,\Pi})\equiv\eval(\CC,\fD)$.
From now on, we let $\FF$, $\CC$ and $\fD$ denote $\FF_{\Sigma,\Pi}$,
  $\CC_{\Sigma,\Pi}$ and $\fD_{\Sigma,\Pi}$, respectively.
By (\ref{normalform}), the new $\FF$ satisfies
\begin{equation}\label{normalnew}
\FF=\bigotimes_{i\in [d]}\hspace{0.06cm} \bcF_{q_i}.
\end{equation}

Before moving forward, we rearrange the prime powers $q_1,q_2,\ldots,q_d$
  and divide them into groups according to different primes.
We need the following notation.

Let $\pp=(p_1,\ldots,p_s)$ be a sequence of primes such that $p_1<p_2<
  \hspace{-0.05cm}\ldots\hspace{-0.05cm}<p_s$
  and $\bft=(t_1,\ldots,t_s)$ be a sequence of positive integers.
Let $\fqq=\{\qq_{i} \hspace{0.07cm}|
  \hspace{0.07cm}i\in [s]\}$ be a collection of $s$ sequences in which
  each $\qq_i$ is a sequence $(q_{i,1},\ldots,q_{i,t_i})$ of powers of $p_i$ such
  that $q_{i,1}\ge \ldots\ge q_{i,t_i}$.
We let $q_i$ denote $q_{i,1}$ for all $i\in [s]$,
\[\mathbb{Z}_{\qq_i}\equiv \prod_{j\in [t_i]}\mathbb{Z}_{q_{i,j}}
= \mathbb{Z}_{q_{i,1}} \times \cdots \times \mathbb{Z}_{q_{i,t_i}},\ \ \text{
  for all $i\in [s]$,}
\]
and\vspace{-0.4cm}
\begin{eqnarray*}
\mathbb{Z}_{\fqq}  \equiv  \prod_{i\in [s],j\in [t_i]}
  \hspace{-0.34cm}\mathbb{Z}_{q_{i,j}} \equiv \prod_{i\in [s]}\mathbb{Z}_{\qq_i}
& \equiv & \mathbb{Z}_{q_{1,1}} \times \cdots \times \mathbb{Z}_{q_{1,t_1}}
 \times\\[-2.2ex]
&&\mathbb{Z}_{q_{2,1}} \times \cdots \times \mathbb{Z}_{q_{2,t_2}}
 \times\\[-1ex]
&        &\ \ \ \ \ \ \ \ \ \ \  \vdots \\[-0.5ex]
&        & \mathbb{Z}_{q_{s,1}} \times \cdots \times \mathbb{Z}_{q_{s,t_s}}
\end{eqnarray*}
be the Cartesian products of the respective finite Abelian groups.
Both $\mathbb{Z}_\fqq$ and $\mathbb{Z}_{\qq_i}$
  are finite Abelian groups, under component-wise operations.
This implies that both $\mathbb{Z}_\fqq$ and $\mathbb{Z}_{\qq_i}$ are
  $\mathbb{Z}$-modules and thus $k\xx$ is well defined for
 all $k\in \mathbb{Z}$
  and $\xx$ in $\mathbb{Z}_\fqq$ or $\mathbb{Z}_{\qq_i}$.
As $\mathbb{Z}$-modules,
we can also refer to their members as ``vectors''.
When we use $\xx$ to denote a vector in
  $\mathbb{Z}_{\fqq}$, we denote its $(i,j)^{th}$ entry
  by $x_{i,j}\in \mathbb{Z}_{q_{i,j}}$. We
  also use $\xx_i$ to denote $(x_{i,j}: j\in [t_i])
  \in \mathbb{Z}_{\qq_i}$, \vspace{0.01cm}so $\xx=(\xx_1,\ldots,\xx_s)$.
Given $\xx,\yy\in \mathbb{Z}_{\fqq}$,
  we let $\xx\pm \yy$ denote the vector in $\mathbb{Z}_{\fqq}$
  whose $(i,j)^{th}$ entry is
 $x_{i,j}\pm y_{i,j}{\pmod{q_{i,j}}}.$
Similarly, for each $i\in [s]$, we can define $\xx\pm \yy$
  for vectors $\xx,\yy\in \mathbb{Z}_{\qq_i}$.\vspace{0.016cm}

By (\ref{normalnew}), there exist $\pp,\bft,\fq$ such that
  $((M,N),\CC,\fD,(\pp,\bft,\fq))$ satisfies the following condition $(\calR)$:\label{CONDITIONR}
\begin{itemize}
\item[$(\calR_1)$]
 $\pp=(p_1,\ldots,p_s)$ is a sequence of primes such that $p_1<\cdots<p_s$;
  $\bft=(t_1,\ldots,t_s)$ is a sequence of positive integers;
  $\fq=\{\qq_{i}\hspace{0.07cm}|\hspace{0.07cm}i\in [s]\}$ is a collection of $s$ sequences,
  in which every $\qq_i$ is a sequence $(q_{i,1},\ldots,q_{i,t_i})$ of
  powers of $p_i$ such that $q_{i,1}\ge \cdots\ge q_{i,t_i}$;\vspace{0.05cm}


\item[$(\calR_2)$] $\CC\in \mathbb{C}^{2m\times 2m}$ is the bipartisation of $\FF\in
  \mathbb{C}^{m\times m}$, and $((M,N),\CC,\fD)$ satisfies
  ($\calU_{1}$)-($\calU_5$);

\item[$(\calR_3)$] There is a bijection $\rho$ from
  $[0:m-1]$ to $\mathbb{Z}_{\fq}$ (so $m=\prod_{i\in [s],j\in [t_i]} q_{i,j}$) such that
\begin{equation}\label{jjj1}
F_{a,b}=\prod_{i\in [s],j\in [t_i]}
\omega_{q_{i,j}}^{x_{i,j}\hspace{0.03cm}y_{i,j}},
  \quad\text{for all\ $a,b\in [0:m-1]$},
\end{equation}
where $(x_{i,j}:i\in [s],j\in [t_i])=\xx=\rho(a)$ and
  $(y_{i,j}:i\in [s],j\in [t_i])=\yy=\rho(b)$.
Note that
  (\ref{jjj1}) above also gives us an expression of $M$ using $\fq$.
It is the product of the largest prime powers $q_{i} = q_{i,1}$
 for each distinct prime $p_i$:
$M= \prod_{i\in [s]} q_{i}.$\vspace{0.03cm}
\end{itemize}

\noindent For convenience, we will from now on directly use $\xx\in \mathbb{Z}_{\fq}$ to
  index the rows and columns of $\FF$:
\begin{equation}\label{new-R3-with-F-in-bold-face-index}
F_{\xx,\yy}\equiv F_{\rho^{-1}(\xx),\rho^{-1}(\yy)}
=\prod_{i\in [s],j\in [t_i]}
\omega_{q_{i,j}}^{x_{i,j}\hspace{0.03cm}y_{i,j}},
  \ \ \ \text{for all\ $\xx,\yy\in \zqt$},
\end{equation}
whenever we have a tuple
  $((M,N),\CC,\fD,(\pp,\bft,\fq))$ that is known to satisfy condition ($\calR$).
We assume that $\FF$ is indexed by $(\xx,\yy)\in \mathbb{Z}_\calQ\times \mathbb{Z}_\calQ$
rather than $(a,b)\in [0:m-1]\times [0:m-1]$, and $(\calR_3)$
refers to (\ref{new-R3-with-F-in-bold-face-index}).
Correspondingly,
to index the entries of matrices $\CC$ and $\DD^{[r]}$, we use
$\{0,1\} \times \mathbb{Z}_{\fqq}$:
$(0,\xx)$ refers to the $\rho^{-1}(\xx)^{th}$ row (or column),
  and $(1,\xx)$ refers to the $(m+\rho^{-1}(\xx))^{th}$ row (or column).\vspace{0.1cm}

\subsubsection{Step 3.3: Affine Support for $\fD$}

Now we have a $4$-tuple $((M,N),\CC,\fD,(\pp,\bft,\fq))$
  that satisfies condition ($\calR$).
In this step, we prove for every $r\in [N-1]$ (recall that $\DD^{[0]}$ is
  already known to be the identity matrix),
  the nonzero entries of the $r^{th}$ matrix $\DD^{[r]}$
  in $\fD$ must have a very nice \emph{coset} structure,
  otherwise $\eval(\CC,\fD)$ is \#P-hard.

For every $r\in [N-1]$, we define $\Lambda_r \subseteq \zqt$
 and $\Delta_r\subseteq \zqt$ as\vspace{0.07cm}
\begin{eqnarray*}
\Lambda_r=\big\{\xx\in \zqt\hspace{0.08cm}\big|\hspace{0.08cm}
  D^{[r]}_{(0,\xx)}\ne 0\big\}\ \ \ \text{and}\ \ \
\Delta_r=\big\{\xx\in \zqt \hspace{0.08cm}\big|\hspace{0.08cm}
  D^{[r]}_{(1,\xx)}\ne 0\big\}.\\[-2.8ex]
\end{eqnarray*}
We let $\cS$ denote the set of $r\in [N-1]$ such that $\Lambda_r\ne \emptyset$
  and $\cT$ denote the set of $r\in [N-1]$ such that $\Delta_r\ne \emptyset$.
We recall the following standard definition of a coset of a group,
specialized to our situation.\vspace{0.1cm}

\begin{definition}
Let $\Phi$ be a nonempty subset of $\mathbb{Z}_{\fqq}$ \emph{(}or $\mathbb{Z}_{
  \qq_i}$ for some $i\in [s]$\emph{)}.
We say $\Phi$ is a \emph{coset} in $\mathbb{Z}_{\fqq}$ \emph{(}or $\mathbb{Z}_{
  \qq_i}$\emph{)} if there exists a
  vector $\xx_0\in \Phi$ such that
$
 \{\xx-\xx_0\hspace{0.05cm} |\hspace{0.05cm} \xx\in \Phi \}
$
is a subgroup of $\mathbb{Z}_{\fqq}$ \emph{(}or $\mathbb{Z}_{\qq_i}$\emph{)}.

Given\vspace{0.004cm} a coset $\Phi$ \emph{(}in $\mathbb{Z}_\calQ$ or $\mathbb{Z}_{\qq_i}$\emph{)},
  we let $\Phi^{\text{\rm lin}}$ denote its corresponding
  subgroup $\{\xx-\xx'\hspace{0.05cm}|\hspace{0.05cm} \xx,\xx'\in \Phi\}$.\hspace{-0.05cm}
Being a subgroup, clearly
 $\Phi^{\text{\rm lin}}=\{\xx-\xx'\hspace{0.05cm}|\hspace{0.05cm} \xx,\xx'\in \Phi\}
= \{\xx-\xx_0\hspace{0.05cm} |\hspace{0.05cm} \xx\in \Phi \}$,
  for any $\xx_0\in \Phi$. \vspace{0.1cm}
\end{definition}

\begin{theorem}\label{bi-step-4}
Let $((M,N),\CC,\fD,(\pp,\bft,\fq))$ be a $4$-tuple that satisfies
  $(\calR)$. Then either $\eval(\CC,\fD)$ is \#P-hard or
  sets $\Lambda_r\subseteq \zqt$ and $\Delta_r\subseteq \zqt$ satisfy the
  following condition $(\calL)$:\vspace{0.07cm}
\begin{itemize}
\item[]\hspace{-0.5cm} $(\calL_1)$ For every $r\in \cS$, $\Lambda_r=\prod_{i=1}^s
  \Lambda_{r,i}$, where for every $i\in [s]$, $\Lambda_{r,i}$
  is a coset in $\mathbb{Z}_{\qq_i}$; and\vspace{-0.05cm}
\item[] \hspace{-0.5cm} $(\calL_2)$ For every $r\in \cT$, $\Delta_r=\prod_{i=1}^s
  \Delta_{r,i}$, where for every $i\in [s]$, $\Delta_{r,i}$
  is a coset in $\mathbb{Z}_{\qq_i}$.\vspace{0.16cm}
\end{itemize}
\end{theorem}

Suppose $\eval(\CC,\fD)$ is not \#P-hard. Then by Theorem \ref{bi-step-4},
  tuple $((M,N),\CC,\fD,(\pp,\bft,\fq))$ satisfies
  not only condition ($\calR$) but also condition ($\calL$).
Actually, by ($\calU_3$), $\fD$ satisfies the following property:\vspace{0.05cm}\label{CONDITIONL}
\begin{itemize}
\item[]\hspace{-0.5cm} $(\calL_3)$ \hspace{0.07cm}For every $r\in \cS$, $\exists\hspace{0.06cm}
  \fa^{[r]}\in \Lambda_r$ such that $D^{[r]}_{(0,\fa^{[r]})}=1$;
 for every $r\in \cT$, $\exists\hspace{0.06cm}\fb^{[r]}\in \Delta_r$,
  $D^{[r]}_{(1,\fb^{[r]})}=1$.\vspace{0.05cm}
\end{itemize}
From now on, when we say condition $(\calL)$, we mean all three conditions
  $(\calL_1)$-$(\calL_3)$.\vspace{0.1cm}

\subsubsection{Step 3.4: Quadratic Structure}

In this final step within
Step 3, we prove that, for every $r\in [N-1]$, the nonzero entries of
  $\DD^{[r]}$ must have a \emph{quadratic}
  structure, otherwise $\eval(\CC,\fD)$ is \#P-hard.
We start with some notation.

Given a vector $\xx$ in $\mathbb{Z}_{\qq_i}$ for some $i\in [s]$,
    we use $\ext_r(\xx)$, where $r\in \cS$, to denote the vector $\xx'\in \zqt$ such that
in the\vspace{-0.03cm} expression $\xx' = (\xx_1', \ldots, \xx_s')
\in \mathbb{Z}_{\fqq} = \prod_{i \in [s]} \mathbb{Z}_{\qq_i}$,
its $i^{th}$ component
    $\xx_i'=\xx$, the vector given in  $\mathbb{Z}_{\qq_i}$,\vspace{-0.09cm}
  and $$\xx_j'=\fa^{[r]}_j,\ \ \text{for all $j\ne i$}.$$
Recall that $\fa^{[r]}$ is a vector we picked from $\Lambda_r$ in condition $(\calL_3)$.
Similarly we let $\ext_r'(\xx)$, where $r\in \cT$,
  denote the vector $\xx'\in \zqt$ such that
  $\xx'_i=\xx$ and $$\xx'_j=\fb^{[r]}_j,\ \ \text{for all $j\ne i$.}$$
Let $\aa$ be a vector in $\mathbb{Z}_{\qq_i}$ for some $i\in [s]$,
  then we use $\widetilde{\aa}$ to denote the vector $\bb\in \zqt$
  such that $\bb_i=\aa$ and $\bb_j=\00$ for all other $j\ne i$.
Also recall that we use $q_k$, where $k\in [s]$, to denote $q_{k,1}$.\vspace{0.05cm}

\begin{theorem}\label{bi-step-5}
Let $((M,N),\CC,\fD,(\pp,\bft,\fq))$ be a tuple that satisfies
  both $(\calR)$ and $(\calL)$ \emph{(}including $(\calL_3)$\emph{)},
  then either $\eval(\CC,\fD)$ is \#P-hard or
  $\fD$ satisfies the following condition $(\calD)$:
\begin{itemize}
\item[]\hspace{-0.5cm}$(\calD_1)$ For every $r\in \cS$, we have\label{CONDITIOND}
\begin{equation}\label{D1eqn-bi-step-5}
D^{[r]}_{(0,\xx)}=D^{[r]}_{(0,\extt_r({\xx_1}))} D^{[r]}_{(0,\extt_r({\xx_2}))} \cdots
   D^{[r]}_{(0,\extt_r({\xx_{s}}))},\ \ \ \text{for all $\xx\in \Lambda_r$.}
\end{equation}

\item[]\hspace{-0.5cm}$(\calD_2)$ For every $r\in \cT$, we have
\begin{equation}\label{D2eqn}
D^{[r]}_{(1,\xx)}=D^{[r]}_{(1,\extt_r'({\xx_1}))} D^{[r]}_{(1,\extt_r'({\xx_2}))} \cdots
  D^{[r]}_{(1,\extt_r'({\xx_{s}}))},\ \ \ \text{for all $\xx\in \Delta_r$.}
\end{equation}

\item[]\hspace{-0.5cm}$(\calD_3)$ For all $r\in \cS$, $k\in [s]$ and $\aa\in
  {\Lambda_{r,k}^{\text{\emph{lin}}}}\subseteq\mathbb{Z}_{\qq_k}$,
  there exist $\bb\in \mathbb{Z}_{\qq_k}$
  and $\alpha\in \mathbb{Z}_{N}$ such that
\begin{equation}\label{targetprove1}
\oo_{N}^\alpha\cdot F_{\xx,\widetilde{\bb}}=D^{[r]}_{(0,\xx+\widetilde{\aa})}\cdot
  \overline{D^{[r]}_{(0,\xx)}},\ \ \ \text{for all\ $\xx\in\Lambda_r$.}
\end{equation}

\item[]\hspace{-0.5cm}$(\calD_4)$ For all $r\in \cT$, $k\in [s]$ and $\aa\in
  {\Delta_{r,k}^{\text{\emph{lin}}}}\subseteq \mathbb{Z}_{\qq_k}$,
  there exist $\bb\in \mathbb{Z}_{\qq_k}$ and $\alpha\in
  \mathbb{Z}_{N}$ such that
\begin{equation}\label{targetprove2}
\oo_{N}^\alpha\cdot F_{\widetilde{\bb},\xx}=D^{[r]}_{(1,\xx+\widetilde{\aa})}\cdot
  \overline{D^{[r]}_{(1,\xx)}},\ \ \ \text{for all\ $\xx\in\Delta_r$.}
\end{equation}
\end{itemize}
Note that in $(\calD_3)$ and $(\calD_4)$,
the expressions on the left-hand-side do not depend
on all other components of $\xx$ except the $k^{th}$ component
$\xx_k$, because all other components of $\widetilde{\bb}$
are $\00$.\vspace{0.065cm}
\end{theorem}

The statements in conditions $(\calD_3)$-$(\calD_4)$
are a technically precise way to express the idea that
there is a quadratic structure on the support of each matrix $\DD^{[r]}$.
We express it in terms of an exponential difference equation.




\subsection{Tractability}

\begin{theorem}\label{tractable-1}
Let $((M,N),\CC,\fD,(\pp,\bft,\fq))$ be a tuple that satisfies
  all three conditions $(\calR),(\calL)$ and\\ $(\calD)$, then
  the problem $\eval(\CC,\fD)$ can be solved in polynomial time.
\end{theorem}

\section{Proof Outline of the Case: $\AA$ is not Bipartite}\label{outlinenonbip}

The definitions and theorems of the case when the fixed matrix $\AA$ is not bipartite are
  similar to, but also have significant differences with, those of the bipartite case.
We will list these theorems.

\subsection{Step 1: Purification of Matrix $\AA$}

We start with $\eval(\AA)$ in which $\AA\in \mathbb{C}^{m\times m}$
  is a symmetric, connected and non-bipartite matrix with algebraic entries.
$\eval(\AA)$ is clearly tractable if $m=1$.
So in the discussion below, we assume $m>1$.

\begin{definition}
Let $\AA\in \mathbb{C}^{m\times m}$ be a symmetric, connected, and
  non-bipartite matrix, then
we say $\AA$ is a \emph{purified} {non-bipartite matrix} if
there exist positive rational numbers $\mu_1,\ldots,\mu_m$
  such that $A_{i,j}/(\mu_i\mu_j)$ is a root of unity for all $i,j\in [m]$.
\end{definition}

In other words,  $\AA$ has the form
\[\AA = \left(\begin{matrix}
\mu_1 \\
& \mu_2 \\
& & \ddots\\
& & & \mu_m
\end{matrix}\right)
\left( \begin{matrix}
\zeta_{1,1} & \zeta_{1,2} & \ldots & \zeta_{1,m} \\
\zeta_{2,1} & \zeta_{2,2} & \ldots & \zeta_{2,m} \\
\vdots      & \vdots      & \ddots & \vdots \\
\zeta_{m,1} & \zeta_{m,2} & \ldots & \zeta_{m,m}
\end{matrix}\right)
\left(\begin{matrix}
\mu_1 \\
& \mu_2 \\
& & \ddots\\
& & & \mu_m
\end{matrix}\right),\vspace{0.05cm}
\]
where $\zeta_{i,j} = \zeta_{j,i}$ are all roots
of unity. We prove the following theorem:\newpage

\begin{theorem}\label{t-step-1}
Let $\AA\in \mathbb{C}^{m\times m}$ be a symmetric, connected, and
  non-bipartite matrix, for some $m>1$.\\
Then either $\eval(\AA)$ is $\#$P-hard or
  there exists a purified non-bipartite matrix\vspace{-0.125cm} $\AA'\in \mathbb{C}^{m\times m}$
  such that $$\eval(\AA)\equiv\eval(\AA').$$
\end{theorem}

\subsection{Step 2: Reduction to Discrete Unitary Matrix}

In this step, we prove the following theorem:

\begin{theorem}\label{t-step-2}
Let $\AA\in \mathbb{C}^{m\times m}$ be a purified non-bipartite matrix,
  then either \emph{1).} $\eval(\AA)$ is tractable; or
  \emph{2).} $\eval(\AA)$ is \#P-hard; or \emph{3).} there exists a triple
  $((M,N),\FF,\fD)$ such that $$\eval(\AA)\equiv \eval(\FF,\fD)$$
  and $((M,N),\FF,\fD)$ satisfies the following conditions $(\calU_1')$-$(\calU_4')$:
\begin{itemize}
\item[]\hspace{-0.5cm}$(\calU_1')$
$M$ and $N$ are positive integers that satisfy $2\hspace{0.04cm}|\hspace{0.04cm}N$
  and $M\hspace{0.04cm}|\hspace{0.04cm} N$. $\FF$
  is an $n\times n$ complex matrix for\\ some $n\ge 1$, and
  $\fD=\{\DD^{[0]},\ldots,\DD^{[N-1]}\}$ is a sequence of $N$
  $n\times n$ diagonal matrices;\vspace{0.005cm}\label{CONDITIONUP}

\item[]\hspace{-0.5cm}$(\calU_2')$
$\FF$ is a \emph{symmetric} $M$-discrete unitary matrix;\vspace{-0.015cm}


\item[]\hspace{-0.5cm}$(\calU_3')$
For all $i\in [n]$, $D_{i}^{[0]}=1$.
For all $r\in [N-1]$, we have
$
\DD^{[r]}\ne \00\ \Longrightarrow\  \exists\hspace{0.06cm} i\in [n],
  \hspace{0.06cm}D_{i}^{[r]}=1;\vspace{-0.015cm}
$

\item[]\hspace{-0.5cm}$(\calU_4')$ For all $r \in [N-1]$ and all $i\in [n]$,
  $D^{[r]}_i\in \mathbb{Q}(\oo_N)$ and $\big|D_i^{[r]}\big|\in \{0,1\}$.\vspace{0.02cm}
\end{itemize}
\end{theorem}

\subsection{Step 3: Canonical Form of $\FF$ and $\fD$}

Now suppose we have a tuple $((M,N),\FF,\fD)$ that satisfies  $(\calU'_1)$-$(\calU'_4)$.
For convenience we still use $m$ to denote the number of rows
  and columns of $\FF$ and each $\DD^{[r]}$ in $\fD$,
  though it should be noted that this new $m$ is indeed the $n$ in Theorem \ref{t-step-2},
  which is different from the $m$ used in the first two steps.

Similar to the bipartite case, we adopt the following convention in
  this step: given an $n\times n$ matrix, we use $[0:n-1]$,
  instead of $[n]$, to index its rows and columns.

We start with the special case when $M=1$.
Because $\FF$ is $M$-discrete unitary, we must have $m=1$ and $\FF=(1)$.
In this case, it is clear that the problem $\eval(\CC,\fD)$ is tractable.
So in the rest of this section, we always assume $M>1$.

\subsubsection{Step 3.1: Entries of $\DD^{[r]}$ are either $0$ or Powers of $\oo_N$}

\begin{theorem}\label{nonbipstep0}
Suppose $((M,N), \FF,\fD)$ satisfies $(\calU_1')$-$(\calU_4')$ and integer $M>1$.
Then either $\eval(\FF,\fD)$ is \#P-hard or $((M,N),\FF,\fD)$
  satisfies the following additional condition $(\calU_5')$:\vspace{-0.01cm}
\begin{enumerate}
\item[]\hspace{-0.5cm}$(\calU_5')$ For all $r\in [N-1]$ and $i\in [0:m-1]$,
  $D^{[r]}_i$ is either zero or a power of $\oo_N$.\label{CONDITIONUP5}
\end{enumerate}
\end{theorem}

\subsubsection{Step 3.2: Fourier Decomposition}\label{reference-for-sec-6.3.2}

Let $q$ be a prime power.
We say $\WW$ is a \emph{non-degenerate} matrix in $\mathbb{Z}_{q}^{2\times 2}$
  if $\WW\xx\ne \00$
for all $\xx\ne \00\in \mathbb{Z}_q^2$.
The following lemma gives some equivalent characterizations of $\WW$
  being non-degenerate.
The proof is elementary, so we omit it here.

\begin{lemma}\label{equivaequiva}
Let $q$ be a prime power and $\WW \in \mathbb{Z}_q^{2\times 2}$.
Then the following statements are equivalent:\vspace{-0.04cm}
\begin{enumerate}
\item $\WW$ is non-degenerate;\ \ \
2. $\xx\mapsto \WW\xx$ is a bijection from $\mathbb{Z}_{q}^2$ to
  $\mathbb{Z}_q^2$;\ \ \
3. $\det(\WW)$ is invertible in $\mathbb{Z}_q$.\vspace{0.08cm}
\end{enumerate}
\end{lemma}



\begin{definition}[Generalized Fourier Matrix]\label{GeneralizedFourierMatrix}
Let $q$ be a prime power and $\WW=(W_{ij})$ be a symmetric non-degenerate matrix
  in $\mathbb{Z}_q^{2\times 2}$.
$\boldsymbol{\calF}_{q,\WW}$ is called
  a \emph{$(q,\WW)$-generalized Fourier matrix}, if it is a
  $q^2\times q^2$ matrix and there is
  a one-to-one correspondence $\rho$ from $[0:q^2-1]$ to $[0:q-1]^2$, such that\vspace{0.14cm}
\begin{equation*}
(\boldsymbol{\calF}_{q,\WW})_{i,j}=\omega_{q}^{W_{11}x_1y_1+ W_{12}x_1y_2
  +W_{21}x_2y_1+ W_{22}x_2y_2},\ \ \ \text{for all $i,j\in [0:q^2-1]$},\vspace{0.14cm}
\end{equation*}
where 
  $\xx=(x_1,x_2)=\rho(i)$ and $\yy=(y_1,y_2)=\rho(j)$.\vspace{0.1cm}
\end{definition}


\begin{theorem}\label{t-step-3}
Suppose $((M,N),\FF,\fD)$ satisfies $(\calU_1')$-$(\calU_5')$,
then either $\eval(\FF,\fD)$ is $\#$P-hard or there exist
  a permutation $\Sigma$ from $[0:m-1]$ to $[0:m-1]$ and
\begin{enumerate}
\item two sequences $\dd=(d_1,\ldots,d_g)$ and $\calW=(\WW^{[1]},\ldots,\WW^{[g]})$,
  for some non-negative $g$
  \emph{(}Note that the $g$ here could be $0$, in which case both $\dd$ and
  $\calW$ are  empty\emph{)}:
  For every $i\in [g]$, $d_i>1$ is a power of $2$, and $\WW^{[i]}$
  is a $2\times 2$ symmetric non-degenerate matrix over $\mathbb{Z}_{d_i}$; and\vspace{-0.04cm}

\item two sequences $\qq=(q_1,\ldots,q_\ell)$ and
  $\kk=(k_1,\ldots,k_\ell)$ \emph{(}Again $\ell$ could be $0$, in which case
  both $\qq$ and $\kk$ are empty\emph{)}, in which for every $i\in [\ell]$, $q_i$ is a
  prime power, $k_i\in \mathbb{Z}_{q_i}$, and $\gcd(q_i,k_i)=1$,
\end{enumerate}
such that
$$
\FF_{\Sigma,\Sigma}=\left(\bigotimes_{i=1}^g {\boldsymbol{\calF}}_{d_i,\WW^{[i]}}\right)
\bigotimes\left(\bigotimes_{i=1}^\ell \boldsymbol{\calF}_{q_i,k_i}\right).\vspace{0.1cm}
$$
\end{theorem}


Suppose there does exist a permutation $\Sigma$ (together with
  $\dd,\calW,\qq$, and $\kk$) such that
  $\FF_{\Sigma,\Sigma}$ satisfies the equation above (otherwise, $\eval(\FF,\fD)$
  is \#P-hard).
Then we apply $\Sigma$ to $\DD^{[r]}$, $r\in [0:N-1]$, to
  get a new sequence $\fD_{\Sigma}$ of $N$ diagonal matrices
  in~which the $r^{th}$ matrix is
$$
\left(\begin{matrix}
D^{[r]}_{\Sigma(0)}\\
& \ddots\\
& & D^{[r]}_{\Sigma(m-1)}\\
\end{matrix}\right).
$$
It is clear that permuting the rows and columns of
  $\FF$ and $\DD^{[r]}$ in $\fD$ by the same permutation $\Sigma$
  does not affect the complexity of $\eval(\FF,\fD)$, so
  $\eval(\FF_{\Sigma,\Sigma},\fD_{\Sigma})\equiv \eval(\FF,\fD)$.
From now on, we simply let $\FF$ and $\fD$ denote
  $\FF_{\Sigma,\Sigma}$ and $\fD_{\Sigma}$, respectively. Thus we have
\begin{equation}\label{juju}
\FF=\left(\bigotimes_{i=1}^g {\boldsymbol{\calF}}_{d_i,\WW^{[i]}}\right)
\bigotimes\left(\bigotimes_{i=1}^\ell \boldsymbol{\calF}_{q_i,k_i}\right).
\end{equation}

Before moving forward to Step 3.3, we rearrange the prime powers in $\dd$ and $\qq$ and
  divide them into groups according to different primes.

By (\ref{juju}), there exist $\dd,\hspace{0.05cm}\cal{W},\hspace{0.05cm}\pp,
  \hspace{0.05cm}\bft,\hspace{0.05cm}\mathcal{Q}$ and $\mathcal{K}$ such that
  tuple $((M,N),\FF,\fD,(\dd, \calW,\pp,\bft,\cal{Q}, \cal{K}))$
  satisfies the following condition $(\calR')$:
\begin{itemize}
\item[$(\calR_1')$]
$\dd=(d_1, \ldots,d_g)$ is a sequence\vspace{0.005cm} of powers of $2$ for some non-negative
  integer $g$, such that if $g>0$, then $d_1\ge \ldots$ $\ge d_g$;
$\calW=(\WW^{[1]}, \ldots,\WW^{[g]})$ is a sequence of matrices. Every $\WW^{[i]}$
  is a symmetric non-degenerate $2\times 2$ matrix over $\mathbb{Z}_{d_i}$
  (Note that $\dd$ and $\calW$ could be empty);\vspace{0.12cm}\label{CONDITIONRP}

$\pp=(p_1,\ldots,p_s)$ is a sequence of $s$ primes, for some $s\ge 1$, such that
  $2=p_1<\ldots <p_s$; $\bft= (t_1,$ $\ldots,t_s)$ is a sequence of integers: $t_1\ge 0$
  and $t_i\ge 1$ for all $i>1$;
  $\mathcal{Q}=\{\qq_i\hspace{0.06cm}|\hspace{0.06cm} i\in [s]\}$ is
  a collection of sequences in which $\qq_i=(q_{i,1},\ldots,q_{i,t_i})$
  is a sequence of powers of $p_i$ such that $q_{i,1}\ge \ldots\ge q_{i,t_i}$ (Only
  $\qq_1$ could be empty. We always fix $p_1$ to be $2$
  even when no powers of $2$ occur in $\calQ$);\vspace{0.13cm}

$\mathcal{K}=\{\kk_i\hspace{0.06cm}|\hspace{0.06cm} i\in [s]\}$
  is a collection of $s$ sequences in which each $\kk_i=(k_{i,1},\ldots,k_{i,t_i})$
  is a sequence of length $t_i$.
For all $i\in [s]$ and $j\in [t_i]$, $k_{i,j}\in [0:q_{i,j}-1]$ and
  $\gcd(k_{i,j},q_{i,j})=\gcd(k_{i,j},p_i)=1$;\vspace{0.06cm}

\item[$(\calR_2')$] $((M,N),\FF,\fD)$ satisfies condition ($\calU'_{1}$)-($\calU'_5$),
  and 
$$
m=\prod_{i\in [g]} (d_i)^2\hspace{0.06cm} \times
  \prod_{i\in [s],j\in [t_i]}q_{i,j}\hspace{0.05cm};\vspace{-0.085cm}
$$

\item[$(\calR_3')$]
There is a one-to-one correspondence $\rho$ from
  $[0:m-1]$ to $\mathbb{Z}^2_{\dd}\times \mathbb{Z}_{\cal{Q}}$, where
$$
\mathbb{Z}^2_{\dd}=\prod_{i\in [g]} \left(\mathbb{Z}_{d_i}\right)^2\ \ \ \text{and}\ \ \
\mathbb{Z}_{\calQ}=\prod_{i\in [s],j\in [t_i]} \mathbb{Z}_{q_{i,j}},
$$
such that (For every $a\in [0:m-1]$, we use $$\big(x_{0,i,j}: i\in [g],j\in \{1,2\}\big)\in
  \mathbb{Z}^2_{\dd}\ \ \ \ \text{and}\
  \ \ \ \big(x_{1,i,j}:i\in [s],j\in [t_i]\big)\in \mathbb{Z}_{\calQ}
  $$
to denote the components of $\xx=\rho(a)\in \mathbb{Z}^2_{\dd}\times \mathbb{Z}_{\calQ}$,
  where $x_{0,i,j}\in \mathbb{Z}_{d_i}$ and $x_{1,i,j}\in \mathbb{Z}_{q_{i,j}}$)
$$
F_{a,b}=\prod_{i\in [g]}\hspace{0.15cm}\oo_{d_i}^{(x_{0,i,1}\hspace{0.08cm}
  x_{0,i,2})\cdot \WW^{[i]} \cdot (y_{0,i,1}\hspace{0.08cm} y_{0,i,2})^T}
\hspace{-0.2cm}\prod_{i\in [s],j\in [t_i]}
  \omega_{q_{i,j}}^{k_{i,j}\cdot x_{1,i,j} y_{1,i,j}},\ \ \
  \text{for all $a,b\in [0:m-1]$,}
$$
where $((x_{0,i,j}),(x_{1,i,j}))=\xx=\rho(a)$ and
  $((y_{0,i,j}),(y_{1,i,j}))=\yy=\rho(b)$.\vspace{0.1cm}
\end{itemize}

\noindent For convenience, we will from now on directly use $\xx\in \mathbb{Z}_{\dd}^2
  \times \mathbb{Z}_{\calQ}$ to\vspace{0.05cm}
  index the rows and columns of $\FF$:
\begin{equation}\label{index}
F_{\xx,\yy}\equiv F_{\rho^{-1}(\xx),\rho^{-1}(\yy)}
=\prod_{i\in [g]}\hspace{0.15cm}\oo_{d_i}^{(x_{0,i,1}\hspace{0.08cm} x_{0,i,2})\cdot \WW^{[i]}
  \cdot (y_{0,i,1}\hspace{0.08cm} y_{0,i,2})^T}
\hspace{-0.2cm}\prod_{i\in [s],j\in [t_i]}
  \omega_{q_{i,j}}^{k_{i,j}\cdot x_{1,i,j}y_{1,i,j}},\ \ \
  \text{for all\ $\xx,\yy$},\vspace{0.05cm}
\end{equation}
whenever we have a tuple
  $((M,N),\FF,\fD,(\dd,\calW,\pp,\bft,\calQ,\calK))$
  that is known to satisfy condition ($\calR'$).
We assume the matrix  $\FF$ is indexed by $(\xx,\yy)$
  rather than $(a,b) \in [0:m-1]^2$, and $(\calR_3')$
  refers to (\ref{index}).

\subsubsection{Step 3.3: Affine Support for $\fD$}\label{sec:Affine-Support}

Now we have a tuple $((M,N),\FF,\fD,(\dd,\calW,\pp,\bft,\calQ,\calK))$
  that satisfies condition ($\calR'$).
In the next step we show for every $r\in [N-1]$ (for $r=0$,
  we already know $\DD^{[0]}$ is the identity matrix),
  the non-zero entries of the $r^{th}$ diagonal matrix $\DD^{[r]}$
  (in $\fD$) must have a coset structure; otherwise $\eval(\FF,\fD)$ is \#P-hard.

\def\calZ{\mathcal{Z}}

For every $r\in [N-1]$, we use $\Gamma_r\subseteq \mathbb{Z}^2_{\dd}\times \mathbb{Z}_{\calQ}$
  to denote the set of $\xx$ such that $$D^{[r]}_{\xx}\ne 0.$$
We also use $\calZ$ to denote the set of $r\in [N-1]$ such that
  $\Gamma_r\ne \emptyset$.\vspace{0.005cm}

\def\hzqt{\hat{\mathbb{Z}}_{\qq_i}}

For convenience, we let $\hat{\mathbb{Z}}_{\qq_i}$, $i\in [s]$,
  denote the following set (or group, more exactly):
When\vspace{-0.01cm} $i>1$, $\hzqt=\mathbb{Z}_{\qq_i}$;
and when $i=1$, $\hat{\mathbb{Z}}_{\qq_1}=\mathbb{Z}^2_{\dd}\times \mathbb{Z}_{\qq_1}$.
This gives us a new way to denote the components of $$\text{$\xx\in \mathbb{Z}_{\dd}^2\times
  \mathbb{Z}_{\calQ}=\prod_{i\in [s]}\hat{\mathbb{Z}}_{\qq_i}$:
  $\xx=(\xx_1,...,\xx_s)$, where $\xx_i\in \hat{\mathbb{Z}}_{\qq_i}$.}$$

\begin{theorem}\label{t-step-4}
Let $((M,N),\FF,\fD,(\dd,\calW,\pp,\bft,\calQ,\calK))$ be a tuple
  that satisfies condition $(\calR')$,
  then either $\eval(\FF,\fD)$ is \#P-hard; or $\fD$ satisfies the
  following condition \emph{(}$\calL_1'$\emph{)}: For every $r\in \calZ$,
\begin{itemize}
\item[]\hspace{-0.5cm} $(\calL_1')$\ \hspace{0.07cm}$\Gamma_r=\prod_{i=1}^{s} \Gamma_{r,i}$,
  where $\Gamma_{r,i}$ is a coset in $\hat{\mathbb{Z}}_{\qq_i}$,
  for all $i\in [s]$.\vspace{0.15cm}\label{CONDITIONLP}
\end{itemize}
\end{theorem}\newpage

Suppose $\eval(\FF,\fD)$ is not \#P-hard, then by Theorem \ref{t-step-4},
  tuple $((M,N),\FF,\fD,(\dd,\calW,\pp,\bft,\calQ,\calK))$ satisfies
  not only ($\calR'$) but also ($\calL_1'$).
By condition ($\calU_3'$), $\fD$ satisfies the following additional property:\vspace{0.05cm}
\begin{itemize}
\item[]\hspace{-0.5cm} $(\calL_2')$\ \hspace{0.07cm}For every $r\in \calZ$, there exists an
  $\fa^{[r]}\in \Gamma_r\subseteq\mathbb{Z}_{\dd}^2\times \mathbb{Z}_{\calQ}
  =\prod_{i\in [s]} \hat{\mathbb{Z}}_{\qq_i}$ such that $D^{[r]}_{\fa^{[r]}}=1$.\vspace{0.05cm}
\end{itemize}
From now on, when we say condition $(\calL')$, we mean both conditions
  $(\calL_{1}')$ and $(\calL_2')$.

\subsubsection{Step 3.4: Quadratic Structure}

In this final step within Step 3 for the non-bipartite case,
  we show that, for any index $r\in [N-1]$, the non-zero entries of
  $\DD^{[r]}$ must have a quadratic
  structure; otherwise $\eval(\FF,\fD)$ is \#P-hard.\vspace{0.005cm}

We need the following notation:
given $\xx$ in $\hat{\mathbb{Z}}_{\qq_i}$ for some $i\in [s]$,
  we let $\ext_r(\xx)$, where $r\in \calZ$, denote the vector $\xx'\in
  \mathbb{Z}_{\dd}^2\times \mathbb{Z}_{\calQ}$ such that
in the expression\vspace{-0.1cm} $$\xx' = (\xx_1', \ldots, \xx_s')
\in \prod_{j \in [s]} \hat{\mathbb{Z}}_{\qq_j},\vspace{-0.2cm}$$
its $i^{th}$ component
    $\xx_i'=\xx$, the vector given in  $\hat{\mathbb{Z}}_{\qq_i}$,
 \vspace{-0.1cm}and
$$\xx_j'=\fa^{[r]}_j,\ \ \ \text{for all $j\ne i$}.\vspace{-0.1cm}$$
Recall that $\fa^{[r]}$ is a vector we picked from $\Gamma_r$ in condition $(\calL_2')$.

Let $\aa$ be a vector in $\hat{\mathbb{Z}}_{\qq_i}$ for some $i\in [s]$.
Then we use $\widetilde{\aa}$ to denote the vector $\bb\in \prod_{j\in [s]}
  \hat{\mathbb{Z}}_{\qq_j}$
  such that $\bb_i=\aa$ and $\bb_j=\00$ for all other $j\ne i$.\vspace{0.05cm}

\begin{theorem}\label{nonbi-step-5}
Suppose $((M,N),\FF,\fD,(\dd,\calW,\pp,\bft,\calQ,\calK))$ satisfies
  conditions $(\calR')$ and $(\calL)$. Then either $\eval(\FF,\fD)$ is \#P-hard or
  $\fD$ satisfies the following condition $(\calD')$:
\begin{itemize}
\item[]\hspace{-0.5cm}$(\calD_1')$ For every $r\in \calZ$, we have\label{CONDITIONDP}
\begin{equation}\label{D1eqn-nonbi-step-5}
D^{[r]}_{\xx}=D^{[r]}_{\extt_r({\xx_1})} D^{[r]}_{\extt_r({\xx_2})} \cdots
   D^{[r]}_{\extt_r({\xx_{s}})},\ \ \ \text{for all $\xx\in \Gamma_r$.}
\end{equation}

\item[]\hspace{-0.5cm}$(\calD_2')$ For all $r\in \calZ$, $k\in [s]$ and $\aa\in
  {\Gamma_{r,k}^{\text{\emph{lin}}}}\subseteq \hat{\mathbb{Z}}_{\qq_k}$,
  there exist $\bb\in \hat{\mathbb{Z}}_{\qq_k}$ and $\alpha\in
  \mathbb{Z}_{N}$ such that
\begin{equation}\label{targetprove3}
\oo_{N}^\alpha\cdot F_{\widetilde{\bb},\xx}=D^{[r]}_{\xx+\widetilde{\aa}}
  \cdot \overline{D^{[r]}_{\xx}},\ \ \ \text{for all\ $\xx\in\Gamma_r$;}
\end{equation}
\end{itemize}
\end{theorem}

Note that in $(\calD_2')$,\vspace{-0.025cm}
  the expressions on the left-hand-side do not depend
  on all other components of $\xx$ except the $k^{th}$ component
  $\xx_k\in \hat{\mathbb{Z}}_{\qq_k}$,
  because all other components of $\widetilde{\bb}$ are $\00$.


\subsection{Tractability}

\begin{theorem}\label{tractable-2}
Let $((M,N),\FF,\fD,(\dd,\calW,\pp,\bft,\calQ,\calK))$ be a tuple\vspace{0.01cm}
  that satisfies all the conditions $(\calR')$,\\ $(\calL')$ and $(\calD')$,
  then the problem $\eval(\FF,\fD)$ can be solved in polynomial time.
\end{theorem}

\section{Proofs of Theorem \ref{bi-step-1} and Theorem \ref{t-step-1}}\label{sec:purification}

In this section, we prove Theorem \ref{bi-step-1} and Theorem \ref{t-step-1}.

Let $\AA=(A_{i.j})$ denote a connected and symmetric $m\times m$ matrix
  in which every entry $A_{i,j}$ is an algebraic number.
(At this moment, we do not make any assumption on whether
  $\AA$ is bipartite or not. $\AA$ could be either bipartite or non-bipartite).
We also let $$\mathscr{A}=\big\{A_{i,j}:i,j\in [m]\big\}$$ denote the finite set of algebraic numbers
  from the entries of $\AA$.
In the first step, we construct a new $m\times m$ matrix $\BB$ from $\AA$,
  which satisfies the following conditions:\vspace{-0.04cm}
\begin{enumerate}
\item $\BB$ is also a connected and symmetric $m\times m$ matrix (so that $\eval(\BB)$
  is well-defined);\vspace{-0.06cm}
\item $\eval(\BB)\equiv \eval(\AA)$; and\vspace{-0.06cm}
\item Every entry of $\BB$ can be expressed as the product
  of a non-negative integer and a root of unity.\vspace{-0.04cm}
\end{enumerate}
We let $\BB'$ be the non-negative matrix
  such that $B_{i,j}'=|B_{i,j}|$.
Then in the second step, we show that,
$$
\eval(\BB') \le \eval(\BB).
$$
Since $\BB'$ is a connected, symmetric
  and non-negative (integer) matrix,
we can apply the dichotomy of Bulatov and
  Grohe~\cite{BulatovGrohe} (see Theorem \ref{basicsharp}) to $\BB'$ and
  show that, either
$\eval(\BB')$ is \#P-hard; or $\BB$ is a (either bipartite or
  non-bipartite, depending on $\AA$) \emph{purified} matrix.
When  $\eval(\BB')$ is \#P-hard,
\[\eval(\BB') \le \eval(\BB) \equiv \eval(\AA),\]
and thus, $\eval(\AA)$
  is also \#P-hard. This proves both Theorem \ref{bi-step-1} and Theorem \ref{t-step-1}.

\subsection{Equivalence between $\eval(\AA)$ and $\COUNT(\AA)$}

Before the construction of the matrix $\BB$,
  we give the definition of a class of counting problems closely related to $\eval(\AA)$.
It has been used in previous work \cite{GoldbergGJT} for
  establishing polynomial-time reductions between different $\eval$ problems.\vspace{0.006cm}

Let $\AA\in \mathbb{C}^{m\times m}$ be a fixed symmetric matrix with algebraic entries,
then the input of the problem $\COUNT(\AA)$ is a pair $(G,x)$,
  where $G=(V,E)$ is an undirected graph and $x\in \mathbb{Q}(\mathscr{A})$.
The output is\vspace{-0.1cm}
$$
\text{\#}_{\AA}(G,x)= \Big|\big\{\text{assignment $\xi:V\rightarrow
  [m]\hspace{0.08cm}\big|\hspace{0.08cm} \text{wt}_{\AA}(\xi)=x$}\big\}\Big|,\label{COUNTA}
$$
a non-negative integer. The following lemma shows that $\eval(\AA)\equiv \COUNT(\AA)$.

\begin{lemma}\label{count}
Let $\AA$ be a symmetric matrix with algebraic entries,
  then $\eval(\AA)\equiv \COUNT(\AA)$.
\end{lemma}
\begin{proof}
To prove $\eval(\AA)\le \COUNT(\AA)$, recall
  that the matrix $\AA$ is considered fixed, with $m$ being a constant.
Let $G=(V,E)$ and $n=|E|$.
We use $X$ to denote the following set of complex numbers:
\begin{equation}\label{definitionreuse}
X=\left\{\prod_{i,j\in [m]}A_{i,j}^{k_{i,j}}\hspace{0.08cm}\Big|\hspace{0.1cm}
  \text{integers $k_{i,j}\ge 0$ and $\sum_{i,j\in [m]}k_{i,j}=n$}
\right\}.
\end{equation}
It is easy to see that $|X|$ is polynomial in $n$,
  being ${n+m^2-1\choose m^2-1}$ counting multiplicity,
  and the elements in $X$ can be enumerated in polynomial time (in $n$).
It then follows from the expression
  in the definition of $\text{wt}_{\AA}(\xi)$ that
  for any $x\notin X$, $\text{\#}_{\AA}(G,x)= 0$.
This gives us the following relation:
\[Z_\AA(G)= \sum_{x\in X} x \cdot \text{\#}_{\AA}(G,x),\ \ \ \text{for any undirected
  graph $G$,}\]
and thus, $\eval(\AA)\le \COUNT(\AA).$


For the other direction,
  we construct, for any $p\in [|X|]$ (Recall that $|X|$ is polynomial in $n$),
  a new undirected graph $G^{[p]}$ from $G$ by replacing
  every edge $uv$ of $G$ with $p$ parallel edges between $u$ and $v$.
It is easy to check that for any assignment $\xi$,
  if its weight over $G$ is $x$, then its weight over $G^{[p]}$ must be $x^p$.
This gives us the following collection of equations: For every $p\in [|X|]$,
$$
Z_\AA(G^{[p]})= \sum_{x\in X} x^p \cdot \text{\#}_{\AA}(G,x),\ \ \ \text{for any undirected
  graph $G$.}
$$
Note that this is a Vandermonde system. Since we can query $\eval(\AA)$ for the
  values of $Z_{\AA}(G^{[p]})$,
  we can solve it and get $\text{\#}_{\AA}(G,x)$ for\vspace{0.0015cm} every non-zero $x\in X$.
To obtain $\text{\#}_{\AA}(G,0)$ (if $0\in X$), we note that
$$
\sum_{x\in X} \text{\#}_{\AA}(G,x) =m^{|V|}.
$$
This gives us a polynomial-time reduction and thus, $\COUNT(\AA)\le \eval(\AA)$.
\end{proof}

\subsection{Step 1.1}\label{generating-sec}

We now show how to build the desired $\BB$ from $\AA$.
We need the following notion of a \emph{generating set}.

\begin{definition}\label{def-generating}
Let $\mathscr{A}=\{a_j\}_{j\in [n]}$ be a set of $n$ non-zero algebraic numbers,
  for some $n \ge 1$.
Then we say $\{g_1,\ldots,g_d\}$, for some integer $d \ge 0$,
 is a \emph{generating set} of $\mathscr{A}$ if
\begin{enumerate}
\item Every $g_i$ is a non-zero algebraic number in $\mathbb{Q}(\mathscr{A})$;\vspace{-0.06cm}
\item For all $(k_1,\ldots,k_d)\in \mathbb{Z}^d$ such that
  $(k_1,\ldots,k_d)\ne \00$, we have
  $$
g_1^{k_1}\cdots g_d^{k_d} \ \text{is not a root of unity.}
  $$
\item For every $a\in \mathscr{A}$, there exists a unique $(k_1,\ldots,k_d)\in
  \mathbb{Z}^d$ such that
$$
\frac{a}{g_1^{k_1}\cdots g_d^{k_d}}\ \text{is a root of unity.}
$$
\end{enumerate}
\end{definition}
Clearly $d=0$ iff the set $\mathscr{A}$ consists of roots of unity only.
The next lemma shows that \emph{every} $\mathscr{A}$
  has a generating set.

\begin{lemma}\label{generating-set-exists}
Let $\mathscr{A}=\{a_j\}_{j\in [n]}$ be a set of non-zero algebraic numbers,
  then it has a generating set.
\end{lemma}

Lemma \ref{generating-set-exists} follows directly from Theorem \ref{lattice}
  in Section \ref{sec:dec}.
Actually, the statement of Theorem \ref{lattice} is stronger: A generating set
  $\{g_1,g_2,\ldots,g_d\}$ can be computed from $\mathscr{A}$ in polynomial time.
More precisely, following the model of computation discussed in Section \ref{complexmodel},
  we let $\alpha$ be a primitive element of $\mathbb{Q}(\mathscr{A})$
  so that $\mathbb{Q}(\mathscr{A})=\mathbb{Q}(\alpha)$
  and let $F(x)$ be a minimal polynomial of $\alpha$.
Then Theorem \ref{lattice} shows that, given the standard representation of the
  $a_j$'s, one can compute the standard representation of $g_1$ $\ldots,g_d\in \mathbb{Q}(\alpha)$ in polynomial time in the input size of the $a_j$'s,
  with $\{g_1,\ldots,g_d\}$ being a generating set of $\mathscr{A}$.
Moreover, for each element $a\in \mathscr{A}$ one can also compute in polynomial time
  the unique tuple of integers $(k_1,\ldots,k_d)$ such that
$$
\frac{a}{g_1^{k_1}\cdots g_d^{k_d}}
$$
is a root of unity.
In addition, if we are also given an approximation $\hat{\alpha}$ of $\alpha$ that uniquely
  determines $\alpha$ as a root of $F(x)$, then we can
  use it to determine which root of unity it is in polynomial time.
Note that in Lemma \ref{generating-set-exists}
  we only need the existence of a generating set $\{g_1,\ldots,g_d\}$.
But later in Section \ref{sec:dec},
  the polynomial-time computability of a generating set will be
  critical to the proof of Theorem \ref{theo-decidability},
  the polynomial-time decidability of the dichotomy theorem.\vspace{0.006cm}

Now we return to the construction of $\BB$, and
  let $\mathscr{A}$ denote the set of all non-zero entries $A_{i,j}$ from $\AA$.
By Lemma~\ref{generating-set-exists},
we know that it has a generating set $\mathscr{G}=\{g_1,\ldots,g_d\}$.
So for each $A_{i,j}$, there exists a unique tuple $(k_1,\ldots,k_d)$
  such that
$$
\frac{A_{i,j}}{g_1^{k_1}\cdots g_d^{k_d}}\ ~~ \text{is a root of unity},
$$
and we denote it by $\zeta_{i,j}$.\vspace{0.006cm}

The matrix $\BB=(B_{i,j})\in \mathbb{C}^{m\times m}$ is constructed as follows. Let
  $p_1<\cdots <p_d$ denote the $d$ smallest primes.
For every $i,j\in [m]$, we define $B_{i,j}$.
If $A_{i,j}=0$, then $B_{i,j}=0$. Suppose  $A_{i,j} \not = 0$. Because
  $\mathscr{G}$ is a generating set,
  we know there exists a unique tuple of integers $(k_1,\ldots,k_d)$ such that
$$
\zeta_{i,j}=\frac{A_{i,j}}{g_1^{k_1}\cdots g_d^{k_d}}\ ~~ \text{is a root of unity}.
$$
Then we set $B_{i,j}$ to be\vspace{-0.1cm}
$$
B_{i,j}=p_1^{k_1}\cdots p_d^{k_d}\cdot \zeta_{i,j}\vspace{0.08cm}
$$

So what we did in constructing $\BB$ is just replacing each
  $g_i$ in $\mathscr{G}$ with a prime $p_i$.
$B_{i,j}$ is well-defined by the uniqueness of
$(k_1,\ldots,k_d)\in\mathbb{Z}^d$ and conversely by taking the prime
factorization of $|B_{i,j}|$ we can recover $(k_1,\ldots,k_d)$
uniquely, and then recover $A_{i,j}$ by
\[A_{i,j} =
 g_1^{k_1}\cdots g_d^{k_d}\cdot  \frac{B_{i,j}}{p_1^{k_1}\cdots p_d^{k_d}}.\]
The next lemma shows that such a replacement does not affect
  the complexity of $\eval(\AA)$.

\begin{lemma}
Let $\AA\in\mathbb{C}^{m\times m}$ be a symmetric and connected matrix with algebraic entries
  and let $\BB$ be the $m\times m$ matrix constructed above,
  then $$\eval(\AA)\equiv \eval(\BB).$$
\end{lemma}
\begin{proof}
By Lemma \ref{count}, it suffices to prove that $\COUNT(\AA)\equiv \COUNT(\BB)$.
Here we only prove one of the two directions:
  $\COUNT(\AA)\le \COUNT(\BB)$. The other direction can be proved similarly.\vspace{0.007cm}

Let $(G,x)$ be an input of $\COUNT(\AA)$, where $G=(V,E)$ and $n=|E|$.
We use $X$ to denote the set of algebraic numbers defined earlier in (\ref{definitionreuse}).
Recall that $|X|$ is polynomial in $n$ since $m$ is a constant,
  and can be enumerated in polynomial time.
Furthermore, if $x\notin X$, then $\#_\AA(G,x)$ must be zero.

Now suppose $x\in X$, then we can find a
particular  sequence of non-negative integers
  $\{k_{i,j}^{*}\}_{i,j\in [m]}$
in polynomial time, such that $\sum_{i,j}k_{i,j}^{*}=n$ and\vspace{-0.3cm}
\begin{equation}\label{ineedu}
x=\prod_{i,j\in [m]} A_{i,j}^{k_{i,j}^{*}}.
\end{equation}
This sequence $\{k_{i,j}^{*}\}_{i,j\in [m]}$
is in general \emph{not unique} for the given $x$.
Using $\{k^*_{i,j}\}$, we define $y$ by
\begin{equation}\label{ineeddu}y = \prod_{i,j\in [m]} B_{i,j}^{k_{i,j}^{*}}.\end{equation}
It is clear that $x=0$ iff $y=0$.
This happens precisely when some $k_{i,j}^{*} >0$ for some entry $A_{i,j} =0$.\vspace{0.006cm}

The reduction  $\COUNT(\AA)\le \COUNT(\BB)$
then follows from the following claim
\begin{equation}\label{count-reduction-x-y}
\#_\AA(G,x)=\#_\BB(G,y).
\end{equation}
To prove this claim, we only need to show that,
  for any assignment $\xi:V\rightarrow [m]$,
$$
\text{wt}_\AA(\xi)= x \ \ \Longleftrightarrow \ \ \text{wt}_\BB(\xi)=y.
$$
We only prove $\text{wt}_\AA(\xi)= x
  \hspace{0.06cm}\Rightarrow\hspace{0.06cm} \text{wt}_\BB(\xi)=y$
  here. The other direction can be proved similarly.

Let $\xi:V\rightarrow [m]$ denote any assignment.
For every $i,j\in [m]$, we use $k_{i,j}$ to denote the
  number of edges $uv\in E$ such that $(\xi(u),\xi(v))=(i,j)$ or $(j,i)$, then for both
  $\AA$ and $\BB$,
\begin{equation}\label{wt-in-A-B-expression}
\text{wt}_\AA(\xi)=\prod_{i,j\in [m]} A_{i,j}^{k_{i,j}}\ \ \ \ \mbox{and}
\ \ \ \ \text{wt}_\BB(\xi)=\prod_{i,j\in [m]}B_{i,j}^{k_{i,j}}.
\end{equation}

For $x=0$, we note that the weight
  $\text{wt}_\AA(\xi)$ is 0 iff for some zero entry $A_{i,j}=0$
we have $k_{i,j} >0$. By the construction of $\BB$, $A_{i,j}=0$ iff $B_{i,j}=0$,
so $\text{wt}_\BB(\xi)$ must also be $0$.

In the following we assume both $x, y \not = 0$,
and we only consider assignments $\xi:V\rightarrow [m]$
such that its $k_{i,j}=0$ for any $A_{i,j}=0$
(equivalently $k_{i,j}=0$ for any $B_{i,j}=0$).
Thus we may consider the products
in (\ref{wt-in-A-B-expression})
are over non-zero entries $A_{i,j}$ and $B_{i,j}$, respectively.

Now we  use the generating set
 $\mathscr{G}=\{g_1,\ldots,g_d\}$ chosen above for
the set $\mathscr{A}$ of all non-zero entries
  $A_{i,j}$ in the matrix $\AA$.  There are integer
exponents $e_{1, (ij)}$, $e_{2, (ij)}$, \ldots, $e_{d, (ij)}$,
such that $$A_{i,j}=\prod_{\ell =1}^d g_\ell^{e_{\ell, (ij)}}\cdot \zeta_{i,j},\ \ \ \
\text{and}\ \ \ \ B_{i,j}=\prod_{\ell =1}^d p_\ell^{e_{\ell, (ij)}}\cdot \zeta_{i,j},
\ \ \ \ \ \text{for all $i,j$ such that $A_{i,j}\ne 0$,} $$
where $\zeta_{i,j}$ is a root of unity.
The expression of $B_{i,j}$ follows from the construction.
By (\ref{ineedu}) and (\ref{wt-in-A-B-expression}),
\[\text{wt}_\AA(\xi)= x  \ \ \Longrightarrow \ \
\prod_{\ell=1}^d g_\ell^{\sum_{i,j}
 (k_{i,j} - k_{i,j}^{*}) \cdot e_{\ell, (ij)} }
\ ~\mbox{is a root of unity}.\]
Here the sum $\sum_{i,j}$ in the exponent
is over all $i, j \in [m]$ where the corresponding
$A_{i,j}$ is non-zero.
This last equation is equivalent to (since $\mathscr{G}$ is a generating set)
\begin{equation}\label{ineedu22}\sum_{i,j} (k_{i,j} - k_{i,j}^{*})\cdot e_{\ell, (ij)} = 0,
~~~~\mbox{for all}\ \ell \in [d],\end{equation}
which in turn implies that
\begin{equation}\label{ineedu2}\prod_{i,j}\big(\zeta_{i,j}\big)^{k_{i,j}}
  =\prod_{i,j}\big(\zeta_{i,j}\big)^{k^*_{i,j}}\hspace{0.04cm}.\end{equation}

It then follows from (\ref{ineeddu}), (\ref{wt-in-A-B-expression}),
  (\ref{ineedu22}) and (\ref{ineedu2}) that $\text{wt}_\BB(\xi)=y$.
%
\end{proof}

\subsection{Step 1.2}

Now we let $\BB'$ denote the $m\times m$ matrix such that
  $B'_{i,j}=|B_{i,j}|$ for all $i,j\in [m]$.
We have (note that Lemma \ref{absolutevalue} actually holds for any
  symmetric matrix $\BB$ and $\BB'$, as long as $B_{i,j}'=|B_{i,j}|$ for all $i,j$)

\begin{lemma}\label{absolutevalue}
$\eval(\BB')\le \eval(\BB)$.
\end{lemma}
\begin{proof}
By Lemma \ref{count}, we only need to show that $\COUNT(\BB')\le \COUNT(\BB)$.

Let $(G,x)$ be an input of $\COUNT(\BB')$.
Since $\BB'$ is non-negative, we have $\#_{\BB'}(G,x)=0$ if $x$ is not real or $x<0$.
Now suppose $x\ge 0$, $G=(V,E)$ and $n=|E| $.
We let $Y$ denote the following set
$$
Y=\left\{ \prod_{i,j\in [m]} B_{i,j}^{k_{i,j}}\hspace{0.08cm}\Big|\hspace{0.12cm}
 \mbox{integers} ~ k_{i,j}\ge 0\ \text{and}
\sum_{i,j\in [m]} k_{i,j}=n\hspace{0.03cm}\right\}.
$$
Again, we know $|Y|$ is polynomial in $n$ and can be enumerated in
  polynomial time in $n$.
Once we have $Y$, we remove all elements in $Y$ whose complex norm
  is not equal to $x$.
We call the subset left $Y_x$.

The lemma then follows directly from the following statement:
$$
\#_{\BB'}(G,x)=\sum_{y\in Y_x} \#_{\BB}(G,y).
$$
This is because, for every assignment $\xi:V\rightarrow [m]$,
$\text{wt}_{\BB'}(\xi)=x$ if and only if $|\text{wt}_{\BB}(\xi)|=x.$
This gives us a polynomial reduction since $Y_x\subseteq Y$,
$|Y_x|$ is polynomially bounded in $n$, and $Y_x$ can be enumerated in
  polynomial time.
\end{proof}

Finally we prove Theorem \ref{bi-step-1} and Theorem \ref{t-step-1}.

\begin{proof}[Proof of Theorem \ref{bi-step-1}]
Let $\AA\in \mathbb{C}^{m\times m}$ be a symmetric, connected and bipartite matrix.
We construct matrices $\BB$ and $\BB'$ as above.
Since we assumed $\AA$ to be connected and bipartite,
  both matrices $\BB$ and $\BB'$ are connected and bipartite.
Therefore, we know there is a permutation $\Pi$ from $[m]$ to itself such that
  $\BB_{\Pi,\Pi}$ is the bipartisation of a $k\times (m-k)$ matrix $\FF$,
  for some $1\le k<m$:
$$
\BB_{\Pi,\Pi}=\begin{pmatrix}\00 & \FF\\ \FF^T & \00\end{pmatrix},
$$
and $\BB_{\Pi,\Pi}'$ is the bipartisation of $\FF'$, where $F'_{i,j}=|F_{i,j}|$
  for all $i\in [k]$ and $j\in [m-k]$.
Since permuting the rows and columns of $\BB$ does not affect the complexity
  of $\eval(\BB)$, we have
\begin{equation}\label{samecomplexity}
\eval(\BB_{\Pi,\Pi}')\le \eval(\BB_{\Pi,\Pi})\equiv \eval(\BB)\equiv \eval(\AA).
\end{equation}
We also know that $\BB_{\Pi,\Pi}'$ is non-negative.
By Bulatov and Grohe's theorem,
  we have the following cases:\vspace{0.1cm}
\begin{flushleft}\begin{enumerate}
\item[--] First, if $\eval(\BB_{\Pi,\Pi}')$ is \#P-hard,
  then by (\ref{samecomplexity}), $\eval(\AA)$ is also \#P-hard.\vspace{-0.05cm}
\item[--] Second, if $\eval(\BB_{\Pi,\Pi}')$ is not \#P-hard
  then the rank of $\FF'$ must be $1$ (it cannot be $0$ since
  $\BB_{\Pi,\Pi}'$ is assumed to be connected and bipartite).
Therefore, there exist non-negative rational numbers
  $\mu_1,\ldots,\mu_k,\ldots,\mu_m$ such that $F_{i,j}'=\mu_i\mu_{j+k}$,
  for all $i\in [k]$ and $j\in [m-k]$.
Moreover, $\mu_i$, for all $i\in [m]$, cannot be $0$ since otherwise 
  $\BB_{\Pi,\Pi}'$ is not connected.\vspace{0.1cm}
\end{enumerate}\end{flushleft}

As every entry of $\BB_{\Pi,\Pi}$ is the product of the corresponding entry of
  $\BB_{\Pi,\Pi}'$ and some root of unity,
  $\BB_{\Pi,\Pi}$ is a purified bipartite matrix.
The theorem is proven since $\eval(\BB)\equiv\eval(\AA)$.\vspace{0.09cm}
\end{proof}

\begin{proof}[Proof of Theorem \ref{t-step-1}]
The proof is similar.

Let $\AA\in \mathbb{C}^{m\times m}$\vspace{-0.012cm} be a
  symmetric, connected and non-bipartite matrix, then we construct $\BB$ and~$\BB'$~as above.
Since $\AA$ is connected and non-bipartite, both $\BB$ and $\BB'$
  are connected and non-bipartite.
Also, $\BB'$~is non-negative. We consider the following two cases.
If $\BB'$ is \#P-hard, then
$ \eval(\BB') \le \eval(\BB)$ $ \equiv  \eval(\AA)$
implies that $\eval(\AA)$
  must also be \#P-hard.
If $\BB'$ is not \#P-hard then it follows from the dichotomy theorem of
  Bulatov and Grohe \cite{BulatovGrohe} that the rank of $\BB$ is $1$
(it cannot be $0$ since we assumed $m>1$, and $\BB$ is connected).
Since $\BB$ is symmetric, it is a purified non-bipartite matrix.
The theorem then follows since $\eval(\BB)\equiv \eval(\AA)$.
\end{proof}

\section{Proof of Theorem \ref{bi-step-2}}

We start by introducing a technique
  for establishing reductions between $\eval(\AA)$
  and $\eval(\CC,\fD)$.
It is inspired by the Twin Reduction Lemma proved in \cite{GoldbergGJT}.

\subsection{Cyclotomic Reduction and Inverse Cyclotomic Reduction}

Let $\AA$ be an $m\times m$  symmetric (but not necessarily bipartite) complex matrix,
and let $(\CC,\fD)$ be a pair that satisfies the
  following condition $(\calT)$:\label{CONDITIONT}
\begin{enumerate}
\item[]\hspace{-0.5cm}($\calT_1$)\hspace{0.07cm}
  $\CC$ is an $n\times n$ symmetric complex matrix;\vspace{-0.04cm}
\item[]\hspace{-0.5cm}($\calT_2$) \hspace{0.07cm}$\fD=\{\DD^{[0]},\ldots,
  \DD^{[N-1]}\}$ is a sequence
  of $N$ $n\times n$ diagonal matrices for some positive integer $N$;\vspace{-0.06cm}
\item[]\hspace{-0.5cm}($\calT_3$)\hspace{0.06cm}
Every diagonal entry $D_a^{[0]}$ in $\DD^{[0]}$ is a positive integer.
Furthermore, for every $a\in [n]$, there\\ exist nonnegative integers
  $\alpha_{a,0},\ldots,\alpha_{a,N-1}$ such that
$$
D_a^{[0]}=\sum_{b=0}^{N-1} \alpha_{a,b}\ \ \ \ \text{and}\ \ \ \
D_a^{[r]}=\sum_{b=0}^{N-1} \alpha_{a,b}\cdot\oo_N^{b\hspace{0.02cm}r},\ \ \ \ \ \text{for all $r\in [N-1]$.}
$$
In particular, we say that the tuple $(\alpha_{a,0},\ldots,\alpha_{a,N-1})$
  \emph{generates} the $a^{th}$ entries of $\fD$.
\end{enumerate}

We show that if $\AA$ and $(\CC,\fD)$ satisfy certain conditions, then
$
\eval(\AA)\equiv \eval(\CC,\fD).
$

\begin{definition}
Let ${\mathscr R}=\{R_{1,0},R_{1,1},\ldots,R_{1,N-1},\ldots,R_{n,0},\ldots,R_{n,N-1}\}$
  be a partition of $[m]$ \emph{(}note that each $R_{a,b}$
  here need not be nonempty\emph{)} such that
\begin{eqnarray*}
&\bigcup_{0\le b\le N-1} R_{a,b}\ne \emptyset,\ \ \ \ \text{for all $a\in [n]$.}&
\end{eqnarray*}
We say $\AA$ can be \emph{generated} by $\CC$ using $\mathscr R$ if
  for all $i,j\in [m]$,
\begin{equation}\label{weneedit}
A_{i,j}=C_{a,a'}\cdot \oo_{N}^{b+b'},\ \ \ \ \text{where
  $i\in R_{a,b}$ and $j\in R_{a',b'}$}.\vspace{0.06cm}
\end{equation}
\end{definition}

Given any pair $(\CC,\fD)$ that satisfies $(\calT)$, we prove the following lemma:

\begin{lemma}[Cyclotomic Reduction Lemma]\label{twinreduction}
Let $(\CC,\fD)$ be a pair that satisfies $(\calT)$, with
  nonnegative integers $\alpha_{a,b}$'s.
Let ${\mathscr R}=\{R_{1,0},\ldots,R_{n,N-1}\}$ be a partition of $[m]$ satisfying
$$
\big|R_{a,b}\big| = \alpha_{a,b}\ \ \ \ \text{and}\ \ \ \
  m = \sum_{a=1}^n \sum_{b=0}^{N-1} \alpha_{a,b} \ge n,
$$
and let $\AA\in \mathbb{C}^{m\times m}$ denote the matrix
  generated by $\CC$ using $\mathscr R$.
Then we have $\eval(\AA)\equiv \eval(\CC,\fD).$
\end{lemma}

\begin{proof}
It suffices to prove for any undirected graph $G=(V,E)$,
$$
Z_\AA(G)=\sum_{\xi:V\rightarrow [m]}\twt_\AA (\xi)\ \ \ \text{and}\ \ \
Z_{\CC,\fD}(G)=\sum_{\eta:V\rightarrow [n]}\twt_{\CC,\fD} (\eta)
$$
are exactly the same.

To prove this, we define a surjective map $\rho$ from $\{\xi\}$,
  the set of all assignments from $V$ to $[m]$, to $\{\eta\}$,
  the set of all assignments from $V$ to $[n]$.
Then we show for every $\eta:V\rightarrow [n]$,
  \begin{equation}\label{guagua}
  \twt_{\CC,\fD}(\eta)=\sum_{\xi:\rho(\xi)=\eta} \twt_{\AA}(\xi).
  \end{equation}
We define $\rho(\xi)$ as follows. Since $\mathscr R$ is a partition of $[m]$,
  for any $v\in V$, there exists a unique pair $(a,b)$
  {such that $\xi(v)\in R_{a,b}.$}
Let $\xi_1(v)=a$ and $\xi_2(v)=b$, then we set
  $\rho(\xi)=\eta\equiv\xi_1$ from $V$ to $[n]$.
It is easy to check that $\rho$ is surjective.\vspace{-0.002cm}

To prove (\ref{guagua}), we write $\twt_\AA(\xi)$ as
$$
\twt_{\AA}(\xi)=\prod_{uv\in E} A_{\xi(u),\xi(v)}=\prod_{uv\in E}
  C_{\eta(u),\eta(v)}\cdot \oo_N^{\xi_2(u)+\xi_2(v)}
  =\prod_{uv\in E} C_{\eta(u),\eta(v)}\cdot \oo_N^{\xi_2(u)}
  \cdot \oo_N^{\xi_2(v)}.
$$
It follows that\vspace{-0.2cm}
\begin{eqnarray*}
\sum_{\xi:\rho(\xi)=\eta} \twt_{\AA}(\xi)&=&
  \prod_{uv\in E} C_{\eta(u),\eta(v)}\times
  \sum_{\xi:\rho(\xi)=\eta}\left(
  \prod_{uv\in E} \oo_{N}^{\xi_2(u)}\cdot \oo_{N}^{\xi_2(v)}\right)\\
  &=&\prod_{uv\in E} C_{\eta(u),\eta(v)}\times
  \sum_{\xi:\rho(\xi)=\eta}\left(
  \prod_{v\in V} \oo_{N}^{\xi_2(v)\cdot \text{deg}(v)} \right)\\
  &=&\prod_{uv\in E} C_{\eta(u),\eta(v)}\times
  \left(\prod_{v\in V} \left(\sum_{b=0}^{N-1}
    \big|R_{\eta(v),b}\big|\cdot \oo_N^{b\cdot \text{deg}(v) }\right) \right)\\[0.3ex]
  &=&\prod_{uv\in E} C_{\eta(u),\eta(v)}\times
  \left(\prod_{v\in V} D_{\eta(v)}^{[\text{deg}(v)\bmod N]}\right)
  \hspace{0.15cm}=\hspace{0.15cm}\twt_{\CC,\fD}(\eta),
\end{eqnarray*}
and the lemma follows.
\end{proof}

By combining Lemma \ref{twinreduction}, Lemma \ref{absolutevalue}, as well as the
  dichotomy theorem of Bulatov and Grohe, we have
  the following handy corollary for dealing with $\eval(\CC,\fD)$:\vspace{0.02cm}

\begin{corollary}[Inverse Cyclotomic Reduction Lemma]\label{inversetwin}
Let $(\CC,\fD)$ be a pair that satisfies condition\\ $(\calT)$.
If $\CC$ has a $2\times 2$ sub-matrix\vspace{-0.08cm}
$$
\left(\begin{matrix} C_{i,k} & C_{i,\ell}\\  C_{j,k} & C_{j,\ell}
\end{matrix}\right)$$  such that all of its four entries are nonzero and
$$
\big|C_{i,k}C_{j,\ell}\big|\ne \big|C_{i,\ell}C_{j,k}\big|,
$$
then the problem $\eval(\CC,\fD)$ is \#P-hard.\vspace{0.02cm}
\end{corollary}
\begin{proof}
By the Cyclotomic\vspace{-0.01cm} Reduction Lemma, we know there exist a symmetric
  $m\times m$ matrix $\AA$, for some positive integer $m$,
  and a partition $\mathscr R$ of $[m]$, where
\begin{equation}\label{killme}
{\mathscr R}=\Big\{R_{a,b}\hspace{0.09cm}\big|\hspace{0.09cm}a\in [n],b\in [0:N-1]\Big\}
 \ \ \ \text{and}\ \ \
\bigcup_{b\in [0:N-1]} R_{a,b}\ne \emptyset,\ \ \ \text{for all $a\in [n]$,}
\end{equation}
such that $\eval(\AA)\equiv\eval(\CC,\fD)$. Moreover, the two matrices
  $\AA$ and $\CC$ satisfy (\ref{weneedit}).

Now suppose there exist $i\ne j,k\ne \ell\in [n]$ such that $|C_{i,k}|,|C_{i,\ell}|,
  |C_{j,k}|$ and $|C_{j,\ell}|$ are non-zero and $|C_{i,k}C_{j,\ell}|\ne
  |C_{i,\ell}C_{j,k}|$.
We arbitrarily pick an integer $i'$ from $\bigcup_{b} R_{i,b}$ (which is
  known to be nonempty), a $j'$ from $\bigcup_b R_{j,b}$,
  a $k'$ from $\bigcup_b R_{k,b}$, and an $\ell'$ from $\bigcup_b R_{\ell,b}$.
Then by (\ref{weneedit}), we have
$$|A_{i',k'}|=|C_{i,k}|,\ \ |A_{i',\ell'}|=|C_{i,\ell}|,\ \
|A_{j',k'}|=|C_{j,k}|,\ \  |A_{j',\ell'}|=|C_{j,\ell}|,\ \ \text{and}\ \
|A_{i',k'}A_{j',\ell'}|\ne |A_{i',\ell'}A_{j',k'}|.$$
Let $\AA'=(|A_{i,j}|)$ for all $i,j\in [m]$, then $\AA'$ has a $2\times 2$ sub-matrix
  of rank $2$ and all its four entries are nonzero.
By the dichotomy of Bulatov and Grohe (Corollary \ref{usefulhahacoro}),
  $\eval(\AA')$ is \#P-hard.
It follows that $\eval(\CC,\fD)$ is \#P-hard, since
  $\eval(\CC,\fD)\equiv \eval(\AA)$ and by Lemma \ref{absolutevalue},
  $\eval(\AA')\le \eval(\AA)$.\vspace{0.03cm}
\end{proof}

By combining Lemma \ref{twinreduction}, Eq.\hspace{0.05cm}(\ref{guagua}),
   and the First Pinning Lemma (Lemma \ref{pinning1}), we have

\begin{corollary}[Third Pinning Lemma]\label{pinning3}
Let $(\CC,\fD)$ be a pair that satisfies $(\calT)$, then
  $$\evalp(\CC,\fD)\equiv \eval(\CC,\fD).$$
In particular, the problem of computing $Z_{\CC,\fD}^{\rightarrow}$ \emph{(}or
  $Z_{\CC,\fD}^{\leftarrow}$\emph{)} is polynomial-time reducible to $\eval(\CC,\fD)$.
\end{corollary}

\begin{proof}
We only need to prove that $\evalp(\CC,\fD)\le \eval(\CC,\fD)$.

By the Cyclotomic Reduction Lemma, we know there exist a symmetric $m\times m$
  matrix $\AA$ for some $m\ge 1$, and a partition $\mathscr R$ of
  $[m]$, such that, $\mathscr R$ satisfies (\ref{killme}) and $\eval(\AA)\equiv \eval(\CC,\fD)$.
$\AA$, $\CC$ and $\mathscr R$ also satisfy (\ref{weneedit}).
By the First Pinning Lemma, we have $\evalp(\AA)\equiv \eval(\AA)\equiv \eval(\CC,\fD)$.
  So we only need to reduce $\evalp(\CC,\fD)$ to $\evalp(\AA)$.

Now let $(G,w,i)$ be an input of $\evalp(\CC,\fD)$, where
  $G$ is an undirected graph, $w$ is a vertex in $G$ and $i\in [n]$.
By (\ref{guagua}), we have
\begin{eqnarray*}
Z_{\CC,\fD}(G,w,i)=\sum_{\eta:\eta(w)=i} \twt_{\CC,\fD}(\eta)
  =\sum_{\xi:\xi_1(w)=i}\twt_\AA(\xi)=\sum_{j\in \cup_{b} R_{i,b}} Z_{\AA}(G,w,j).
\end{eqnarray*}
This gives us a polynomial-time reduction from $\evalp(\CC,\fD)$ to $\evalp(\AA)$.
\end{proof}

Notice that, compared to the Second Pinning Lemma, the Third Pinning
  Lemma does not require the matrix $\CC$ to be the bipartisation
  of a unitary matrix.
It only requires $(\CC,\fD)$ to satisfy $(\calT)$.

\subsection{Step 2.1}

Let $\AA$ be a purified bipartite matrix.
Then after collecting its entries of equal norm in decreasing order by
  permuting the rows and columns of $\AA$,
  there exist a positive integer $N$ and four
  sequences $\boldsymbol{\mu},\boldsymbol{\nu},\mm$ and $\nn$ such that
  $(\AA,(N,\boldsymbol{\mu},\boldsymbol{\nu},\mm,\nn))$ satisfies the following condition:\label{CONDITIONS1}
\begin{enumerate}
\item[($\calS_1$)] Matrix $\AA$ is the bipartisation of an $m\times n$ matrix $\BB$
  so $\AA$ is $(m+n)\times (m+n)$.
$\boldsymbol{\mu}=\{\mu_1,\ldots,\mu_s\}$ and
  $\boldsymbol{\nu}=\{\nu_1,\ldots,\nu_t\}$ are two sequences of
  positive rational numbers, of lengths $s\ge 1$ and $t\ge 1$, respectively.
$\boldsymbol{\mu}$ and $\boldsymbol{\nu}$ satisfy $\mu_1>\mu_2>\ldots >\mu_s$
  and $\nu_1>\nu_2>\ldots>\nu_t$.
$\mm=\{m_1,...,m_s\}$ and $\nn=\{n_1,...,n_t\}$ are two sequences of positive
  integers such that, $m=\sum m_i$ and $n=\sum n_i$.
The rows of $\BB$ are indexed by $\xx=(x_1,x_2)$ where $x_1\in [s]$
  and $x_2\in [m_{x_1}]$, while the columns of $\BB$ are indexed by
  $\yy=(y_1,y_2)$ where $y_1\in [t]$ and $y_2\in [n_{y_1}]$.
For all $\xx,\yy$, we have
$$
B_{\xx,\yy}=B_{(x_1,x_2),(y_1,y_2)}=\mu_{x_1}\nu_{y_1} S_{\xx,\yy},
$$
where $\SS=\{S_{\xx,\yy}\}$ is an $m\times n$ matrix in which
  every entry is a power of $\oo_N$.
\begin{equation*}
\BB = \left(\begin{matrix}
\mu_1\II_{m_1}\\
& \hspace{-0.15cm}\mu_2\II_{m_2}\hspace{-0.15cm} \\
& & \hspace{-0.15cm}\ddots\hspace{-0.15cm}\\
& & & \mu_s\II_{m_s}
\end{matrix}\right)
\left( \begin{matrix}
\SS_{(1,*),(1,*)} & \SS_{(1,*),(2,*)} & \ldots & \SS_{(1,*),(t,*)} \\
\SS_{(2,*),(1,*)} & \SS_{(2,*),(2,*)} & \ldots & \SS_{(2,*),(t,*)} \\
\vdots & \vdots & \ddots & \vdots \\
\SS_{(s,*),(1,*)} & \SS_{(s,*),(2,*)} & \ldots & \SS_{(s,*),(t,*)}
\end{matrix}\right)
\left(\begin{matrix}
\nu_1\II_{n_1} \\
& \hspace{-0.15cm}\nu_2\II_{n_2}\hspace{-0.15cm} \\
& & \hspace{-0.15cm}\ddots\hspace{-0.15cm}\\
& & & \nu_t\II_{n_t}
\end{matrix}\right),
\end{equation*}
where $\II_{k}$ denotes the $k\times k$ identity matrix.
\end{enumerate}

We let\vspace{-0.1cm}
$$I\equiv\bigcup_{i\in [s]} \{\hspace{0.03cm}(i,j)\hspace{0.05cm}\big|
  \hspace{0.05cm}j\in [m_i]\hspace{0.03cm}\}\ \ \ \text{and}\ \ \ J
  \equiv\bigcup_{i\in [t]} \{\hspace{0.03cm}(i,j)\hspace{0.05cm}\big|
  \hspace{0.05cm}j\in [n_i]\hspace{0.03cm}\},$$respectively.
We use $\{0\}\times I$ to index the first $m$ rows (or columns) of $\AA$,
  and $\{1\}\times J$ to index the last $n$ rows (or columns) of $\AA$.
Given $\xx\in I$ and $j\in [t]$, we let\vspace{0.01cm}
  $$\SS_{\xx,(j,*)}=\big(S_{\xx,(j,1)},\ldots,S_{\xx,(j,n_j)}\big)\in \mathbb{C}^{n_j}$$
denote the $j^{th}$ block of the $\xx^{th}$ row vector of $\SS$.
Similarly, given $\yy\in J$ and $i\in [s]$, we let $$\SS_{(i,*),\yy}
= \big(S_{(i,1),\yy},\ldots,S_{(i,m_i),\yy}\big)\in \mathbb{C}^{m_i}$$
denote the $i^{th}$ block of the $\yy^{th}$ column vector of $\SS$.

\begin{figure}
\center
\includegraphics[height=3.1cm]{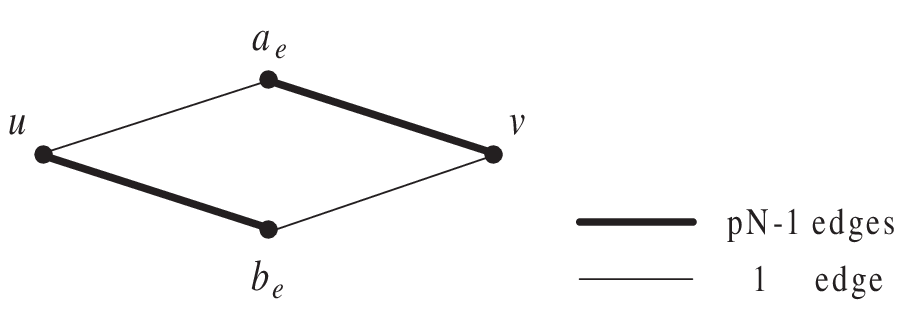}
\caption{Gadget for constructing graph $G^{[p]}$, $p\ge 1$.}\label{figure_2}
\end{figure}

\begin{lemma}\label{hahajaja}
Suppose\vspace{0.005cm} $(\AA,(N,\boldsymbol{\mu},\boldsymbol{\nu}
  ,\mm,\nn))$ satisfies $(\calS_1)$, then either
  $\eval(\AA)$ is \#P-hard or $(\AA,(N,\boldsymbol{\mu},$ $\boldsymbol{\nu},\mm,\nn))$ satisfies
  the following two conditions:\label{CONDITIONS23}
\begin{enumerate}
\item[]$\hspace{-0.6cm}(\calS_2)$ For all $\xx,\xx'\in I$, either there exists
  an integer $k$ such that $\SS_{\xx,*}=\oo_N^k\cdot \SS_{\xx',*}$ or
  for every $j\in [t]$,
$$
\big\langle \SS_{\xx,(j,*)},\SS_{\xx',(j,*)}\big\rangle =0;
$$
\item[]\hspace{-0.6cm}$(\calS_3)$
For all $\yy,\yy'\in J$, either there exists an integer $k$ such that
  $\SS_{*,\yy}=\oo_N^k\cdot \SS_{*,\yy'}$ or for every
  $i\in [s]$,
$$
\big\langle \SS_{(i,*),\yy},\SS_{(i,*),\yy'}\big\rangle =0.
$$
\end{enumerate}
\end{lemma}

\begin{proof}
Assume $\eval(\AA)$ is not \#P-hard. We only prove ($\calS_2$) here.
($\calS_3$) can be proved similarly.

Let $G=(V,E)$ be an undirected graph. For each $p\ge 1$, we construct
  a new graph $G^{[p]}$ by replacing every edge $uv$ in $E$ with
  a gadget which is shown in Figure \ref{figure_2}.

More exactly, we define graph $G^{[p]}=(V^{[p]},E^{[p]})$ as follows:
$$
V^{[p]}=V\cup\big\{a_e,b_e\hspace{0.05cm}\big|\hspace{0.05cm}e\in E\big\}
$$
and $E^{[p]}$ contains exactly the following edges: For each $e=uv\in E$,
\begin{enumerate}
\item one edge between $(u,a_e)$ and $(b_e,v)$;\vspace{-0.08cm}
\item $(pN-1)$ edges between $(a_e,v)$ and $(u,b_e)$.
\end{enumerate}
The construction of $G^{[p]}$, for each $p\ge 1$, gives us an $(m+n)\times (m+n)$ matrix
  $\AA^{[p]}$ such that
$$
Z_{\AA^{[p]}}(G)=Z_\AA(G^{[p]}),\ \ \ \text{for all undirected graphs $G$.}
$$
Thus, we have $\eval(\AA^{[p]})\le \eval(\AA)$,
  and $\eval(\AA^{[p]})$ is also not \#P-hard.\vspace{0.0025cm}

The entries of $\AA^{[p]}$ are as follows. First,
$$
A^{[p]}_{(0,\uu),(1,\vv)}=A^{[p]}_{(1,\vv),(0,\uu)}=0,\ \ \ \text{for
  all $\uu \in I$ and $\vv \in J$.}
$$
So $\AA^{[p]}$ is a block diagonal matrix with $2$ blocks of
  $m\times m$ and $n\times n$, respectively.
The entries in the upper-left $m\times m$ block are
\begin{eqnarray*}
A^{[p]}_{(0,\uu),(0,\vv)}&=&\left(\sum_{\aa\in J}
  A_{(0,\uu),(1,\aa)}(A_{(0,\vv),(1,\aa)})^{pN-1}\right)
  \left(\sum_{\bb\in J}
  (A_{(0,\uu),(1,\bb)})^{pN-1}A_{(0,\vv),(1,\bb)}\right)\\[0.6ex]
  &=&\left(\sum_{\aa\in J}
  B_{\uu,\aa}(B_{\vv,\aa})^{pN-1}\right)
  \left(\sum_{\bb\in J}
  (B_{\uu,\bb})^{pN-1}B_{\vv,\bb}\right)
\end{eqnarray*}
for all $\uu,\vv\in I$. The first factor of the last expression is
$$
\sum_{\aa\in J} \mu_{u_1}\nu_{a_1}S_{\uu,\aa} (\mu_{v_1}\nu_{a_1})^{pN-1}\overline{S_{\vv,\aa}}
=\mu_{u_1}\mu_{v_1}^{pN-1} \sum_{\aa\in J} \nu_{a_1}^{pN}
  S_{\uu,\aa}\overline{S_{\vv,\aa}}
=\mu_{u_1}\mu_{v_1}^{pN-1} \sum_{i\in [t]} \nu_{i}^{pN} \langle
  \SS_{\uu,(i,*)},\SS_{\vv,(i,*)}\rangle.
$$
Similarly, we have for the second factor
$$ \sum_{\bb\in J}
  (B_{\uu,\bb})^{pN-1}B_{\vv,\bb}
=\mu_{u_1}^{pN-1} \mu_{v_1} \sum_{i\in [t]} \nu_i^{pN}
  \overline{\langle
  \SS_{\uu,(i,*)},\SS_{\vv,(i,*)}\rangle}.
$$
As a result,\vspace{-0.16cm}
$$
A^{[p]}_{(0,\uu),(0,\vv)}=(\mu_{u_1}\mu_{v_1})^{pN}\left|
  \sum_{i\in [t]} \nu_i^{pN} \langle
  \SS_{\uu,(i,*)},\SS_{\vv,(i,*)}\rangle \right|^2.
$$
It is clear that the upper-left $m\times m$ block of $\AA^{[p]}$
  is a nonnegative real matrix.
Similarly one can prove that the same holds for its
  lower-right $n\times n$ block, so $\AA^{[p]}$ is a nonnegative real matrix.

Now let $\uu\ne \vv$ be two indices in $I$ (note that if $|I|=1$,
  then $(\calS_2)$ is trivially true), then we have
$$A^{[p]}_{(0,\uu),(0,\uu)}A^{[p]}_{(0,\vv),(0,\vv)}=(\mu_{u_1}\mu_{v_1})^{2pN}\left(
\sum_{i\in [t]} n_i\cdot \nu_i^{pN} \right)^4,$$
which is positive, and
$$
A^{[p]}_{(0,\uu),(0,\vv)}A^{[p]}_{(0,\vv),(0,\uu)}=
  (\mu_{u_1}\mu_{v_1})^{2pN} \left|\sum_{i\in [t]} \nu_i^{pN} \langle
  \SS_{\uu,(i,*)},\SS_{\vv,(i,*)}\rangle \right|^4.
$$
Since $\eval(\AA^{[p]})$ is not \#P-hard,
  by the dichotomy theorem of Bulatov and Grohe (Corollary \ref{usefulhahacoro}),
$$
\left|\sum_{i\in [t]} \nu_i^{pN} \langle
  \SS_{\uu,(i,*)},\SS_{\vv,(i,*)}\rangle\right|
$$
is either $0$ or $\sum_{i\in [t]}n_i\cdot \nu_i^{pN}$.\vspace{0.006cm}

Now suppose vectors $\SS_{\uu,*}$ and $\SS_{\vv,*}$
  are linearly dependent, then
because entries of $\SS$ are all powers of $\oo_N$,
  there must exist an integer $k\in [0:N-1]$ such that
$\SS_{\uu,*}=\oo_N^k\cdot \SS_{\vv,*}$,
and we are done.

Otherwise, assuming $\SS_{\uu,*}$ and
  $\SS_{\vv,*}$ are linearly independent,
  we have
\begin{equation}\label{temptemp}
\left|\sum_{i\in [t]} \nu_i^{pN}\cdot \langle
  \SS_{\uu,(i,*)},\SS_{\vv,(i,*)}\rangle\right|<\sum_{i\in [t]}n_i\cdot \nu_i^{pN},
  \ \ \ \text{for any $p\ge 1$.}
\end{equation}
This is because,\vspace{-0.062cm} if the left-hand side is equal to the right-hand side,
  then $|\langle \SS_{\uu,(i,*)},\SS_{\vv,(i,*)}\rangle|=n_i$
  for all $i\in [t]$ and thus,
  $\SS_{\uu,(i,*)}=\oo_N^{k_i}\cdot \SS_{\vv,(i,*)}$ for some $k_i\in [0:N-1]$.
Moreover, these $k_i$'s must be the same since we assumed (\ref{temptemp}) is
  an equality:
  $$\left|\sum_{i\in [t]} \nu_i^{pN}n_i\cdot \oo_N^{k_i}\right|=\sum_{i\in [t]}n_i\cdot
  \nu_i^{pN}.$$
As a result, $\SS_{\uu,*}$ and $\SS_{\vv,*}$ are linearly dependent, which
  contradicts the assumption.
By (\ref{temptemp}), we have
$$
\sum_{i\in [t]} \nu_i^{pN} \langle
  \SS_{\uu,(i,*)},\SS_{\vv,(i,*)}\rangle=0,\ \ \ \text{for all $p\ge 1$.}
$$
Since $\nu_1>\ldots >\nu_t$ is strictly decreasing, by using the
  Vandermonde matrix, we have
$$
\langle
  \SS_{\uu,(i,*)},\SS_{\vv,(i,*)}\rangle=0,\ \ \ \text{for all $i\in [t]$}.
$$
This finishes the proof of $(\calS_2)$.
\end{proof}

We then have the following corollary: 

\begin{corollary}\label{fullrank}
For all $i\in [s]$ and $j\in [t]$, the rank of the $(i,j)^{th}$
  block matrix $\SS_{(i,*),(j,*)}$ of $\SS$ has\\ exactly the same rank as $\SS$.
\end{corollary}
\begin{proof}
Without loss of generality, we prove $\text{rank}(\SS_{(1,*),(1,*)})=\text{rank}(\SS)$.

First, we use Lemma \ref{hahajaja} to show that
$$
\text{rank}\left(\begin{matrix} \SS_{(1,*),(1,*)}\\\SS_{(2,*),(1,*)}\\ \vdots
  \\ \SS_{(s,*),(1,*)}\end{matrix}\right)=\text{rank}(\SS).
$$
To see this, we take any $h=\text{rank}(\SS)$ rows of $\SS$
  which are linearly independent.
Since any two of them $\SS_{\xx,(*,*)}$ and $\SS_{\yy,(*,*)}$
  are linearly independent, by condition $(\calS_2)$, the two subvectors
  $\SS_{\xx,(1,*)}$ and $\SS_{\yy,(1,*)}$ are orthogonal.
Therefore, the corresponding $h$ rows of the matrix on the left-hand side
  are pairwise orthogonal, and the left-hand side is at least $h$.
Of course it cannot be larger than $h$, so it is equal to $h$.

By using condition ($\calS_3$), we can similarly show that
$$
\text{rank}(\SS_{(1,*),(1,*)})=\text{rank}\left(\begin{matrix} \SS_{(1,*),(1,*)}
  \\\SS_{(2,*),(1,*)}\\ \vdots \\ \SS_{(s,*),(1,*)}\end{matrix}\right).
$$
As a result, we have $\text{rank}(\SS_{(1,*),(1,*)})=\text{rank}(\SS)$.
\end{proof}

Now suppose $h=\text{rank}(\SS)$, then by Corollary \ref{fullrank},
  there must exist indices $1\le i_1<\ldots<i_h\le m_1$
  and $1\le j_1<\ldots<j_h\le n_1$, such that, the $\{(1,i_1),\ldots,(1,i_h)\}\times
  \{(1,j_1),\ldots,(1,j_h)\}$ sub-matrix of $\SS$ has full rank $h$.
Without loss of generality (if this is not true, we can apply an
  appropriate permutation $\Pi$ to the rows and columns of $\AA$
  so that the new $\SS$ has this property) we assume $i_k=k$ and $j_k=k$
  for all $k\in [h]$.
We use $\HH$ to denote this $h\times h$ matrix: $H_{i,j}=S_{(1,i),(1,j)}$.\vspace{0.007cm}

By Corollary \ref{fullrank} and Lemma \ref{hahajaja},
  for every index $\xx\in I$, there exist
  two unique integers $j\in [h]$ and $k\in [0:N-1]$ \vspace{-0.1cm}such that
\begin{equation}\label{case1}
\SS_{\xx,*} = \oo_N^k \cdot \SS_{(1,j),*}.
\end{equation}
This gives us a partition of index set $\{0\}\times I$:
$$
{\mathscr R}_0 =\big\{R_{(0,i,j),k}\hspace{0.05cm}\Big|\hspace{0.05cm}
  i\in [s],j\in [h],k\in [0:N-1]\big\},
$$
as follows: For every $\xx\in I$, $(0,\xx)\in R_{(0,i,j),k}$ if
  $i=x_1$ and $\xx,j,k$ satisfy (\ref{case1}).
By Corollary \ref{fullrank}, we have
  $$\bigcup_{k\in [0:N-1]} R_{(0,i,j),k}\ne \emptyset,\ \ \
  \text{for all $i\in [s]$ and $j\in [h]$.}$$

Similarly, for every $\yy\in J$, there exist two unique integers $j\in [h]$
  and $k\in [0:N-1]$ such that
\begin{equation}\label{case2}
\SS_{*,\yy}= \oo_N^k \cdot \SS_{*,(1,j)},
\end{equation}
and we partition $\{1\}\times J$ into
$$
{\mathscr R}_1 =\big\{R_{(1,i,j),k}\hspace{0.05cm}\Big|\hspace{0.05cm}
  i\in [t],j\in [h],k\in [0:N-1]\big\},
$$
as follows: For every $\yy\in J$, $(1,\yy)\in R_{(1,i,j),k}$ if
  $i=y_1$ and $\yy,j,k$ satisfy (\ref{case2}).
Again by Corollary \ref{fullrank},
$$ \bigcup_{k\in [0:N-1]} R_{(1,i,j),k}\ne \emptyset,\ \ \
  \text{for all $ i \in [t]$ and $j\in [h]$.}$$

Now we define $(\CC,\fD)$ and use the Cyclotomic Reduction
  Lemma (Lemma \ref{twinreduction}) to show that
$$\eval(\CC,\fD)\equiv\eval(\AA).$$
First, $\CC$ is an $(s+t)h\times (s+t)h$ matrix which is the bipartisation
  of an $sh\times th$ matrix $\FF$.
We use set $I'\equiv [s]\times [h]$ to index the rows of $\FF$, and $J'\equiv[t]\times [h]$
  to index the columns of $\FF$.
We have
$$
F_{\xx,\yy}=\mu_{x_1}\nu_{y_1}H_{x_2,y_2}=\mu_{x_1}\nu_{y_1}S_{(1,x_2),(1,y_2)},
\ \ \ \text{for all $\xx\in I'$, $\yy\in J'$},$$
or equivalently,\vspace{-0.26cm}
$$
\FF = \left(\begin{matrix}
\mu_1\II \\
& \mu_2\II \\
& & \ddots\\
& & & \mu_s\II
\end{matrix}\right)
\left( \begin{matrix}
\HH & \HH & \ldots & \HH \\
\HH & \HH & \ldots & \HH \\
\vdots & \vdots & \ddots & \vdots \\
\HH & \HH & \ldots & \HH
\end{matrix}\right)
\left(\begin{matrix}
\nu_1\II \\
& \nu_2\II \\
& & \ddots\\
& & & \nu_t\II
\end{matrix}\right),
$$
where $\II$ is the $h\times h$ identity matrix.
We use $(\{0\}\times I')\cup (\{1\}\times J')$ to
  index the rows/columns of $\CC$. 

Second, $\fD=\{\DD^{[0]},...,\DD^{[N-1]}\}$ is a sequence of $N$ diagonal
  matrices of the same size as $\CC$.
We use $\{0\}\times I'$ to index the first $sh$ diagonal entries, and
  $\{1\}\times J'$ to index the last $th$ diagonal entries. Then the
  $(0,\xx)^{th}$ entries of $\fD$ are generated by $(|R_{(0,x_1,x_2),0}|,\ldots,
  |R_{(0,x_1,x_2),N-1}|)$ and the $(1,\yy)^{th}$ entries of $\fD$ are
  generated by $(|R_{(1,y_1,y_2),0}|,\ldots, |R_{(1,y_1,y_2),N-1}|)$:
\begin{eqnarray*}
D^{[r]}_{(0,\xx)}=\sum_{k=0}^{N-1} \big|R_{(0,x_1,x_2),k}\big|\cdot \oo_N^{kr}\ \ \ \ \text{
  and}\ \ \ \
D^{[r]}_{(1,\yy)}=\sum_{k=0}^{N-1} \big|R_{(1,y_1,y_2),k}\big|\cdot \oo_N^{kr},
\end{eqnarray*}
for all $r\in [0:N-1],\xx=(x_1,x_2)\in I'$ and $\yy=(y_1,y_2)\in J'$.\vspace{0.007cm}

The following lemma is a direct application of
  the Cyclotomic Reduction Lemma (Lemma \ref{twinreduction}).

\begin{lemma}
$\eval(\AA)\equiv \eval(\CC,\fD)$.
\end{lemma}
\begin{proof}
First we show that $\AA$ can be generated from $\CC$ using
  ${\mathscr R}_0\cup {\mathscr R}_1$.

Let $\xx,\xx'\in I$, $(0,\xx)\in R_{(0,x_1,j),k}$ and $(0,\xx')\in$
  $R_{(0,x_1',j'),k'}$, then
we have $$A_{(0,\xx),(0,\xx')}= C_{(0,x_1,j),(0,x_1',j')}=0,$$ since
  $\AA$ and $\CC$ are the bipartisations of $\BB$ and $\FF$, respectively.
As a result,
$$
A_{(0,\xx),(0,\xx')}=C_{(0,x_1,j),(0,x_1',j')}\cdot \oo_N^{k+k'}
$$
holds trivially. Clearly, this is also true for the lower-right $n\times n$ block of $\AA$.

Let $\xx\in I$, $(0,\xx)\in R_{(0,x_1,j),k}$, $\yy\in J$, and
  $(1,\yy)\in R_{(1,y_1,j'),k'}$ for some $j,k,j',k'$, then by (\ref{case1})-(\ref{case2}),
\begin{equation*}
A_{(0,\xx),(1,\yy)}=\mu_{x_1}\nu_{y_1} S_{\xx,\yy }
  =\mu_{x_1}\nu_{y_1} S_{(1,j), \yy }\cdot \oo_N^{k}
  =\mu_{x_1}\nu_{y_1} S_{( 1,j),( 1,j')}\cdot \oo_N^{k+k'}
  =C_{(0,x_1,j),(1,y_1,j')}\cdot \oo_N^{k+k'}.
\end{equation*}
A similar equation holds for the lower-left block of $\AA$, so
  it can be generated from $\CC$ using ${\mathscr R}_0\cup {\mathscr R}_1$.

On the other hand, the construction of $\fD$ implies that $\fD$ can be generated from
  partition ${\mathscr R}_0\cup {\mathscr R}_1$.
The lemma then follows directly from the Cyclotomic Reduction Lemma.
\end{proof}

\subsection{Step 2.2}

We first summarize what we have proved in Step 2.1.
We showed that the problem $\eval(\AA)$ is either \#P-hard or equivalent
  to $\eval(\CC,\fD)$, where $(\CC,\fD)$ satisfies the following condition
  ({\sl Shape}):\label{SHAPECONDITION}
\begin{enumerate}
\item[] \hspace{-0.6cm}({\sl Shape}$_1$): $\CC\in \mathbb{C}^{m\times m}$ (note that this $m$
  is different from the $m$ used in Step 2.1) is
  the bipartisation of an $sh\times th$ matrix $\FF$ (thus $m=(s+t)h$).
$\FF$ is an $s\times t$ block matrix and we use $I=[s]\times [h]$, $J=[t]\times [h]$
  to index the rows and columns of $\FF$, respectively.

\item[] \hspace{-0.6cm}({\sl Shape}$_2$): There are two sequences $\boldsymbol{\mu}
  =\{ \mu_1> \ldots>\mu_s>0\}$ and
  $\boldsymbol{\nu}=\{ \nu_1>\ldots>\nu_t>0\}$ of rational numbers together with
  an $h\times h$ full-rank matrix $\HH$ whose entries are all powers of $\oo_N$, for
  some positive integer $N$.
For all $\xx\in I$ and $\yy\in J$, we have\vspace{-0.1cm}
$$
F_{\xx,\yy}=\mu_{x_1}\nu_{y_1}H_{x_2,y_2}.\vspace{-0.1cm}
$$

\item[] \hspace{-0.6cm}({\sl Shape}$_3$): $\fD=\{\DD^{[0]},\ldots,\DD^{[N-1]}\}$ is a sequence of
  $m\times m$ diagonal matrices.
  $\fD$ satisfies $(\calT_3)$, so
  $$
  D^{[r]}_{(0,\xx)}=\overline{D^{[N-r]}_{(0,\xx)}}\ \ \text{and}\ \
  D^{[r]}_{(1,\yy)}=\overline{D^{[N-r]}_{(1,\yy)}},\ \ \ \text{for all $r\in [N-1]$,
    $\xx\in [s]\times [h]$ and $\yy\in [t]\times [h]$.}
  $$
\end{enumerate}
We use $(\{0\}\times I)\cup(\{1\}\times J)$ to index the rows and columns
  of matrices $\CC$ and $\DD^{[r]}$.

Now in Step 2.2, we prove 
  the following lemma:

\begin{lemma}\label{shapecondition}
Either $\eval(\CC,\fD)$ is \#P-hard, or
  $\HH$ and $\DD^{[0]}$ satisfy the following two conditions:\vspace{0.04cm}
\begin{enumerate}
\item[]\hspace{-0.4cm}\emph{({\sl Shape}$_4$)}: \emph{$\frac{1}{\sqrt{h}}\cdot \HH$ is a
  unitary matrix, i.e., $\langle \HH_{i,*},\HH_{j,*}\rangle
  =\langle \HH_{*,i},\HH_{*,j}\rangle =0$ for all $i\ne j\in [h]$.}\vspace{-0.05cm}

\item[]\hspace{-0.4cm}\emph{({\sl Shape}$_5$)}: $\DD^{[0]}$ satisfies $D^{[0]}_{(0,\xx)}=
  D^{[0]}_{(0,(x_1,1))}$ for all $\xx\in I$, and
  $D^{[0]}_{(1,\yy)}=D^{[0]}_{(1,(y_1,1))}$ for all $\yy\in J$.\vspace{0.03cm}
\end{enumerate}
\end{lemma}
\begin{proof}
We rearrange the diagonal entries of $\DD^{[0]}$ indexed by $\{1\}\times J$
  into a $t\times h$ matrix $\XX$:
$$
X_{i,j}=D^{[0]}_{(1,(i,j))},\ \ \ \text{for all $i\in [t]$ and $j\in [h]$,}
$$
and its diagonal entries indexed by $\{0\}\times I$ into an $s\times h$ matrix $\YY$:
$$
Y_{i,j}=D^{[0]}_{(0,(i,j))},\ \ \ \text{for all $i\in [s]$ and $j\in [h]$.}
$$
Note that by condition $(\calT_3)$, all entries of $\XX$ and $\YY$
  are positive integers.

The proof has two stages: First, we show in Lemma \ref{stage1}
  that, either $\eval(\CC,\fD)$ is \#P-hard, or
\begin{eqnarray}\label{assumption11}
&\langle \HH_{i,*}\circ\overline{\HH_{j,*}},\XX_{k,*}\rangle=0,\ \ \ \text{for all
  $k\in [t]$ and $i\ne j\in [h]$,}&\text{and}\\[0.6ex]
\label{assumption21}
&\langle \HH_{*,i}\circ\overline{\HH_{*,j}},\YY_{k,*}\rangle=0,\ \ \
\text{for all $k\in [s]$ and $i\ne j\in [h]$.}&
\end{eqnarray}
We use $U$ to denote the set of $h$-dimensional
  vectors that are orthogonal to
$$\HH_{1,*}\circ \overline{\HH_{2,*}},\ \HH_{1,*}\circ
  \overline{\HH_{3,*}},\ \ldots,\ \HH_{1,*}\circ \overline{\HH_{h,*}}.$$
The above set of $h-1$ vectors is linearly independent.
This is because
$$\sum_{i=2}^{h} a_i\big(\HH_{1,*}\circ \overline{\HH_{i,*}}\big)=\HH_{1,*}\circ
  \left(\sum_{i=2}^h a_i\overline{\HH_{i,*}}\right),$$
and\vspace{0.003cm} if $\sum_{i=2}^{h} a_i (\HH_{1,*}\circ \overline{\HH_{i,*}})=\00$,
  then $\sum_{i=2}^h a_i\overline{\HH_{i,*}}=\00$
  since all entries of $\HH_{1,*}$ are nonzero.
Because $\HH$ has full rank, we have $a_i=0$, $i=2,\ldots,h$.
As a result, $U$ is a linear space of dimension $1$ over $\mathbb{C}$.

In the second stage, we show in Lemma \ref{stage2} that, assuming (\ref{assumption11}) and
  (\ref{assumption21}), either
\begin{eqnarray}\label{assumption12}
&\langle \HH_{i,*}\circ\overline{\HH_{j,*}},(\XX_{k,*})^2\rangle =0,\ \ \ \text{for
  all $k\in [t]$ and $i\ne j\in [h]$,}&\text{and}\\[0.5ex]
\label{assumption22}
&\langle \HH_{*,i}\circ\overline{\HH_{*,j}},(\YY_{k,*})^2\rangle =0,\ \ \ \text{for all
  $k\in [s]$ and $i\ne j\in [h]$,}&
\end{eqnarray}
or $\eval(\CC,\fD)$ is \#P-hard. Here we use $(\XX_{k,*})^2$
  to denote $\XX_{k,*}\circ\XX_{k,*}$.\vspace{0.005cm}

(\ref{assumption11}) and (\ref{assumption12}) then
  imply that both $\XX_{k,*}$ and $(\XX_{k,*})^2$ are in $U$
  and thus, they are linearly dependent (since the dimension of $U$ is $1$).
On the other hand, by $(\calT_3)$, every entry in $\XX_{k,*}$ is a positive
  integer. Therefore, $\XX_{k,*}$ must have the form $u\cdot \11$,
  for some positive integer $u$.
The same argument works for $\YY_{k,*}$ and the latter must also
  have the form $u'\cdot \11$.
By (\ref{assumption11}) and (\ref{assumption21}), this further implies that
$$
\langle \HH_{i,*},\HH_{j,*}\rangle =0\ \ \text{and}\ \
\langle \HH_{*,i},\HH_{*,j}\rangle =0,\ \ \ \text{for all $i\ne j\in [h]$.}
$$
This finishes the proof of Lemma \ref{shapecondition}.\end{proof}

Now we proceed to the two stages of the proof.
In the first stage, we prove the following lemma:
\begin{lemma}\label{stage1}
Either matrices $\HH$, $\XX$ and $\YY$ satisfy \emph{(\ref{assumption11})} and
  \emph{(\ref{assumption21})}, or $\eval(\CC,\fD)$ is \#P-hard.
\end{lemma}
\begin{proof}
Suppose problem $\eval(\CC,\fD)$ is not \#P-hard, otherwise we are already done.
We let $\fD^*$ denote a sequence of $N$ $m\times m$ diagonal
  matrices in which every matrix is a copy of $\DD^{[0]}$ (as in $\fD$):\vspace{-0.1cm}
$$
\fD^*=\{\DD^{[0]},\ldots,\DD^{[0]}\}.\vspace{-0.1cm}
$$
It is easy to check that $\fD^*$ satisfies condition $(\calT_3)$.

Let $G=(V,E)$ be an undirected graph. For each $p\ge 1$,
  we build a new graph $G^{[p]}=(V^{[p]},E^{[p]})$ in the
  same way as we did in the proof of Lemma \ref{hahajaja}.
This gives us an $m\times m$ matrix $\CC^{[p]}$ such that
$$
Z_{\CC^{[p]},\fD^*}(G)=Z_{\CC,\fD}(G^{[p]}),\ \ \ \text{for all undirected graphs $G$,}
$$
and thus, $\eval(\CC^{[p]},\fD^*)\le \eval(\CC,\fD)$,
  and $\eval(\CC^{[p]},\fD^*)$ is also not \#P-hard.

Matrix $\CC^{[p]}$ is a block matrix which has the same
  block dimension structure as $\CC$.
The upper-right and lower-left blocks of $\CC^{[p]}$ are zero matrices.
For $\xx,\yy\in I$, we have
$$
C^{[p]}_{(0,\xx),(0,\yy)}=\left(\sum_{\aa\in J}F_{\xx,\aa}(F_{\yy,\aa})^{pN-1}
  X_{a_1,a_2}\right)
  \left(\sum_{\bb\in J} (F_{\xx,\bb})^{pN-1}F_{\yy,\bb}X_{b_1,b_2}\right).
$$
By ({\sl Shape}$_2$) and the fact that all entries of $\XX$
  are positive integers, we can rewrite the first factor as
$$
\mu_{x_1}(\mu_{y_1})^{pN-1}\sum_{\aa\in J} (\nu_{a_1})^{pN}
  H_{x_2,a_2}\overline{H_{y_2,a_2}}X_{a_1,a_2}=
  \mu_{x_1}(\mu_{y_1})^{pN-1}\sum_{a \in [t]} (\nu_a)^{pN} \langle
  \HH_{x_2,*}\circ\overline{\HH_{y_2,*}},\XX_{a,*}\rangle.
$$
Similarly, we have\vspace{-0.1cm}
$$
(\mu_{x_1})^{pN-1}\mu_{y_1}\sum_{a\in [t]} (\nu_a)^{pN}
  \overline{\langle
  \HH_{x_2,*}\circ\overline{\HH_{y_2,*}},\XX_{a,*}\rangle}
$$
for the second factor. Since $\nu_a>0$ for all $a$, we have\vspace{-0.05cm}
\begin{equation}\label{star}
C^{[p]}_{(0,\xx),(0,\yy)}=(\mu_{x_1}\mu_{y_1})^{pN} \left|
  \sum_{a \in [t]} (\nu_a)^{pN} \langle
  \HH_{x_2,*}\circ\overline{\HH_{y_2,*}},\XX_{a,*}\rangle\right|^2,
\end{equation}
so the upper-left block of $\CC^{[p]}$ is a nonnegative real matrix.
Similarly one can show that the same holds for its lower-right block,
  so $\CC^{[p]}$ is a nonnegative real matrix.

Now for any $\xx\ne \yy\in I$, we have
$$
C^{[p]}_{(0,\xx),(0,\xx)}=(\mu_{x_1})^{2pN}
  \left( \sum_{a\in [t]}(\nu_a)^{pN}\sum_{b\in [h]} X_{a,b}\right)^2\ \text{and}\ \
C^{[p]}_{(0,\yy),(0,\yy)}=(\mu_{y_1})^{2pN}
  \left( \sum_{a\in [t]}(\nu_a)^{pN}\sum_{b\in [h]} X_{a,b}\right)^2\hspace{-0.2cm},
$$
which are positive, and\vspace{-0.2cm}
$$
C^{[p]}_{(0,\xx),(0,\xx)}C^{[p]}_{(0,\yy),(0,\yy)}
  =(\mu_{x_1}\mu_{y_1})^{2pN}\left( \sum_{a\in [t]}(\nu_a)^{pN}\sum_{b\in [h]} X_{a,b}\right)^4>0.
$$
Since $\eval(\CC^{[p]},\fD^*)$ is not \#P-hard
  and $(\CC^{[p]},\fD^*)$ satisfies $(\calT)$, by the Inverse Cyclotomic Reduction Lemma
  (Corollary \ref{inversetwin}), we have
$$
\text{either\ \ }\big(C^{[p]}_{(0,\xx),(0,\yy)}\big)^2
  =C^{[p]}_{(0,\xx),(0,\xx)}C^{[p]}_{(0,\yy),
  (0,\yy)}\ \ \text{or }\ C^{[p]}_{(0,\xx),(0,\yy)}=0.
$$
We claim that if the former is true, then we must have $x_2=y_2$.
This is because, in this case, we have
$$
\left|\sum_{a \in [t]} (\nu_a)^{pN} \langle
  \HH_{x_2,*}\circ\overline{\HH_{y_2,*}},\XX_{a,*}\rangle\right|=\sum_{a\in [t]}(\nu_a)^{pN}
  \sum_{b\in [h]}X_{a,b},
$$
and the norm of $\langle\vspace{-0.04cm}
  \HH_{x_2,*}\circ\overline{\HH_{y_2,*}},\XX_{a,*}\rangle$ must be $\sum_{b\in [h]}X_{a,b}$.
However the inner product is a sum of $X_{a,b}$'s weighted by
  roots of unity, so the entries of $\HH_{x_2,*}\circ \overline{\HH_{y_2,*}}$
  must be the same root of unity.
Thus, $\HH_{x_2,*}$ and $\HH_{y_2,*}$ are linearly dependent.
Since $\HH$ is a matrix of full rank, we conclude that $x_2=y_2$.\vspace{-0.045cm}

In other words, if $x_2\ne y_2$, then we have $
C^{[p]}_{(0,\xx),(0,\yy)}=0$ and thus, $$\sum_{a\in [t]} (\nu_a)^{pN}
  \langle\HH_{x_2,*}\circ \overline{\HH_{y_2,*}},\XX_{a,*}\rangle=0,\ \ \
  \text{for all $p\ge 1$ and all $x_2\ne y_2$,}
$$
since the argument has nothing to do with $p$.
By using the Vandermonde matrix, we have
  $$\langle\HH_{x_2,*}\circ \overline{\HH_{y_2,*}},\XX_{a,*}\rangle=0,\ \ \
  \text{for all $a\in [t]$ and all $x_2\ne y_2$.}$$
This finishes the proof of (\ref{assumption11}).
(\ref{assumption21}) can be proved similarly.
\end{proof}

In the second stage, we prove the following lemma:
\begin{lemma}\label{stage2}
Suppose matrices $\HH$, $\XX$ and $\YY$ satisfy both
  \emph{(\ref{assumption11})} and \emph{(\ref{assumption21})}. Then
  either they also satisfy \emph{(\ref{assumption12})}
  and \emph{(\ref{assumption22})}, or $\eval(\CC,\fD)$ is \#P-hard.
\end{lemma}
\begin{proof}
We will only prove (\ref{assumption22}). (\ref{assumption12}) can be proved similarly.
Again, we let $\fD^*$  denote a sequence of $N$ $m\times m$
  diagonal matrices in which every matrix is a copy of $\DD^{[0]}$
  ($\fD^*$ satisfies $(\calT_3)$).

\begin{figure}
\center
\includegraphics[height=6.6cm]{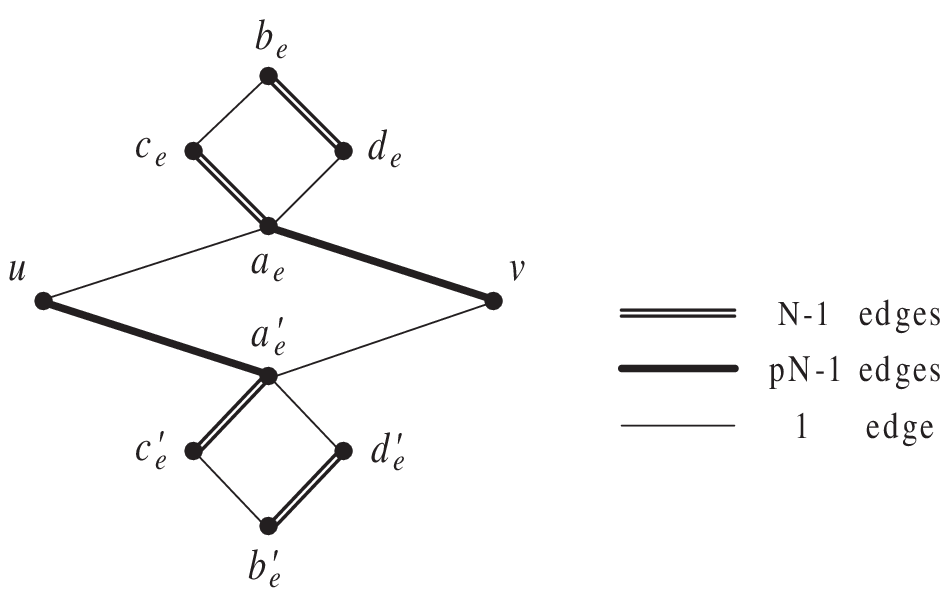}
\caption{Gadget for constructing $G^{(p)}$, $p\ge 1$.}\label{figure_3}
\end{figure}

Before starting the proof we note the following property of
  the matrix $\CC^{[1]}$ which we used in the proof of Lemma \ref{stage1} since we need it
  to prove (\ref{assumption22}) here: When $x_2=y_2$, by (\ref{star}), we have
$$
C^{[1]}_{(0,\xx),(0,\yy)}=(\mu_{x_1}\mu_{y_1})^{N} \left(
  \sum_{a \in [t]} (\nu_a)^{N}\sum_{b\in [h]}{X_{a,b}} \right)^2,
$$
and is equal to $0$ when $x_2\ne y_2$. We use $L$ to denote the
  second factor on the right-hand side, which is independent of $\xx$ and $\yy$,
  so the right-hand side becomes $(\mu_{x_1}\mu_{y_1})^{N}\cdot L$.\vspace{0.005cm}

Additionally, because of (\ref{assumption21}), we have $\YY_{k,*}$ and
  $\YY_{1,*}$ are linearly dependent for every $k$. Thus there exists
  a positive rational number $\lambda_k$ such that
\begin{equation}\label{hahahaha}
\YY_{k,*}=\lambda_k\cdot \YY_{1,*},\ \ \ \text{for all $k\in [s]$.}
\end{equation}
Because of this, we only need to prove (\ref{assumption22}) for the case when $k=1$.

Now we start the proof of (\ref{assumption22}).
Suppose $\eval(\CC,\fD)$ is not \#P-hard.
We use $G=(V,E)$ to denote an undirected graph,
  then for each $p\ge 1$, we build a new graph $G^{(p)}=(V^{(p)},E^{(p)})$
  by replacing every edge $e=uv\in E$ with a gadget
  which is shown in Figure \ref{figure_3}.

More exactly, we define $G^{(p)}=(V^{(p)},E^{(p)})$ as follows:
$$
V^{(p)}=V\cup \big\{a_e,b_e,c_e,d_e,a_e',b_e',c_e',d_e'\hspace{0.1cm}\Big|
  \hspace{0.1cm} e\in E\big\},
$$
and $E^{(p)}$ contains exactly the following edges: For
  every edge $e=uv\in E$,
\begin{enumerate}
\item One edge between $(u,a_e),(a_e',v),(c_e,b_e),(d_e,a_e),
  (c_e',b_e')$ and $(d_e',a_e')$;
\item $pN-1$ edges between $(a_e,v)$ and $(u,a_e')$;
\item $N-1$ edges between $(a_e,c_e),(b_e,d_e),(a_e',c_e')$
  and $(b_e',d_e')$.
\end{enumerate}
It is easy to check that the degree of every vertex in $G^{(p)}$
  is a multiple of $N$.

Moreover, the construction of $G^{(p)}$ gives us
  a new $m\times m$ matrix $\RR^{(p)}$ which is symmetric
  since the gadget is symmetric, such that
$$
Z_{\RR^{(p)},\fD^*}(G)=Z_{\CC,\fD}(G^{(p)}),\ \ \ \text{for all undirected graphs $G$}
$$
and thus, $\eval(\RR^{(p)},\fD^*)\le \eval(\CC,\fD)$, and
  $\eval(\RR^{(p)},\fD^*)$ is also not \#P-hard.

The matrix $\RR^{(p)}$ is a block matrix which has the
  same block dimension structure as $\CC$.
The upper-right and lower-left blocks of $\RR^{(p)}$ are zero matrices.
The entries in its lower-right block are as follows:
\begin{equation*}
R^{(p)}_{(1,\xx),(1,\yy)}=\left(\sum_{\aa,\bb\in I }
  F_{\aa,\xx}(F_{\aa,\yy})^{pN-1} C^{[1]}_{(0,\aa),(0,\bb)}
  Y_{a_1,a_2}Y_{b_1,b_2}\right)
\left(\sum_{\aa,\bb\in I }
  (F_{\aa,\xx})^{pN-1} F_{\aa,\yy} C^{[1]}_{(0,\aa),(0,\bb)}
  Y_{a_1,a_2}Y_{b_1,b_2}\right)
\end{equation*}
for $\xx,\yy\in J$.
Firstly, by (\ref{hahahaha}),
  we have $Y_{a_1,a_2}Y_{b_1,b_2}=\lambda_{a_1}\lambda_{b_1}
  Y_{1,a_2}Y_{1,b_2}.$
Secondly, we have
$$
C^{[1]}_{(0,\aa),(0,\bb)}=0,\ \ \ \text{whenever $a_2\ne b_2$.}
$$
As a result, we can simplify the first factor to be
\begin{eqnarray*}
&&\nu_{x_1}(\nu_{y_1})^{pN-1} L\cdot \sum_{\aa,\bb\in I,a_2=b_2} (\mu_{a_1})^{pN}
  H_{a_2,x_2}\overline{H_{a_2,y_2}}(\mu_{a_1}\mu_{b_1})^N\lambda_{a_1}\lambda_{b_1}
  Y_{1,a_2}Y_{1,b_2}\\[0.8ex]
&&=\hspace{0.07cm} \nu_{x_1}(\nu_{y_1})^{pN-1} L\cdot \sum_{a_1,b_1\in [s]} (\mu_{a_1})^{(p+1)N}
  (\mu_{b_1})^{N}\lambda_{a_1}\lambda_{b_1}\sum_{a_2\in [h]}
  H_{a_2,x_2}\overline{H_{a_2,y_2}} (Y_{1,a_2})^2
  \\[0.4ex]
&&=\hspace{0.07cm} \nu_{x_1}(\nu_{y_1})^{pN-1} L'\cdot
  \langle \HH_{*,x_2}\circ \overline{\HH_{*,y_2}}\hspace{0.03cm},(\YY_{1,*})^2\rangle,
\end{eqnarray*}
where $$L'=L\hspace{-0.15cm}\sum_{a_1,b_1\in [s]} (\mu_{a_1})^{(p+1)N}
  (\mu_{b_1})^{N}\lambda_{a_1}\lambda_{b_1}$$ is a positive
  number that is independent from $\xx, \yy$.
Similarly the second factor can be simplified to be
$$
(\nu_{x_1})^{pN-1}\nu_{y_1}L'\cdot \overline{
  \langle \HH_{*,x_2}\circ\overline{\HH_{*,y_2}},(\YY_{1,*})^2\rangle}.
$$
As a result, we have\vspace{-0.2cm}
$$
R^{(p)}_{(1,\xx),(1,\yy)}=(L')^2\cdot (\nu_{x_1}\nu_{y_1})^{pN}\cdot \Big|
  \langle \HH_{*,x_2}\circ \overline{\HH_{*,y_2}}\hspace{0.03cm},(\YY_{1,*})^2\rangle\Big|^2.
$$
Thus the lower-right block of $\RR^{(p)}$ is non-negative.
Similarly one can prove that the same holds for its upper-left block,
  so $\RR^{(p)}$ is non-negative.\vspace{-0.007cm}

We now apply Corollary \ref{inversetwin} to $(\RR^{(p)},\fD^*)$.
Since $\eval(\RR^{(p)},\fD^*)$ is not \#P-hard, we have
$$
\text{either\ \ }\big(R^{(p)}_{(1,\xx),(1,\yy)}\big)^2
  =R^{(p)}_{(1,\xx),(1,\xx)}R^{(p)}_{(1,\yy),
  (1,\yy)}\ \ \text{or }\ R^{(p)}_{(1,\xx),(1,\yy)}=0,\ \ \ \text{for any
  $\xx\ne\yy\in J$.}
$$
We claim that if the former is true, then we must have $x_2=y_2$.
This is because, in this case, $$\Big|\langle \HH_{*,x_2}\circ
  \overline{\HH_{*,y_2}},(\YY_{1,*})^2\rangle\Big|=\sum_{i\in [h]}
  Y_{1,i}^2.\vspace{-0.14cm}$$
However, the left-hand side is a sum of $(Y_{1,i})^2$'s, which are positive
  integers, weighted by roots of unity.
To sum to a number of norm $\sum_{i\in [h]}Y_{1,i}^2$
  the entries of $\HH_{*,x_2}\circ \overline{\HH_{*,y_2}}$ must
  be the same root of unity.
As a result, $\HH_{*,x_2}$ and $\HH_{*,y_2}$ are linearly dependent.
Since $\HH$ is of full rank, we conclude that $x_2=y_2$.
In other words, we have shown that
$$
\langle \HH_{*,x_2}\circ \overline{\HH_{*,y_2}}\hspace{0.03cm},(\YY_{1,*})^2\rangle=0,
  \ \ \ \text{for all $x_2\ne y_2$.}
$$
By combining it with (\ref{hahahaha}), we have finished the proof
  of (\ref{assumption22}).
\end{proof}
\input{Rankone.tex}

\section{Proofs of Theorem \ref{step30} and Theorem \ref{bi-step-3}}

Let $((M,N),\CC,\fD)$ be a tuple that satisfies $(\calU_1)$-$(\calU_4)$
  and $\FF\in \mathbb{C}^{m\times m}$ be the upper-right block of $\CC$.
In this section,
  we index the rows and columns of an $n\times n$ matrix with $[0:n-1]$.

\subsection{The Group Condition}

We first prove that either $\FF$ satisfies the following
  condition or $\eval(\CC,\fD)$ is $\#$P-hard:

\begin{figure}
\center
\includegraphics[height=6cm]{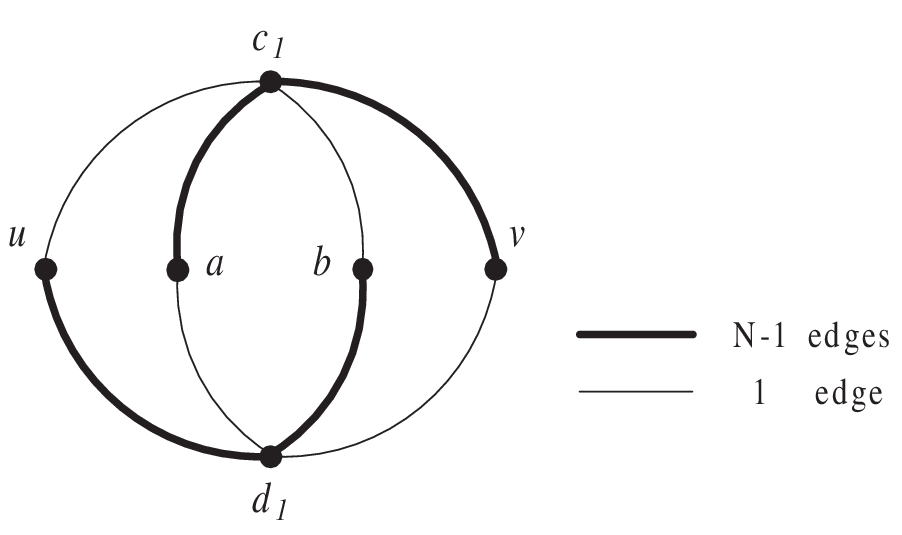}
\caption{The gadget for $p=1$ (Note that the subscript $e$ is suppressed).}\label{figure_5}
\end{figure}

\begin{lemma}\label{groupcondition1}
Let $((M,N),\CC,\fD)$ be a tuple that satisfies $(\calU_1)$-$(\calU_4)$,
  then either $\FF$ satisfies the following group condition \emph{(\GC)}:\vspace{0.025cm}
\begin{enumerate}
\item {\emph{(row-\GC)}}: $\forall\hspace{0.06cm} i,j\in [0:m-1]$,
  $\exists\hspace{0.06cm} k\in [0:m-1]$ such that
  $\FF_{k,*}=\FF_{i,*}\circ \FF_{j,*}$;\vspace{-0.02cm}\label{CONDITIONGC}

\item {\emph{(column-\GC)}}: $\forall\hspace{0.06cm} i,j\in [0:m-1]$,
  $\exists\hspace{0.06cm} k\in [0:m-1]$ such that
  $\FF_{*,k}=\FF_{*,i}\circ \FF_{*,j}$,\vspace{0.025cm}
\end{enumerate}
or $\eval(\CC,\fD)$ is $\#$P-hard.
\end{lemma}
\begin{proof}
Suppose $\eval(\CC,\fD)$ is not $\#$P-hard.

Let $G=(V,E)$ be an undirected graph.
For every integer $p\ge 1$, we construct a new graph $G^{[p]}$ by replacing
  every edge $uv\in E$ with a gadget.
The gadget for $p=1$ is shown in Figure \ref{figure_5}.\vspace{0.012cm}


More exactly, we define $G^{[p]}=(V^{[p]},E^{[p]})$ as
\begin{equation*}
V^{[p]}=V\cup\big\{a_e,b_e,c_{e,1},\ldots,c_{e,p},d_{e,1},\ldots,d_{e,p}
  \hspace{0.07cm}\big|\hspace{0.07cm}e\in E\big\},
\end{equation*}
and $E^{[p]}$ contains exactly the following edges: For each $e=uv\in E$,
and for every $1 \le i \le p$,
\begin{enumerate}
\item {One edge between $(u,c_{e,i})$, $(c_{e,i},b_e)$,
  $(d_{e,i},a_{e})$, and $(d_{e,i},v)$;}\vspace{-0.06cm}
\item {$N-1$ edges between $(c_{e,i},v)$, $(c_{e,i},a_e)$,
  $(d_{e,i},b_{e})$, and $(d_{e,i},u)$.}
\end{enumerate}
It is easy to check that the degree of every vertex in $G^{[p]}$
is a multiple of $N$, so
$${Z}_{\CC,\fD}(G^{[p]}) = {Z}_{\CC}(G^{[p]}),$$
since $\fD$ satisfies $(\calU_3)$.\vspace{0.01cm}
On the other hand, the way we build $G^{[p]}$ gives us,
 for every $p\ge 1$, a
symmetric  matrix $\AA^{[p]}\in \mathbb{C}^{2m\times 2m}$
  which only depends on $\CC$, such that
$$Z_{\AA^{[p]}}(G)=Z_{\CC}(G^{[p]})=Z_{\CC,\fD}(G^{[p]}), \ \ \text{for all $G$}.$$
As a result, we have\vspace{0.01cm} $\eval(\AA^{[p]})\le \eval(\CC,\fD)$ and thus,
  $\eval(\AA^{[p]})$ is not $\#$P-hard for all $p\ge 1$.

The $(i,j)^{th}$ entry of $\AA^{[p]}$, where $i,j\in [0:2m-1]$, is
\begin{eqnarray*}
A_{i,j}^{[p]}&=&\sum_{a=0}^{2m-1}\sum_{b=0}^{2m-1}
  \left(\sum_{c=0}^{2m-1}C_{i,c}\overline{C_{a,c}}C_{b,c}\overline{C_{j,c}}
  \right)^p
  \left(\sum_{d=0}^{2m-1}\overline{C_{i,d}}C_{a,d}\overline{C_{b,d}}C_{j,d}
  \right)^p.\\[0.6ex]
&=&\sum_{a=0}^{2m-1}\sum_{b=0}^{2m-1}
  \left|\sum_{c=0}^{2m-1}C_{i,c}\overline{C_{a,c}}C_{b,c}\overline{C_{j,c}}
  \right|^{2p}.
\end{eqnarray*}
To derive the first equation, we use the fact that $M\hspace{0.03cm}|
  \hspace{0.03cm}N$ and thus, e.g.,
  $(C_{a,c})^{N-1} =\overline{C_{a,c}}$ since
  $C_{a,c}$ is a power of $\oo_M$.
Note that $\AA^{[p]}$ is a symmetric non-negative matrix.
Furthermore, it is easy to check that
$$A_{i,j}^{[p]}=0,\ \ \text{$\forall\hspace{0.06cm}i\in [0:m-1],
 \forall  j\in [m,2m-1]$; ~~~~~and}\ \ A_{i,j}^{[p]}=0,\ \ \text{$\forall
  \hspace{0.06cm}i\in [m,2m-1], \forall j\in [0:m-1]$}.
$$
For $i,j\in [0:m-1]$, we have\vspace{-0.23cm}
\begin{eqnarray}\nonumber
&&A_{i,j}^{[p]}=\sum_{a=0}^{m-1}\sum_{b=0}^{m-1}\left|
\langle \FF_{i,*}\circ \overline{\FF_{j,*}}, \FF_{a,*}\circ
  \overline{\FF_{b,*}}\rangle
\right|^{2p},\ \ \text{and}\\[0.3ex]
 \label{groupcondition3}
&&A_{i+m,j+m}^{[p]}=\sum_{a=0}^{m-1}\sum_{b=0}^{m-1}\left|
\langle \FF_{*,i}\circ \overline{\FF_{*,j}}, \FF_{*,a}\circ
  \overline{\FF_{*,b}}\rangle
\right|^{2p}.
\end{eqnarray}
It is clear that all these entries are
  positive real numbers (by taking $a=i$ and $b=j$).
Now let us focus on the upper-left $m\times m$ block of $\AA^{[p]}$.
Since it is a non-negative symmetric matrix, we can apply
  the dichotomy theorem of Bulatov and Grohe.

On the one hand, for the special case when $j=i\in [0:m-1]$, we have
$$
A_{i,i}^{[p]}=\sum_{a=0}^{m-1}\sum_{b=0}^{m-1}
  \left|\langle \11,\FF_{a,*}\circ \overline{\FF_{b,*}}\rangle \right|^{2p}=
\sum_{a=0}^{m-1}\sum_{b=0}^{m-1}
  \left|\langle\FF_{a,*},\FF_{b,*}\rangle\right|^{2p}.
$$
As $\FF$ is a discrete unitary matrix, we have
  $A_{i,i}^{[p]}=m\cdot m^{2p}$.
On the other hand, assuming $\eval(\CC,\fD)$ is not \#P-hard,
  then by using Bulatov and Grohe's dichotomy theorem (Corollary \ref{usefulhahacoro}),
  we have
$$
A_{i,i}^{[p]}\cdot A_{j,j}^{[p]}=A_{i,j}^{[p]}\cdot A_{j,i}^{[p]}
  =(A_{i,j}^{[p]})^2,\ \ \ \text{for all $i\ne j\in [0:m-1]$,}
$$
and thus $A_{i,j}^{[p]}=m^{2p+1}$ for all $i,j\in [0:m-1]$.

Now we use this condition to show that $\FF$ satisfies (row-\GC).
We introduce the following notation: For $i,j\in [0:m-1]$, let\vspace{-0.2cm}
$$
X_{i,j}=\Big\{|\langle \FF_{i,*}\circ\overline{\FF_{j,*}},\FF_{a,*}
  \circ\overline{\FF_{b,*}}\rangle| ~\Big|~ a,b\in [0:m-1]\Big\}.
$$
Clearly set $X_{i,j}$ is finite for all $i,j$, with cardinality
$|X_{i,j}| \le m^2$.
Each $x \in X_{i,j}$ satisfies $0 \le x \le m$.
For each $x\in X_{i,j}$, we let $s_{i,j}(x)$ denote
  the number of pairs $(a,b) \in [0:m-1] \times [0:m-1]$ such that
$$
|\langle \FF_{i,*}\circ\overline{\FF_{j,*}},\FF_{a,*}
  \circ\overline{\FF_{b,*}}\rangle| =x.
$$
We can now rewrite $A_{i,j}^{[p]}$ as
\begin{equation}\label{group-cond-Vandermonde}
A_{i,j}^{[p]}=\sum_{x\in X_{i,j}} s_{i,j}(x)\cdot x^{2p},
\end{equation}
and is equal to $m^{2p+1}$ for all $p\ge 1$.
Also note that $s_{i,j}(x)$, for all $x\in X_{i,j}$,
  do not depend on $p$, and
\begin{equation}\label{group-cond-Vandermonde2}
\sum_{x\in X_{i,j}} s_{i,j}(x)=m^2.
\end{equation}
We can view (\ref{group-cond-Vandermonde}) and (\ref{group-cond-Vandermonde2})
  as a linear system of equations
in the unknowns $s_{i,j}(x)$.
Fix $i,j$, then there are $|X_{i,j}|$ many variables
$s_{i,j}(x)$, one for each distinct value  $x \in X_{i,j}$.
Equations in (\ref{group-cond-Vandermonde}) are
indexed  by $p \ge 1$. If we choose (\ref{group-cond-Vandermonde2})
  and (\ref{group-cond-Vandermonde}) for $p = 1, \ldots, |X_{i,j}|-1$,
this linear system has an $|X_{i,j}| \times |X_{i,j}|$ Vandermonde matrix
$( (x^2)^{p} )$, with row index $p$ and column index $x \in X_{i,j}$.
It  has full rank. Note that by setting $(a,b)=(i,j)$ and $(i',j)$,
  where $i'\ne i$, respectively, we get $m\in X_{i,j}$ and $0\in X_{i,j}$, respectively.
Moreover, $s_{i,j}(0)=m^2-m$, $s_{i,j}(m)=m$, and all other $s_{i,j}(x) = 0$ is a solution
  to the linear system.
Therefore this must be the unique solution.
As a result, we have $X_{i,j}=\{0,m\}$, $$s_{i,j}(m)=m\ \ \text{and}\ \
  s_{i,j}(0)=m^2-m,\ \ \text{for all $i,j\in [0:m-1]$}.
$$
This implies that for all $i,j,a,b\in [0:m-1]$,
  $|\langle\FF_{i,*}\circ\overline{\FF_{j,*}},
  \FF_{a,*}\circ\overline{\FF_{b,*}}\rangle| $ is either $m$ or $0$.\vspace{0.008cm}

Finally, we prove (row-\GC).
Set $j=0$. Because $\FF_{0,*}=\11$, the all-1 vector, we have
$$|\langle\FF_{i,*}\circ\11,
  \FF_{a,*}\circ\overline{\FF_{b,*}} \rangle| =|\langle\FF_{i,*}\circ \FF_{b,*},
  \FF_{a,*}\rangle|\in \{0,m\},\ \ \ \text{for all $i,a,b \in [0:m-1]$.}$$
As $\{\FF_{a,*}, a\in [0:m-1]\}$ is an orthogonal basis,
where each $\|\FF_{a,*}\|^2 = m$, by Parseval, we have
\[\sum_a \left|  \langle \FF_{i,*}\circ\FF_{b,*},\FF_{a,*}\rangle
\right|^2 =
m \cdot\|\FF_{i,*}\circ\FF_{b,*}\|^2.\]
Since every entry of $\FF_{i,*}\circ\FF_{b,*}$ is a root of unity,
$\|\FF_{i,*}\circ\FF_{b,*}\|^2 = m$.  Hence
\[\sum_a \left|  \langle \FF_{i,*}\circ\FF_{b,*},\FF_{a,*}\rangle
\right|^2
= m^2.\]
As a result, for all $i,b\in [0:m-1]$, there exists a unique $a$ such that
$
\left|
  \langle \FF_{i,*}\circ\FF_{b,*},\FF_{a,*}\rangle
  \right|=m.
$

By property ($\calU_2$), every entry of $\FF_{i,*}$, $\FF_{b,*}$, and $\FF_{a,*}$ is
  a root of unity.
The inner product $
\langle \FF_{i,*}\circ\FF_{b,*},\FF_{a,*}\rangle$
is a sum of $m$ terms each of complex norm 1.  To sum
to a complex number of norm $m$,
each term must be a complex number of unit norm with the {\it same}
argument, i.e., they are the same complex number $e^{i\theta}$.
Thus, $\FF_{i,*}\circ \FF_{b,*} = e^{i\theta}\cdot \FF_{a,*}$.
We assert that in fact $e^{i\theta} = 1$,
and $\FF_{i,*}\circ \FF_{b,*}=\FF_{a,*}$. This is because
   $\FF_{i,1}=\FF_{a,1}=\FF_{b,1}=1$.
This proves the group condition (row-\GC).\vspace{0.008cm}

One can prove (column-\GC)  similarly using (\ref{groupcondition3}) and
  the lower-right $m\times m$ block of $\AA^{[p]}$.\vspace{0.015cm}
\end{proof}

We prove the following property concerning
  discrete unitary matrices that satisfy (\GC):
(Given an $n\times n$ matrix $\AA$, we let $A^R$ denote the set of its row vectors
  $\{\AA_{i,*}\}$, and $A^C$ denote the set of its column
  vectors $\{\AA_{*,j}\}$. For general matrices, it is possible
  that $|A^R|, |A^C|<n$, since $\AA$ might have duplicate rows or columns.
  However, if $\AA$ is $M$-discrete unitary, then it is clear that $|A^R|=|A^C|=n$.)\vspace{0.02cm}

\begin{property}\label{gcproperty1}
Let $\AA\in \mathbb{C}^{n\times n}$ be an $M$-discrete unitary matrix
  that satisfies \emph{(}\GC\emph{)}.
Then both $A^R$ and $A^C$ are finite Abelian groups \emph{(}of order $n$\emph{)}
  under the Hadamard product.
\end{property}
\begin{proof}
The Hadamard product $\circ$ gives a binary operation
on both $A^R$ and $A^C$. The group condition (\GC)
states that  both sets $A^R$ and $A^C$ are closed under this
operation, and it is clearly associative and commutative.
Being discrete unitary, the all-1 vector $\11$ belongs to
both $A^R$ and $A^C$, and serves as the identity element.
This operation also satisfies the cancelation law:
if $x \circ y = x \circ z$ then $y=z$.
From general group theory, a finite set with these
properties already forms a group.  But here we can be more
specific about the inverse of an element.
For each $\AA_{i,*}$,
the inverse should clearly be $\overline{\AA_{i,*}}$.
  By (\GC), there exists a $k\in [0:m-1]$ such
  that $\AA_{k,*}=(\AA_{i,*})^{M-1}=\overline{\AA_{i,*}}$.
The second equation is because $A_{i,j}$, for all $j$,
  is a power of $\oo_M$.
\end{proof}

\subsection{Proof of Theorem \ref{step30}}

In this section, we prove Theorem \ref{step30}.
Suppose $\eval(\CC,\fD)$ is not \#P-hard (otherwise we are already done),
  then by Lemma \ref{groupcondition1}, \vspace{-0.02cm}$((M,N),\CC,\fD)$ satisfies
  not only $(\calU_1)$-$(\calU_4)$, but also  (\GC).
Let us fix $r$ to be any index in $[N-1]$.
We will prove $(\calU_5)$ for $D^{[r]}_{i}$ where $i\in [m:2m-1]$.
The proof for the first half of $\DD^{[r]}$ is similar.
For simplicity, we let $\DD$ be the $m$-dimensional vector such that
$$
D_i=D^{[r]}_{m+i},\ \ \ \text{for all $i\in [0:m-1]$}.
$$
We also need the following notation: Let $K=\{i\in [0:m-1]\hspace{0.08cm}|\hspace{0.08cm}
  D_i\ne 0\}$.

If $|K|=0$, then there is nothing to prove; If $|K|=1$, then
  by $(\calU_3)$, the only non-zero entry in $\DD$ must be $1$.
So we assume $|K|\ge 2$.

We claim that $D_i$, for every $i\in K$, must be a root of unity
  otherwise problem $\eval(\CC,\fD)$ is \#P-hard, which contradicts the assumption.
Actually, the lemma below shows that, such a claim is all we need
  to prove Theorem \ref{step30}:

\begin{lemma}\label{fieldlemma}
If $D\in \mathbb{Q}(\oo_N)$
  is a root of unity, 
then $D$ must be a power of $\oo_N$. \emph{(}$N$ is \emph{even} by \emph{($\calU_1$)}.\emph{)}
\end{lemma}

\def\ZZ{\mathbf{Z}}

We delay the proof to the end of the section.
Now we use it to show that every $D_i$, $i\in K$, is a root of unity.
Suppose for a contradiction that this is not true.
We start by proving the following lemma about $\ZZ=(Z_0,\ldots,Z_{m-1})$,
  where $Z_i=(D_i)^N$ for all $i$:

\begin{lemma}\label{lemma-approach-1}
Assume there exists some $k\in K$ such that $Z_k$ is not a root of unity,
  then there exists an
  infinite integer sequence $\{P_n\} $
  such that, when $n\rightarrow \infty$, the vector sequence $((Z_k)^{P_n} :k
\in K)$ approaches to, but never equals to, the all-one vector of
dimension $|K|$.
\end{lemma}
\begin{proof}
Since $Z_k$, for $k\in K$, has norm $1$, there exists a real number
  $\theta_k\in [0,1)$ such that, $Z_k=e^{2\pi i\theta_k}.$
We will treat $\theta_k$ as a number in the $\mathbb{Z}$-module
  $\mathbb{R}_{\bmod 1}$, i.e., real numbers modulo $1$.
By the assumption we know that at least one of the $\theta_k$'s, $k\in K$,
  is irrational.

This lemma follows from the well-known
Dirichlet's Box Principle.
For completeness, we include a proof here.
Clearly, for any positive integer $P$, $((Z_{k})^{P} : k \in K)$ does not
equal to the all-one vector of dimension $|K|$;
Otherwise, every $\theta_k$ is rational, contradicting the assumption.


Let $n^* = n^{|K|}+1$, for some positive integer $n>1$.
We consider $(L\cdot\theta_k:k\in K)$ for all $L\in [n^*]$.
We divide the unit cube $[0,1)^{|K|}$
  into $n^*-1$ sub-cubes of the following form
\[\left[\frac{a_1}{n}, \frac{a_1+1}{n}\right)
\times  \cdots \times \left[\frac{a_{|K|}}{n}, \frac{a_{|K|}+1}{n}\right),\]
where $a_k \in \{0, \ldots, n-1\}$ for all $k\in |K|$.
By cardinality,
  there exist $L\ne L'\in [n^*]$ such that $$\big(L\cdot \theta_k \bmod 1: k\in K\big) \ \ \
  \text{and}\ \ \ \big(L'\cdot \theta_k\bmod 1: k\in K\big)$$ fall in the same sub-cube.
Assume $L>L'$, then by setting $P_n=L-L'\ge 1$, we have
\[\big|P_n\cdot \theta_k\bmod 1\big|=
  \left|(L  -L')\cdot \theta_k \bmod 1 \right| \le \frac{1}{n},\ \ \ \text{for all $k\in K$.}\]

It is clear that by repeating the procedure for every $n$,
  we get an infinite sequence $\{P_n\}$
  such that $$\left((Z_k)^{P_n}=e^{2\pi i(P_n\cdot \theta_k)}:
  k\in K\right)$$ approaches to, but never equals to, the
all-one vector of dimension $|K|$.\vspace{0.06cm}
\end{proof}


\begin{figure}
\center
\includegraphics[height=6.3cm]{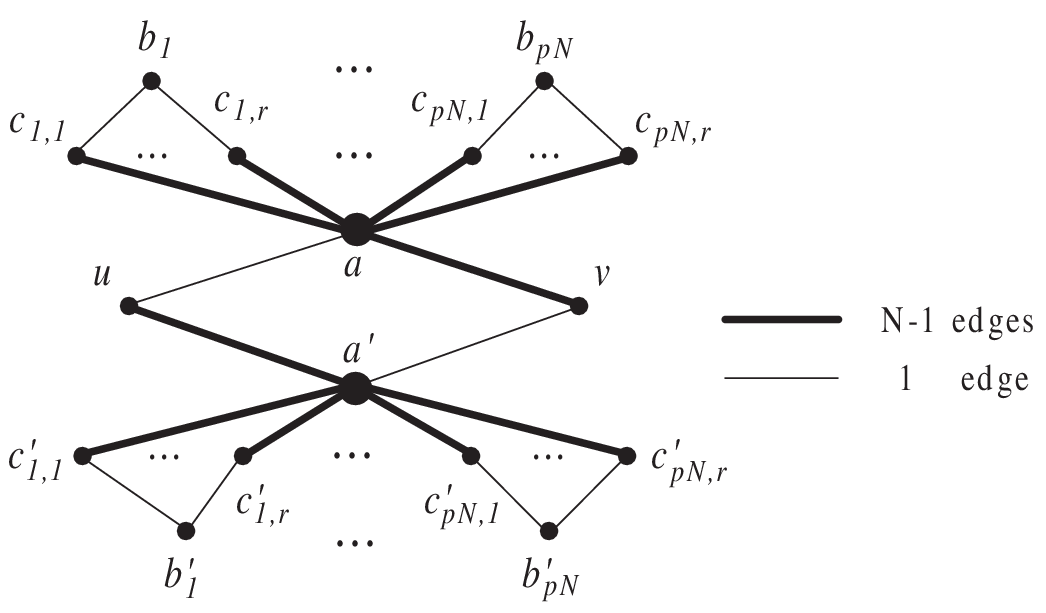}
\caption{The gadget for $p=1$ (Note that the subscript $e$ is suppressed).}\label{figure_6}
\end{figure}

Let $G=(V,E)$ be an undirected graph. Then for each $p\ge 1$, we build
  a graph $G^{[p]}$ by replacing every edge $e=uv\in E$
  with a gadget which is shown in Figure \ref{figure_6}.
Recall that $r\in [N-1]$ is fixed.\vspace{0.01cm}

More exactly, we define $G^{[p]}=(V^{[p]},E^{[p]})$ as follows:
$$
V^{[p]}=V\cup\big\{a_e,b_{e,i},c_{e,i,j},a_e',b_{e,i}',c_{e,i,j}'\hspace{0.08cm}\big|
  \hspace{0.08cm} e\in E,i\in [pN], j\in [r]\big\},
$$
and $E^{[p]}$ contains the following edges: For each edge $e=uv\in E$,
\begin{enumerate}
\item One edge between $(u,a_e)$ and $(v,a_e')$;\vspace{-0.05cm}
\item $N-1$ edges between $(a_e,v)$ and $(u,a_e')$;\vspace{-0.05cm}
\item One edge between $(c_{e,i,j},b_{e,i})$ and $(c_{e,i,j}',b_{e,i}')$,
  for all $i\in [pN]$ and $j\in [r]$;\vspace{-0.05cm}
\item $N-1$ edges between $(a_e,c_{e,i,j})$ and $(a_e',c_{e,i,j}')$,
  for all $i\in [pN]$ and $j\in [r]$.
\end{enumerate}
It is easy to check that the degree of every vertex in $G^{[p]}$ is
  a multiple of $N$ except $b_{e,i}$ and $b_{e,i}'$, which have
  degree $r\bmod N$.

As the gadget is symmetric, the construction gives us
  a symmetric $2m\times 2m$ matrix $\AA^{[p]}$ such that
$$
Z_{\AA^{[p]}}(G)=Z_{\CC,\fD}(G^{[p]}),\ \ \ \text{for any
  undirected graph $G$,}
$$
and thus, $\eval(\AA^{[p]})\le \eval(\CC,\fD)$, and $\eval(\AA^{[p]})$ is also not \#P-hard.

The entries of $\AA^{[p]}$ are as follows: First, for all $u,v\in [0:m-1]$,
$$
A^{[p]}_{u,m+v}=A^{[p]}_{m+u,v}=0.
$$
The entries in the upper-left $m\times m$ block of $\AA^{[p]}$ are
\begin{eqnarray*}
A^{[p]}_{u,v}&=&\left(\sum_{a\in [0:m-1]} F_{u,a}\overline{F_{v,a}}
  \left(\sum_{b\in [0:m-1]} D^{[r]}_{m+b} \left(
  \sum_{c\in [0:m-1]} F_{c,b}\overline{F_{c,a}}
  \right)^r  \right)^{pN}
\right)\\[0.5ex] &&\times \left(\sum_{a\in [0:m-1]} \overline{F_{u,a}}{F_{v,a}}
  \left(\sum_{b\in [0:m-1]} D^{[r]}_{m+b} \left(
  \sum_{c\in [0:m-1]} F_{c,b}\overline{F_{c,a}}
  \right)^r  \right)^{pN}
\right),\end{eqnarray*}
for all $u,v\in [0:m-1]$. Since $\FF$ is discrete unitary,
$$\sum_{c\in [0:m-1]} F_{c,b}\overline{F_{c,a}}=\langle \FF_{*,b},\FF_{*,a}\rangle
  =0,$$
unless $a=b$. As a result, the equation can be simplified to be
$$
A^{[p]}_{u,v}=L_p\cdot \left(\sum_{k\in K}
  \big(D_k\big)^{pN} F_{u,k}\overline{F_{v,k}}\right)\left(
  \sum_{k\in K} \big(D_k\big)^{pN} \overline{F_{u,k}}{F_{v,k}}\right),
  \ \ \ \text{for all $u,v\in [0:m-1]$.}
$$
where $L_p$ is a positive constant that is independent of $u$ and $v$.

Assume for a contradiction that not all the $D_k$'s, $k\in K$,
  are roots of unity, then by Lemma \ref{lemma-approach-1} we know there exists a sequence
  $\{P_n\}$ such that $((D_k)^{NP_n}: k \in K) $ approaches
to, but never equals to, the all-$1$ vector of dimension $|K|$, when $n\rightarrow \infty$.
Besides, by ($\calU_3$)\vspace{0.005cm} we know there exists an $i\in K$ such that $D_i=1$.
Now consider $G^{[P_n]}$ with parameter $p=P_n$ from this sequence.
We have
$$ A^{[P_n]}_{u,u} =L_{P_n}\cdot \left(\sum_{k\in K}(D_k)^{NP_n}\right)^2,\ \ \ \text{
  for any $u\in [0:m-1]$}.$$
We let $T_n$ denote the second factor on the right-hand side,
  then $|T_n|$ could be arbitrarily close to $|K|^2$
  if we choose $n$ large enough. 
By using the dichotomy theorem of Bulatov and Grhoe (and Lemma \ref{absolutevalue})
  together with the assumption that problem $\eval(\AA^{[P_n]})$ is not \#P-hard,
  we know\vspace{0.005cm} the norm of every entry of $\AA^{[P_n]}$ in its
  upper-left block is either $0$ or $L_{P_n}\cdot |T_n|$.\vspace{0.005cm}

Now we focus on the first row by fixing $u=0$.
Since $\FF_{0,*}=\11$, we have
$$ A^{[P_n]}_{0,v} =L_{P_n}\cdot \left(\sum_{k\in K}(D_k)^{NP_n} \overline{F_{v,k}}
  \right)\left(\sum_{k\in K}(D_k)^{NP_n} {F_{v,k}}\right) ,\ \ \ \text{
  for any $v\in [0:m-1]$}.$$
By Property \ref{gcproperty1}, $F^R=\{\FF_{v,*}\}$ is a group under the Hadamard product.
We let
$$S = \{v \in [0:m-1]\hspace{0.07cm}\big|\hspace{0.1cm}
  \forall\hspace{0.08cm} i, j \in K, F_{v,i} = F_{v, j}\},$$
and denote $\{\FF_{v,*}:v \in S\}$ by $F^S$,
Then it is clear that $F^S$ is a subgroup of $F^R$.
Also note that, $0\in S$ since $\FF_{0,*}$ is the all-one
  vector of dimension $m$.\vspace{-0.047cm}

\def\bchi{\boldsymbol{\chi}}

For any $v\notin S$, when $n$ is sufficiently large, we have
$|A^{[P_n]}_{0,v}|<|A^{[P_n]}_{0,0}|$.
This is because when $n \rightarrow \infty$, $$T_n\rightarrow |K|^2\ \ \ \text{but}\ \ \
  \left(\sum_{k\in K}(D_k)^{NP_n} \overline{F_{v,k}}\right)\left(
  \sum_{k\in K}(D_k)^{NP_n} {F_{v,k}}\right) \rightarrow
  \left(\sum_{k\in K}\overline{F_{v,k}}\right)\left(\sum_{k\in K} {F_{v,k}}\right),
$$
which\vspace{-0.028cm} has norm strictly smaller than $|K|^2$ (since $v\notin S$).
So when $n$ is sufficiently large, $A^{[P_n]}_{0,v}$ must be $0$ for all $v\notin S$.
We denote $((D_k)^{NP_n}:k\in [0:m-1])$ by $\DD^n$, then for $v\notin S$
  and sufficiently large
  $n$, \begin{equation}\label{uuuu}\text{either \ $\langle \DD^n, \FF_{v,*} \rangle =0$
\  \ or\ \ $\langle \DD^n,\overline{ \FF_{v,*}} \rangle =0$.}\end{equation}


Next, we focus on the characteristic vector $\bchi$ (of dimension $m$) of $K$:
  $\chi_k=1$ if $k\in K$ and $\chi_k=0$ elsewhere.
By (\ref{uuuu}) and the definition of $S$, we have
\begin{equation}\label{blaalb}
\langle \bchi,\FF_{v,*}\rangle=0,\ \ \text{for all $v\notin S$}\ \ \
\text{and}\ \ \ \left|\langle \bchi,\FF_{v,*}\rangle\right|=|K|,\ \ \ \text{for all $v\in S$}.
\end{equation}
To prove the first equation, we note that by Eq.\hspace{0.04cm}(\ref{uuuu}),
  either there is an infinite subsequence $\{\DD^n\}$ that satisfies
  $\langle \DD^n,\FF_{v,*}\rangle=0$ or there is an infinite subsequence that
  satisfies $\langle \DD^n,\overline{\FF_{v,*}}\rangle=0$.
Since $\DD^n\rightarrow \bchi$ when $n\rightarrow \infty$,
  we have either $\langle \bchi,\FF_{v,*}\rangle =0$ or $\langle\bchi,\overline{\FF_{v,*}}\rangle = 0$.
The second case still gives us $\langle \bchi,\FF_{v,*}\rangle=0$
  since $\bchi$ is real.
The second equation in (\ref{blaalb}) follows directly from the definition of $S$.
As a result, we have $$
\bchi= \frac{1}{m}\sum_{v\in S} \langle \bchi, \FF_{v,*} \rangle\cdot  \FF_{v,*}.$$

Now we assume the expression of vector $\DD^n$, under the orthogonal
  basis $\{\FF_{v,*}\}$, is
\begin{equation*}
\DD^n=\sum_{i=0}^{m-1} x_{i,n} \FF_{i,*},\ \ \ \text{where}\ \
 x_{i,n}=\frac{1}{m} \langle \DD^n, \FF_{i,*} \rangle.
\end{equation*}
If for some $n$ we have $x_{i,n}=0$ for all $i\notin S$,
  then we are done because by the definition of $S$,
  every $\FF_{i,*}$, $i\in S$, is a constant over $K$ and thus,
  the vector $\DD^n$ is a constant over $K$.
Since we know there exists an $i\in K$ such that $D_i=1$,
  every $D_j$, $j\in K$, must be a root of unity.\vspace{0.005cm}

Suppose this is not the case.
Then (here consider those sufficiently large
    $n$ so that (\ref{uuuu}) holds)
$$\bchi=\DD^n\circ \overline{\DD^n}=
\left(\sum_{i} x_{i,n} \FF_{i,*}\right)\circ \left(\sum_j
  \overline{x_{j,n}} \overline{\FF_{j,*}}\right)=\sum_v y_{v,n} \FF_{v,*},
\ \ \text{where}\ \ y_{v,n}=\hspace{-0.3cm}\sum_{\FF_{i,*} \circ \overline{\FF_{j,*}}
  =\FF_{v,*}}\hspace{-0.3cm} x_{i,n} \overline{x_{j,n}}.$$
The last equation uses the fact that $F^R$ is a group under the
  Hadamard product (so for any $i,j$ there exists a unique $v$ such that
  $\FF_{v,*}=\FF_{i,*}\circ\overline{\FF_{j,*}}$).

Since the Fourier expansion of $\bchi$ under $\{\FF_{v,*}\}$ is unique, we have $y_{v,n}=0$,
  for any $v\not \in S$.
Because $\DD^n\rightarrow \bchi$,
  by (\ref{blaalb}),
we know that when $n\rightarrow \infty$, $x_{i,n}$, for any $i\notin S$,
  can be arbitrarily close to $0$,
while $|x_{i,n}|$ can be arbitrarily close to ${|K|}/{m}$,
  for any $i \in S$. So there exists a sufficiently large $n$ such that
$$|x_{i,n}|<\frac{4|K||S|}{5m^2},\ \ \ \text{for all $i\notin S$},\ \
\text{and}\ \ \ |x_{i,n}|>\frac{4|K|}{5m},\ \ \ \text{for all $i \in S$.}$$
We pick such an $n$ and will use it to reach a contradiction.
Since we assumed that for any $n$ (which is of course also true for this particular
  $n$ we picked here), there exists at least one index $i \notin S$
  such that $x_{i,n} \neq 0$, we can choose a $w\notin S$
  that maximizes $|x_{i,n}|$ among all $i\notin S$. Clearly, $|x_{w,n}|$ is positive.

We consider the expression of $y_{w,n}$ using $x_{i,n}$.
We divide the summation into two parts:
the \emph{main} terms $x_{i,n}\overline{x_{j,n}}$ in which either
  $i\in S$ or $j\in S$ and the remaining terms in which $i,j\notin S$
(note that if $\FF_{w,*}=\FF_{i,*}\circ \overline{\FF_{j,*}}$,
  then  $i$ and $j$ cannot be both in $S$.
Otherwise, since $F^S$ is a subgroup, we have $w\in S$ which contradicts
  the assumption that $w \not \in S$.)\vspace{0.1cm}
\begin{eqnarray*}
\mbox{The main terms of }\hspace{0.05cm} y_{w,n} \hspace{-0.17cm}&=&\hspace{-0.17cm}
  \frac{1}{m^2}\sum_{j\in S} \langle \DD^{n}, \FF_{w,*}\circ \FF_{j,*}\rangle
 \overline{\langle \DD^{n}, \FF_{j,*}\rangle}+ \frac{1}{m^2}\sum_{i\in S}
 \langle \DD^{n}, \FF_{i,*}\rangle
 \overline{\langle \DD^{n}, \FF_{i,*} \circ\overline{\FF_{w,*}}\rangle}\vspace{0.1cm}
\end{eqnarray*}

Note that $x_{0,n}=(1/m)\langle \DD^n,\FF_{0,*}\rangle$
  and $\FF_{0,*}=\11$.
Also note that (by the definition of $S$), when $j\in S$,
  $F_{j,k}=\alpha_j$ for all $k\in K$, for some complex number $\alpha_j$ of
  norm $1$.
Since $\DD^n$ is only non-zero on $K$,\vspace{0.06cm}
$$\langle \DD^{n}, \FF_{w,*}\circ \FF_{j,*}\rangle
 \overline{\langle \DD^{n}, \FF_{j,*}\rangle}= \langle \DD^n,\alpha_j\FF_{w,*}\rangle
 \overline{\langle \DD^n,\alpha_j\11\rangle}=m\overline{x_{0,n}}
 \cdot \langle \DD^n,\FF_{w,*}\rangle.\vspace{0.06cm}
$$
Similarly, we can simplify the other sum so that
\[
\text{The main terms of }\hspace{0.05cm} y_{w,n}\hspace{0.05cm} =\hspace{0.05cm}
  \frac{ |S|}{m} \Big(\overline{x_{0,n}} \langle\DD^n, \FF_{w,*}\rangle +
  x_{0,n} \langle \overline{\DD^n}, \FF_{w,*}\rangle \Big).
 \]
By (\ref{uuuu}) we have either $\langle \DD^n, \FF_{w,*}\rangle$ or
  $\langle \overline{\DD^n}, \FF_{w,*}\rangle$ is $0$.
Since we assumed that $x_{w,n}= \frac{1}{m}\langle \DD^n,\FF_{w,*}\rangle\ne 0$, the latter
  has to be $0$. Therefore, the sum of the main terms of $y_{w,n}$
is equal to $\overline{x_{0,n}} x_{w,n} |S|$. As $0\in S$,
$$
\Big|\overline{x_{0,n}} x_{w,n} |S|\Big|\ge \frac{4|K||S|}{5m}|x_{w,n}|.
$$

Now we consider the remaining terms. Below we show that the sum of all these terms
  cannot have a norm as large as $|\overline{x_{0,n}}x_{w,n}|S||$ and thus,
  $y_{w,n}$ is non-zero and we get a contradiction.
To prove this, it is easy to check that the number of remaining terms
  is at most $m$, and the norm of each of them is
  $$|x_{i,n}\overline{x_{j,n}}|\le
  |x_{w,n}|^2<\frac{4|K||S|}{5m^2} |x_{w,n}|,$$
since $i,j\notin S$. So the norm of their sum
  is $<\frac{4|K||S|}{5m}|x_{w,n}|$.
This finishes the proof of Theorem \ref{step30}.\vspace{0.025cm}

\begin{proof}[Proof of Lemma \ref{fieldlemma}]
Assume $D=\oo_M^{k}$, for some positive integers $k$\vspace{-0.005cm} and $M$ with $\gcd(k,M)=1$.
Since $D\in \mathbb{Q}(\oo_N)$, we have $\oo_M^k \in \mathbb{Q}(\omega_N)$.
By $\gcd(k,M)=1$, we have $\omega_M \in \mathbb{Q}(\omega_N)$ and
\[\mathbb{Q}(\omega_N)=\mathbb{Q}(\omega_N,\omega_M)=\mathbb{Q}(\omega_{\lcm(M,N)}).\]
The degree of the field extension is $[\mathbb{Q}(\omega_N) : \mathbb{Q}] = \phi(N$),
  the Euler function \cite{Euler}.\vspace{0.006cm}

When $N\hspace{0.06cm} |\hspace{0.06cm} N'$, and $\phi(N) = \phi(N')$,
by expanding according to the prime factorization for $N$,
we can get (and indeed this is all there is to be had) that
if $N$ is even, then $N'= N$; if $N$ is odd, then $N' = N$ or $N'=2N$.
Since by ($\calU_1$) $N$ is even, we have $\lcm(M,N)=N$, $M\hspace{0.06cm}
  |\hspace{0.06cm}N$, and $D$ is a power of $\oo_N$.
\end{proof}

\subsection{Decomposing $\FF$ into Fourier Matrices}

Assume that $((M,N),\CC,\fD)$ satisfies not \vspace{0.01cm}only conditions $(\calU_1)$--$(\calU_5)$ but
  also the group condition (\GC); since otherwise $\eval(\CC,\fD)$ is \#P-hard.

To prove a decomposition for $\FF$ (recall that $\FF$ is the
  upper-right $m\times m$ block matrix of $\CC$)
 we first show that if $M=pq$ and $\gcd(p,q)=1$,
  then up to a permutation of rows and columns, $\FF$ must be
  the tensor product of two smaller matrices,
both of which are discrete unitary and satisfy (\GC).
Note that $p,q$ here are not necessarily primes or prime powers.

\begin{lemma}\label{decomp1}
Let $\FF\in \mathbb{C}^{m\times m}$ be an $M$-discrete unitary matrix that
  satisfies \emph{(}\GC\emph{)}.
Moreover, $M=pq$, $p,q>1$ and $\gcd(p,q)=1$.
Then there exist two permutations $\Pi,\Sigma:[0:m-1]\rightarrow [0:m-1]$
  such that
$$
\FF_{\Pi,\Sigma}=\FF'\otimes \FF'',
$$
where $\FF'$ is $p$-discrete unitary, $\FF''$ is $q$-discrete
  unitary, and both of them satisfy \emph{(}\GC\emph{)}.
\end{lemma}
\begin{proof}
By Property \ref{gcproperty1}, both $F^R$ and $F^C$ are
  finite Abelian groups.
Since $\FF$ is $M$-discrete unitary, the order of any vector
  in $F^R$ and $F^C$ is a divisor of $M$.

By the fundamental theorem of Abelian groups,
  there is a group isomorphism
$$
\rho: F^R\rightarrow \mathbb{Z}_{g_1}\times\cdots \times \mathbb{Z}_{g_h}\equiv \mathbb{Z}_\gg,
$$  where $g_1,\ldots,g_h$ are prime powers, and
  $g_i\hspace{0.05cm}|\hspace{0.05cm} M$ for all $i$.
As $\gcd(p,q)=1$, without loss of generality, we may assume
  there exists an integer $h'$ such that
  $g_i\hspace{0.05cm}|\hspace{0.05cm} p$ for all $i\in [h']$ and
  $g_i\hspace{0.05cm}|\hspace{0.05cm} q$ for all other $i$.

We use $\rho^{-1}$ to define the following two subsets of $F^R$:
$$
S^p=\{\rho^{-1}(\xx) \;\big|\; \xx\in \mathbb{Z}_\gg,\hspace{0.06cm}
  \text{$x_i=0$ for all $i>h'$}\}\ \ \text{and}
\ \ S^q=\{\rho^{-1}(\xx) \;\big|\;\xx\in \mathbb{Z}_\gg,\hspace{0.06cm}
\text{$x_i=0$ for all $i\le h'$}\}.
$$
Then it is easy to show the following four properties:
\begin{enumerate}
\item Both $S^p$ and $S^q$ are subgroups of $F^R$;\vspace{-0.02cm}
\item $S^p=\{\uu\in F^R\hspace{0.07cm}|\hspace{0.07cm}(\uu)^p=\11\}$ {and}
$S^q=\{\vv\in F^R\hspace{0.07cm}|\hspace{0.07cm}(\vv)^q=\11\}$;\vspace{-0.02cm}
\item Let $m'=|S^p|$, $m''=|S^q|$, then $m=m'\cdot m''$,
  $\gcd(m',q)=1$, $\gcd(m'',p)=1$, $\gcd(m',m'')=1$;\vspace{-0.02cm}
\item $(\uu,\vv)\mapsto \uu\circ \vv$ is a group isomorphism
  from $S^p\oplus S^q$ onto $F^R$.
\end{enumerate}
Let\vspace{0.01cm} $S^p=\{\uu_0=\11,\uu_1,\ldots,\uu_{m'-1}\}$ {and}
  $S^q=\{\vv_0=\11,\vv_1,\ldots,\vv_{m''-1}\}$. Then by 4)
  there is a one-to-one correspondence $f:i\mapsto (f_1(i),f_2(i))$
  from $[0:m-1]$ to $[0:m'-1]\times [0:m''-1]$ such that
\begin{equation}\label{haha1}
\FF_{i,*}=\uu_{f_1(i)}\circ \vv_{f_2(i)},\ \ \text{for all $i\in [0:m-1]$.}
\end{equation}

Next we apply the fundamental theorem to $F^C$.
We use the group isomorphism, in the same way, to define two subgroups $T^p$ and $T^q$
  with four corresponding properties:
\begin{enumerate}
\item Both $T^p$ and $T^q$ are subgroups of $F^C$;\vspace{-0.02cm}
\item $T^p=\{\ww\in F^C\hspace{0.06cm}|\hspace{0.06cm}(\ww)^p=\11\}$
  {and} $T^q=\{\rr\in F^C\hspace{0.06cm}|\hspace{0.06cm}(\rr)^q=\11\}$;\vspace{-0.02cm}
\item $m=|T^p|\cdot |T^q|$,
  $\gcd(|T^p|,q)=1$, $\gcd(|T^q|,p)=1$, and $\gcd(|T^p|,|T^q|)=1$;\vspace{-0.02cm}
\item $(\ww,\rr)\mapsto \ww\circ \rr$ is a group isomorphism
  from $T^p\oplus T^q$ onto $F^C$.
\end{enumerate}
By comparing item 3) in both lists,
we have $|T^p|=|S^p|=m'$ and $|T^q|=|S^q|=m''$.

Let
$
T^p=\{\ww_0=\11,\ww_1,\ldots,\ww_{m'-1}\}$  and
$T^q=\{\rr_0=\11,\rr_1,\ldots,\rr_{m''-1}\}.
$
Then by item 4),
we have a one-to-one correspondence $g$
  from $[0:m-1]$ to $[0:m'-1]\times [0:m''-1]$ and
\begin{equation}\label{haha2}
\FF_{*,j}=\ww_{g_1(j)}\circ \rr_{g_2(j)},\ \ \text{for all $j\in [0:m-1]$.}
\end{equation}

Now we are ready to permute the rows and columns of $\FF$ to get a
  new matrix $\GG$ that is the tensor product
  of two smaller matrices.
We use $(x_1,x_2)$, where $x_1\in [0:m'-1],
  x_2\in [0:m''-1]$, to index the rows and columns of $\GG$.
We use $\Pi(x_1,x_2)=f^{-1}(x_1,x_2)$, from $[0:m'-1]\times [0:m''-1]$
  to $[0:m-1]$, to permute the rows of $\FF$ and
  $\Sigma(y_1,y_2)=g^{-1}(y_1,y_2)$ to permute the columns of $\FF$, respectively.
As a result, we get $\GG=\FF_{\Pi,\Sigma}$ where $$G_{(x_1,x_2),(y_1,y_2)}=
  F_{\Pi(x_1,x_2),\Sigma(y_1,y_2)},\ \ \text{for all $x_1,y_1\in [0:m'-1]$ and
  $x_2,y_2\in [0:m''-1]$}.$$
By (\ref{haha1}), and using the fact that
$\uu_0=\11$ and $\vv_0=\11$,
we have
$$
\GG_{(x_1,x_2),*}=\GG_{(x_1,0),*}\circ \GG_{(0,x_2),*},\ \ \ \text{for all
  $x_1\in [0:m'-1]$ and $x_2\in [0:m''-1]$.}$$
Similarly by (\ref{haha2}) and $\ww_0=\11$ and $\rr_0=\11$, we have $$
\GG_{*,(y_1,y_2)}=\GG_{*,(y_1,0)}\circ \GG_{*,(0,y_2)},\ \ \ \text{for all
  $y_1\in [0:m'-1]$ and $y_2\in [0:m''-1]$.}
$$
Therefore, applying both relations, we have\vspace{-0.02cm}
$$
G_{(x_1,x_2),(y_1,y_2)}=G_{(x_1,0),(y_1,0)}\cdot
G_{(x_1,0),(0,y_2)}\cdot G_{(0,x_2),(y_1,0)}\cdot G_{(0,x_2),(0,y_2)}.
$$

We claim
\begin{equation}\label{claimhaha}
G_{(x_1,0),(0,y_2)}=1\ \ \ \text{and}\ \ \ G_{(0,x_2),(y_1,0)}=1.
\end{equation}
Then we have
\begin{equation}\label{decompostionG-tensor}
G_{(x_1,x_2),(y_1,y_2)}= G_{(x_1,0),(y_1,0)}\cdot G_{(0,x_2),(0,y_2)}.\vspace{0.1cm}
\end{equation}

To prove the first equation in (\ref{claimhaha}), we realize that
  it appears as an entry
  in both $\uu_{x_1}$ and $\rr_{y_2}$.
Then by item 2) for $S^p$ and $T^q$, both of its $p${th} and $q${th} powers
  are $1$.
Thus it has to be $1$.
The other equation in (\ref{claimhaha}) can be proved the same way.\vspace{0.005cm}

As a result, we have obtained our tensor product
decomposition $\GG=\FF'\otimes \FF''$,
where $$\FF'=\Big(F'_{x,y}\equiv G_{(x,0),(y,0)}\Big)\ \
  \text{and}\ \ \FF''=\Big(F''_{x,y}\equiv G_{(0,x),(0,y)}\Big).$$

The only thing left is to show that $\FF'$ and $\FF''$
  are discrete unitary, and satisfy (\GC).
Here we only prove it for $\FF'$. The proof for $\FF''$ is the same.
To see $\FF'$ is discrete unitary, for all $x\ne y\in [0:m'-1]$,
\begin{eqnarray*}
0\hspace{-0.15cm}&=&\hspace{-0.15cm}
  \langle \GG_{(x,0),*},\GG_{(y,0),*}\rangle\hspace{0.1cm}=\hspace{0.1cm}
  \sum_{z_1,z_2} G_{(x,0),(z_1,z_2)}\overline{
  G_{(y,0),(z_1,z_2)}}\\
\hspace{-0.15cm}&=&\hspace{-0.15cm}
\sum_{z_1,z_2} G_{(x,0),(z_1,0)}G_{(0,0),(0,z_2)}\overline{
  G_{(y,0),(z_1,0)}G_{(0,0),(0,z_2)}}\\[0.8ex]
\hspace{-0.15cm}&=&\hspace{-0.15cm}m''\cdot \langle \FF'_{x,*},\FF'_{y,*}\rangle.
\end{eqnarray*}
Here we used the factorization (\ref{decompostionG-tensor})
and $\uu_0 = \11$ and $\vv_0 = \11$.
Similarly, we can prove that $\FF'_{*,x}$ and $\FF'_{*,y}$ are orthogonal for $x\ne y$.
$\FF'$ also satisfies (\GC) because both $S^p$ and $T^p$ are groups
  and thus, closed under the Hadamard product.
Finally, $\FF'$ is exactly $p$-discrete unitary: First, by definition, we have
$$
pq=M=\text{lcm}\big\{\text{order of $G_{(x_1,x_2),(y_1,y_2)}$}:\xx,\yy\big\}=
\text{lcm}\big\{\text{order of $G_{(x_1,0),(y_1,0)}\cdot
  G_{(x_2,0),(y_2,0)}$}:\xx,\yy\big\};
$$
Second, the order of $G_{(x_1,0),(y_1,0)}$ divides $p$
  and the order of $G_{(x_2,0),(y_2,0)}$ divides $q$.
As a result, we have
$$
p=\text{lcm} \big\{\text{order of $G_{(x,0),(y,0)}$}:x,y\big\}
$$
and by definition, $\FF'$ is a $p$-discrete unitary matrix.\vspace{0.02cm}
\end{proof}


Next we prove Lemma~\ref{decomp2} which deals with the case\vspace{0.02cm}
when $M$ is a prime power. 

\begin{property}\label{gcproperty2}
Let $\AA$ be an $M$-discrete unitary matrix that satisfies the group condition
  \emph{(}\GC\emph{)}. If $M$ is a prime power, then \vspace{0.02cm}
  one of its entries is equal to $\oo_M$.
\end{property}


\begin{proof}
Since $M$ is a prime power, some entry of $\AA$ has order exactly $M$
as a root of unity. Hence it has the form $\oo_M^k$ for some $k$
relatively prime to $M$. Then by the group condition \emph{(}\GC\emph{)}
all powers of $\oo_M^k$ also appear as entries of $\AA$,
in particular $\oo_M$.
\end{proof}

\begin{lemma}\label{decomp2}
Let $\FF\in \mathbb{C}^{m\times m}$ be an $M$-discrete unitary matrix that
  satisfies \emph{(}\GC\emph{)}.
Moreover, $M=p^k$ is a prime power for some $k\ge 1$.
Then there exist two permutations $\Pi$ and $\Sigma$ such that
$$
\FF_{\Pi,\Sigma}=\boldsymbol{\mathcal{F}}_{M}\otimes\hspace{0.04cm} \FF',
$$
where $\FF'$ is an $M'$-discrete unitary matrix, $M'=p^{k'}$ for some
  $k'\le k$, and $\FF'$ satisfies \emph{(}\GC\emph{)}.
\end{lemma}

\begin{proof}
By Property \ref{gcproperty2}, there exist $a$ and $b$
  such that $F_{a,b}=\oo_M$.
Thus both the order of $\FF_{a,*}$ (in $F^R$) and the order of
  $\FF_{*,b}$ (in $F^C$) are $M$.
Let\vspace{-0.07cm}
$$S_1=\big\{\11,\FF_{a,*},(\FF_{a,*})^2,\ldots,(\FF_{a,*})^{M-1}\big\}\vspace{-0.07cm}$$
denote the subgroup of $F^R$ generated by $\FF_{a,*}$.
Since the order of $\FF_{a,*}$ is $M$, we have $|S_1|=M$.\vspace{0.005cm}

Let $S_2$ denote the subset of $F^R$ such that $\uu\in S_2$ iff
 its $b^{th}$ entry $u_b=1$.
It is easy to see that $S_2$ is a subgroup of $F^R$.
Moreover, one can show that $(\ww_1,\ww_2)\mapsto \ww_1\circ \ww_2$
  is a group isomorphism from $S_1\oplus S_2$ onto $F^R$.
As a result, $|S_2|=m/M$ which we denote by $n$.

Let $S_2=\{\uu_0=\11,\uu_1,\ldots,\uu_{n-1}\}$, then
  there exists a one-to-one correspondence $f$ from $[0:m-1]$
  to $[0:M-1]\times [0:n-1]$, where
$i \mapsto f(i) = (f_1(i), f_2(i))$, such that\vspace{-0.02cm}
\begin{equation}\label{baba1}
\FF_{i,*}=(\FF_{a,*})^{f_1(i)}\circ \uu_{f_2(i)},
  \ \ \text{for all $i\in [0:m-1]$}.\vspace{-0.02cm}
\end{equation}
In particular, we have $f(a)=(1,0)$.

Similarly, we use $T_1$ to denote the subgroup of $F^C$
  generated by $\FF_{*,b}$ ($|T_1|=M$), and $T_2$ to denote the subgroup
  of $F^C$ that contains all the $\vv\in F^C$ such that $v_a=1$.
$(\ww_1,\ww_2)\mapsto \ww_1\circ\ww_2$
  also gives us a~natural group isomorphism from $T_1 \oplus T_2$ onto $F^C$,
  so $|T_2|=m/M=n$\vspace{0.005cm}

Let $T_2=\{\vv_0=\11,\vv_1,\ldots,\vv_{n-1}\}$, then
  there exists a one-to-one correspondence $g$ from $[0:m-1]$
  to $[0:M-1]\times [0:n-1]$, where
$j \mapsto g(j) = (g_1(j), g_2(j))$, such that\vspace{-0.02cm}
\begin{equation}\label{baba2}
\FF_{*,j}=(\FF_{*,b})^{g_1(j)}\circ \vv_{g_2(j)},
  \ \ \text{for all $j\in [0:m-1]$}.\vspace{-0.02cm}
\end{equation}
In particular, we have $g(b)=(1,0)$.

Now we are ready to permute the rows and columns of $\FF$ to get a new
  $m\times m$ matrix $\GG$.
Again we use $(x_1,x_2)$, where $x_1\in [0:M-1]$ and $x_2\in [0:n-1]$,
  to index the rows and columns of matrix $\GG$.
We use $\Pi(x_1,x_2)=f^{-1}(x_1,x_2)$, from
  $[0:M-1]\times [0:n-1]$ to $[0:m-1]$, to permute the rows
  and $\Sigma(y_1,y_2)=g^{-1}(y_1,y_2)$ to permute the columns of $\FF$,
  respectively.
As a result, we get $\GG=\FF_{\Pi,\Sigma}$.\vspace{0.005cm}

By equations (\ref{baba1}) and (\ref{baba2}),
and $\uu_0 = \11$ and $\vv_0 = \11$, we have
$$
\GG_{(x_1,x_2),*}=(\GG_{(1,0),*})^{x_1}\circ \GG_{(0,x_2),*}\ \ \text{and}\ \
\GG_{*,(y_1,y_2)}=(\GG_{*,(1,0)})^{y_1}\circ \GG_{*,(0,y_2)}.
$$
Applying them in succession, we get
\begin{eqnarray*}\label{succession-G}
G_{(x_1,x_2),(y_1,y_2)}=(G_{(1,0),(y_1,y_2)})^{x_1}
G_{(0,x_2),(y_1,y_2)}
=(G_{(1,0),(1,0)})^{x_1y_1}(G_{(1,0),(0,y_2)})^{x_1}
(G_{(0,x_2),(1,0)})^{y_1}G_{(0,x_2),(0,y_2)}.
\end{eqnarray*}
We can check that $G_{(1,0),(1,0)}=F_{a,b} =\omega_M$. Indeed, by $f(a)=(1,0)$ and $g(b)=(1,0)$,
  we have
$$
G_{(1,0),(1,0)}=F_{\Pi(1,0),\Sigma(1,0)}=F_{f^{-1}(1,0),g^{-1}(1,0)}=F_{a,b}=\oo_M.
$$
By (\ref{baba2}), and a similar reasoning, we have
$$G_{(1,0),(0,y_2)}= F_{a,g^{-1}(0,y_2)}= (F_{a,b})^0\cdot v_{y_2,a}=v_{y_2,a}=1,$$
where $v_{y_2,a}$ denotes the $a^{th}$ entry of $\vv_{y_2}$,
which is $1$ by the definition of $T_2$.
By (\ref{baba1}), we also have
$$G_{(0,x_2),(1,0)} =F_{f^{-1}(0,x_2),b}
= (F_{a,b})^0 \cdot u_{x_2,b} = u_{x_2,b} = 1,$$
where $u_{x_2,b}$ denotes the $b^{th}$ entry of $\uu_{x_2}$,
which is 1 by the definition of $S_2$.

Combining all these equations, we have
\begin{equation}\label{haharefer}
G_{(x_1,x_2),(y_1,y_2)}=\oo_M^{x_1y_1}\cdot G_{(0,x_2),(0,y_2)}.
\end{equation}
As a result,
  $\GG=\bcF_M\otimes \FF'$, where $\FF'=(F'_{x,y}\equiv G_{(0,x),(0,y)})$
  is an $n\times n$ matrix.\vspace{0.005cm}

To see $\FF'$ is discrete unitary, by (\ref{haharefer}), we have
$$
0=\langle \GG_{(0,x),*},\GG_{(0,y),*}\rangle=M\cdot\langle \FF'_{x,*},\FF'_{y,*}\rangle,
\ \ \ \text{for any $x\ne y\in [0:n-1]$.}
$$
Similarly we can prove that $\FF'_{*,x}$ and $\FF'_{*,y}$ are
  orthogonal for $x\ne y$.
$\FF'$ also satisfies the group condition
because both $S_2$ and $T_2$ are groups and
  thus, closed under the Hadamard product.
More precisely, for (row-\GC), suppose
$\FF'_{x,*}$ and $\FF'_{y,*}$ are two rows of $\FF'$.
The corresponding  two rows
$\GG_{(0,x),*}$ and $\GG_{(0,y),*}$ in $\GG$ are permuted versions of $\uu_x$
and $\uu_y$, respectively.
We have, by (\ref{haha1}),
\[F'_{x,z} = F_{f^{-1}(0,x),
  g^{-1}( 0,z)}= u_{x,g^{-1}(0,z)}
\ \ \ \text{and}\ \ \
F'_{y,z} = F_{f^{-1}(0,y),g^{-1}(0,z)}
  = u_{y,g^{-1}(0,z)}.
\]
Since $S_2$ is a group, we have some $w\in [0:n-1]$
  such that $\uu_x \circ \uu_y = \uu_w$, and thus
\[F'_{x,z} \cdot F'_{y,z} = u_{w,g^{-1}(0,z)} = F'_{w, z}.\]
The verification of (column-\GC) is similar.
Finally, it is also easy to see that $\FF'$ is $p^{k'}$-discrete unitary,
for some integer $k' \le k$.
\end{proof}

Theorem \ref{bi-step-3} then follows from Lemma \ref{decomp1}
  and Lemma \ref{decomp2}.

\section{Proof of Theorem \ref{bi-step-4}}

Let $((M,N),\CC,\fD,(\qq,\bft,\fq))$ be a $4$-tuple that satisfies
  condition ($\calR$).
Also assume that $\eval(\CC,\fD)$ is not $\#$P-hard (since otherwise,
  we are done).
For every $r$ in $\cT$ (recall that $\cT$ is the set of $r\in [N-1]$ such
  $\Delta_r\ne \emptyset$), we show that $\Delta_r$ must be a coset in $\zqt$.
Condition ($\calL_2$) then follows from the following lemma which we
  will prove at the end of this section.
Condition ($\calL_1$) about $\Lambda_r$ can be proved similarly.

\begin{lemma}\label{cosetlemma}
Let $\Phi$ be a coset in $G_1\oplus G_2$, where $G_1$ and $G_2$
  are finite Abelian groups such that $$\gcd\big(|G_1|,|G_2|\big)=1.$$
Then for both $i=1,2$, there exists a coset $\Phi_i$ in $G_i$
  such that $\Phi=\Phi_1\times \Phi_2$.
\end{lemma}

\begin{figure}
\center
\includegraphics[height=8.3cm]{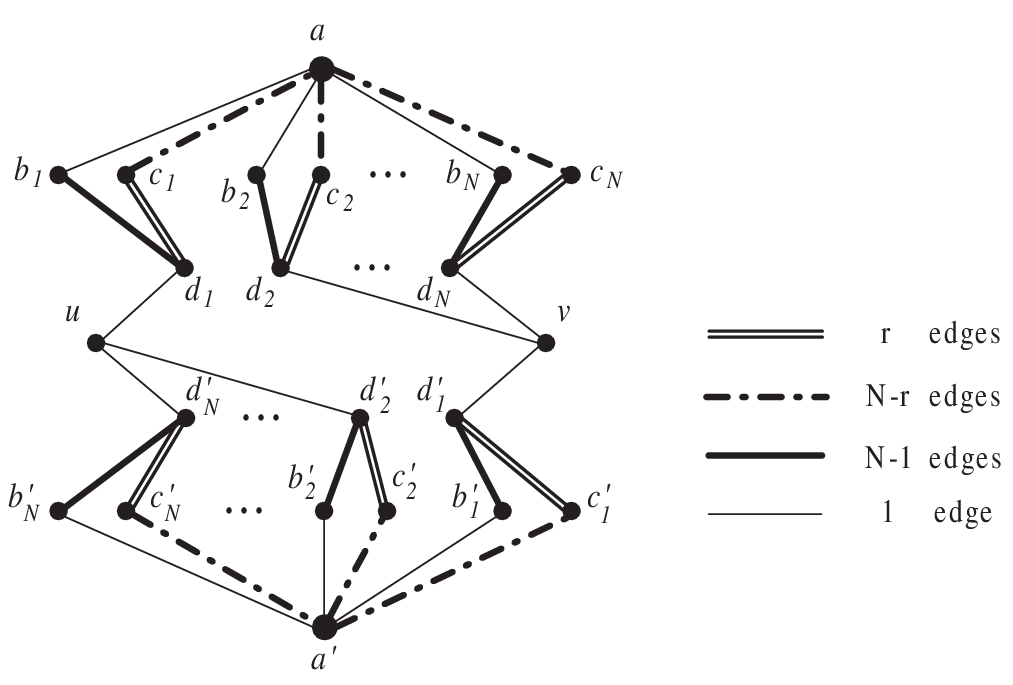}
\caption{The gadget for constructing graph $G'$ (Note that the subscript $e$ is suppressed).}
\label{figure_7}
\end{figure}

Let $G=(V,E)$ be an undirected graph.
We build a new graph $G'$ by replacing
  every $e=uv\in E$ with the gadget as shown in Figure \ref{figure_7}.
More exactly, we define $G'=(V',E')$ as
\begin{equation*}
V'=V\cup\big\{a_e,b_{e,i},c_{e,i},d_{e,i},
  a_e',b_{e,i}',c_{e,i}',d_{e,i}' \hspace{0.06cm}
  \big|\hspace{0.06cm} e\in E\text{\ and\ }i\in [N]\big\}
\end{equation*}
and $E'$ contains exactly the following edges: For each $e=uv\in E$,
\begin{enumerate}
\item {One edge between $(u,d_{e,1})$, $(v,d_{e,1}')$, $(u,d_{e,i}')$
  and $(v,d_{e,i})$ for all $i\in [2:N]$;}\vspace{-0.03cm}
\item {For every $i\in [N]$, one edge between $(a_e,b_{e,i})$, $N-1$ edges
  between $(b_{e,i},d_{e,i})$;}\vspace{-0.03cm}
\item {For every $i\in [N]$, $N-r$ edges between $(a_e,c_{e,i})$,
  $r$ edges between $(c_{e,i},d_{e,i})$;}\vspace{-0.03cm}
\item {For every $i\in [N]$, one edge between $(a_e',b_{e,i}')$,
  $N-1$ edges between $(b_{e,i}',d_{e,i}')$;}\vspace{-0.03cm}
\item {For every $i\in [N]$, $N-r$ edges between $(a_e',c_{e,i}')$,
  $r$ edges between $(c_{e,i}',d_{e,i}')$.}
\end{enumerate}
It is easy to check that the degree of $d_{e,i}$
  and $d_{e,i}'$, for all $e\in E,i\in [N]$, is exactly $r\hspace{-0.05cm}\pmod{
  \hspace{-0.03cm}N}$ while
  all other vertices in $V'$ have degree $0\pmod N$.
It is also noted that the graph fragment which defines the gadget
is bipartite, with all $u, v, b_{e,i},c_{e,i}, b_{e,i}',c_{e,i}'$
on one side and all $a_e, a_e',d_{e,i},d_{e,i}'$ on the other side.\vspace{0.007cm}

The way we construct $G'$ gives us a $2m\times 2m$
  matrix $\AA$ such that
$$
Z_{\AA}(G)=Z_{\CC,\fD}(G'),\ \ \text{for all $G$,}
$$
and thus, $\eval(\AA)\le \eval(\CC,\fD)$, and $\eval(\AA)$ is also not \#P-hard.
We use $\{0,1\}\times \zqt$ to index the rows
  and columns of $\AA$.
Then for all $\uu,\vv\in \zqt$, we have
$$A_{(0,\uu),(1,\vv)}=A_{(1,\uu),(0,\vv)}=0.$$
This follows from the bipartiteness of the gadget.

We now analyze the upper-left $m\times m$ block of $\AA$.
For $\uu,\vv\in \zqt$, we have
\begin{eqnarray*}
A_{(0,\uu),(0,\vv)}=&&\hspace{-0.5cm}
  \left(\sum_{ \aa,\dd_{1},...,\dd_N\in \zqt}F_{\uu,\dd_1}\prod_{i=2}^N
  F_{\vv,\dd_i} \left(\prod_{i=1}^N\left(\sum_{\bb_i\in\zqt}
  F_{\bb_i,\aa}\overline{F_{\bb_i,\dd_i}}\right)
  \left(\sum_{\cc_i\in \zqt}F_{\cc_i,\aa}^{N-r}F_{\cc_i,\dd_i}^r\right)\right)
  \prod_{i=1}^N D^{[r]}_{(1,\dd_i)}\right)\\[1ex]
\times &&\hspace{-0.5cm}\left(\sum_{
  \aa,\dd_{1},...,\dd_N\in \zqt}F_{\vv,\dd_1} \prod_{i=2}^N
  F_{\uu,\dd_i} \left(\prod_{i=1}^N\left(\sum_{\bb_i\in\zqt}
  F_{\bb_i,\aa}\overline{F_{\bb_i,\dd_i}}\right)
  \left(\sum_{\cc_i\in \zqt}F_{\cc_i,\aa}^{N-r}F_{\cc_i,\dd_i}^r\right)\right)
  \prod_{i=1}^N D^{[r]}_{(1,\dd_i)}\right).
\end{eqnarray*}
Note that in deriving this equation, we used the fact
  that $M\hspace{0.04cm}|\hspace{0.04cm}N$ and entries of $\FF$ are all powers of $\oo_M$.

Since $\FF$ is discrete unitary,
$$
\sum_{\bb_i\in\zqt}
  F_{\bb_i,\aa}\overline{F_{\bb_i,\dd_i}}=\langle \FF_{*,\aa},\FF_{*,\dd_i}\rangle
$$
is $0$ unless $\dd_i=\aa$.
When $\dd_i=\aa$ for every $i \in [N]$,
the inner product $\langle \FF_{*,\aa},\FF_{*,\dd_i}\rangle
= m$, and likewise so are the sums over $\cc_i$.
Also the product\vspace{-0.2cm}
 $$\prod_{i\in [N]} D^{[r]}_{(1,\dd_i)} = \left(D^{[r]}_{(1,\aa)}\right)^N\vspace{-0.25cm}
= 1,$$ when each $\dd_i=\aa \in \Delta_r$, and 0 otherwise.
 This is because by $(\calU_5)$, $D^{[r]}_{(1,\aa)}$ is a power of $\oo_N$
  when $\aa \in \Delta_r$, and 0 otherwise.

 As a result, we have
\begin{equation}\label{complicated}
A_{(0,\uu),(0,\vv)}=\left(\sum_{\aa\in \Delta_r} F_{\uu,\aa}\overline{F_{\vv,\aa}}\cdot m^{2N}
\right)\times \left(\sum_{\aa\in \Delta_r} F_{\vv,\aa}\overline{F_{\uu,\aa}}
  \cdot m^{2N}\right)
=m^{4N}\left|\sum_{\aa\in \Delta_r}F_{\uu,\aa}\overline{F_{\vv,\aa}}\right|^2.
\end{equation}
By using condition $(\calR_3)$, we can further simplify (\ref{complicated}) to be
\begin{equation}\label{simplified}
A_{(0,\uu),(0,\vv)}=m^{4N}\left|\sum_{\aa\in \Delta_r}F_{\uu-\vv,\aa}\right|^2
=m^{4N}\Big|\langle \chi,\FF_{\uu-\vv,*}\rangle\Big|^2,
\end{equation}
where $\chi$ is a $0$-$1$ characteristic vector such
  that $\chi_{\aa}=0$ if $\aa\notin \Delta_r$
  and $\chi_{\aa}=1$ if $\aa\in \Delta_r$, for all $\aa\in \zqt$.

Since $\FF$ is discrete unitary, it is easy to show that
$$0\le A_{(0,\uu),(0,\vv)}\le m^{4N}|\Delta_r|^2\ \ \ \text{and}\ \ \
  A_{(0,\uu),(0,\uu)}=m^{4N}|\Delta_r|^2,\ \ \ \text{for all $\uu,\vv\in\zqt$.}$$
As $r\in \cT$, we have $|\Delta_r| \ge 1$ and let $n$ denote $|\Delta_r|$.
Using the dichotomy theorem of Bulatov and Grohe
  (Corollary \ref{useful1}) together with the assumption that $\eval(\AA)$
  is not \#P-hard, we have
$$
A_{(0,\uu),(0,\vv)}\in\{0,m^{4N}n^2\},\ \ \text{for all\ $\uu,\vv\in \zqt$}.
$$
As a result, we have for all $\uu\in \zqt$,
\begin{equation}\label{only-0-orn}
\Big|\langle \chi,\FF_{\uu,*}\rangle\Big|\in \{0,n\}.
\end{equation}
The inner product $\langle \chi,\FF_{\uu,*}\rangle$
is a sum of $n$ terms, each term a power of $\oo_M$.
To sum to a complex number of norm $n$, each term must have exactly the same
argument; any misalignment will result in a
complex number of norm $< n$,
which is the maximum possible. This
implies that
\begin{equation}\label{possible-values-chi-by-F}
\langle \chi,\FF_{\uu,*}\rangle\in \big\{0,n,n\oo_M,n\oo_M^2,\ldots,n\oo_M^{M-1}\big\}.
\end{equation}

Next, let $\fa$ denote a vector in $\Delta_r$.
We use $\Phi$ to denote $\fa+\langle\Delta_r-\fa\rangle$, where
$$
\Delta_r-\fa \equiv \big\{\xx-\fa\hspace{0.05cm}\big|
  \hspace{0.05cm}\xx\in \Delta_r\big\}
$$
and $\langle\Delta_r-\fa\rangle$ is the subgroup generated by
  $\Delta_r-\fa$.
Clearly $\Delta_r \subseteq \Phi$.
We want to prove that $\Delta_r$ is equal to
$ \Phi$, which by definition is a coset in $\zqt$.
This statement, together with Lemma~\ref{cosetlemma},
will  finish the proof of Theorem~\ref{bi-step-4}.

To this end we use $\kappa$ to denote the characteristic
  vector of $\Phi$: $\kappa_\xx=0$ if $\xx\notin \Phi$ and
  $\kappa_\xx=1$ if $\xx\in \Phi$.
We will show for every $\uu\in\zqt$,
\begin{equation}\label{target}
\langle \kappa,\FF_{\uu,*}\rangle=\frac{|\Phi|}{|\Delta_r|}
  \hspace{0.08cm}\langle\chi,\FF_{\uu,*}\rangle.
\end{equation}
Since $\FF$ is discrete unitary, $\{\FF_{\uu,*},\uu\in \zqt\}$ is an orthogonal basis.
From (\ref{target}) we have
$$
\kappa=\frac{|\Phi|}{|\Delta_r|}\hspace{0.05cm}\chi,
$$
which implies $\kappa=\chi$ (since both of them are $0$-$1$ vectors)
  and thus, $\Delta_r=\Phi$ is a coset in $\zqt$.

We now prove (\ref{target}). We make the following Observations 1) and 2):\vspace{0.1cm}
\begin{enumerate}
\item If $|\langle \chi,\FF_{\uu,*}\rangle|=n$, then there exists
  an $\alpha\in \mathbb{Z}_M$ such that $F_{\uu,\xx}=\oo_M^\alpha$
  for all $\xx\in \Delta_r$;\vspace{0.05cm}
\item Otherwise, (which is equivalent to $\langle \chi,\FF_{\uu,*}\rangle =0$
from (\ref{only-0-orn})), there exist $\yy$ and
  $\zz$ in $\Delta_r$ such\\ that $F_{\uu,\yy}\ne F_{\uu,\zz}$.\vspace{0.1cm}
\end{enumerate}
Observation 1) has already been noted when we proved (\ref{possible-values-chi-by-F}).
Observation 2) is obvious since if $F_{\uu,\yy}\hspace{-0.05cm} =$ $F_{\uu,\zz}$ for all
  $\yy,\zz\in \Delta_r$, then clearly $\langle \chi,\FF_{\uu,*}\rangle \ne 0$.\vspace{0.007cm}

Equation (\ref{target}) then follows from the following two lemmas.\vspace{0.12cm}

\begin{lemma}\label{kkkkbbbb}
If there exists an $\alpha$ such that $F_{\uu,\xx}=\oo_M^{\alpha}$ for all
  $\xx\in \Delta_r$, then $F_{\uu,\xx}=\oo_M^{\alpha}$ for all
  $\xx\in \Phi$.
\end{lemma}
\begin{proof}
Let $\xx$ be a vector in $\Phi$,
  then there exist $\xx_1,\ldots,\xx_k\in \Delta_r$
  and $h_1,\ldots,h_k\in \{\pm 1\}$ for some
$k \ge 0$, such that $\xx=\fa+\sum_{i=1}^k h_i(\xx_i-\fa)$.
By using ($\calR_3$) together with the assumption
  that $F_{\uu,\fa}=F_{\uu,\xx_i}=\oo_M^\alpha$,
$$
F_{\uu,\xx}=F_{\uu,\fa+\sum_i h_i(\xx_i-\fa)}
  =F_{\uu,\fa}\prod_{i} F_{\uu,h_i(\xx_i-\fa)}
  =F_{\uu,\fa}\prod_{i} \left(F_{\uu, \xx_i}\overline{F_{\uu,\fa}}\right)^{h_i}
  =\oo_M^\alpha.\vspace{-0.6cm}
$$
\end{proof}

\begin{lemma}\label{kkkbb}
If there exist $\yy,\zz\in \Phi$ such that $F_{\uu,\yy}\ne F_{\uu,\zz}$,
  then $\sum_{\xx\in \Phi} F_{\uu,\xx}=0$.
\end{lemma}
\begin{proof}
Let $l$ be the smallest positive integer such that
  $l(\yy-\zz)=\00$, then
$l$ exists because $\zqt$ is a finite group,
and $l>1$ because $\yy\ne \zz$.
We use $c$ to denote $F_{\uu,\yy}\overline{F_{\uu,\zz}}$.
By using condition ($\calR_3$) together with the
  assumption, we have $c^l=F_{\uu,l(\yy-\zz)}=1$ but $c\ne 1$.\vspace{0.01cm}

We define the following equivalence relation over $\Phi$: For $\xx,\xx'\in \Phi$,
$$
\xx \sim\xx'\ \ \text{if there exists an integer $k$ such that
  $\xx-\xx'=k(\yy-\zz)$.}
$$
For every $\xx\in \Phi$, its equivalence class contains the following
  $l$ vectors:
$$
\xx,\ \xx+(\yy-\zz),\ \ldots,\ \xx+(l-1)(\yy-\zz),
$$
as $\Phi$ is a coset in $\zqt$. We conclude that $\sum_{\xx\in \Phi}F_{\uu,\xx}=0$
  since for every class, we have (by using ($\calR_3$))
$$
\sum_{i=0}^{l-1}F_{\uu,\xx+i(\yy-\zz)}=F_{\uu,\xx}\sum_{i=0}^{l-1}c^i
  =F_{\uu,\xx}\frac{1-c^l}{1-c}=0.\vspace{-0.6cm}
$$
\end{proof}

Now (\ref{target}) can be proved as follows:
 If $|\langle \chi,\FF_{\uu,*}\rangle|=n ~~(= |\Delta_r|)$, then
by Observation 1) and Lemma~\ref{kkkkbbbb} $$\left|\langle \kappa,\FF_{\uu,*}\rangle\right|=
|\Phi|.$$
If  $|\langle \chi,\FF_{\uu,*}\rangle| \not = n ~~(= |\Delta_r|)$,
then $\langle \chi,\FF_{\uu,*}\rangle = 0$.
By Observation 2) and $\Delta_r \subseteq \Phi$,
Lemma~\ref{kkkbb} implies $$\langle \kappa,\FF_{\uu,*}\rangle= 0.$$

This concludes that $\Delta_r$ is a coset in $\mathbb{Z}_{\fq}$.
To get the decomposition $(\calL_2)$
  for $\Delta_r=\prod_{i=1}^s\Delta_{r,i}$, we use Lemma \ref{cosetlemma}.

\subsection{Proof of Lemma \ref{cosetlemma}}

First, we show that if $\uu=(u_1,u_2) \in \Phi$
and $\vv=(v_1,v_2)\in \Phi$, for $u_i, v_i \in G_i$, $i=1,2$, then
  $(u_1,v_2)\in\Phi$.\vspace{0.008cm}

On the one hand, since $\gcd(|G_1|,|G_2|)=1$, there exists
  an integer $k$ such that $|G_1|\hspace{0.07cm}\big|\hspace{0.05cm}k$ and
  $k\equiv 1\pmod{|G_2|}$.
On the other hand, since $\Phi$ is a coset, we have $\uu+k(\vv-\uu)\in \Phi$.\vspace{-0.1cm}
Since $$u_1+k(v_1-u_1)=u_1\ \ \text{and}\ \ u_2+k(v_2-u_2)=v_2,\vspace{-0.1cm}$$
  we conclude that $(u_1,v_2)\in \Phi$.

This implies the existence of subsets $\Phi_1\subseteq G_1$ and
  $\Phi_2\subseteq G_2$ such that $\Phi=\Phi_1\times \Phi_2$.
Namely we let
$$\Phi_1 = \big\{ x \in G_1 \hspace{0.06cm}|\hspace{0.06cm}\exists\hspace{0.05cm} y
\in G_2, (x,y) \in \Phi\big\}\ \ \ \ \text{and}\ \ \ \
 \Phi_2  = \big\{y  \in G_2 \hspace{0.06cm}|\hspace{0.06cm}
  \exists\hspace{0.05cm} x \in G_1, (x,y) \in \Phi\big\}.$$
It is easy to check that both $\Phi_1$ and $\Phi_2$ are cosets
  (in $G_1$ and $G_2$, respectively), and $\Phi=\Phi_1\times \Phi_2$.

\subsection{Some Corollaries of Theorem~\ref{bi-step-4}}

Now that we have proved Theorem~\ref{bi-step-4}, we know that
unless the problem is $\#$P-hard, we may assume
that condition $(\calL)$ holds. Thus $\Lambda_r$ and
$\Delta_r$ are cosets.

\begin{lemma}\label{lemma8-4}
Let $\HH$ be the $m \times |\Delta_r|$  submatrix
obtained from  $\FF$ by restricting to the columns indexed by $\Delta_r$.
Then for any two rows $\HH_{\uu,*}$ and $\HH_{\vv,*}$, where $\uu, \vv \in \zqt$,
either there exists some $\alpha \in \mathbb{Z}_M$ such that
$\HH_{\uu,*} = \omega_M^\alpha\cdot \HH_{\vv,*}$, or
$\langle \HH_{\uu,*}, \HH_{\vv,*} \rangle = 0$.\vspace{0.028cm}

Similarly we denote by $\GG$ the $|\Lambda_r| \times m$ submatrix
obtained from  $\FF$ by restricting to the rows indexed by $\Lambda_r$.
Then for any two columns $\GG_{*,\uu}$ and $\GG_{*,\vv}$, where $\uu, \vv \in \zqt$,
either there exists an $\alpha \in \mathbb{Z}_M$ such that
$\GG_{*,\uu} = \omega_M^\alpha\cdot \GG_{*,\vv}$, or
$\langle \GG_{*,\uu}, \GG_{*,\vv} \rangle = 0$.\vspace{0.13cm}
\end{lemma}

\begin{proof}
The rows of $\HH$ are restrictions of $\FF$.
Any two rows $\HH_{\uu,*}, \HH_{\vv,*}$  satisfy
$$\HH_{\uu,*} \circ \overline{\HH_{\vv,*}} = \FF_{\uu-\vv,*}\hspace{0.03cm} |_{\Delta_r}
\hspace{-0.1cm}
= \HH_{\uu-\vv,*},$$ which is a row in $\HH$.
If this $\HH_{\uu-\vv,*}$ is a constant, namely $\omega_M^\alpha$
for some  $\alpha \in \mathbb{Z}_M$, then
$\HH_{\uu,*} = \omega_M^\alpha\hspace{0.05cm} \HH_{\vv,*}$ holds.
Otherwise, Lemma~\ref{kkkbb} says
$\langle \HH_{\uu,*}, \HH_{\vv,*} \rangle = 0$.\vspace{0.005cm}

The proof for $\GG$ is exactly the same.
\end{proof}

As part of a discrete unitary matrix $\FF$,
all columns $\{\HH_{*,\uu}\hspace{0.06cm}|\hspace{0.06cm} \uu \in \Delta_r\}$ of
$\HH$ must be orthogonal and thus $\text{rank}(\HH)=|\Delta_r|$.
We denote by $n$ the cardinality $|\Delta_r|$.
There must be $n$ linearly independent rows in $\HH$.
We may start with $\fb_0=\00$, and assume the following
$n$ vectors
$\fb_0=\00,\fb_1,\ldots,\fb_{n-1}\in \zqt$ are the
indices of  a set of linearly independent rows.
By Lemma~\ref{lemma8-4}, these must be orthogonal as row vectors
(over $\mathbb{C}$).
Since the rank of the matrix $\HH$ is exactly $n$,
it is clear that all other rows must be a multiple of
these rows, since the only alternative is to be orthogonal to them all,
by Lemma~\ref{lemma8-4} again,
which is~absurd.
A symmetric statement for $\GG$ also holds.




\section{Proof of Theorem \ref{bi-step-5}}

Let $((M,N),\CC,\fD,(\pp,\bft,\fq))$ be a tuple that
  satisfies both conditions $(\calR)$ and $(\calL)$ (including $(\calL_3)$).
We also assume that $\eval(\CC,\fD)$ is not \#P-hard.
By $(\calL)$, we have
$$
\Lambda_r=\prod_{i=1}^s \Lambda_{r,i}\ \ \ \text{for every $r\in \cS$,\ \ \ \ and}\ \ \ \
\Delta_r=\prod_{i=1}^s \Delta_{r,i}\ \ \ \text{for every $r\in \cT$,}
$$
where both $\Lambda_{r,i}$ and $\Delta_{r,i}$ are cosets
  in $\mathbb{Z}_{\qq_i}$.

Let $r$ be an integer in $\cS$. Below we will prove ($\calD_1$) and $(\calD_3)$
  for $\Lambda_r$.
The other parts of the theorem, that is, $(\calD_2)$ and
  $(\calD_4)$, can be proved similarly.\vspace{0.005cm}

Let $\GG$ denote the $|\Lambda_r|\times m$ submatrix
  of $\FF$ whose row set is $\Lambda_r\subseteq \zqt$.
We start with the following simple lemma about $\GG$.
  In this section we denote by $n$ the cardinality $|\Lambda_r|\ge 1$.
A symmetric statement also holds for the $m \times |\Delta_r|$
submatrix of $\FF$ whose column set is $\Delta_r$, where we replace
$n = |\Lambda_r|$ by $|\Delta_r|$, which could be different.

\begin{lemma}\label{useful1}
There exist
  vectors $\fb_0=\00,\fb_1,\ldots,\fb_{n-1}\in \zqt$ such that
\begin{enumerate}
\item $\{\GG_{*,\fb_i}\hspace{0.07cm}\big|
  \hspace{0.07cm}i\in [0:n-1]\}$ forms an orthogonal basis;
\item For all $\bb\in \zqt$, there exist $i\in [0:n-1]$ and $\alpha\in \mathbb{Z}_M$
  such that $\GG_{*,\bb}=\omega_M^\alpha\cdot \GG_{*,\fb_i}$; and
\item Let $A_i$ denote the set of $\bb\in \zqt$ such that
  $\GG_{*,\bb}$ is linearly dependent with $\GG_{*,\fb_i}$,\vspace{-0.1cm}
  then $$|A_0|=|A_1|=\ldots=|A_{n-1}|=\frac{m}{n}.\vspace{-0.2cm}$$
\end{enumerate}
\end{lemma}
\begin{proof}
By Lemma~\ref{lemma8-4}, and the discussion following Lemma~\ref{lemma8-4}
  (the symmetric statements regarding $\Lambda_r$ and $\GG$),
there exist vectors $\fb_0=\00,\fb_1,\ldots,\fb_{n-1}\in \zqt$
such that Properties 1) and 2) hold.
%

We now prove property 3).
By condition ($\calR_3$),
fixing $\fb_i$, for any $i$, there is
a one-to-one correspon\-dence between $A_i$ and $A_0$,
by $\bb \mapsto \bb-\fb_i$.
This is clear from $\GG_{\bb-\fb_i, *} =
\GG_{\bb, *} \circ \overline{\GG_{\fb_i, *}}$.
Hence  we have $A_0=\{\bb-\fb_i\hspace{0.08cm}|
  \hspace{0.08cm}\bb \in A_i\}$ for all sets $A_i$.
It then follows that $|A_0|=|A_1|=\ldots=|A_{n-1}|=m/n$.\vspace{0.06cm}
\end{proof}

\begin{figure}
\center
\includegraphics[height=16cm]{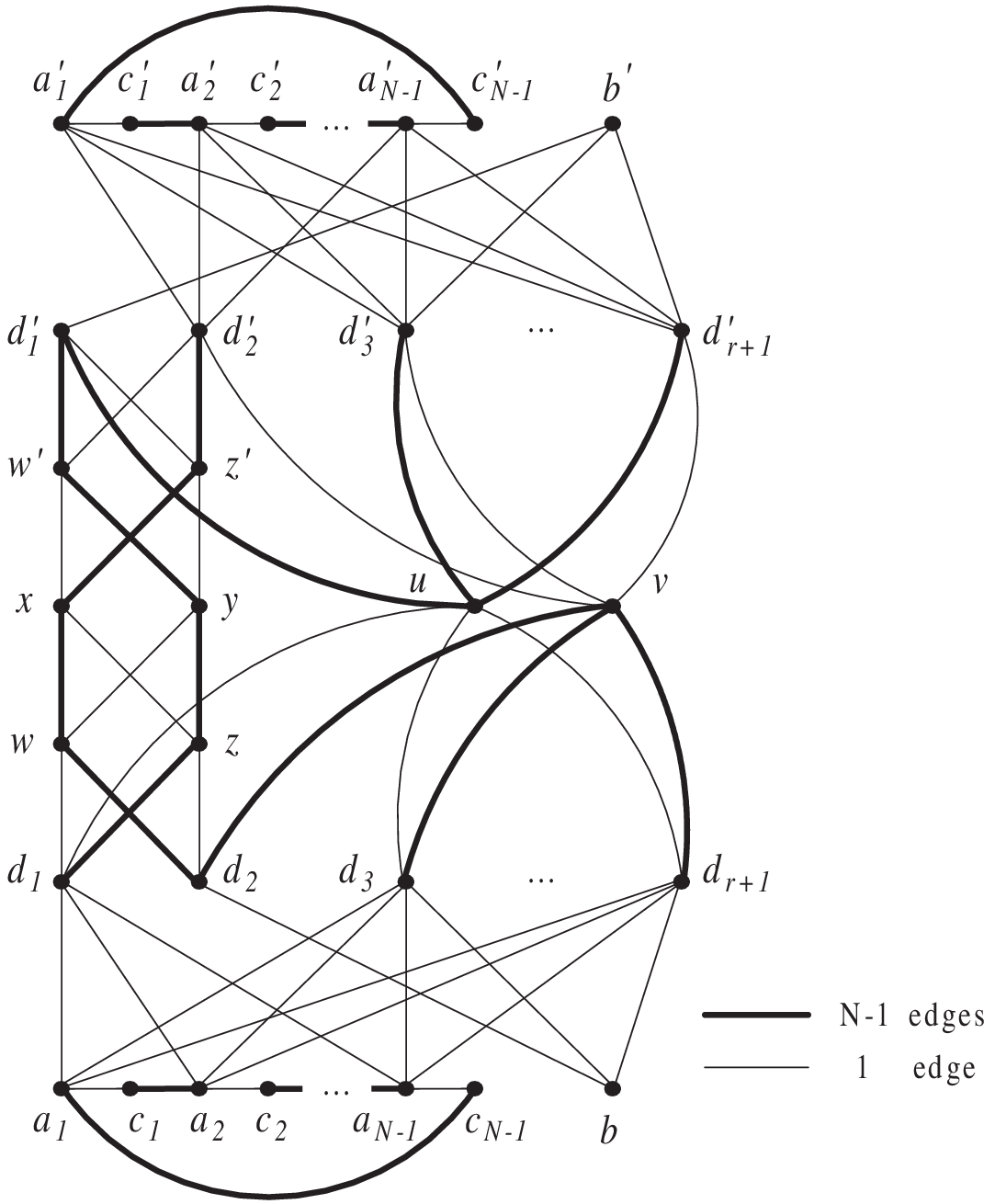}\vspace{0.25cm}
\caption{The gadget for constructing $G^{[1]}$ (Note that the subscript $e$ is suppressed).}\label{figure_8}
\end{figure}

Now let $G=(V,E)$ be an undirected graph.
For every positive integer $p$,
  we can build a new graph $G^{[p]}$ from $G$ by
  replacing every edge $e=uv\in E$ with a gadget.
  We will need $G^{[2]}$ in the proof.  But
it is more convenient to describe $G^{[1]}$ first
and illustrate it only with the case $p=1$.  (The picture for $G^{[2]}$ will
be too cumbersome to draw.)
The gadget for $G^{[1]}$ is shown in Figure \ref{figure_8}.\vspace{0.008cm}

More exactly, we have $G^{[1]}=(V^{[1]},E^{[1]})$ where
\begin{equation*}
V^{[1]}=V\cup\big\{x_{e},y_e,a_{e,i},a_{e,i}',b_e,b_e',c_{e,i},
  c_{e,i}',d_{e,j},d_{e,j}',w_{e},w_e',z_e,z_e'\ \big|\hspace{0.1cm}  e\in E,
  i\in [N-1],j\in [r+1]\big\},
\end{equation*}
and $E^{[1]}$ contains exactly the following edges: For every edge $e=uv\in E$,
\begin{enumerate}
\item {one edge between $(u,d_{e,j})$ for all $j\in [r+1]-\{2\}$};\vspace{-0.06cm}
\item {$N-1$ edges between $(v,d_{e,j})$ for all $j\in [r+1]-\{1\}$};\vspace{-0.06cm}
\item {one edge between $(d_{e,1},w_e)$, $(d_{e,2},z_{e})$, $(w_e,y_e)$
  and $(z_e,x_e)$;}\vspace{-0.06cm}
\item {$N-1$ edges between $(d_{e,1},z_e)$, $(d_{e,2},w_e)$, $(w_e,x_e)$
  and $(z_e,y_e)$;}\vspace{-0.06cm}
\item {one edge between $(a_{e,i},d_{e,j})$ for all $i\in [N-1]$ and $j\in [r+1]-
  \{2\}$;}\vspace{-0.06cm}
\item {one edge between $(b_e,d_{e,j})$ for all $j\in [r+1]-\{1\}$;}\vspace{-0.06cm}
\item {$N-1$ edges between $(c_{e,N-1},a_{e,1})$ and
  $(c_{e,i},a_{e,i+1})$ for all $i\in [N-2]$;}\vspace{-0.06cm}
\item {one edge between $(a_{e,i},c_{e,i})$ for all $i\in [N-1]$};\vspace{-0.06cm}
\item {$N-1$ edges between $(u,d_{e,j}')$ for all $j\in [r+1]-\{2\}$;}\vspace{-0.06cm}
\item {one edge between $(v,d_{e,j}')$ for all $j\in [r+1]-\{1\}$;}\vspace{-0.06cm}
\item {one edge between $(d_{e,1}',z_e')$, $(d_{e,2}',w_e')$,
  $(w_e',x_e)$ and $(z_e',y_e)$;}\vspace{-0.06cm}
\item {$N-1$ edges between $(d_{e,1}',w_e')$, $(d_{e,2}',z_e')$,
  $(w_e',y_e)$ and $(z_e',x_e)$;}\vspace{-0.06cm}
\item {one edge between $(a_{e,i}',d_{e,j}')$ for all $i\in [N-1]$ and $j\in [r+1]
  -\{1\}$;}\vspace{-0.06cm}
\item {one edges between $(b_e',d_{e,j}')$ for all $j\in [r+1]-\{2\}$;}\vspace{-0.06cm}
\item {$N-1$ edges between $(c_{e,N-1}',a_{e,1}')$ and $(c_{e,i}',
  a_{e,i+1}')$ for all $i\in [N-2]$;}\vspace{-0.06cm}
\item {one edge between $(a_{e,i}',c_{e,i}')$ for all $i\in [N-1]$}.
\end{enumerate}
As indicated earlier, the graph we really need in the proof is $G^{[2]}$.
The gadget for $G^{[2]}$ can be built from the one for $G^{[1]}$ in Figure \ref{figure_8}
  as follows: First, we make a new copy of the subgraph spanned
  by vertices $$\big\{u,v,x,y,w,z,d_j,a_i,c_i,b\mid i\in [N-1],j\in [r+1]\big\}.$$
All vertices are new except $x,y,u$ and $v$.
Second, make a new copy of the subgraph spanned by
  $$\big\{u,v,x,y,w',z',d_j',a_i',c_i',b'\mid i\in [N-1],j\in [r+1]\big\}.$$
Again all vertices are new except $x,y,u$ and $v$.
In this way, we get a new gadget and we use it to build $G^{[2]}$ by
  replacing every edge $e=uv\in E$ with this gadget.

It is easy to verify that the degree of every vertex in $G^{[2]}$
  is a $0\pmod N$ except both copies of $a_{e,i},$ $a_{e,i}',b_e$ and $b_e'$ whose
  degree is $r\pmod N$.
The construction gives us a $2m\times 2m$ matrix $\AA$ such that
$$
Z_{\AA}(G)=Z_{\CC,\fD}(G^{[2]}),\ \ \text{for any undirected graph $G$,}
$$
and thus, $\eval(\AA)$ ($\le \eval(\CC,\fD)$)
  (right now it is not clear whether $\AA$ is
  a symmetric matrix, which we will prove later)
  is not \#P-hard.
We index the rows (columns) of $\AA$ in the same way as
  we do for $\CC$: The first $m$ rows (columns) are
  indexed by $\{0\}\times \zqt$ and the last $m$ rows (columns)
  are indexed by $\{1\}\times \zqt$.
Since $\CC$ is the bipartisation of $\FF$, we have
$
A_{(0,\uu),(1,\vv)}=A_{(1,\uu),(0,\vv)}=0,$ for all
  $\uu,\vv\in \zqt$.\vspace{0.02cm}

We now analyze the upper-left $m\times m$ block of $\AA$.
For $\uu,\vv\in \zqt$, we have
$$
A_{(0,\uu),(0,\vv)}=\sum_{\xx,\yy\in \zqt} A_{\uu,\vv,\xx,\yy}^2 B_{\uu,\vv,\xx,\yy}^2,
\vspace{-0.4cm}
$$
where\vspace{-0.2cm}
\begin{eqnarray*}
A_{\uu,\vv,\xx,\yy}&=&\sum_{\aa_1,\ldots,\aa_{N-1},\bb\in \Lambda_r,
  \dd_1,\dd_2\in \zqt} D^{[r]}_{(0,\bb)}\prod_{i=1}^{
  N-1} D^{[r]}_{(0,\aa_i)}
  \left(\sum_{\ww\in \zqt} F_{\ww,\dd_1}F_{\ww,\yy}
  \overline{F_{\ww,\dd_2}F_{\ww,\xx}}\right)\\[0.6ex]
  &\times&\hspace{-0.2cm} \left(\sum_{\zz\in \zqt} F_{\zz,\dd_2}
  F_{\zz,\xx}\overline{F_{\zz,\dd_1}F_{\zz,\yy}}\right)
  \left(\prod_{i=1}^{N-2} \sum_{\cc_i\in \zqt} F_{\aa_i,\cc_i}
  \overline{F_{\aa_{i+1},\cc_i}}\right) \left(\sum_{\cc_{N-1}\in \zqt}
  F_{\aa_{N-1},\cc_{N-1}}\overline{F_{\aa_1,\cc_{N-1}}}\right)\\[1ex]
  &\times&\hspace{-0.2cm} \left(\prod_{i=3}^{r+1}\hspace{0.07cm}\sum_{\dd_i\in \zqt}
  F_{\uu,\dd_i}F_{\bb,\dd_i}
  \overline{F_{\vv,\dd_i}} \prod_{j=1}^{N-1}F_{\aa_j,\dd_i}\right)
  F_{\uu,\dd_1}\left(\prod_{j=1}^{N-1}F_{\aa_j,\dd_1}\right)
  \overline{F_{\vv,\dd_2}}F_{\bb,\dd_2},\ \ \ \text{and}
\end{eqnarray*}
\begin{eqnarray*}
B_{\uu,\vv,\xx,\yy}&=&\sum_{\aa_1,\ldots,\aa_{N-1},\bb\in \Lambda_r,\dd_1,\dd_2\in \zqt}
  D^{[r]}_{(0,\bb)}\prod_{i=1}^{N-1}D^{[r]}_{(0,\aa_i)}\left(\sum_{\ww\in \zqt}
  F_{\ww,\dd_2}F_{\ww,\xx}\overline{F_{\ww,\dd_1}F_{\ww,\yy}}\right)\\[0.6ex]
  &\times&\hspace{-0.3cm} \left(\sum_{\zz\in\zqt}F_{\zz,\dd_1}F_{\zz,\yy}\overline{F_{\zz,\dd_2}
  F_{\zz,\xx}}\right)
  \left(\prod_{i=1}^{N-2} \sum_{\cc_i\in \zqt} F_{\aa_i,\cc_i}
  \overline{F_{\aa_{i+1},\cc_i}}\right) \left(\sum_{\cc_{N-1}\in \zqt}
  F_{\aa_{N-1},\cc_{N-1}}\overline{F_{\aa_1,\cc_{N-1}}}\right)\\[1ex]
  &\times&\hspace{-0.3cm} \left(\prod_{i=3}^{r+1}\hspace{0.07cm} \sum_{\dd_i\in \zqt}
  F_{\vv,\dd_i}F_{\bb,\dd_i}\overline{F_{\uu,\dd_i}}
  \prod_{j=1}^{N-1}F_{\aa_j,\dd_i}\right)
  F_{\vv,\dd_2}\left(\prod_{j=1}^{N-1}F_{\aa_j,\dd_2}\right)
  \overline{F_{\uu,\dd_1}}F_{\bb,\dd_1}.
\end{eqnarray*}
We simplify $A_{\uu,\vv,\xx,\yy}$ first. Since $\FF$ is discrete unitary
  and satisfies ($\calR_3$), we have
$$
\sum_{\ww\in \zqt} F_{\ww,\dd_1}F_{\ww,\yy}\overline{F_{\ww,\dd_2}F_{\ww,\xx}}
=\langle \FF_{*,\dd_1+\yy},\FF_{*,\dd_2+\xx}\rangle
$$
is zero unless $\dd_1-\dd_2=\xx-\yy$.
When this equation holds, the inner product
$\langle \FF_{*,\dd_1+\yy},\FF_{*,\dd_2+\xx}\rangle = m$.
Also when $\dd_1-\dd_2=\xx-\yy$ the sum
$\sum_{\zz\in \zqt} F_{\zz,\dd_2} F_{\zz,\xx}
\overline{F_{\zz,\dd_1}F_{\zz,\yy}} = m$ as well.
Similarly,
$$
\sum_{\cc_i\in \zqt} F_{\aa_i,\cc_i}
  \overline{F_{\aa_{i+1},\cc_i}}=\langle \FF_{\aa_i,*},\FF_{\aa_{i+1},*}\rangle
$$
is zero unless $\aa_i=\aa_{i+1}$, for $i=1, \ldots, N-2$.
Also, we have  $$\sum_{\cc_{N-1}\in \zqt}
F_{\aa_{N-1},\cc_{N-1}}\overline{F_{\aa_1,\cc_{N-1}}} =
\langle \FF_{\aa_{N-1},*},\FF_{\aa_{1},*} \rangle$$
is zero unless $\aa_{N-1} = \aa_{1}$.
When $\aa_1=  \ldots = \aa_{N-1}$,
all these inner products are equal to $m$.
So now we may assume $\dd_1-\dd_2=\xx-\yy$ and all $\aa_i$'s are equal,
call it $\aa$,
in the sum for $A_{\uu,\vv,\xx,\yy}$.

Let $\xx-\yy=\mathbf{s}$, then $A_{\uu,\vv,\xx,\yy}$ is equal to
\begin{equation}
m^{N+1}\hspace{-0.2cm} \sum_{\aa,\bb\in \Lambda_r,
  \dd_2 \in \zqt}\hspace{-0.2cm} D_{(0,\bb)}^{[r]}\overline{
  D_{(0,\aa)}^{[r]}}\hspace{0.05cm} \left(\prod_{i=3}^{r+1}\hspace{0.06cm}
  \sum_{\dd_i\in \zqt} F_{\uu,\dd_i}
  F_{\bb,\dd_i}\overline{F_{\vv,\dd_i} F_{\aa,\dd_i}}\right)
  F_{\uu,\dd_2+\mathbf{s}}F_{\bb,\dd_2}
  \overline{F_{\vv,\dd_2}F_{\aa,\dd_2+\mathbf{s}}}.
\end{equation}
Again,\vspace{0.02cm} $$\sum_{\dd_i\in \zqt} F_{\uu,\dd_i}
  F_{\bb,\dd_i}\overline{F_{\vv,\dd_i} F_{\aa,\dd_i}}=\langle \FF_{\uu+\bb,*},
  \FF_{\vv+\aa,*}\rangle=0$$ unless $\uu+\bb=\vv+\aa$.
When $\uu+\bb=\vv+\aa$, the inner product
$\langle \FF_{\uu+\bb,*}, \FF_{\vv+\aa,*}\rangle= m$.
As a result, if $$\vv-\uu\notin \lin{\Lambda_r}\equiv\{\xx-\xx'\mid \xx,\xx'\in \Lambda_r\},$$
  then $A_{\uu,\vv,\xx,\yy}=0$ since $\aa, \bb \in \Lambda_r$
  and $\bb-\aa\in \lin{\Lambda_r}$.\vspace{0.008cm}

For every vector $\hh\in \lin{\Lambda_r}$ (e.g., $\hh=\vv-\uu$),
  we define a $|\Lambda_r|$-dimensional vector $\TT^{[\hh]}$ as
follows:
$$
T^{[\hh]}_{\xx}=D^{[r]}_{(0,\xx+\hh)}\overline{D^{[r]}_{(0,\xx)}},
  \ \ \ \text{for all $\xx\in \Lambda_r$.}
$$
By $(\calL)$, $\Lambda_r$ is a coset in $\zqt$, so
for any $\xx\in \Lambda_r$, we also have $\xx+\hh\in \Lambda_r$.
Therefore every entry of $\TT^{[\hh]}$ is non-zero
and is a power of $\oo_N$.

Now we use $\TT^{[\vv-\uu]}$ to express $A_{\uu,\vv,\xx,\yy}$.
Suppose $\vv-\uu\in \lin{\Lambda_r}$, then
\begin{eqnarray*}
A_{\uu,\vv,\xx,\yy}\hspace{-0.15cm}
&=&\hspace{-0.15cm}
  m^{N+r} \sum_{\aa\in \Lambda_r, \dd_2 \in \zqt,\bb=\aa+\vv-\uu}
  D_{(0,\bb)}^{[r]}\overline{
  D_{(0,\aa)}^{[r]}} F_{\uu,\dd_2+\mathbf{s}}F_{\bb,\dd_2}
  \overline{F_{\vv,\dd_2}F_{\aa,\dd_2+\mathbf{s}}}\\[1.2ex]
&=&\hspace{-0.15cm} m^{N+r+1} \sum_{\aa\in \Lambda_r} D_{(0,\aa+\vv-
  \uu)}^{[r]}\overline{
  D_{(0,\aa)}^{[r]}} F_{\uu,\mathbf{s}} \overline{F_{\aa,\mathbf{s}}}
      \hspace{0.09cm}= \hspace{0.09cm}
   m^{N+r+1} F_{\uu,\xx-\yy} \langle \TT^{[\vv-\uu]},\GG_{*,\xx-\yy}\rangle.\vspace{0.04cm}
\end{eqnarray*}
Here we used ($\calR_3$) in the second equality, and we recall
the definition of $\mathbf{s} = \xx-\yy$.\vspace{0.005cm}

Similarly, when $\vv-\uu\notin \lin{\Lambda_r}$, we have
  $B_{\uu,\vv,\xx,\yy}=0$; and when $\vv-\uu\in \lin{\Lambda_r}$,\vspace{0.06cm}
\begin{eqnarray*}
B_{\uu,\vv,\xx,\yy}\hspace{-0.15cm}&=&\hspace{-0.15cm}
  m^{N+r}\sum_{\bb\in \Lambda_r,\dd_2\in \zqt,\aa=\bb+\vv-\uu} D^{[r]}_{(0,\bb)}
  \overline{D^{[r]}_{(0,\aa)}} F_{\vv,\dd_2}F_{\bb,\dd_2+\xx-\yy}\overline{
  F_{\aa,\dd_2}F_{\uu,\dd_2+\xx-\yy}}\\[1.2ex]
&=&\hspace{-0.15cm} m^{N+r}\sum_{\bb\in \Lambda_r,
  \dd_2\in \zqt} D^{[r]}_{(0,\bb)}\overline{D^{[r]}_{(0,\bb+\vv-\uu)}}F_{\bb,\xx-\yy}
  \overline{F_{\uu,\xx-\yy}}\hspace{0.09cm}
=\hspace{0.09cm} m^{N+r+1} \overline{F_{\uu,\xx-\yy} \langle
  \TT^{[\vv-\uu]},\GG_{*,\xx-\yy}\rangle}.\vspace{0.06cm}
\end{eqnarray*}
To summarize, when $\vv-\uu\notin \lin{\Lambda_r}$, $A_{(0,\uu),(0,\vv)}=0$;
  and when $\vv-\uu\in \lin{\Lambda_r}$,
\begin{equation}\label{exp-1}
A_{(0,\uu),(0,\vv)}=m^{4(N+r+1)}\sum_{\xx,\yy\in \zqt} \left|\langle \TT^{[\vv-\uu]},
  \GG_{*,\xx-\yy}\rangle\right|^4
  =m^{4N+4r+5} \sum_{\bb\in \zqt} \left|\langle \TT^{[\vv-\uu]},
  \GG_{*,\bb}\rangle\right|^4.
\end{equation}

We now show that $\AA$ is a symmetric non-negative matrix.
Let $\aa=\vv-\uu\in \lin{\Lambda_r}$. Then by ($\calR_3$),
  we have for every $\bb\in \zqt$,
\begin{eqnarray*}
\left|\langle \TT^{[-\aa]},\GG_{*,-\bb}\rangle\right|
\hspace{-0.13cm}&=&\hspace{-0.13cm}\left|\sum_{\xx\in \Lambda_r}
  D_{(0,\xx-\aa)}^{[r]}\overline{D_{(0,\xx)}^{[r]} G_{\xx,-\bb}}\right|
  \hspace{0.08cm}=\hspace{0.08cm}
  \left|{\sum_{\xx\in \Lambda_r} D_{(0,\xx)}^{[r]}
\overline{D_{(0,\xx-\aa)}^{[r]}
  G_{\xx,\bb}}}\right|\\[0.6ex]
&=&\hspace{-0.13cm}\left|{\sum_{\yy\in \Lambda_r}
  D_{(0,\yy+\aa)}^{[r]}\overline{D_{(0,\yy)}^{[r]}G_{\yy,\bb}F_{\aa,\bb}}}\right|
\hspace{0.08cm}=\hspace{0.08cm} \left|{\sum_{\yy\in \Lambda_r}
  D_{(0,\yy+\aa)}^{[r]}\overline{D_{(0,\yy)}^{[r]}G_{\yy,\bb}}}\right|
  \hspace{0.08cm}=\hspace{0.08cm}\left|\langle \TT^{[\aa]},\GG_{*,\bb}\rangle\right|,
\end{eqnarray*}
where the second equation is by conjugation, the third equation is
by the substitution $\xx = \yy + \aa$
and the fourth equation is because $F_{\aa,\bb}$ is a root of unity.
It then follows that $A_{(0,\uu),(0,\vv)}=A_{(0,\vv),(0,\uu)}$.
The lower-right block can be proved similarly.
Hence $\AA$ is symmetric.\vspace{0.005cm}

Next, we further simplify (\ref{exp-1}) using Lemma \ref{useful1}:
\begin{equation}\label{exp-2}
A_{(0,\uu),(0,\vv)}=\frac{m^{4N+4r+6}}{n}\cdot\hspace{0.06cm}\sum_{i=0}^{n-1}\left|
  \langle\TT^{[\vv-\uu]},\GG_{*,\fb_i}\rangle\right|^4.
\end{equation}
For the special case when $\uu=\vv$, we know exactly what
  $A_{(0,\uu),(0,\uu)}$ is: Since $\TT^{[\00]}=\11=\GG_{*,\fb_0}$,
  we have $$\langle \TT^{[\00]}, \GG_{*,\fb_0}
\rangle = n;$$
By Lemma \ref{useful1}, $\{\GG_{*,\fb_0}, \ldots,
  \GG_{*,\fb_{n-1}}\}$ is an orthogonal basis,
hence $$\sum_{i=0}^{n-1} \left| \langle\TT^{[\00]}, \GG_{*,\fb_i}\rangle
\right|^4 = n^4\ \ \ \text{and}\ \ \
  A_{(0,\uu),(0,\uu)}=L\cdot n^4,\ \ \ \text{where
  $L\equiv m^{4N+4r+6}/n$.}$$

Our next goal is to prove (\ref{imp-need}). Note that
  if $|\Lambda_r^{\text{lin}}|=1$ then (\ref{imp-need})
  is trivially true.
So below we assume $|\Lambda_r^{\text{lin}}|>1$.
Because $\AA$ is symmetric and non-negative,
we can apply  the dichotomy theorem of Bulatov and Grohe.
For any pair $\uu\ne \vv $ such that $\uu-\vv\in \lin{\Lambda_r}$,
  we consider the following $2 \times 2$ submatrix\vspace{0.01cm}
$$
\left(
\begin{matrix}
A_{(0,\uu),(0,\uu)} & A_{(0,\uu),(0,\vv)} \\
A_{(0,\vv),(0,\uu)} & A_{(0,\vv),(0,\vv)}
\end{matrix}
\right)
$$
of $\AA$. Since $\eval(\AA)$ is assumed to be not \#P-hard,
  by Corollary \ref{usefulhahacoro}, we have
$$A_{(0,\uu),(0,\vv)}=A_{(0,\vv),(0,\uu)}\in \big\{0,L\cdot n^4\big\},$$
and thus from (\ref{exp-2}), we get
\begin{equation}\label{nonzero}
\sum_{i=0}^{n-1}\left|\langle \TT^{[\vv-\uu]},\GG_{*,\fb_i}\rangle\right|^4\in \big\{0,n^4\big\},
\ \ \ \text{for all $\uu,\vv$ such that $\uu-\vv\in \Lambda_r^{\text{lin}}$.}
\end{equation}

However, the sum in (\ref{nonzero}) cannot be zero. This is because the following:
By Lemma~\ref{useful1}, $\{\GG_{*,\fb_i} \hspace{0.06cm}|\hspace{0.06cm} i\in [0:n-1]\}$
is an orthogonal basis, with each $\|\GG_{*,\fb_i}\|^2 = n$. Then by Parseval,
$$
\sum_{i=0}^{n-1} \left|\langle \TT^{[\vv-\uu]},\frac{\GG_{*,\fb_{i}} }{
\|\GG_{*,\fb_i}\|}
\rangle\right|^2= \|\TT^{[\vv-\uu]}\|^2 =
n,
$$
since each entry of $\TT^{[\vv-\uu]}$ is a root of unity.
Hence, $\sum_{i=0}^{n-1} |\langle \TT^{[\vv-\uu]}, \GG_{*,\fb_{i}}
\rangle |^2 = n^2$.
This\vspace{0.005cm} shows that for  some $0 \le i < n$,
$ |\langle \TT^{[\vv-\uu]}, \GG_{*,\fb_{i}} \rangle | \not = 0$,
and therefore, the sum in (\ref{nonzero}) is non-zero, and thus in fact
\[\sum_{i=0}^{n-1} \left|\langle \TT^{[\vv-\uu]}, \GG_{*,\fb_{i}}
\rangle\right|^4 = n^4,\ \ \ \text{for all $\uu,\vv$ such that
  $\uu-\vv\in \Lambda_r^{\text{lin}}$.}\]

If we temporarily denote $x_i = |\langle \TT^{[\vv-\uu]}, \GG_{*,\fb_{i}}
\rangle |$, for $0 \le i < n$, then each $x_i \ge 0$.
We have both
$$\sum_{i=0}^{n-1} x_i^2 = n^2\ \ \ \ \text{and}\ \ \ \ \sum_{i=0}^{n-1} x_i^4 = n^4 .$$
By taking the square, we have $$n^4 = \left( \sum_{i=0}^{n-1} x_i^2 \right)^2
= \sum_{i=0}^{n-1} x_i^4 +\hspace{0.06cm} \text{non-negative cross terms}.$$
It follows that
all cross terms must be zero.  Thus, there exists
a unique term $x_i \not = 0$. Moreover, this $x_i$ must equal to $n$ while
  all other $x_j = 0$.
We conclude that,
for all $\uu$ and $\vv\in \zqt$ such that $\uu-\vv\in \lin{\Lambda_{r}}$,
  there exists a unique $i\in [0:n-1]$ such that
$$
\left|\langle \TT^{[\vv-\uu]},\GG_{*,\fb_i}\rangle\right|=n.
$$

Again apply the argument that $
\langle \TT^{[\vv-\uu]},\GG_{*,\fb_i}\rangle$ is a sum of $n$
terms, each of which is a root of unity, we can conclude the following:
For all $\aa\in \lin{\Lambda_r}$, there exist $\bb\in \zqt$ and
  $\alpha\in \mathbb{Z}_N$ such that
\begin{equation}\label{imp-need}
\TT^{[\aa]}=\oo_N^{\alpha}\cdot \GG_{*,\bb}.
\end{equation}

Below we use (\ref{imp-need}) to prove $(\calD_3)$.
Note that, if $s=1$, then $(\calD_3)$ follows directly from (\ref{imp-need}).
So below we assume $s>1$.
First, (\ref{imp-need}) implies the following useful lemma:

\begin{lemma}\label{ratio-of-T}
Let $\aa$ be a vector in ${\Lambda_{r,k}^{\text{\emph{lin}}}}$ for some
  $k\in [s]$. Then for every $\cc\in \Lambda_{r,\ell}^{\text{\emph{lin}}}$, where $\ell\ne k$,
$$
{T_{\xx+\wt{\cc}}^{[\wt{\aa}]}}\Big/{T_\xx^{[\wt{\aa}]}}
~~~~\mbox{for all $\xx \in \Lambda_{r}$,}
$$
is a power of $\oo_{q_{\ell}}$.
\emph{(}Recall we use $q_\ell$ to denote $q_{\ell,1}$.
Also note that for every $\xx \in \Lambda_{r}$, the translated
point $\xx+\wt{\cc}$ is in $\Lambda_{r}$, so
$\TT^{[\wt{\aa}]}$ is defined at both $\xx$ and $\xx+\wt{\cc}$.
Since they are roots of unity, one can divide one by the other.\emph{)}
\end{lemma}
\begin{proof}
By (\ref{imp-need}), there exists a vector $\bb\in \zqt$
  such that\vspace{-0.05cm}
$$
{T_{\xx+\wt{\cc}}^{[\wt{\aa}]}}\Big/{T_\xx^{[\wt{\aa}]}}=G_{\xx+\wt{\cc},{\bb}}
  \Big/ G_{\xx,{\bb}}=F_{\wt{\cc},{\bb}},\vspace{-0.05cm}
$$
which, by $(\calR_3)$, must be a power of $\oo_{q_{\ell}}$.
\end{proof}

Let\vspace{-0.09cm}
$\aa$ still denote an arbitrary vector in
 ${\Lambda_{r,k}^{\text{{lin}}}}$, and
$\cc\in \Lambda_{r,\ell}^{\text{{lin}}}$,
where $\ell\ne k$ and $\ell, k \in [s]$.
By writing out the definition of
$T^{[\hh]}_{\xx}$ in term of ${D^{[r]}_{*}}$, we have
$$
T^{[\wt{\cc}]}_{\xx+\wt{\aa}} \cdot T^{[\wt{\aa}]}_{\xx}=
  T_{\xx}^{[\wt{\aa}+\wt{\cc}]}=T^{[\wt{\aa}]}_{\xx+\wt{\cc}}
\cdot T^{[\wt{\cc}]}_{\xx},\vspace{-0.1cm}
$$
and thus,\vspace{-0.05cm}
$$
T^{[\wt{\cc}]}_{\xx+\wt{\aa}}\Big/T^{[\wt{\cc}]}_{\xx}=
  T^{[\wt{\aa}]}_{\xx+\wt{\cc}}\Big/T^{[\wt{\aa}]}_{\xx}.
$$
By Lemma~\ref{ratio-of-T}, the left
hand side of the equation is a power of $\oo_{q_{k}}$,
  while the right hand side of the equation is a power of $\oo_{q_{\ell}}$.
Since $k\ne \ell$, ${\rm gcd}(q_{k}, q_{\ell}) =1$, we have
\begin{equation}\label{jujutemp}
T^{[\wt{\aa}]}_{\xx+\wt{\cc}}\Big/T^{[\wt{\aa}]}_{\xx}=1,\ \ \ \text{for all
  $\cc\in \lin{\Lambda_{r,\ell}}$\hspace{0.06cm} such that $\ell\ne k$}.\vspace{-0.15cm}
\end{equation}
This implies that $T^{[\wt{\aa}]}_{\xx}$, as a function of $\xx$,
  only depends on $\xx_k\in \Lambda_{r,k}$.
It then follows from (\ref{imp-need}) that
$$
T^{[\wt{\aa}]}_{\xx}=T^{[\wt{\aa}]}_{\ext_r(\xx_k)}=
  \oo_N^\alpha \cdot G_{\ext_r(\xx_k),\bb}=\oo_N^{\alpha+\beta}
  \cdot F_{\widetilde{\xx_k},\widetilde{\bb_k}}=
  \oo_N^{\alpha+\beta} \cdot F_{\xx,\widetilde{\bb_k}},\ \ \ \text{for any
  $\xx\in \Lambda_r$,}
$$
and for some constants $\alpha,\beta\in \mathbb{Z}_N$
  and $\bb_k\in \mathbb{Z}_{\qq_k}$ that are independent of $\xx$.
This proves condition $(\calD_3)$.

Finally we prove $(\calD_1)$ from $(\calD_3)$.\vspace{-0.025cm}

Recall that, in condition ($\calL_3$), we have
$D^{[r]}_{(0,\fa^{[r]})}=1$. Let $\fa^{[r]}
= (\aa_1, \aa_2, \ldots, \aa_s) \in \Lambda_{r}$, then
\begin{eqnarray*}
D^{[r]}_{(0,\xx)}\hspace{-0.15cm} &=&\hspace{-0.15cm} D^{[r]}_{(0,(\xx_1,\xx_2, \ldots, \xx_s))}
            \overline{D^{[r]}_{(0,(\aa_1,\aa_2, \ldots, \aa_s))}}\\[0.5ex]
 &=&\hspace{-0.15cm}
\left( D^{[r]}_{(0,(\xx_1,\xx_2, \ldots, \xx_{s-1}, \xx_s))}
\overline{D^{[r]}_{(0,(\xx_1,\xx_2, \ldots, \xx_{s-1}, \aa_s))}} \right)\\[0.5ex]
&&
\times
\left( D^{[r]}_{(0,(\xx_1,\xx_2, \ldots, \xx_{s-1}, \aa_s))}
\overline{D^{[r]}_{(0,(\xx_1, \ldots, \xx_{s-2}, \aa_{s-1}, \aa_s))}} \right)\\[-0.8ex]
&&
 \ \ \ \ \ \ \ \ \ \ \vdots \\[0.5ex]
&& \times
\left( D^{[r]}_{(0,(\xx_1, \aa_2, \ldots, \aa_s))}
\overline{D^{[r]}_{(0,(\aa_1,\aa_2, \ldots, \aa_s))}} \right),\ \ \ \text{for any
  $\xx\in \Lambda_r$.}
\end{eqnarray*}

We consider the $k^{th}$ factor
\[D^{[r]}_{(0,(\xx_1,\ldots, \xx_{k-1}, \xx_{k}, \aa_{k+1},
\ldots, \aa_s))}
\overline{D^{[r]}_{(0,(\xx_1,\ldots, \xx_{k-1}, \aa_{k}, \aa_{k+1},
\ldots, \aa_s))}
}.\]
By (\ref{jujutemp}) this factor is independent of all other components
in the starting point
$(\xx_1,\ldots, \xx_{k-1}, \aa_{k},\aa_{k+1},$ $\ldots, \aa_s)$
{\it except} the  $k^{th}$ component $\aa_{k}$.
In particular we can replace all other components,
as long as we stay within $\Lambda_{r}$.
We choose to replace the first $k-1$ components
$\xx_i$ by $\aa_i$, then\vspace{0.06cm}
\begin{eqnarray*}
&& D^{[r]}_{(0,(\xx_1,\ldots, \xx_{k-1}, \xx_{k}, \aa_{k+1},
\ldots, \aa_s))}
\overline{D^{[r]}_{(0,(\xx_1,\ldots, \xx_{k-1}, \aa_{k}, \aa_{k+1},
\ldots, \aa_s))}
}\\[1.3ex]
&&=\hspace{0.03cm}
 D^{[r]}_{(0,(\aa_1,\ldots, \aa_{k-1}, \xx_{k}, \aa_{k+1},
\ldots, \aa_s))}
\overline{D^{[r]}_{(0,(\aa_1,\ldots, \aa_{k-1}, \aa_{k}, \aa_{k+1},
\ldots, \aa_s))}
}\hspace{0.03cm}=\hspace{0.03cm} D^{[r]}_{(0,\ext_r({\xx_k}))}
\overline{ D^{[r]}_{(0,\fa^{[r]})} }
\hspace{0.03cm}=\hspace{0.03cm} D^{[r]}_{(0,\ext_r({\xx_k}))}.
\end{eqnarray*}
$(\calD_1)$ is proved.




\section{Tractability: Proof of Theorem \ref{tractable-1}}\label{tractabilitysec}

Let $((M,N),\CC,\fD,(\pp,\bft,\fq))$ be a tuple that satisfies
  all the three conditions $(\calR),(\calL)$ and $(\calD)$.
In this section, we reduce $\eval(\CC,\fD)$ to the following problem:

\begin{quote}
$\eval(q)$: Let $q=p^k$ be a prime power for some prime $p$ and
  positive integer $k$. The input of $\eval(q)$ is a
  quadratic polynomial $f(x_1,x_2,\ldots,x_n)=\sum_{i,j\in [n]}a_{i,j}x_ix_j$,
  where $a_{i,j}\in \mathbb{Z}_q$ for all $i,j$;
  and the output is
$$
Z_q(f)=\sum_{x_1,\ldots,x_n\in \mathbb{Z}_q} \oo_q^{f(x_1,\ldots,x_n)}.
$$
\end{quote}

We postpone the proof of the following theorem to the end of this section.

\begin{theorem}\label{pinyan}
Problem $\eval(q)$ can be solved in polynomial time \emph{(}in $n$: the number
  of variables\emph{)}.
\end{theorem}

The reduction goes as follows:
First, we use conditions $(\calR)$, $(\calL)$, and $(\calD)$ to show that
  $\eval(\CC,\fD)$ can be decomposed into $s$ smaller problems
  (recall $s$ is the number of primes in the sequence $\pp$):
  $$\eval(\CC^{[1]},\fD^{[1]}),\ldots,\eval(\CC^{[s]},\fD^{[s]}).$$
If every $\eval(\CC^{[i]},\fD^{[i]})$ is tractable, then so is $\eval(\CC,\fD)$.
Second, for each problem $\eval(\CC^{[i]},\fD^{[i]})$ where
  $i\in [s]$, we reduce it to $\eval(q)$ for some prime\vspace{0.005cm} power $q$
  which will become clear later, and thus,
  by Theorem \ref{pinyan}, all $\eval(\CC^{[i]},\fD^{[i]})$'s
  can be solved in polynomial time.

\subsection{Step 1}

For every integer $i\in [s]$, we define a $2m_i\times 2m_i$ matrix $
  \CC^{[i]}$ where $m_i=|\mathbb{Z}_{\qq_i}|$:
  $\CC^{[i]}$ is the bipartisation of the following $m_i\times m_i$
  matrix $\FF^{[i]}$, where
  (we index the rows and columns of $\FF^{[i]}$ using
  $\xx\in \mathbb{Z}_{\qq_i}$ and index the rows and columns
  of $\CC^{[i]}$ using $\{0,1\}\times
  \mathbb{Z}_{\qq_i}$)
\begin{equation}\label{step1-1}
F^{[i]}_{\xx,\yy}=\prod_{j\in [t_i]} \oo_{q_{i,j}}^{
  x_{j}y_{j}},\ \ \ \text{for all $\xx=(x_1,\ldots,x_{t_i}),\yy
  =(y_1,\ldots,y_{t_i})\in \mathbb{Z}_{\qq_i}$}.
\end{equation}
Here we use $x_{j}$, where $j\in [t_i]$, to
  denote the $j^{th}$ entry of $\xx$ in $\mathbb{Z}_{q_{i,j}}$.
It then follows from $(\calR_3)$ that
\begin{equation}\label{tensor-prod-F-diff-primes}
F_{\xx,\yy}=F^{[1]}_{\xx_1,\yy_1}\cdot F^{[2]}_{\xx_2,\yy_2} \cdots
   F^{[s]}_{\xx_s,\yy_s},\ \ \ \text{for all $\xx,\yy\in \zqt$}.
\end{equation}

On the other hand, for each integer $i\in [s]$, we define a sequence
  of $N$ $2m_i\times 2m_i$ diagonal matrices $$\fD^{[i]}=\{\DD^{[i,0]},\ldots,
  \DD^{[i,N-1]}\}:$$ $\DD^{[i,0]}$ is the $2m_i\times 2m_i$ identity matrix; and for every
  $r\in [N-1]$, we set\vspace{0.06cm}
\begin{eqnarray*}
&\DD^{[i,r]}_{(0,*)}=\00,\ \text{if $r\notin \calS$,}\ \ \ \ \text{and}\ \ \ \
D^{[i,r]}_{(0,\xx)}=D^{[r]}_{(0,\ext_r(\xx))}\ \text{for all $\xx\in \mathbb{Z}_{\qq_i}$,
  if $r\in \calS$;}&\text{and}\\[1.1ex]
&\DD^{[i,r]}_{(1,*)}=\00,\ \text{if $r\notin \calT$,}\ \ \ \ \text{and}\ \ \ \
D^{[i,r]}_{(1,\xx)}=D^{[r]}_{(1,\ext_r'(\xx))}\ \text{for all
  $\xx\in \mathbb{Z}_{\qq_i}$,\ if $r\in \calT$}.& \\[-3ex]
\end{eqnarray*}
By conditions $(\calD_1)$ and $(\calD_2)$, we have
\begin{equation}\label{step1-2}
D^{[r]}_{(b,\xx)}=D^{[1,r]}_{(b,\xx_1)}\cdots  D^{[s,r]}_{(b,\xx_s)},\ \ \
\text{for all $b\in \{0,1\}$ and  $\xx\in \zqt$.}
\end{equation}
Eq.\hspace{0.03cm}(\ref{step1-2}) is valid for all $\xx\in \zqt$:
For example for $b=0$ and $\xx\in\mathbb{Z}_{\fq}-\Lambda_r$,
  the left-hand side is $0$ because $\xx\notin \Lambda_r$.
The right-hand side is also $0$,\vspace{-0.055cm} since there exists an index $i\in [s]$
  such that $\xx_i\notin \Lambda_{r,i}$ and thus, $\ext_r(\xx_i)\notin \Lambda_r$,
  and $D^{[i,r]}_{(0,\xx_i)}=0$.
It then follows from (\ref{step1-1}), (\ref{step1-2}) and the following lemma that\vspace{-0.1cm}
$$
\eval(\CC^{[i]},\fD^{[i]})\ \text{is in polynomial time for all $i\in [s]$}\
\Longrightarrow\ \eval(\CC,\fD)\ \text{is in polynomial time}.\vspace{0.08cm}
$$

\begin{lemma}\label{tensorproduct2}
Suppose we have the following:
For each $i\in \{0,1,2\}$,\vspace{0.02cm} $\FF^{[i]}$ is an $m_i\times m_i$ complex matrix,
  for some positive integers $m_i$;
  $\CC^{[i]}$ is the bipartisation of $\FF^{[i]}$; and
${\frak D}^{[i]}=\{\DD^{[i,0]},\ldots,\DD^{[i,N-1]}\}$ is
  a sequence of $N$ $2m_i\times 2m_i$ diagonal matrices for
  some positive integer $N$, where\vspace{-0.1cm}
$$
\DD^{[i,r]}=\left(\begin{matrix} \PP^{[i,r]} & \\
  & \QQ^{[i,r]}\end{matrix}\right)\vspace{-0.1cm}
$$
and both $\PP^{[i,r]}$ and $\QQ^{[i,r]}$ are $m_i\times m_i$
   diagonal matrices.

For each $i\in \{0,1,2\}$, $(\CC^{[i]},\fD^{[i]})$ satisfies
   \emph{({\sl Pinning})}. Moreover,\vspace{0.05cm}
  $m_0=m_1\cdot m_2$,
$$\FF^{[0]}=\FF^{[1]}\otimes \FF^{[2]},\ \ \PP^{[0,r]}=
  \PP^{[1,r]}\otimes \PP^{[2,r]},\ \
  \QQ^{[0,r]}=\QQ^{[1,r]}\otimes \QQ^{[2,r]},\ \
  \ \text{for all $r\in [0:N-1]$. }$$
Then if both $\eval(\CC^{[1]},\fD^{[1]})$ and $\eval(\CC^{[2]},\fD^{[2]})$
  are tractable, $\eval(\CC^{[0]},\fD^{[0]})$ is also tractable.
\end{lemma}
\begin{proof}
By the Second Pinning\vspace{-0.02cm} Lemma (Lemma \ref{pinning2}), we can compute
  $Z^{\rightarrow}_{\CC^{[i]},\fD^{[i]}}$
  and $Z^{\leftarrow}_{\CC^{[i]},\fD^{[i]}}$, for both $i=1$ and $2$, in polynomial time.
The lemma then follows from Lemma \ref{tensorproduct}.
\end{proof}

We now use condition ($\calD_4$) to prove the following lemma
  about $\DD^{[i,r]}_{(1,*)}$, where $r\in \calT$.
\begin{lemma}\label{useitonce2}
For any $r\in \calT$, $i\in [s]$ and $\aa\in \Delta_{r,i}^{\text{\rm lin}}$,
  there exist $\bb\in {\mathbb{Z}}_{\qq_i}$ and $\alpha\in \mathbb{Z}_N$ such that
$$
D^{[i,r]}_{(1,\xx+\aa)}\cdot \overline{D^{[i,r]}_{(1,\xx)}}=
  \oo_N^\alpha\cdot F^{[i]}_{\bb,\xx},\ \ \ \text{for all $\xx\in \Delta_{r,i}$.}
$$
\end{lemma}
\begin{proof}
By the definition of $\DD^{[i,r]}$, we have
$$
D^{[i,r]}_{(1,\xx+\aa)}\cdot \overline{D^{[i,r]}_{(1,\xx)}}=
  D^{[r]}_{(1,\ext_r'(\xx+\aa))}\cdot \overline{
  D^{[r]}_{(1,\ext_r'(\xx))}}=D^{[r]}_{(1,\ext_r'(\xx)+\widetilde{\aa})}\cdot \overline{
  D^{[r]}_{(1,\ext_r'(\xx))}}.
$$
Recall that we use $\widetilde{\aa}$ to denote the vector $\xx\in \mathbb{Z}_\calQ$
  such that $\xx_i=\aa$ and $\xx_j=0$ for all other $j\ne i$.

Then by condition ($\calD_4$), we know there exist $\bb\in {\mathbb{Z}}_{\qq_i}$ and
  $\alpha\in \mathbb{Z}_N$ such that
$$
D^{[i,r]}_{(1,\xx+\aa)}\cdot \overline{D^{[i,r]}_{(1,\xx)}}=\oo_N^\alpha\cdot
  F_{\widetilde{\bb},\ext_r'(\xx)}=\oo_N^{\alpha}\cdot  F^{[i]}_{\bb,\xx},
  \ \ \ \text{for all $\xx\in \Delta_{r,i},$}
$$
and the lemma is proven.\vspace{-0.15cm}
\end{proof}

One can also prove a similar lemma for $\DD^{[i,r]}_{(0,*)}$, $r\in \calS$,
  using condition ($\calD_3$).

\subsection{Step 2}

For convenience, in this subsection  we abuse the notation slightly and
  use $\eval(\CC,\fD)$ to
 denote one of the subproblems we defined in the last step:
  $\eval(\CC^{[i]},\fD^{[i]})$, $i\in [s]$.
Then by using conditions $(\calR),(\calL)$ and $(\calD)$, we summarize
  the properties of this new $(\CC,\fD)$ that we need in the reduction as
  follows:
\begin{itemize}
\item[($\calF_1$)] There exist a prime $p$ and a sequence $\boldsymbol{\pi}=\{\pi_1\ge\pi_2
  \ge \ldots \ge \pi_h\}$ of powers of the same $p$.
$\FF$ is an $m\times m$ complex matrix, where $m=\pi_1\pi_2\ldots \pi_h$,
  and $\CC$ is the bipartisation of $\FF$.
We let $\pi$ denote $\pi_1$.
We use $\mathbb{Z}_{\boldsymbol{\pi}}\equiv \mathbb{Z}_{\pi_1}\times \cdots
  \times \mathbb{Z}_{\pi_h}$ to index the rows and columns of $\FF$, then
$$
F_{\xx,\yy}=\prod_{i\in [h]} \oo_{\pi_i}^{x_{i}y_{i}},
\ \ \ \text{for all $\xx=(x_1,\ldots,x_h)$ and $\yy
  =(y_1,\ldots,y_h)\in \mathbb{Z}_{\boldsymbol{\pi}}$,}\vspace{-0.08cm}
$$
where we use $x_{i}$ to denote the $i^{th}$ entry of $\xx$ in $\mathbb{Z}_{\pi_i}$, $i\in [h]$.\label{CONDITIONF14}

\item[($\calF_2$)] $\fD=\{\DD^{[0]},\ldots,\DD^{[N-1]}\}$ is a sequence
  of $N$ $2m\times 2m$ diagonal matrices, for some positive integer $N$
  with $ \pi \hspace{0.06cm}|\hspace{0.06cm}N$.
$\DD^{[0]}$ is the identity matrix, and every diagonal entry of $\DD^{[r]}$,
   $r\in [N-1]$, is either $0$ or a power of $\oo_N$.
We use $\{0,1\}\times \mathbb{Z}_{\boldsymbol{\pi}}$ to index the
  rows and columns of matrices $\CC$ and $\DD^{[r]}$.
(The condition $\pi\hspace{0.05cm}|\hspace{0.05cm}N$
  is from the condition $M\hspace{0.05cm}|\hspace{0.05cm}N$
in $(\calU_1)$, and the expression of $M$ in terms of the prime
powers, stated after $(\calR_3)$. The $\pi$ here is one of the $q_{i}
= q_{i,1}$ there.)

\item[($\calF_3$)] For each $r\in [0:N-1]$, we let $\Lambda_r$ and $\Delta_r$ denote
$$
\Lambda_r=\{\xx\in \mathbb{Z}_{\boldsymbol{\pi}}\hspace{0.07cm}\big|
  \hspace{0.07cm} D^{[r]}_{(0,\xx)}\ne 0\}\ \ \ \text{and}\ \ \
\Delta_r=\{\xx\in \mathbb{Z}_{\boldsymbol{\pi}}\hspace{0.07cm}\big|
  \hspace{0.07cm} D^{[r]}_{(1,\xx)}\ne 0\}.
$$
We let $\cS$ denote the set of $r$ such that $\Lambda_r\ne \emptyset$,
  and $\cT$ denote the set of $r$ such that $\Delta_r\ne \emptyset$.
Then for every $r\in \cS$, $\Lambda_r$ is a coset in $\mathbb{Z}_{\boldsymbol{\pi}}$;
  and for every $r\in \cT$, $\Delta_r$ is a coset in $\mathbb{Z}_{\boldsymbol{\pi}}$.
Moreover, for\vspace{-0.01cm} every $r\in \cS$ (and $r\in \cT$, resp.),
  there exists a vector $\fa^{[r]}\in \Lambda_r$ (and $\fb^{[r]}\in
  \Delta_r$, resp.) such that
$$D_{(0,\fa^{[r]})}^{[r]}=1\ \ \text{$\Big($and
  $D_{(1,\fb^{[r]})}^{[r]}=1\text{,\ resp.}\Big)$}.\vspace{-0.1cm}$$

\item[$(\calF_4)$] For all $r\in \cS$ and $\aa\in \lin{\Lambda_r}$,
  there exist $\bb\in \mathbb{Z}_{\boldsymbol{\pi}}$ and $\alpha\in \mathbb{Z}_N$ such that
$$
D^{[r]}_{(0,\xx+\aa)}\overline{D^{[r]}_{(0,\xx)}}=\oo_{N}^\alpha\cdot
  \FF_{\xx,\bb}, \ \ \ \text{for all $\xx\in \Lambda_r$};
$$
For all $r\in \cT$ and $\aa\in \lin{\Delta_r}$,
  there exist $\bb\in \mathbb{Z}_{\boldsymbol{\pi}}$ and $\alpha\in \mathbb{Z}_N$ such that
$$
D^{[r]}_{(1,\xx+\aa)}\overline{D^{[r]}_{(1,\xx)}}=\oo_N^\alpha\cdot
  \FF_{\bb,\xx},\ \ \ \text{for all $\xx\in \Delta_r$}.\vspace{-0.1cm}
$$
\end{itemize}

Now let $G$ be a connected graph. Below we will reduce the computation
  of $Z_{\CC,\fD}(G)$ to $\eval(\hq)$,
\begin{center}
where $\hq=\pi$ if $p\ne 2$;\ \ \ and $\hq=2 \pi$ if $p=2$.
\end{center}
Given $a\in \mathbb{Z}_{\pi_i}$ for some $i\in [h]$,
  we let $\widehat{a}$ denote an element in $\mathbb{Z}_{\hq}$
  such that $\widehat{a}\equiv a\pmod{\pi_i}$.
As $\pi_h\hspace{0.05cm}|$ $\pi_{h-1}\hspace{0.05cm}|\hspace{0.05cm}\ldots
  |\hspace{0.08cm}\pi_1 = \pi \hspace{0.08cm}|\hspace{0.05cm} \hq$,
this lifting of $a$ is certainly feasible.
For definiteness, we can choose $a$ itself if we consider
$a$ to be an integer between $0$ and $\pi_i-1$.

First, if $G$ is not bipartite, then $Z_{\CC,\fD}(G)$ is trivially $0$.
So from now on in this section,
we assume $G=(U\cup V, E)$ to be bipartite: every
  edge $uv\in E$ has one vertex in $U$ and one vertex in $V$.

Let $u^*$ be a vertex in $U$, then we can decompose $Z_{\CC,\fD}(G)$ into
$$
Z_{\CC,\fD}(G)=Z_{\CC,\fD}^\rightarrow(G,u^*)+Z_{\CC,\fD}^\leftarrow(G,u^*).
$$
We will give a reduction from the computation of
  $Z_{\CC,\fD}^\rightarrow(G,u^*)$ to $\eval(\hq)$.
The other part concerning $Z^\leftarrow $
 can be proved similarly.

We use $U_r$, where $r\in [0:N-1]$, to denote the set of vertices
  in $U$ whose degree is $r \pmod N$,
  and $V_\rho$ to denote the set of vertices in $V$ whose degree is
 $\rho\pmod N$.
We further decompose $E$ into $\bigcup_{i,j}E_{i,j}$ where $E_{i,j}$
contains
  the edges between $U_i$ and $V_j$.

It is clear that if $U_r\ne \emptyset$ for some $r\notin \cS$
  or if $V_\rho \ne \emptyset$ for some $\rho \notin \cT$,
  then $Z_{\CC,\fD}^{\rightarrow}(G)=0$.
Therefore we assume $U_r=\emptyset$ for all $r \not \in \cS$ and
  $V_\rho=\emptyset$ for all $\rho \not \in \cT$.
In this case, we have\vspace{0.02cm}
\begin{equation}\label{expand1}
Z_{\CC,\fD}^\rightarrow (G,u^*)=
\sum_{(f,g)}\hspace{0.1cm}
\left[  \prod_{r\in \cS} \left(
  \prod_{u \in U_r} D^{[r]}_{(0,\xx_{u })}\right) \cdot
  \prod_{\rho\in \cT}\left(
  \prod_{v \in V_\rho } D^{[r]}_{(1,\yy_{v })}\right) \right]
\cdot
\left[
  \prod_{(r, \rho) \in \cS \times \cT}
\prod_{u v \in E_{r,\rho}} F_{\xx_{u},\yy_{v}} \right].\vspace{0.02cm}
\end{equation}
Here the sum ranges over all pairs $(f,g)$,
where $$f = (f_r;r \in \cS)
\in \prod_{r \in \cS} \left( U_r \rightarrow \Lambda_r \right)\ \ \ \text{and}
\ \ \ g = (g_\rho;\rho \in \cT)
\in \prod_{\rho \in \cT} \left(V_\rho \rightarrow
\Delta_\rho \right),\vspace{-0.08cm}$$
such that $f (u ) = \xx_{u }$ and $g (v ) = \yy_{v }$.

The following lemma gives us
  a convenient way to do summation over a coset.\vspace{0.03cm}

\begin{lemma}\label{quiteuseful}
Let\vspace{-0.006cm} $\Phi$ be a coset in $\mathbb{Z}_{\boldsymbol{\pi}}$
  and $\cc = (c_1, \ldots, c_h)$ be a vector in $\Phi$,
then there exist a positive integer $s$ and an $s\times h$ matrix
  $\AA$ over $\mathbb{Z}_{\hq}$
such that the following map $\tau : (\mathbb{Z}_{\hq})^s \rightarrow
\mathbb{Z}_{\pi_1} \times \cdots \times \mathbb{Z}_{\pi_h}$
\begin{equation}\label{stars1}
\tau(\xx)=\big(\tau_1(\xx),\ldots,\tau_h(\xx)\big),\ \text{where}\
  \tau_j(\xx)=\big(\hspace{0.04cm}\xx\AA_{*,j}+{\widehat{c}}_j
  \hspace{-0.15cm}\pmod{\pi_j}\hspace{0.04cm}\big)\in \mathbb{Z}_{\pi_j}
    \ \text{for all $j\in [h]$},
\end{equation}
is a {\em uniform} map from $\mathbb{Z}_{\hq}^s$ onto $\Phi$.
This uniformity means that for all
  $\bb,\bb'\in \Phi$, the number of $\xx\in \mathbb{Z}_{\hq}^s$
  such that $\tau(\xx)=\bb$ is the same as the number of
  $\xx$ such that $\tau(\xx)=\bb'$.\vspace{0.03cm}
\end{lemma}
\begin{proof}
By the fundamental theorem of finite Abelian groups,  there
  is a group isomorphism $f$ from $\mathbb{Z}_\gg$
onto $\lin{\Phi}$,
  where  $\gg=(g_1, \ldots,g_s)$ is a sequence of powers
  of $p$ and satisfies $\hq \ge  \pi =\pi_1\ge g_1\ge \ldots \ge g_s$, for some $s\ge 1$.
$\mathbb{Z}_\gg\equiv \mathbb{Z}_{g_1}\times\ldots\times
  \mathbb{Z}_{g_s}$ is a $\mathbb{Z}_{\hq}$-module. This is clear, since
as a $\mathbb{Z}$-module, any multiple of $\hq$
  annihilates $\mathbb{Z}_\gg$.
Thus $f$ is also a $\mathbb{Z}_{\widehat{\pi}}$-module isomorphism.\vspace{0.005cm}

Let  $\aa_i = f(\ee_i)\in \lin{\Phi}$, for each $i\in [s]$,
where
  $\ee_i\in \mathbb{Z}_\gg$ is the vector whose $i^{th}$ entry is $1$ and all
  other entries are $0$.
  Let $\aa_i = (a_{i,1},
\ldots, a_{i,h}) \in \mathbb{Z}_{\boldsymbol{\pi}}$ where $a_{i,j} \in
\mathbb{Z}_{\pi_j}$, $i \in [s]$, $j \in [h]$. Let $\widehat{\aa}_i =
(\widehat{a}_{i,1}, \ldots, \widehat{a}_{i,h}) \in
(\mathbb{Z}_{\hq})^h$ be a lifting of $\aa_i$ component-wise. Similarly let
$\widehat{\cc}$ be a lifting of $\cc$ component-wise. Then we claim
that $\AA=(\widehat{a}_{i,j})$ and $\widehat{\cc}$ together
  give us a~\emph{uniform} map $\tau$ from $\mathbb{Z}_{\hq}^s$ to
  $\Phi$ defined in (\ref{stars1}).\vspace{0.005cm}

To prove that $\tau$ is uniform,
we consider the linear part of the map $\tau': \mathbb{Z}_{\hq}^s
\rightarrow \lin{\Phi}$,
$$\tau'(\xx) = (\tau'_1(\xx), \ldots, \tau'_h(\xx)),\  \text{
where $\tau'_j(\xx) = (\xx \AA_{*,j}\hspace{-0.25cm}
  \pmod{\pi_j}) \in \mathbb{Z}_{\pi_j}$, for all $j \in [h]$.}$$
Clearly we only need to show that $\tau'$ is a uniform map.

Let $\sigma$ be the natural projection from $\mathbb{Z}_{\hq}^s$
  to $\mathbb{Z}_\gg$:\vspace{-0.1cm}
\[\xx = (x_1, \ldots, x_s) \mapsto \big(x_1\hspace{-0.27cm}
  \pmod{g_1}, \ldots, x_s\hspace{-0.27cm} \pmod{g_s}\big).\vspace{-0.1cm}\]
$\sigma$ is certainly a uniform map, being a
surjective homomorphism. Thus every vector $\bb \in \mathbb{Z}_\gg$
has exactly $|\ker \sigma | = \hq^s/(g_1 \cdots g_s)$ many
preimages.
We show that the map $\tau'$ factors through $\sigma$ and $f$:
$\tau' = f \circ \sigma$.
Since $f$ is an isomorphism, this implies that $\tau'$ is also a uniform map.

Since $g_i \ee_i = \00$ in $\mathbb{Z}_\gg$,\vspace{-0.1cm}
  the following is a valid expression in the $\mathbb{Z}_{\hq}$-module for $\sigma(\xx)$
\begin{eqnarray*}&
\big(x_1\hspace{-0.27cm} \pmod{g_1}, \ldots, x_s\hspace{-0.27cm} \pmod{g_s}\big)
= \sum_{i=1}^s x_i \ee_i.
\end{eqnarray*}
Apply $f$ as a $\mathbb{Z}_{\hq}$-module homomorphism\vspace{-0.2cm}
\[f(\sigma(\xx)) = \sum_{i=1}^s x_i f(\ee_i),\vspace{-0.15cm}\]
which has its $j^{th}$ entry $\sum_{i=1}^s x_i a_{i,j}.$
This is an expression in the $\mathbb{Z}_{\hq}$-module $\mathbb{Z}_{\pi_j}$,
which is the same as
\[\sum_{i=1}^s \big(x_i\hspace{-0.27cm} \pmod{\pi_j}\big)\cdot a_{i,j}
=\sum_{i=1}^s x_i \widehat{a}_{i,j}\hspace{-0.23cm} \pmod{\pi_j}
= \tau'_j(\xx).\vspace{-0.55cm}\]
\end{proof}

By applying Lemma \ref{quiteuseful} to coset $\Lambda_r$,
  we know for every $r\in \cS$, there exist a
  positive integer $s_r$ and an $s_r\times h$ matrix $\AA^{[r]}$ over
  $\mathbb{Z}_{\hq}$ which give us a uniform map $\lambda^{[r]}(\xx)$
  from $\mathbb{Z}_{\hq}^{s_r}$ to $\Lambda_r$, where
\begin{equation}\label{stars2}
\lambda^{[r]}_i(\xx)=\big(\hspace{0.04cm}\xx\AA^{[r]}_{*,i} +\widehat{\frak a}^{[r]}_i
  \hspace{-0.2cm}\pmod{\pi_i}\hspace{0.04cm}\big),\ \ \ \text{for all $i\in [h]$
  and $\xx\in \mathbb{Z}_{\widehat{\pi}}^{s_r}$.}
\end{equation}
Similarly\vspace{-0.012cm} for every $r\in \cT$, there exist a positive integer $t_r$ and
  an $t_r\times h$ matrix $\BB^{[r]}$ over $\mathbb{Z}_{\hq}$ which
  give us a uniform map $\delta^{[r]}$ from $\mathbb{Z}_{\hq}^{t_r}$ to $\Delta_r$, where
\begin{equation}\label{stars3}
\delta^{[r]}_i(\yy)=\big(\hspace{0.04cm}\yy\BB^{[r]}_{*,i}+
  \widehat{\frak b}^{[r]}_i\hspace{-0.2cm} \pmod{\pi_i}\hspace{0.04cm}\big),\ \ \ \text{for all
  $i\in [h]$ and $\yy\in \mathbb{Z}_{\widehat{\pi}}^{t_r}$.}
\end{equation}
Using $(\calF_3)$, we have
\begin{equation}\label{zeropoint}
D^{[r]}_{(0,\lambda^{[r]}(\00))} =1,\ \text{when $r\in \cS$;\ \ and}\
D^{[r]}_{(1,\delta^{[r]}(\00))}=1,\ \text{when $r\in \cT$.}
\end{equation}
Because both $\lambda^{[r]}$ and $\delta^{[r]}$ are uniform,
  and we know the multiplicity of each map (cardinality of inverse images),
  to compute (\ref{expand1}), it suffices to compute the following
\begin{equation}\label{stars4}
\sum_{ 
  (\xx_u),(\yy_v)}\hspace{0.1cm}
  \prod_{r\in \cS} \left(
  \prod_{u\in U_r} D^{[r]}_{(0,\lambda^{[r]}(\xx_u))}\right)
  \prod_{r\in \cT}\left(
  \prod_{v\in V_r} D^{[r]}_{(1,\delta^{[r]}(\yy_v))}\right)
  \prod_{r_1\in \cS,r_2\in \cT}\left(\prod_{uv\in E_{r_1,r_2}} F_{\lambda^{[r_1]}(\xx_u),
  \delta^{[r_2]}(\yy_v)}\right),
\end{equation}
where the sum is over pairs of sequences
$$
\Big(\xx_u;u\in \bigcup_{r\in \calS} U_r\Big)\in \prod_{r\in \calS}
  \left(\mathbb{Z}_{\hq}^{s_r}\right)^{|U_r|}\ \ \ \text{and}\ \ \
\Big(\yy_v;v\in \bigcup_{r\in \calT} V_r\Big)\in \prod_{r\in \calT}
  \left(\mathbb{Z}_{\hq}^{t_r}\right)^{|V_r|}.
$$

If we can show for all $r\in \cS$, there is a quadratic polynomial
  $f^{[r]}$ over $\mathbb{Z}_{\hq}$ such that
\begin{equation}\label{condition1}
D^{[r]}_{(0,\lambda^{[r]}(\xx))}=\oo_{\hq}^{f^{[r]}(\xx)},\ \ \ \text{for all
  $\xx\in \mathbb{Z}_{\hq}^{s_{r}};$}
\end{equation}
and for all $r\in \cT$, there is a quadratic polynomial $g^{[r]}$ over $\mathbb{Z}_{\hq}$
  such that
\begin{equation}\label{conditionyy}
D^{[r]}_{(1,\delta^{[r]}(\yy))}=\oo_{\hq}^{g^{[r]}(\yy)},\ \ \ \text{for all
  $\yy\in \mathbb{Z}_{\hq}^{t_r};$}
\end{equation}
and for all $r_1\in \cS$ and $r_2\in \cT$, there is a quadratic polynomial
$f^{[r_1,r_2]}$ over $\mathbb{Z}_{\hq}$ such that
\begin{equation}\label{condition3}
F_{\lambda^{[r_1]}(\xx),\delta^{[r_2]}(\yy)}=
  \oo_{\hq}^{f^{[r_1,r_2]}(\xx,\yy)},\ \ \ \text{for all $\xx\in
  \mathbb{Z}_{\hq}^{s_{r_1}}$ and $\yy \in \mathbb{Z}_{\hq}^{t_{r_2}}$},
\end{equation}
then we can reduce the computation of the summation in (\ref{stars4}) to problem $\eval(\hq)$.

We start by proving the existence of the quadratic polynomial $f^{[r_1,r_2]}$.
Let $r_1\in \cS$ and $r_2\in \cT$ then by ($\calF_1$), the following
  map $f^{[r_1,r_2]}$ satisfies (\ref{condition3}):
$$
f^{[r_1,r_2]}(\xx,\yy)=\sum_{i\in [h]} \left(\frac{\hq}{\pi_i}\right)\cdot
  \lambda^{[r_1]}_i(\xx)\cdot \delta^{[r_2]}_i(\yy)
  =\sum_{i\in [h]} \left(\frac{\hq}{\pi_i}\right) \left(\xx\AA^{[r_1]}_{*,i}+
    \widehat{\frak a}^{[r_1]}_i
  \right)\left(\yy\BB^{[r_2]}_{*,i}+\widehat{\frak b}^{[r_2]}_i\right).
$$
Note\vspace{-0.04cm} that the presence of the integer $\hq/\pi_i$ is crucial to be able to
  substitute the mod ${\pi_i}$ expressions for $\lambda_i^{[r_1]}(\xx)$ in
  (\ref{stars2}) and $\delta_i^{[r_2]}(\yy)$ in (\ref{stars3}) respectively, as if they
  were mod ${\hq}$ expressions.
Now it is clear that $f^{[r_1,r_2]}$ is indeed a quadratic polynomial over
  $\mathbb{Z}_{\hq}$.\vspace{0.002cm}

Next, we prove the existence of a quadratic polynomial $f^{[r]}$ for $\Lambda_r$,
  $r\in \calS$,
  in (\ref{condition1}), which is a little more complicated.
One can prove the same result for (\ref{conditionyy}) similarly.

Let $r\in S$ and $\ee_i$ denote the vector in $\mathbb{Z}_{\hq}^{s_r}$ whose
  $i^{th}$ entry is $1$ and all other entries are $0$.
Then by ($\calF_4$),
  for each $i\in [s_r]$, there exist $\alpha_i\in \mathbb{Z}_N$
  and $\bb_i=(b_{i,1},\ldots,b_{i,h})\in \mathbb{Z}_{\boldsymbol{\pi}}$,
  where $b_{i,j}\in \mathbb{Z}_{\pi_j}$, such that
\begin{equation}\label{qutient}
D^{[r]}_{(0,\lambda^{[r]}(\xx+\ee_i))}\overline{
  D^{[r]}_{(0,\lambda^{[r]}(\xx))}}=\oo_N^{\alpha_i}
  \prod_{j\in [h]} \oo_{\pi_j}^{b_{i,j}\cdot \lambda^{[r]}_j(\xx)},\ \ \
  \text{for all $\xx\in \mathbb{Z}_{\hq}^{s_r}$.}
\end{equation}
We have this equation because $\lambda^{[r]}(\xx+\ee_i)-\lambda^{[r]}(\xx)$
  is indeed a vector in $\mathbb{Z}_{\boldsymbol{\pi}}$ that is independent of $\xx$.
To see this, its $j^{th}$ entry in $\lambda^{[r]}(\xx+\ee_i)-\lambda^{[r]}(\xx)$
  is $$\ee_i\AA^{[r]}_{*,j}= A^{[r]}_{i,j} \pmod{\pi_j},$$ and thus the
  displacement vector $\lambda^{[r]}(\xx+\ee_i)-\lambda^{[r]}(\xx)$
  is independent of $\xx$,\vspace{0.005cm}
  and is in $\Lambda_r^{\text{lin}}$  by definition.
This is the $\aa\in \Lambda_r^{\text{lin}}$
  in the statement of $(\calF_4)$ which we applied.

Before moving forward, we show that $\oo_N^{\alpha_i}$ must be a power of $\oo_{\hq}$.
This is because
\begin{equation}\label{checkbla1}
1=\prod_{j=0}^{\hq-1} D^{[r]}_{(0,\lambda^{[r]}((j+1)\ee_i))}
  \overline{D^{[r]}_{(0,\lambda^{[r]}(j\ee_i))}}
=(\oo_N^{\alpha_i})^{\hq} \prod_{k\in [h]}
  \oo_{\pi_k}^{b_{i,k}\cdot[\lambda_k^{[r]}(0\ee_i)+...+\lambda_k^{[r]}((\hq-1)\ee_i)]}.
\end{equation}
For each $k\in [h]$, the exponent of $\oo_{\pi_k}$ is $b_{i,k}Q_k\in \mathbb{Z}_{\pi_k}$
  where $Q_k$ is the following summation:
\begin{equation}\label{checkbla2}
\sum_{j=0}^{\hq-1}\lambda_k^{[r]}(j\ee_i)=\sum_{j=0}^{\hq-1} \Big(
  (j\ee_i)\AA^{[r]}_{*,k}+\widehat{\frak a}^{[r]}_k\hspace{-0.18cm}
  \pmod{\pi_k}\hspace{0.02cm}\Big)
  = \left(\sum_{j=1}^{\hq-1}j\ee_i\right)\AA^{[r]}_{*,k}
  \hspace{-0.18cm}\pmod{\pi_k}=0.
\end{equation}
The last equality comes from $J\equiv\sum_{j=1}^{\hq-1}j=0 \pmod{\pi_k}$, and this
  is due to our definition of $\hq$.
When $p$ is odd, $J$ is a multiple of $\hq$ and $\pi_k\hspace{0.06cm}|\hspace{0.06cm}\hq$;
When $p=2$, $J$ is a multiple of $\hq/2$.
However in this case, we have $\hq/2=\pi_1$ and $\pi_k\hspace{0.06cm}|\hspace{0.06cm}
  \pi_1$.\vspace{0.02cm}

As a result, $(\oo_N^{\alpha_i})^{\hq}=1$ and
  $\oo_{N}^{\alpha_i}$ is a power of $\oo_{\hq}$.
So there exists $\beta_i\in \mathbb{Z}_{\hq}$ for each $i\in [s_r]$
  such that
\begin{equation}\label{stars5}
D^{[r]}_{(0,\lambda^{[r]}(\xx+\ee_i))}\overline{
  D^{[r]}_{(0,\lambda^{[r]}(\xx))}}=\oo_{\hq}^{\beta_i}
  \prod_{j\in [h]} \oo_{\pi_j}^{b_{i,j}\cdot \lambda^{[r]}_j(\xx)},\ \ \
  \text{for all $\xx\in \mathbb{Z}_{\hq}^{s_r}$.}
\end{equation}
It follows that every non-zero entry of
  $\DD^{[r]}$ is a power of $\oo_{\hq}$.
This uses $(\calF_3)$:
  the $(0,\fa^{[r]})^{th}$ entry of $\DD^{[r]}$ is $1$, and the fact that $\lambda^{[r]}$
  is surjective to $\Lambda_r$: any point in $\Lambda_r$
  is connected to the normalizing point $\fa^{[r]}$ by a sequence of moves
  $\lambda^{[r]}(\xx)\rightarrow \lambda^{[r]}(\xx+\ee_i)$, for $i\in [s_r]$.\vspace{0.005cm}

Now we\vspace{-0.0123cm} know there is a function $f^{[r]}$: $\mathbb{Z}_{\widehat{\pi}}^{s_r}
  \rightarrow \mathbb{Z}_{\widehat{\pi}}$ \hspace{0.075cm}satisfying (\ref{condition1}).
We want to show that we can take a quadratic polynomial $f^{[r]}$ for this purpose.\vspace{0.05cm}
To see this, by (\ref{stars5}),
  we have for every $i\in [s_r]$,
\begin{equation}\label{finally}
f^{[r]}(\xx+\ee_i)-f^{[r]}(\xx)=\beta_i+\sum_{j\in [h]}
 \left( \left(\frac{\hq}{\pi_j}\right) b_{i,j}\right)\cdot \lambda^{[r]}_j(\xx)
  =\beta_i+\sum_{j\in [h]}
  \left(\frac{\hq}{\pi_j}\right) \widehat{b}_{i,j}\cdot
  \big(\xx\AA^{[r]}_{*,j}+\widehat{\frak a}^{[r]}_j\big).\vspace{0.05cm}
\end{equation}
We should remark that, originally $b_{i,j}$ is in $\mathbb{Z}_{\pi_j}$;
  however with the integer multiplier $(\hq/\pi_j)$, the quantity
  $(\hq/\pi_j)\cdot {b}_{i,j}$ is now considered in $\mathbb{Z}_{\hq}$.
Furthermore, $$\widehat{b}_{i,j}\equiv b_{i,j}\hspace{-0.15cm}\pmod{\pi_j}
  \ \ \ \text{implies that}\ \
 \left(\frac{\hq}{\pi_j}\right)\widehat{b}_{i,j}\equiv
  \left(\frac{\hq}{\pi_j}\right)b_{i,j}\hspace{-0.15cm}\pmod{\hq}.$$
Thus the expression in (\ref{finally}) happens in $\mathbb{Z}_{\hq}$.
It means for any $i\in [s_r]$, there
  exist $c_{i,0},c_{i,1},\ldots,c_{i,s_r}\in \mathbb{Z}_{\hq}$,
\begin{equation}\label{bbb}
f^{[r]}(\xx+\ee_i)-f^{[r]}(\xx)=c_{i,0}+\sum_{j\in [s_r]} c_{i,j}x_j.\vspace{-0.15cm}
\end{equation}
Since $D^{[r]}_{(0,\lambda^{[r]}(\00))}=1$,
  $f^{[r]}(\00)$ is $0$.
The case when the prime $p$ is odd follows from the lemma below.

\begin{lemma}\label{quadraticlemma1}
Let $f$ be a map from $\mathbb{Z}_\pi^s$, for some positive integer $s\ge 1$, to
  $\mathbb{Z}_\pi$, and $\pi$ is a power of an odd prime.
Suppose for every $i\in [s]$, there exist $c_{i,0},c_{i,1},\ldots,c_{i,s}
  \in \mathbb{Z}_\pi$ such that
$$
f(\xx+\ee_i)-f(\xx)=c_{i,0}+\sum_{j\in [s]} c_{i,j}x_j,\ \ \text{
  for all $\xx\in \mathbb{Z}_\pi^s$},
$$
and $f(\00)=0$.
Then there exist $a_{i,j},a_i\in \mathbb{Z}_\pi$ such that
  $$f(\xx)=\sum_{i\le j\in [s]} a_{i,j}x_{i}x_{j}
  +\sum_{i\in [s]} a_i x_i,\ \ \ \text{for all $\xx\in \mathbb{Z}_\pi^s$.}$$
\end{lemma}
\begin{proof}
First note that $f$ is uniquely determined by the conditions
  on $f(\xx+\ee_i)-f(\xx)$ and $f(\00)$.
Second we show that $c_{i,j}=c_{j,i}$ for all $i,j\in [s]$;
  otherwise $f$ does not exist, contradicting the assumption.

On the one hand, we have
$$
f(\ee_i+\ee_j)=f(\ee_i+\ee_j)-f(\ee_j)+f(\ee_j)-f(\00)
  =c_{i,0}+c_{i,j}+c_{j,0}.
$$
On the other hand,
$$
f(\ee_i+\ee_j)=f(\ee_i+\ee_j)-f(\ee_i)+f(\ee_i)-f(\00)
  =c_{j,0}+c_{j,i}+c_{i,0}.
$$
As a result, we have $c_{i,j}=c_{j,i}$.

Finally, we set $a_{i,j}=c_{i,j}$ for all $i<j\in [s]$;
  $$a_{i,i}=c_{i,i}\big/2,\ \ \ \text{for all $i\in [s]$};$$
(Here $c_{i,i}/2$ is well defined because
  $\pi$ is odd) and $a_i=c_{i,0}-a_{i,i}$ for all $i\in [s]$.
We now claim that $$g(\xx)=\sum_{i\le j\in [s]}a_{i,j}x_ix_j+
  \sum_{i\in [s]}a_ix_i$$ satisfies both conditions
  and thus, $f=g$.
To see this, we check the case when $i=1$ and the other cases are similar:
\begin{eqnarray*}
g(\xx+\ee_1)-g(\xx)=
  2a_{1,1}x_1+\sum_{j>1} a_{1,j}x_j+(a_{1,1}+a_1)
  =c_{1,1}x_1+\sum_{j>1} c_{1,j} x_j+ c_{1,0}.\\[-5.5ex]
\end{eqnarray*}
\end{proof}

The case when $p=2$ is a little more complicated.
We first claim for every $i\in [s]$, the constant
  $c_{i,i}$ in (\ref{bbb}) must be even.
This is because
\begin{equation*}
0=f^{[r]}(\hq\ee_i)-f^{[r]}((\hq-1)\ee_i)+\ldots
  +f^{[r]}(\ee_i)-f^{[r]}(\00)
  =\hq\cdot c_{i,0}+c_{i,i}(\hq-1+\hq-2+\ldots+1+0).
\end{equation*}
This equality happens in $\mathbb{Z}_{\hq}$. So
$$
c_{i,i}\frac{\hq}{2}(\hq-1)=0\hspace{-0.13cm} \pmod{\hq}.
$$
When $\hq-1$ is odd we have $2\hspace{0.05cm}|\hspace{0.05cm} c_{i,i}$.
It follows from the lemma below that $f^{[r]}$ is a
  quadratic polynomial.

\begin{lemma}\label{quadraticlemma2}
Let $\pi$ be a power of $2$ and $f$ be a map from $\mathbb{Z}_\pi^s$ to
  $\mathbb{Z}_\pi$, for some positive integer $s\ge 1$.
Suppose for every $i\in [s]$, there exist $c_{i,0},c_{i,1},\ldots,c_{i,s}
  \in \mathbb{Z}_\pi$, where $2\hspace{0.06cm}|\hspace{0.07cm} c_{i,i}$, such that
$$
f(\xx+\ee_i)-f(\xx)=c_{i,0}+\sum_{j\in [s]} c_{i,j}x_j,\ \ \text{
  for all $\xx\in \mathbb{Z}_\pi^s$},
$$
and $f(\00)=0$.
Then there exist $a_{i,j},a_i\in \mathbb{Z}_\pi$ such that\vspace{-0.05cm}
  $$f(\xx)=\sum_{i\le j\in [s]} a_{i,j}x_{i}x_{j}
  +\sum_{i\in [s]} a_i x_i,\ \ \ \text{for all $\xx\in \mathbb{Z}_\pi^s$.}\vspace{-0.05cm}
$$
\end{lemma}
\begin{proof}
The proof of Lemma \ref{quadraticlemma2} is
  essentially the same as in Lemma \ref{quadraticlemma1}.
The only thing to notice is that,
  because $2\hspace{0.06cm}|\hspace{0.07cm} c_{i,i}$,
  $a_{i,i}=c_{i,i}/2$ is well defined (in particular, when $c_{i,i}=0$,
  we set $a_{i,i}=0$).
\end{proof}

\subsection{Proof of Theorem \ref{pinyan}}

\input{Tractability.tex}

\section{Proof of Theorem \ref{t-step-2}}

Let $\AA$ be a symmetric, non-bipartite and purified matrix.
After collecting its entries of equal norm in decreasing order (by
  permuting the rows and columns of $\AA$),
there exist a positive integer $N$, and two
  sequences $\boldsymbol{\kappa}$ and $\mm$ such that
  $(\AA,(N,\boldsymbol{\kappa},\mm))$ satisfies the following condition:
\begin{enumerate}
\item[($\calS_1'$)] Matrix $\AA$ is an $m\times m$ symmetric matrix.\label{CONDITIONSP}
$\boldsymbol{\kappa}=\{\kappa_1,\kappa_2,\ldots,\kappa_s\}$ is a sequence of
  positive rational numbers of length $s\ge 1$ such that
  $\kappa_1>\kappa_2>\ldots>\kappa_s>0$.
$\mm=\{m_1,\ldots,m_s\}$ is a sequence of positive
  integers such that $m=\sum m_i$.
The\vspace{0.005cm} rows (and columns) of $\AA$ are indexed by $\xx=(x_1,x_2)$ where $x_1\in [s]$
  and $x_2\in [m_{x_1}]$.
For all $\xx,\yy$, we have
$$
A_{\xx,\yy}=A_{(x_1,x_2),(y_1,y_2)}=\kappa_{x_1}\kappa_{y_1} S_{\xx,\yy},
$$
where $\SS=\{S_{\xx,\yy}\}$ is an $m\times m$ symmetric matrix in which
  every entry is a power of $\oo_N$:\vspace{0.03cm}
\begin{equation*}
\AA= \left(\begin{matrix}
\kappa_1\II_{m_1}\\
& \hspace{-0.15cm}\kappa_2\II_{m_2}\hspace{-0.15cm} \\
& & \hspace{-0.15cm}\ddots\hspace{-0.15cm}\\
& & & \kappa_s\II_{m_s}
\end{matrix}\right)
\left( \begin{matrix}
\SS_{(1,*),(1,*)} & \SS_{(1,*),(2,*)} & \ldots & \SS_{(1,*),(s,*)} \\
\SS_{(2,*),(1,*)} & \SS_{(2,*),(2,*)} & \ldots & \SS_{(2,*),(s,*)} \\
\vdots & \vdots & \ddots & \vdots \\
\SS_{(s,*),(1,*)} & \SS_{(s,*),(2,*)} & \ldots & \SS_{(s,*),(s,*)}
\end{matrix}\right)
\left(\begin{matrix}
\kappa_1\II_{m_1} \\
& \hspace{-0.15cm}\kappa_2\II_{m_2}\hspace{-0.15cm} \\
& & \hspace{-0.15cm}\ddots\hspace{-0.15cm}\\
& & & \kappa_s\II_{m_s}
\end{matrix}\right),\vspace{0.1cm}
\end{equation*}
where $\II_{m_i}$ is the $m_i\times m_i$ identity matrix.
\end{enumerate}
We use $I$ to denote
$$
I=\big\{(i,j)\hspace{0.07cm}\big|\hspace{0.07cm}i\in [s],j\in [m_i]\big\}.
$$

The proof of Theorem \ref{t-step-2}, just like the one of Theorem \ref{bi-step-2},
  consists of five steps.
All the proofs, as one will see, use the following strategy:
We construct, from the $m\times m$ matrix $\AA$, its bipartisation $\AA'$ (which
  is a $2m\times 2m$ symmetric matrix).
Then we just apply the lemmas for the bipartite case to $\AA'$,
  and show that $\AA'$ is either \#P-hard or has certain properties.
Finally, we use these properties of $\AA'$ to derive properties of $\AA$.\vspace{0.016cm}

We need the following lemma:

\begin{lemma}\label{biparbipar}Let $\AA$ be a symmetric matrix, and
  $\AA'$ be its bipartisation, then $\eval(\AA')\le \eval(\AA)$.
\end{lemma}
\begin{proof}
Suppose $\AA$ is an $m\times m$ matrix. Let $G$ be a connected undirected graph.
If $G$ is not bipartite, then $Z_{\AA'}(G)$ is trivially $0$, since
  $\AA'$ is the bipartisation of $\AA$.
Otherwise, we assume $G=(U\cup V,E)$ to be a bipartite and connected graph,
  and $u^*$ be a vertex in $U$. It is easy to show that
$$
Z_{\AA}(G,u^*,i)=Z_{\AA'}(G,u^*,i)=Z_{\AA'}(G,u^*,m+i),\ \ \ \text{for any $i\in [m]$.}
$$
It then follows that $Z_{\AA'}(G)=2\cdot Z_{\AA}(G)$, and
  $\eval(\AA')\le \eval(\AA)$.
\end{proof}

\subsection{Step 2.1}

\begin{lemma}\label{step221-nonbip}
Suppose $(\AA,(N,\boldsymbol{\kappa},\mm))$ satisfies $(\calS_1')$, then either
  $\eval(\AA)$ is \#P-hard or $(\AA,(N,\boldsymbol{\kappa},\mm))$ satisfies
  the following condition:
\begin{enumerate}
\item[]\hspace{-0.6cm}$(\calS_2')$ For all $\xx,\xx'\in I$, either there exists
  an integer $k$ such that $\SS_{\xx,*}=\oo_N^k\cdot \SS_{\xx',*}$; or
  for every $j\in [s]$,\vspace{-0.1cm}
$$
\langle \SS_{\xx,(j,*)},\SS_{\xx',(j,*)}\rangle =0.\vspace{-0.2cm}
$$
\end{enumerate}
\end{lemma}
\begin{proof}
Suppose $\eval(\AA)$ is not \#P-hard.

Let $\AA'$ denote the bipartisation of $\AA$. Then
  by Lemma \ref{biparbipar}, $\eval(\AA')\le \eval(\AA)$, and $\eval(\AA')$ is also not \#P-hard.
It is easy to check that $(\AA',(N,\boldsymbol{\kappa},\boldsymbol{\kappa},
  \mm,\mm))$ satisfies condition $(\calS_1)$,
so by Lemma \ref{hahajaja} together with the assumption that $\AA'$
  is not \#P-hard (also note that the $\SS$ matrix in Lemma \ref{hahajaja}
  is exactly the same $\SS$ we have here), $\SS$ satisfies $(\calS_2)$
  which is exactly the same as $(\calS_2')$ here (note that in Lemma \ref{hahajaja},
  $\SS$ also need to satisfy $(\calS_3)$, but since $\SS$ is symmetric here,
  $(\calS_3)$ is the same as $(\calS_2)$).
\end{proof}

We also have the following corollary.
The proof is exactly the same as the one of Corollary \ref{fullrank}.\vspace{0.04cm}

\begin{corollary}\label{fullrank-nonbip}
For all $i,j\in [s]$, the $(i,j)^{th}$
  block matrix $\SS_{(i,*),(j,*)}$ has the same rank as $\SS$.\vspace{0.04cm}
\end{corollary}

Next, we apply the Cyclotomic Reduction Lemma
  on $\AA$ to build a pair $(\FF,\fD)$ such that\vspace{-0.07cm}
$$\eval(\AA)\equiv\eval(\FF,\fD).\vspace{-0.07cm}$$

Let $h=\text{rank}(\SS)$. By Corollary\vspace{0.008cm} \ref{fullrank-nonbip},
  it can be easily proved that there exist $1\le i_1<\ldots<i_h\le m_1$
  such that, the $\{(1,i_1),\ldots,(1,i_h)\}\times
  \{(1,i_1),\ldots,(1,i_h)\}$ submatrix of $\SS$ has full rank $h$ (using
  the fact that $\SS$ is symmetric).
Without loss of generality (if this is not the case, we can apply an
  appropriate permutation $\Pi$ to the rows and columns of $\AA$
  so that the new $\SS$ has this property), we assume $i_k=k$
  for all $k\in [h]$.
We use $\HH$ to denote this $h\times h$
  symmetric matrix: $H_{i,j}=S_{(1,i),(1,j)}$.\vspace{0.012cm}

By Corollary \ref{fullrank-nonbip} and Lemma \ref{step221-nonbip},
  for any index $\xx\in I$, there exist
  two unique integers $j\in [h]$ and $k\in [0:N-1]$ such that\vspace{-0.3cm}
\begin{equation}\label{t-case1-bipa}
\SS_{\xx,*} = \oo_N^k \cdot \SS_{(1,j),*}\ \ \ \text{and}\ \ \
  \SS_{*,\xx}=\oo_N^k \cdot \SS_{*,(1,j)}.
\end{equation}
This gives us a partition of the index set $I$
$$
{\mathscr R} =\big\{R_{(i,j),k}\hspace{0.05cm}\big|\hspace{0.05cm}
  i\in [s],j\in [h],k\in [0:N-1]\big\},
$$
as follows: For every $\xx\in I$, $\xx\in R_{(i,j),k}$ iff
  $i=x_1$ and $\xx,j,k$ satisfy (\ref{t-case1-bipa}).
By Corollary \ref{fullrank-nonbip}, we have
  $$\bigcup_{k\in [0:N-1]} R_{(i,j),k}\ne \emptyset,\ \ \
  \text{for all $i\in [s]$ and $j\in [h]$.}$$

Now we define $(\FF,\fD)$ and use the Cyclotomic Reduction
  Lemma together with $\mathscr{R} $ to show that
$$\eval(\FF,\fD)\equiv\eval(\AA).$$
First, $\FF$ is an $sh\times sh$ matrix.
We use $I'\equiv [s]\times [h]$ to index the rows and columns of $\FF$.
Then
$$
F_{\xx,\yy}=\kappa_{x_1}\kappa_{y_1}H_{x_2,y_2}=\kappa_{x_1}\kappa_{y_1}S_{(1,x_2),(1,y_2)},
  \ \ \ \text{ for all $\xx,\yy\in I'$.}
$$
or equivalently,\vspace{-0.2cm}
$$
\FF = \left(\begin{matrix}
\kappa_1\II \\
& \kappa_2\II \\
& & \ddots\\
& & & \kappa_s\II
\end{matrix}\right)
\left( \begin{matrix}
\HH & \HH & \ldots & \HH \\
\HH & \HH & \ldots & \HH \\
\vdots & \vdots & \ddots & \vdots \\
\HH & \HH & \ldots & \HH
\end{matrix}\right)
\left(\begin{matrix}
\kappa_1\II \\
& \kappa_2\II \\
& & \ddots\\
& & & \kappa_s\II
\end{matrix}\right),
$$
where $\II$ is the $h\times h$ identity matrix.\vspace{0.003cm}

Second, $\fD=\{\DD^{[0]},\ldots,\DD^{[N-1]}\}$\vspace{-0.004cm} is a sequence of $N$ diagonal
  matrices of the same size as $\FF$.
We use $I'$ to index its diagonal entries.
The $\xx^{th}$ entries of $\fD$ are generated by $(|R_{(x_1,x_2),0}|,\ldots,
  |R_{(x_1,x_2),N-1}|)$:
$$
D^{[r]}_{\xx}=\sum_{k=0}^{N-1} \big|R_{(x_1,x_2),k}\big|\cdot \oo_N^{kr},
\ \ \ \text{for all $r\in [0:N-1],\xx\in I'$.}
$$\newpage

The following lemma is a direct application of the Cyclotomic Reduction Lemma
  (Lemma \ref{twinreduction}).\vspace{0.16cm}

\begin{lemma}
$\eval(\AA)\equiv \eval(\FF,\fD)$.\vspace{0.16cm}
\end{lemma}
\begin{proof}
First we show that matrix $\AA$ can be generated from $\FF$ using\vspace{-0.02cm}
  ${\mathscr R}$.
Let $\xx,\yy\in I$, $\xx \in R_{(x_1,j),k}$ and
  $\yy\in R_{(y_1,j'),k'}$ for some $j,k,j',k'$, then by (\ref{t-case1-bipa}),
\begin{equation*}
A_{\xx,\yy}=\kappa_{x_1}\kappa_{y_1} S_{\xx,\yy }
  =\kappa_{x_1}\kappa_{y_1} S_{(1,j), \yy }\cdot \oo_N^{k}
  =\kappa_{x_1}\kappa_{y_1} S_{( 1,j),( 1,j')}\cdot \oo_N^{k+k'}
  =F_{(x_1,j),(y_1,j')}\cdot \oo_N^{k+k'}.
\end{equation*}

On the other hand, the construction of $\fD$ implies that $\fD$ can be generated from the
  partition ${\mathscr R}$.
The lemma then follows directly from the Cyclotomic Reduction Lemma.
\end{proof}

\subsection{Steps 2.2 and 2.3}

Now we get a pair $(\FF,\fD)$ that satisfies the following condition ({\sl Shape}$'$):
\begin{enumerate}
\item[] \hspace{-0.7cm}({\sl Shape}$_1'$): $\FF\in \mathbb{C}^{m\times m}$ (note that this $m$
  is different from the $m$ used in Step 2.1)
  is a symmetric $s\times s$ block matrix and we use $I=[s]\times [h]$
  to index its rows and columns.\label{SHAPECONDITIONP}

\item[] \hspace{-0.7cm}({\sl Shape}$_2'$): There is a sequence  $\boldsymbol{\kappa}
  =\{\kappa_1>\ldots >\kappa_s>0\}$ of rational numbers together with
  an $h\times h$ matrix $\HH$ of full rank, whose entries are all powers of $\oo_N$, for
  some positive integer $N$.
We have
$$
F_{\xx,\yy}=\kappa_{x_1}\kappa_{y_1}H_{x_2,y_2},\ \ \ \text{for all $\xx,\yy\in I$}.
$$

\item[] \hspace{-0.7cm}({\sl Shape}$_3'$): $\fD=\{\DD^{[0]},\ldots,\DD^{[N-1]}\}$ is a sequence of $N$
  $m\times m$ diagonal matrices.
  $\fD$ satisfies $(\calT_3)$, so
  $$
  D^{[r]}_{\xx}=\overline{D^{[N-r]}_{\xx}},\ \ \ \text{for all $r\in [N-1]$,
    and $\xx\in I$.}
  $$
\end{enumerate}

\noindent Now suppose $\eval(\FF,\fD)$ is not \#P-hard.\vspace{0.006cm}

We build the following pair $(\CC,\hat{\fD})$:
$\CC$ is the bipartisation of $\FF$ and $\hat{\fD}=\{\hat{\DD}^{[0]},\ldots,\hat{\DD}^{[N-1]}\}$,
  where
$$
\hat{\DD}^{[r]}=\left(\begin{matrix}
\DD^{[r]}\\ & \DD^{[r]}
\end{matrix}\right),\ \ \ \text{for all $r\in [0:N-1]$.}
$$
The proof of the following lemma is the same as the one of Lemma \ref{biparbipar}.\vspace{0.04cm}

\begin{lemma}\label{checkcheck2}
$\eval(\CC,\hat{\fD})\le \eval(\FF,\fD)$.\vspace{0.04cm}
\end{lemma}

By Lemma \ref{checkcheck2} above, we have $\eval(\CC,\hat{\fD})\le \eval(\FF,\fD)$,
  and $\eval(\CC,\hat{\fD})$ is also not \#P-hard.
Using ({\sl Shape}$_1'$)-({\sl Shape}$_3'$), one can check that
  $(\CC,\hat{\fD})$ satisfies ({\sl Shape}$_1$)-({\sl Shape}$_3$).
Therefore, by Lemma \ref{shapecondition} and Lemma \ref{horrible},
  $(\CC,\hat{\fD})$ must also satisfy ({\sl Shape}$_4$)-({\sl Shape}$_6$).
Since $(\CC,\hat{\fD})$ is built from $(\FF,\fD)$, we have the latter
  must satisfy the following conditions:\vspace{0.06cm}
\begin{enumerate}
\item[]\hspace{-0.7cm}{({\sl Shape}$_4'$)}: \emph{$\frac{1}{\sqrt{h}}\cdot \HH$ is unitary:
  $\langle \HH_{i,*},\HH_{j,*}\rangle
  =\langle \HH_{*,i},\HH_{*,j}\rangle =0$ for all $i\ne j\in [h]$;}\vspace{-0.02cm}

\item[]\hspace{-0.7cm}{({\sl Shape}$_5'$)}: $D^{[0]}_{\xx}=
  D^{[0]}_{(x_1,1)}$ for all $\xx\in I$;\vspace{-0.02cm}

\item[]\hspace{-0.7cm}{({\sl Shape}$_6'$): For every $r\in [N-1]$, there
  exist two diagonal matrices: $\KK^{[r]}\in \mathbb{C}^{s\times s}$
  and $\LL^{[r]}\in \mathbb{C}^{h\times h}$.\\
The norm of every diagonal entry in $\LL^{[r]}$ is
  either $0$ or $1$. We have}\vspace{-0.04cm}
$$
\DD^{[r]} =\KK^{[r]} \otimes \LL^{[r]},\ \ \ \text{for any $r\in [N-1]$.}\vspace{-0.04cm}$$
Moreover, for any $r\in [N-1]$,\vspace{-0.04cm}
\begin{eqnarray*}
\KK^{[r]} =\00 \ \Longleftrightarrow \ \LL^{[r]} =\00\ \ \ \ \text{and}\ \ \ \
  \LL^{[r]} \ne \00 \ \Longrightarrow\ \exists\hspace{0.05cm}i\in [h],\ L^{[r]}_i =1.
\end{eqnarray*}
\end{enumerate}\newpage
In particular, ({\sl Shape}$_5'$) means by setting\vspace{-0.04cm}
$$ K^{[0]}_i=D^{[0]}_{(i,1)}\ \ \ \text{and}\ \ \ L^{[0]}_j=1,\ \ \ \text{for all
  $i\in [s]$ and $j\in [h]$.}\vspace{-0.04cm}$$
we have $\DD^{[0]}=\KK^{[0]}\otimes \LL^{[0]}$, where $\LL^{[0]}$ is
  the $h\times h$ identity matrix.
By $(\calT_3)$ in ({\sl Shape}$_3'$), every entry of $\KK^{[0]}$ is a positive integer.

\subsection{Step 2.4}

Suppose $(\FF,\fD)$ satisfies conditions ({\sl Shape}$_1'$)-({\sl Shape}$_6'$).
By ({\sl Shape}$_2'$), we have $\FF=\MM\otimes \HH$, where $\MM$
  is the $s\times s$ matrix of rank $1$:
  $M_{i,j}=\kappa_i\kappa_j$ for all $i,j\in [s]$.\vspace{0.005cm}

We now decompose
  $\eval(\FF,\fD)$ into two problems $\eval(\MM,{\frak K})$ and $\eval(\HH,{\frak L})$,
  where
$$
{\frak K}=\{\KK^{[0]},\ldots,\KK^{[N-1]}\},\ \ \ \text{and}\ \ \
{\frak L}=\{\LL^{[0]},\ldots,\LL^{[N-1]}\}.
$$
The proof of the following lemma is essentially the same as the one of Lemma \ref{jajajaja}:

\begin{lemma}
$\eval(\FF,\fD)\equiv \eval(\HH,\calL)$.
\end{lemma}

\subsection{Step 2.5}

We normalize the matrix $\HH$ in the same way we did for the bipartite case
  and obtain a new pair that 1). satisfies conditions $(\calU_1')$--$(\calU_4')$; and 2).
  is polynomial-time equivalent to $\eval(\HH,\calL)$.

\section{Proofs of Theorem \ref{nonbipstep0} and Theorem \ref{t-step-3}}

Let $((M,N),\FF,\fD)$ be a triple that satisfies ($\calU_1'$)-($\calU_4'$).
We prove Theorem \ref{nonbipstep0} and \ref{t-step-3} in this section.\vspace{0.005cm}

We first prove that, if $\FF$ does not satisfy the group condition (\GC),
  then $\eval(\FF,\fD)$ is $\#$P-hard.
This is done by applying Lemma \ref{groupcondition1} (for the bipartite
  case) to the bipartisation $\CC$ of $\FF$:\vspace{0.04cm}


\begin{lemma}\label{groupcondition2}
Let $((M,N),\FF,\fD)$ be a triple that satisfies conditions $(\calU_1')$-$(\calU_4')$,
  then either the matrix $\FF$ satisfies the group condition \emph{(\GC)},
  or $\eval(\FF,\fD)$ is \#P-hard.\vspace{0.04cm}
\end{lemma}
\begin{proof}
Suppose $\eval(\FF,\fD)$ is not \#P-hard.\vspace{0.01cm}

Let $\CC$ and ${\frak E}=\{\EE^{[0]}, \ldots,\EE^{[N-1]}\}$ denote the
  bipartisations of $\FF$ and $\fD$, respectively:
$$
\CC=\left( \begin{matrix}\00 & \FF \\ \FF &\00\end{matrix}\right),\ \ \ \text{and}\ \
\ \EE^{[r]} =\left( \begin{matrix}\DD^{[r]}& \00 \\ \00 & \DD^{[r]}\end{matrix}\right),\ \ \
\text{for all $r\in [0:N-1]$.}$$
By using $(\calU_1')$-$(\calU_4')$, one can\vspace{0.007cm} show that $((M,N),\CC,{\frak E})$
  satisfies $(\calU_1)$-$(\calU_4)$.
Furthermore, by Lemma \ref{checkcheck2}, we have
  $\eval(\CC,{\frak E})\le \eval(\FF,\fD)$ and thus,
  $\eval(\CC,{\frak E})$ is also not \#P-hard.
It then follows from Lemma \ref{groupcondition1} that
  $\FF$ satisfies the group condition (\GC).
\end{proof}

\subsection{Proof of Theorem \ref{nonbipstep0}}

We prove Theorem \ref{nonbipstep0}, again, by using $\CC$ and ${\frak E}$: the bipartisations
  of $\FF$ and $\fD$, respectively.\vspace{0.004cm}

Suppose $\eval(\FF,\fD)$ is not \#P-hard.
On the one hand, $\eval(\CC,{\frak E})\le \eval(\FF,\fD)$
  and $\eval(\CC,{\frak E})$ is also not \#P-hard.
On the other hand, $((M,N),\CC,{\frak E})$ satisfies conditions $(\calU_1)$-$(\calU_4)$.
As a result, by Theorem \ref{step30}, ${\frak E}$ must satisfy ($\calU_5$):
Every entry of $\EE^{[r]}$, $r\in [N-1]$, is either $0$ or a power of $\oo_N$.
It then follows directly that every entry of $\DD^{[r]}$, $r\in [N-1]$,
  is either $0$ or a power of $\oo_N$.

\subsection{Proof of Theorem \ref{t-step-3}}

In this section, we prove Theorem \ref{t-step-3}.

However, we can not simply reduce it, using pair $(\CC,{\frak E})$, to the bipartite
  case (Theorem \ref{bi-step-3}).
The reason is because, in Theorem \ref{t-step-3},
  we are only allowed to permute the rows and columns symmetrically,
  while in Theorem \ref{bi-step-3}, one can use two different permutations
  to permute the rows and columns.
But as we will see below,
  for most of the lemmas we need here, their proofs are exactly the same
  as those for the bipartite case.
The only exception is the counterpart of Lemma \ref{decomp2},
  in which we have to bring in the generalized Fourier matrices
(see Definitions~\ref{FourierMatrix} and \ref{GeneralizedFourierMatrix}).

Suppose $\FF$ satisfies (\GC) (otherwise we already
  know that $\eval(\FF,\fD)$ is \#P-hard).\vspace{0.004cm}

We let $F^R$ denote the set of row vectors $\{\FF_{i,*} \}$ of $\FF$ and
  $F^C$ denote the set of column vectors $\{\FF_{*,j}\}$ of $\FF$.
Since $\FF$ satisfies (\GC), by Property \ref{gcproperty1},
  both $F^R$ and $F^C$ are finite Abelian groups
  of order $m$, under the Hadamard product.

We start the proof by proving a symmetric version of Lemma \ref{decomp1},
  stating that
  when $M=pq$~and $\gcd(p,q)=1$ (note that $p$ and $q$ are not necessarily primes),
  $\FF$ (after an appropriate permutation)
  is the tensor pro\-duct of two smaller discrete unitary matrices, both
  of which satisfy the group condition.

\begin{lemma}\label{sys-decomp1}
Let $\FF\in \mathbb{C}^{m\times m}$ be a symmetric $M$-discrete
  unitary matrix that satisfies \emph{(}\GC\emph{)}. \hspace{-0.05cm}Moreover,
  $M=pq$, $p,q>1$ and $\gcd(p,q)=1$.
Then there is a permutation $\Pi:[0:m-1]\rightarrow [0:m-1]$
  such that
$$
\FF_{\Pi,\Pi}=\FF'\otimes \FF'',
$$
where $\FF'$ is a symmetric $p$-discrete unitary matrix,
  $\FF''$ is a symmetric $q$-discrete
  unitary matrix, and both of them satisfy \emph{(}\GC\emph{)}.
\end{lemma}
\begin{proof}
The proof is almost the same as the one of Lemma \ref{decomp1}.
The only thing to notice is that, as $\FF$ is symmetric,
  the two correspondences $f,g$ that we defined in the
  proof of Lemma \ref{decomp1}, from $[0:m-1]$ to
  $[0:m'-1]\times [0:m''-1]$, are exactly the same. As a result,
  the row permutation $\Pi$ and the column permutation $\Sigma$
  that we apply on $\FF$ are the same.
\end{proof}

As a result, we only need to deal with the case when $M=p^\beta$ is a prime power.

\begin{lemma}\label{t-decomp1}
Let $\FF\in \mathbb{C}^{m\times m}$ be a symmetric $M$-discrete unitary matrix
  that satisfies \emph{(}\GC\emph{)}. \hspace{-0.05cm}Moreover $M=p^{\beta}$ is
  a prime power, $p\ne 2$, and $\beta\ge 1$.
Then there must exist an integer $k\in [0:m-1]$ such that
  $F_{k,k}=\oo_M^{\alpha_{k,k}}$ and $p\nmid \alpha_{k,k}$.
\end{lemma}
\begin{proof}
For $i,j\in [0:m-1]$, we let $\alpha_{i,j}$ denote the integer in $[0:M-1]$
  such that $F_{i,j}=\oo_M^{\alpha_{i,j}}$.\vspace{0.005cm}

Assume the lemma is not true, that is, $p\hspace{0.06cm}|\hspace{0.06cm}
  \alpha_{k,k}$ for all $k$.
Since $\FF$ is $M$-discrete unitary, there must exist
  $i\ne j\in [0:m-1]$ such that $p\nmid \alpha_{i,j}$.
Without loss of generality, we assume $p\nmid \alpha_{2,1}=\alpha_{1,2}$.

As $\FF$ satisfies (\GC), there must exist a $k\in [0:m-1]$
  such that $\FF_{k,*}=\FF_{1,*}\circ \FF_{2,*}$.
However,
$$
\omega_M^{\alpha_{k,k}}=F_{k,k}=F_{1,k}F_{2,k}=F_{k,1}F_{k,2}
=F_{1,1}F_{2,1}F_{1,2}F_{2,2}=\omega_M^{\alpha_{1,1}+\alpha_{2,2}+2\alpha_{1,2}},
$$
and $\alpha_{k,k}\equiv \alpha_{1,1}+\alpha_{2,2}+2\alpha_{1,2}\hspace{-0.06cm}\pmod{M}$
  implies that $0\equiv 0+0+2\alpha_{1,2}\hspace{-0.06cm}\pmod{p}$.
Since $p\ne 2$ and $p\nmid \alpha_{1,2}$ we get a contradiction.
\end{proof}

The next lemma is the symmetric version of Lemma \ref{decomp2} showing
  that when there exists a diagonal entry $F_{k,k}$ such that
  $p\nmid \alpha_{k,k}$, then $\FF$ is the tensor product
  of a Fourier matrix and a discrete unitary matrix.
Note that this lemma also applies to the case when $p=2$.
So the only case left is when $p=2$ but
  $2\hspace{0.06cm}|\hspace{0.06cm} \alpha_{i,i}$ for all $i\in [0:m-1]$.

\begin{lemma}\label{t-decomp2}
Let $\FF\in \mathbb{C}^{m\times m}$ be a symmetric $M$-discrete unitary matrix
  that satisfies \emph{(}\GC\emph{)}. \hspace{-0.05cm}Moreover, $M=p^{\beta}$ is
  a prime power.
If there exists a $k\in [0:m-1]$ such that $F_{k,k}=\oo_M^{\alpha}$
  and $p\nmid \alpha$, then one can find a permutation $\Pi$ such that
$$
\FF_{\Pi,\Pi}=\boldsymbol{\calF}_{M,\alpha}\otimes \FF',
$$
where $\FF'$ is a symmetric $M'$-discrete unitary matrix, $M'=p^{\beta'}$
  for some $\beta'\le \beta$, and $\FF'$ satisfies \emph{(}\GC\emph{)}.
\end{lemma}
\begin{proof}
The proof is exactly the same as the one of Lemma \ref{decomp2} by
  setting $a=k$ and $b=k$.
The only thing to notice is that, as $\FF$ is symmetric,
  the two correspondences $f$ and $g$ that we defined in the proof of Lemma \ref{decomp2}
  are the same.
As a result, the row permutation $\Pi$ and the column permutation $\Sigma$ that we apply
  on $\FF$ are the same.
Also note that, since $F_{k,k}=\oo_M^\alpha$, (\ref{haharefer}) becomes
$$
G_{(x_1,x_2),(y_1,y_2)}=\oo_M^{\alpha x_1y_1}\cdot G_{(0,x_2),(0,y_2)}.
$$
This explains why we need to use Fourier matrix $\boldsymbol{\calF}_{M,\alpha}$ here.
\end{proof}

Finally, we deal with the case when $p=2$ and $2\hspace{0.06cm}
  |\hspace{0.06cm} \alpha_{i,i}$ for all $i\in [0:m-1]$.

\begin{lemma}\label{t-decomp3}
Let $\FF\in \mathbb{C}^{m\times m}$ be a symmetric $M$-discrete unitary matrix
  that satisfies condition \emph{(}\GC\emph{)}.
Moreover, $M=2^{\beta}$ and
  $2\hspace{0.06cm}|\hspace{0.06cm} \alpha_{i,i}$ for all $i\in [0:m-1]$.
Then one can find a permutation $\Pi$ together with a symmetric non-degenerate matrix
  $\WW$ in $\mathbb{Z}_M^{2\times 2}$
  \emph{(see Section~\ref{reference-for-sec-6.3.2} and
        Definition~\ref{GeneralizedFourierMatrix})} such that
$$
\FF_{\Pi,\Pi}=\boldsymbol{\calF}_{M,\WW}\otimes \FF',
$$
where $\FF'$ is a symmetric $M'$-discrete unitary matrix, $M'=2^{\beta'}$
  for some $\beta'\le \beta$, and $\FF'$ satisfies \emph{(}\GC\emph{)}.
\end{lemma}
\begin{proof}
By Property \ref{gcproperty2}, there exist two integers $a\ne b$
  such that $F_{a,b}=F_{b,a}=\oo_M$.
Let $F_{a,a}=\oo^{\alpha_a}$ and $F_{b,b}=\oo^{\alpha_b}$.
The assumption of the lemma implies that $2\hspace{0.06cm}|\hspace{0.06cm} \alpha_a,\alpha_b$.

We let $S^{a,b}$ denote the following subset of $F^R$:\vspace{-0.04cm}
$$
S^{a,b}=\{\uu\in F^R\hspace{0.08cm}\big|\hspace{0.08cm}u_a=u_b=1\}.\vspace{-0.04cm}
$$
It is easy to see that $S^{a,b}$ is a subgroup of $F^R$.
On the other hand, let $S^a$ denote the subgroup
  of $F^R$ that is generated by $\FF_{a,*}$,
  and $S^b$ denote the subgroup generated by $\FF_{b,*}$:\vspace{-0.03cm}
$$
S^a=\{(\FF_{a,*})^0,(\FF_{a,*})^1,\ldots,(\FF_{a,*})^{M-1}\}\ \ \ \text{and}\ \ \
S^b=\{(\FF_{b,*})^0,(\FF_{b,*})^1,\ldots,(\FF_{b,*})^{M-1}\}.\vspace{-0.03cm}
$$
We have $|S^a|=|S^b|=M$, because $F_{a,b}=\oo_M$.
It is clear that $(\uu_1,\uu_2,\uu_3)\mapsto
  \uu_1\circ\uu_2\circ\uu_3$ is a group homomorphism from
  $S^a\oplus S^b\oplus S^{a,b}$ to $F^R$.
We now prove that it is a surjective group isomorphism.

Toward this end, we first note that the matrix $\WW$, where\vspace{-0.05cm}
$$
\WW=\left(\begin{matrix}
\alpha_a & 1\\
1 & \alpha_b
\end{matrix}\right),\vspace{-0.05cm}
$$
is non-degenerate. This follows from Lemma \ref{equivaequiva},
  since $\det(\WW)=\alpha_a\alpha_b-1$ is odd.


First, we show that $(\uu_1,\uu_2,\uu_3)\mapsto
  \uu_1\circ\uu_2\circ\uu_3$ is surjective.
This is because for any $\uu\in F^R$, there exist integers
  $k_1$ and $k_2$ such that (since $\WW$ is non-degenerate,
  by Lemma \ref{equivaequiva}, $\xx\mapsto\WW\xx$ is a bijection) $$ {u_a}=
  F_{a,a}^{k_1}\cdot F_{b,a}^{k_2}=\oo_M^{\alpha_ak_1+k_2}\ \ \
  \text{and}\ \ \ {u_b}=F_{a,b}^{k_1}\cdot F_{b,b}^{k_2}=\oo_M^{k_1+\alpha_bk_2},$$
and thus, $\uu\circ \overline{\FF_{a,*}^{k_1}}\circ \overline{\FF_{b,*}^{k_2}}\in S^{a,b}$.
It then follows that $\uu=\FF_{a,*}^{k_1}\circ \FF_{b,*}^{k_2}\circ \uu_3$
  for some $\uu_3\in S^{a,b}$.

Second, we show that it is injective.
Suppose this is not true. Then there exist
  $k_1,k_2,k_1',k_2'\in \mathbb{Z}_M$, and $\uu,\uu'\in S^{a,b}$ such that
  $(k_1,k_2,\uu) \ne (k_1',k_2',\uu')$ but
$$
(\FF_{a,*})^{k_1}\circ (\FF_{b,*})^{k_2}\circ \uu=
(\FF_{a,*})^{k_1'}\circ (\FF_{b,*})^{k_2'}\circ \uu'.
$$
If $k_1=k_1'$ and $k_2=k_2'$, then $\uu=\uu'$, which
  contradicts with our assumption.
Therefore, we may assume that $\boldsymbol{\ell}=(\ell_1,\ell_2)^T
 =(k_1-k_1',k_2-k_2')^T\ne \00$.
By restricting on the $a^{th}$ and $b^{th}$ entries, we get\vspace{0.007cm}
  $\WW\boldsymbol{\ell}=\00$.
This contradicts with the fact that $\WW$ is non-degenerate.\vspace{-0.005cm}

Now we know that $(\uu_1,\uu_2,\uu_3)\mapsto \uu_1\circ\uu_2\circ \uu_3$
  is a group isomorphism from $S^a\oplus S^b\oplus S^{a,b}$ to $F^R$.
As a result, $|S^{a,b}|=m/M^2$ which we denote by $n$.
Let $S^{a,b}=\{\vv_0=\11,\vv_1,\ldots,\vv_{n-1}\}$, then there
  exists a one-to-one correspondence $f$ from
  $[0:m-1]$ to $[0:M-1]\times [0:M-1]\times [0:n-1]$,
  $f(i)=(f_1(i),$ $f_2(i),f_3(i))$, such that\vspace{-0.06cm}
\begin{equation}\label{huhu1}
\FF_{i,*}=(\FF_{a,*})^{f_1(i)}\circ (\FF_{b,*})^{f_2(i)}\circ \vv_{f_3(i)},
\ \ \ \text{for all $i\in [0:m-1]$.}
\end{equation}
Since $\FF$ is symmetric, this also implies that
\begin{equation}\label{huhu2}
\FF_{*,j}=(\FF_{*,a})^{f_1(j)}\circ (\FF_{*,b})^{f_2(j)}\circ \vv_{f_3(j)},
\ \ \text{for all $j\in [0:m-1]$.}
\end{equation}
Note that $f(a)=(1,0,0)$ and $f(b)=(0,1,0)$.

Finally we permute the rows and columns of $\FF$ to obtain a new matrix $\GG$.
For convenience, we use $(x_1,x_2,x_3)$ and $(y_1,y_2,y_3)$, where
  $x_1,x_2,y_1,y_2\in [0:M-1]$ and $x_3,y_3\in [0:n-1]$,
  to\vspace{0.005cm} index the rows and columns of $\GG$, respectively.
We permute $\FF$ using $\Pi(x_1,x_2,x_3)=f^{-1}(x_1,x_2,x_3)$:
\begin{equation}\label{basicstuff}
G_{(x_1,x_2,x_3),(y_1,y_2,y_3)}=F_{\Pi(x_1,x_2,x_3),\Pi(y_1,y_2,y_3)}.
\end{equation}
Then by (\ref{huhu1}) and (\ref{huhu2}),
\begin{eqnarray*}
\GG_{(x_1,x_2,x_3),*}=(\GG_{(1,0,0),*})^{x_1}\circ (\GG_{(0,1,0),*})^{x_2}\circ
\GG_{(0,0,x_3),*}&&\text{and}\\[0.8ex]
\GG_{*,(y_1,y_2,y_3)}=(\GG_{*,(1,0,0)})^{y_1}\circ (\GG_{*,(0,1,0)})^{y_2}\circ
\GG_{*,(0,0,y_3)}.
\end{eqnarray*}
As a result,\vspace{-0.1cm}
$$
G_{(x_1,x_2,x_3),(y_1,y_2,y_3)}=(G_{(1,0,0),(y_1,y_2,y_3)})^{x_1}\cdot
  (G_{(0,1,0),(y_1,y_2,y_3)})^{x_2}\cdot G_{(0,0,x_3),(y_1,y_2,y_3)}.\vspace{0.04cm}
$$
We analyze the three factors. First, we have $G_{(1,0,0),(y_1,y_2,y_3)}$ is equal to
$$
(G_{(1,0,0),(1,0,0)})^{y_1}\cdot (G_{(1,0,0),(0,1,0)})^{y_2}
  \cdot G_{(1,0,0),(0,0,y_3)}=F_{a,a}^{y_1}\cdot F_{a,b}^{y_2}\cdot
  v_{y_3,a}=\omega_M^{\alpha_ay_1+y_2},
$$
where $v_{y_3,a}$ denotes the $a^{th}$ entry of $\vv_{y_3}$.
Similarly, $G_{(0,1,0),(y_1,y_2,y_3)}=\omega_M^{y_1+\alpha_by_2}$.
Second,
$$
G_{(0,0,x_3),(y_1,y_2,y_3)}=(G_{(0,0,x_3),(1,0,0)})^{y_1}\cdot
(G_{(0,0,x_3),(0,1,0)})^{y_2}\cdot G_{(0,0,x_3),(0,0,y_3)}.
$$
By (\ref{basicstuff}) and (\ref{huhu2})\vspace{-0.1cm} we
have $$G_{(0,0,x),(1,0,0)}=F_{\Pi(0,0,x),\Pi(1,0,0)}=
  F_{\Pi(0,0,x),a}\hspace{0.06cm}.\vspace{0.08cm}$$
Then by (\ref{huhu1}), $F_{\Pi(0,0,x),a}=v_{x,a}=1$.
  Similarly, we have $G_{(0,0,x),(0,1,0)}=v_{x,b}=1$.
Therefore,
$$
G_{(x_1,x_2,x_3),(y_1,y_2,y_3)}=\omega_M^{\alpha_ax_1y_1+x_1y_2+x_2y_1+\alpha_bx_2y_2}
  \cdot G_{(0,0,x_3),(0,0,y_3)}.
$$
In other words, we have
$$
\GG=\boldsymbol{\calF}_{M,\WW}\otimes \FF', \ \text{where $\WW$ is non-degenerate
and $\FF'\equiv\left(F'_{i,j}=G_{(0,0,i),(0,0,j)}\right)$
  is symmetric.}$$

The only thing left is to show $\FF'$ is discrete unitary
  and satisfies (\GC).
$\FF'$ satisfies (\GC) because $S^{a,b}$ is a group and thus, closed
  under the Hadamard product.
To see $\FF'$ is discrete unitary, we have
$$
0=\langle \GG_{(0,0,i),*},\GG_{(0,0,j),*}\rangle
  =M^2\cdot \langle \FF'_{i,*},\FF'_{j,*}\rangle,\ \ \ \text{for any $i\ne j\in [0:n-1]$.}
$$
Since $\FF'$ is symmetric, columns $\FF'_{*,i}$ and $\FF'_{*,j}$ are also orthogonal.
\end{proof}

Theorem \ref{t-step-3} then follows from Lemma \ref{t-decomp1},
  Lemma \ref{t-decomp2}, and Lemma \ref{t-decomp3}.

\section{Proofs of Theorem \ref{t-step-4} and Theorem \ref{nonbi-step-5}}

Suppose $((M,N),\FF,\fD,(\dd,\calW,\pp,\bft,\calQ,\calK))$ satisfies $(\calR')$.
We first prove Theorem \ref{t-step-4}:\hspace{-0.02cm} either $\eval(\FF,\fD)$ is \#P-hard or
  $\fD$ satisfies conditions $(\calL_1')$ and $(\calL_2')$.\vspace{0.008cm}

Suppose $\eval(\FF,\fD)$ is not \#P-hard.
We use $(\CC,{\frak E})$ to denote the bipartisation of $(\FF,\fD)$.
The plan is to show that $(\CC,{\frak E})$ (together with appropriate
  $\pp',\bft'$ and $\fq'$) satisfies condition $(\calR)$.

To see this is the case we permute $\CC$ and ${\frak E}$
  using the following permutation $\Sigma$.
We index the rows (and columns) of $\CC$ and $\EE^{[r]}$ using
  $\{0,1\}\times \mathbb{Z}^2_{\dd}\times \mathbb{Z}_{\calQ}.$
We set $\Sigma(1,\yy)=(1,\yy)$ for all $\yy\in \mathbb{Z}_{\dd}^2\times \mathbb{Z}_{\calQ}$
  (that is, $\Sigma$ fixes pointwise the second half of the rows and columns),
  and $\Sigma(0,\xx)=(0,\xx')$, where $\xx'$ satisfies\vspace{0.04cm}
$$
x_{0,i,1}=W^{[i]}_{1,1}x'_{0,i,1}+W^{[i]}_{2,1}x'_{0,i,2},\ \ \
x_{0,i,2}=W^{[i]}_{1,2}x'_{0,i,1}+W^{[i]}_{2,2}x'_{0,i,
  2},\ \ \ \text{for all $i\in [g]$,}\vspace{0.1cm}
$$
and\vspace{-0.1cm}
$$x_{1,i,j}=k_{i,j}\cdot x'_{1,i,j},\ \ \ \text{for all $i\in [s]$ and $j\in [t_i]$.}$$
See ($\calR'$)  for definition of these symbols.\vspace{-0.005cm}

Before proving properties of $ \CC_{\Sigma,\Sigma}$ and ${\frak E}_{\Sigma}$,
  we need to verify that $\Sigma$ is indeed a permutation.
This follows from the fact that $\WW^{[i]}$, for every $i\in [g]$,
  is non-degenerate over $\mathbb{Z}_{d_i}$, and $k_{i,j}$, for all $i\in [s]$ and $j\in [t_i]$,
  satisfies $\gcd(k_{i,j},q_{i,j})=1$ (so the $\xx'$ above is unique).
We use $\Sigma_0$ to denote the $(0,*)$-part of $\Sigma$\vspace{0.01cm}
  and $I$ to denote the identity map: $$\Sigma(0,\xx)=(0,\Sigma_0(\xx))
  =(0,\xx'),\ \ \ \text{for all $\xx\in \mathbb{Z}_{\dd}^2\times \mathbb{Z}_{\calQ}$.}$$

Now we can write $\CC_{\Sigma,\Sigma}$ and ${\frak E}_{\Sigma}=\{\EE^{[0]}_{\Sigma}
  ,\ldots,\EE^{[N-1]}_\Sigma\}$ as
\begin{equation}\label{everything}
\CC_{\Sigma,\Sigma}=\left(\begin{matrix}\00\hspace{0.15cm} & \FF_{\Sigma_0,I}\\
\FF_{I,\Sigma_0} & \00\end{matrix}\right)\ \ \ \text{and}\ \ \
\EE^{[r]}_{\Sigma}=\left(\begin{matrix}\DD^{[r]}_{\Sigma_0} & \00\\
\00& \DD^{[r]}\end{matrix}\right),\ \ \ \text{for all $r\in [0:N-1]$.}
\end{equation}
We make the following observations:
\begin{enumerate}
\item[] \hspace{-0.6cm} Observation $1$: $\eval(\CC_{\Sigma,\Sigma},{\frak E}_{\Sigma})
  \equiv \eval(\CC,{\frak E})\le \eval(\FF,\fD)$,
  thus $\eval(\CC_{\Sigma,\Sigma},{\frak E}_{\Sigma})$ is not \#P-hard;\vspace{-0.08cm}
\item[] \hspace{-0.6cm} Observation $2$: $\FF_{\Sigma_0,I}$ satisfies (letting $\xx'=\Sigma_0(\xx)$)
\begin{eqnarray*}
\big(\FF_{\Sigma_0,I}\big)_{\xx,\yy}=F_{\Sigma_0(\xx),\yy}
  =F_{\xx',\yy}\hspace{-0.16cm}&=&\hspace{-0.18cm}\prod_{i\in [g]}\hspace{0.15cm}
\oo_{d_i}^{(x'_{0,i,1}\hspace{0.07cm} x'_{0,i,2})\cdot \WW^{[i]}
  \cdot (y_{0,i,1}\hspace{0.07cm} y_{0,i,2})^T}
\hspace{-0.2cm}\prod_{i\in [s],j\in [t_i]}
  \omega_{q_{i,j}}^{k_{i,j}\cdot x'_{1,i,j} y_{1,i,j}}\\[1.4ex]
  &=&\hspace{-0.18cm}\prod_{i\in [g]}\hspace{0.15cm}
\oo_{d_i}^{x_{0,i,1}y_{0,i,1}+ x_{0,i,2} y_{0,i,2} }
\hspace{-0.2cm}\prod_{i\in [s],j\in [t_i]}
  \omega_{q_{i,j}}^{x_{1,i,j} y_{1,i,j}}.
\end{eqnarray*}
\end{enumerate}

By Observation 2, it\vspace{0.003cm} is easy to show that
  $\CC_{\Sigma,\Sigma}$ and ${\frak E}_{\Sigma}$
  (together with appropriate $\qq',\bft',\fq'$) satisfy condition $(\calR)$.
Since $\eval(\CC_{\Sigma,\Sigma},{\frak E}_\Sigma)$, \vspace{0.002cm}by Observation 1,
  is not \#P-hard, it follows from Theorem \ref{bi-step-4} and (\ref{everything})
  that $\DD^{[r]}$, for all $r$, satisfy conditions $(\calL_2)$ and $(\calL_3)$.
This proves Theorem \ref{t-step-4} since $(\calL_1')$
  and $(\calL_2')$ follow directly from $(\calL_2)$ and $(\calL_3)$, respectively.\vspace{0.01cm}

We continue to prove Theorem \ref{nonbi-step-5}.\vspace{-0.004cm}
Suppose $\eval(\FF,\fD)$ is not \#P-hard, then the argument above shows that
  $(\CC_{\Sigma,\Sigma},{\frak E}_{\Sigma})$\vspace{0.01cm} (with appropriate $\pp',\bft'
  ,\fq'$) satisfies both $(\calR)$ and $(\calL)$.
Since by Observation $1$, $\eval(\CC_{\Sigma,\Sigma},{\frak E}_{\Sigma})$
  is not \#P-hard, by Theorem \ref{bi-step-5} and (\ref{everything}),
  $\DD^{[r]}$  satisfies  $(\calD_2)$ and $(\calD_4)$ for all $r\in \calZ$. \vspace{0.021cm}

Condition $(\calD_1')$ follows directly from $(\calD_2)$.
To prove $(\calD_2')$,
  we let $\FF'$ denote $\FF_{\Sigma_0,I}$.\vspace{-0.018cm}

By $(\calD_4)$, for any $r\in \calZ$, $k\in [s]$ and $\aa\in \Gamma_{r,k}^{\text{lin}}$,
  there exist $\bb\in \hat{\mathbb{Z}}_{\qq_k}$ and $\alpha\in \mathbb{Z}_N$ such that
$$
\oo_N^{\alpha}\cdot F_{\widetilde{\bb},
  \xx}'=D^{[r]}_{\xx+\widetilde{\aa}}\cdot \overline{
  D^{[r]}_{\xx}},\ \ \ \text{for all $\xx\in \Gamma_r$,\ \ where\ \ \
$\FF_{\widetilde{\bb},*}'=
  \FF_{\Sigma_0(\widetilde{\bb}),*}$.}$$
Also note that $\Sigma_0$ works within each prime factor, so there exists
  a $\bb'\in \hat{\mathbb{Z}}_{\qq_k}$ such that
  $\Sigma_0(\widetilde{\bb})=\widetilde{\bb'}$, and ($\calD_2'$) follows.

\section{Tractability: Proof of Theorem \ref{tractable-2}}

In this section, we prove Theorem \ref{tractable-2}.
The proof is almost the same as the one of Theorem \ref{tractable-1}
  for the bipartite case.

Let $((M,N),\FF,\fD,(\dd,\cal{W},\pp,\bft,\calQ,\calK))$ be a tuple that satisfies
    ($\calR'$),($\calL'$) and ($\calD'$).
The proof has the following two steps.
In the first step, we use ($\calR'$), ($\calL'$) and ($\calD'$)
  to decompose the problem $\eval(\FF,\fD)$
  into a collection of $s$ subproblems (Recall $s$ is the length of the sequence $\pp$):
$$
\eval(\FF^{[1]},\fD^{[1]}),\ldots,\eval(\FF^{[s]},\fD^{[s]}),
$$
such that, if every $\eval(\FF^{[i]},\fD^{[i]})$, $i\in [s]$, is tractable, then $\eval(\FF,\fD)$ is also
  tractable.
In the second step, we reduce $\eval(\FF^{[i]},\fD^{[i]})$, for every $i\in [s]$,
  to problem $\eval(\pi)$ for some prime power $\pi$.
Recall that $\eval(\pi)$ is the following problem: Given a quadratic polynomial
  $f(x_1,\ldots,x_n)$ over $\mathbb{Z}_\pi$,
  compute
$$
Z_\pi(f)=\sum_{x_1,\ldots,x_n\in \mathbb{Z}_\pi} \oo_{\pi}^{f(x_1,\ldots,x_n)}.
$$
By Theorem \ref{pinyan}, we have for any prime power $\pi$, problem $\eval(\pi)$ can be solved in polynomial time.
As a result, $\eval(\FF^{[i]},\fD^{[i]})$ is tractable for all $i\in [s]$, and so is
  $\eval(\FF,\fD)$.

\subsection{Step 1}

Fix $i$ to be any index in $[s]$.
We start by defining $\FF^{[i]}$ and $\fD^{[i]}$.
First recall the definition of  $\hat{\mathbb{Z}}_{\qq_i}$
 from Section~\ref{sec:Affine-Support}.

For any $\xx\in \hat{\mathbb{Z}}_{\qq_i}$,\vspace{-0.08cm}
  we use $\widetilde{\xx}$ to denote the vector $\yy\in
  \mathbb{Z}_{\dd}^2\times \mathbb{Z}_{\calQ}=\prod_{j=1}^s \hat{\mathbb{Z}}_{\qq_j}$
  such that
$$\yy_i=\xx\ \text{and}\ \yy_j=\00\ \text{for all $j\ne i$},\ \ \
  \text{where\ $\yy=(\yy_1,\ldots,\yy_s)$\ and\
  $\yy_j\in \hat{\mathbb{Z}}_{\qq_j}$}.$$
Then we define $\FF^{[i]}$.
$\FF^{[i]}$ is an $m_i\times m_i$ symmetric matrix,
  where $m_i= |\hat{\mathbb{Z}}_{\qq_i}|$.
We use $\hat{\mathbb{Z}}_{\qq_i}$ to index the rows and columns of $\FF^{[i]}$.
Then\vspace{-0.18cm}
$$
F^{[i]}_{\xx,\yy}=F_{\widetilde{\xx},\widetilde{\yy}},\ \ \ \text{for all
  $\xx,\yy\in \hat{\mathbb{Z}}_{\qq_i}$.}
$$
By condition ($\calR_3'$), it is easy to see that $\FF,\FF^{[1]},\ldots,\FF^{[s]}$ satisfy
\begin{equation}\label{dec1}
\FF=\FF^{[1]}\otimes \ldots\otimes \FF^{[s]}.
\end{equation}

Next, we define $\fD^{[i]}$.
$\fD^{[i]}=\{\DD^{[i,0]},\ldots,\DD^{[i,N-1]}\}$ is a sequence
  of $m_i\times m_i$ diagonal matrices:
$\DD^{[i,0]}$ is the $m_i\times m_i$ identity matrix; and for every $r\in [N-1]$, the
  $\xx^{th}$ entry, where $\xx\in \hat{\mathbb{Z}}_{\qq_i}$, of $\DD^{[i,r]}$ is
$$
D^{[i,r]}_{\xx}=D^{[r]}_{\ext_r(\xx)}.
$$
By condition ($\calD_1'$), we have
\begin{equation}\label{dec2}
\DD^{[r]}=\DD^{[1,r]}\otimes \ldots \otimes \DD^{[s,r]},\ \ \ \text{for
  all $r\in [0:N-1]$.}
\end{equation}

It then follows from (\ref{dec1}) and (\ref{dec2}) that
$$
Z_{\FF,\fD}(G)=Z_{\FF^{[1]},\fD^{[1]}}(G)\times \ldots\times
  Z_{\FF^{[s]},\fD^{[s]}}(G),\ \ \ \text{for all undirected graphs $G$.}
$$
As a result, we have the following lemma:\vspace{0.03cm}
\begin{lemma}
If $\eval(\FF^{[i]},\fD^{[i]})$ is tractable for all $i\in [s]$, then $\eval(\FF,\fD)$
  is also tractable.
\end{lemma}

We can use condition ($\calD_2'$) to prove the following lemma
  about the matrix $\DD^{[i,r]}$ (recall $\calZ$ is the set of $r\in [N-1]$
  such that $\DD^{[r]}\ne \00$, and $\Gamma_{r,i}$ is a coset in $\hat{\mathbb{Z}}_{\qq_i}$ for every $i\in [s]$,
  such that, $\Gamma_r=\prod_{i\in [s]}\Gamma_{r,i}$):
\begin{lemma}\label{useitonce}
Let $r\in \calZ$. Then for any $i\in [s]$, $\aa\in \Gamma_{r,i}^{\text{\rm lin}}$,
  there exist $\bb\in \hat{\mathbb{Z}}_{\qq_i}$ and $\alpha\in \mathbb{Z}_N$ such that
$$
D^{[i,r]}_{\xx+\aa}\cdot \overline{D^{[i,r]}_{\xx}}=\oo_N^\alpha\cdot F^{[i]}_{\bb,\xx},
  \ \ \ \text{for all $\xx\in \Gamma_{r,i}$.}
$$
\end{lemma}
\begin{proof}
By the definition of $\DD^{[i,r]}$, we have
$$
D^{[i,r]}_{\xx+\aa}\cdot \overline{D^{[i,r]}_{\xx}}=D^{[r]}_{\ext_r(\xx+\aa)}\cdot \overline{
  D^{[r]}_{\ext_r(\xx)}}=D^{[r]}_{\ext_r(\xx)+\widetilde{\aa}}\cdot \overline{
  D^{[r]}_{\ext_r(\xx)}}.
$$
Then by condition ($\calD_2'$), we know there exist $\bb\in \hat{\mathbb{Z}}_{\qq_i}$ and
  $\alpha\in \mathbb{Z}_N$ such that
$$
D^{[i,r]}_{\xx+\aa}\cdot \overline{D^{[i,r]}_{\xx}}=\oo_N^\alpha\cdot
  F_{\widetilde{\bb},\ext_r(\xx)}=\oo_N^{\alpha}\cdot  F^{[i]}_{\bb,\xx},
  \ \ \ \text{for all $\xx\in \Gamma_{r,i}$,}
$$
and the lemma is proven.
\end{proof}

\subsection{Step 2}

\def\bpi{\boldsymbol{\pi}}

Now we let $\eval(\FF,\fD)$ denote one of the subproblems $\eval(\FF^{[i]},\fD^{[i]})$
  we defined in the last step.
By conditions $(\calR')$, ($\calL'$), ($\calD'$) and Lemma \ref{useitonce},
  we summarize the properties of $(\FF,\fD)$ as follows.
We will use these properties to show that $\eval(\FF,\fD)$ is tractable.
\begin{enumerate}
\item[($\calF_1'$)] There exist a prime $p$ and a sequence $\boldsymbol{\pi}
  =(\pi_1\ge \pi_2\ge \ldots \ge \pi_h)$ of powers of $p$.\label{CONDITIONFP}
  $\FF$ is an $m\times m$ symmetric matrix, where $m= \pi_1\pi_2\ldots\pi_h$.
We let $\pi$ denote $\pi_1$ and use $\mathbb{Z}_{\bpi}\equiv\mathbb{Z}_{\pi_1}\times
  \ldots\times \mathbb{Z}_{\pi_h}$
to index the rows and columns of $\FF$.
We also let $\calT$ denote the set of pairs $(i,j)\in [h]\times [h]$
  such that $\pi_i=\pi_j$.
Then there exist $c_{i,j}\in \mathbb{Z}_{\pi_i}=\mathbb{Z}_{\pi_j}$
  for all $(i,j)\in \calT$ such that $c_{i,j}=c_{j,i}$ and
$$
F_{\xx,\yy}=\prod_{(i,j)\in \calT} \oo_{\pi_i}^{c_{i,j}x_iy_j},\ \ \ \text{
  for all $\xx=(x_1,\ldots,x_h),
  \yy=(y_1,\ldots,y_h)\in \mathbb{Z}_{\bpi}$},
$$
where we use $x_i\in \mathbb{Z}_{\pi_i}$ to denote the $i^{th}$ entry of $\xx$
(The reason we express $\FF$ in this very general form is to unify the
  proofs for the two slightly different cases:
  $(\FF^{[1]},\fD^{[1]})$ and $(\FF^{[i]},\fD^{[i]})$, $i\ge 2$);

\item[($\calF_2'$)] $\fD=\{\DD^{[0]},\ldots,\DD^{[N-1]}\}$ is a sequence of
  $N$ $m\times m$ diagonal matrices, for some positive integer $N$ with $\pi\hspace{0.06cm}
  |\hspace{0.06cm}N$.
$\DD^{[0]}$ is the identity matrix; and every diagonal entry of $\DD^{[r]}$,
   $r\in [N-1]$, is either $0$ or a power of $\oo_N$.
We also use $\mathbb{Z}_{\bpi}$ to index the diagonal entries of $\DD^{[r]}$;

\item[($\calF_3'$)] For every $r\in [0:N-1]$, we let $\Gamma_r$ denote
  the set of $\xx\in \mathbb{Z}_{\bpi}$ such that $D^{[r]}_\xx\ne 0$,
  and let $\calZ$ denote the set of $r$ such that $\Gamma_r\ne \emptyset$.
For every $r\in \calZ$, $\Gamma_r$ is a coset in $\mathbb{Z}_{\bpi}$.
Moreover,\vspace{-0.04cm} for every $r\in \calZ$, there exists a vector
  $\fa^{[r]}\in \Gamma_r$ such that $D^{[r]}_{\fa^{[r]}}=1$;

\item[($\calF_4'$)] For all $r\in \calZ$ and $\aa\in \Gamma_r^{\text{lin}}$,
  there exist $\bb\in \mathbb{Z}_{\bpi}$ and $\alpha\in \mathbb{Z}_N$ such that
$$
D^{[r]}_{\xx+\aa}\cdot \overline{D^{[r]}_{\xx}}=\oo_N^{\alpha}\cdot
  F_{\bb,\xx},\ \ \ \text{for all $\xx\in \Gamma_r$.}
$$
\end{enumerate}

Now let $G$ be an undirected graph. Below we will reduce
  the computation of $Z_{\FF,\fD}(G)$ to $\eval(\hpi)$,\vspace{-0.05cm}
$$
\text{where}\ \hpi=\pi\ \text{if $p\ne 2$,\ \ and}\ \hpi=2\pi\ \text{if $p=2$.}
$$
Given $a\in \mathbb{Z}_{\pi_i}$ for some $i\in [h]$, we use $\widehat{a}$ to
  denote an element in $\mathbb{Z}_{\hpi}$ such that $\widehat{a}\equiv
  a\pmod{\pi_i}$.
For definiteness we can choose $a$ itself if we consider $a$
  to be an integer between $0$ and $\pi_i-1$.

Let $G=(V,E)$. We let $V_r$, $r\in [0:N-1]$, denote
  the set of vertices in $V$ whose degree is $r \bmod N$.
We further decompose $E$ into $\cup_{i\le j\in [0:N-1]} E_{i,j}$,
  where $E_{i,j}$ contains the edges between $V_i$ and $V_j$.\vspace{0.025cm}

It is clear that if $V_r\ne \emptyset$ for some $r\notin \calZ$, then
  $Z_{\FF,\fD}(G)$ is trivially $0$.
As a result, we assume $V_r=\emptyset$ for all $r\notin \calZ$.
In this case, we have
$$
Z_{\FF,\fD}(G)=\sum_{\xi}\hspace{0.05cm}\left[\prod_{r\in \calZ}
\left(\prod_{v\in V_r} D^{[r]}_{\xx_{v}}\right)\right]\left[
\prod_{r\le r'\in \calZ}\left(\prod_{uv\in E_{r,r'}}
F_{\xx_{u},\xx_{v}}\right)\right],
$$
where the sum ranges over all assignments $\xi=(\xi_r:V_r\rightarrow
  \Gamma_r\hspace{0.08cm}|\hspace{0.08cm}r\in \calZ)$
  such that $\xi (v )=\xx_{v }$.

Next by using Lemma \ref{quiteuseful},
  we know for every $r\in \calZ$, there exist a positive integer $s_r$
  and an $s_r\times h$ matrix $\AA^{[r]}$ over $\mathbb{Z}_{\hpi}$
  which gives us a \emph{uniform} map $\gamma^{[r]}$ (see Lemma \ref{quiteuseful}
  for definition) from $\mathbb{Z}_{\hpi}^{s_r}$ to $\Gamma_r$:
$$
\gamma^{[r]}_i(\xx)=\left(\xx\AA^{[r]}_{*,i}+\widehat{\frak a}^{[r]}_i
  \hspace{-0.15cm}\pmod{\pi_i}\right), \ \ \ \text{for all $i\in [h]$.}
$$
Recall that for every $r\in \calZ$,
 $\fa^{[r]}$ is a vector in $\Gamma_r$ such that $D^{[r]}_{\fa^{[r]}}=1$.
Thus, $$D^{[r]}_{\gamma^{[r]}(\00)}=1.$$
Since $\gamma^{[r]}$ is uniform,  and we know the multiplicity of
this map, in order to compute $Z_{\FF,\fD}(G)$,
 it suffices to compute
$$
\sum_{(\xx_v)}\hspace{0.05cm}\left[\prod_{r\in \calZ}\left(\prod_{v\in V_r}
  D^{[r]}_{\gamma^{[r]}(\xx_{v})}\right)\right]\left[\prod_{r\le r'\in \calZ}
  \left(\prod_{uv\in E_{r,r'}} F_{\gamma^{[r]}(\xx_{u}),
  \gamma^{[r']}(\xx_{v})}\right)\right],
$$
where the sum is over $$\big(\xx_{v}\in \mathbb{Z}_{\hpi}^{s_r}:v \in V_r,r\in \calZ\big)=
  \prod_{r\in \calZ}(\mathbb{Z}_{\hpi}^{s_r})^{|V_r|}.$$

If we can show for every $r\in \calZ$, there is a quadratic polynomial
  $f^{[r]}$ over $\mathbb{Z}_{\hpi}$, such that,
\begin{equation}\label{finaleq1}
D^{[r]}_{\gamma^{[r]}(\xx)}=\oo_{\hpi}^{f^{[r]}(\xx)},\ \ \ \text{for all $\xx\in
  \mathbb{Z}_{\hpi}^{s_r}$,}
\end{equation}
and for all $r\le r'\in \calZ$, there is a quadratic polynomial $f^{[r,r']}$
  over $\mathbb{Z}_{\hpi}$, such that,
\begin{equation}\label{finaleq2}
F_{\gamma^{[r]}(\xx),\gamma^{[r']}(\yy)}=\oo_{\hpi}^{
  f^{[r,r']}(\xx,\yy)},\ \ \ \text{for all $\xx\in \mathbb{Z}_{\hpi}^{s_r}$
    and $\yy\in \mathbb{Z}_{\hpi}^{s_{r'}}$},
\end{equation}
then we can reduce the computation of $Z_{\FF,\fD}(G)$ to $\eval(\hpi)$ and finish the proof.

First, we prove the existence of the quadratic polynomial $f^{[r,r']}$.
By condition ($\calF_1'$), the following function $f^{[r,r']}$ satisfies (\ref{finaleq2}):
$$
f^{[r,r']}(\xx,\yy)=
  \sum_{(i,j)\in \calT} \left(\frac{\hpi}{\pi_i}\right)\cdot c_{i,j}\cdot \gamma_i^{[r]}(\xx)
    \cdot \gamma_j^{[r']}(\yy)= \sum_{(i,j)\in \calT}\widehat{c}_{i,j}
    \left(\frac{\hpi}{\pi_i}\right)
    \left(\xx\AA^{[r]}_{*,i}+\widehat{\frak a}^{[r]}_i\right)\left(
    \yy\AA^{[r']}_{*,j}+\widehat{\frak a}^{[r']}_j\right).
$$
Note that $(i,j)\in \calT$ implies that $\pi_i=\pi_j$ and thus,\vspace{0.0cm}
  $$\gamma_i^{[r]}(\xx),\hspace{0.06cm}
  \gamma_j^{[r']}(\yy)\in \mathbb{Z}_{\pi_i}=\mathbb{Z}_{\pi_j}.$$
The presence of $\hpi/\pi_i$ is crucial to be able to substitute the
  $\bmod\hspace{0.06cm} \pi_i$\vspace{-0.08cm} expressions
  for $\gamma_i^{[r]}(\xx)$ and $\gamma_j^{[r']}(\yy)$,
  as if they were $\bmod\hspace{0.06cm} \hpi$ expressions.
It is clear that $f^{[r,r']}$ is a quadratic polynomial over $\mathbb{Z}_{\hpi}$.\vspace{-0.06cm}

Next we prove the existence of the quadratic polynomial $f^{[r]}$.
Let us fix $r$ to be an index in $\calZ$. We use $\ee_i$, $i\in [s_r]$, to denote the
  vector in $\mathbb{Z}_{\hpi}^{s_r}$ whose $i^{th}$ entry is $1$
  and all other entries are $0$.
By $(\calF_4')$, we know for every $i\in [s_r]$,
  there exist $\alpha_i\in \mathbb{Z}_N$ and $\bb_i=(b_{i,1},...,b_{i,h})\in \mathbb{Z}_{\bpi}$,
  where $b_{i,j}\in \mathbb{Z}_{\pi_j}$, such that
$$
D^{[r]}_{\gamma^{[r]}(\xx+\ee_i)}\cdot \overline{D^{[r]}_{\gamma^{[r]}(\xx)}}
  =\oo_N^{\alpha_i}\cdot \prod_{j\in [h]} \oo_{\pi_j}^{b_{i,j}\cdot \gamma^{[r]}_j(\xx)},
  \ \ \ \text{for all $\xx\in \mathbb{Z}_{\hpi}^{s_r}$.}
$$
We have this equation because $\gamma^{[r]}(\xx+\ee_i)-\gamma^{[r]}(\xx)$
  is a vector in $\mathbb{Z}_{\bpi}$ that is independent of $\xx$.\vspace{0.007cm}

By the same argument we used in the proof of Theorem \ref{tractable-1}
  ((\ref{checkbla1}) and (\ref{checkbla2}), more exactly),
  one can show that $\oo_N^{\alpha_i}$ must be a power of $\oo_{\hpi}$,
  for all $i\in [s_r]$.
As a result, there exists $\beta_i\in \mathbb{Z}_{\hpi}$ such that
\begin{equation}\label{finaleq3}
D^{[r]}_{\gamma^{[r]}(\xx+\ee_i)}\cdot \overline{D^{[r]}_{\gamma^{[r]}(\xx)}}
  =\oo_{\hpi}^{\beta_i}\cdot \prod_{j\in [h]} \oo_{\pi_j}^{b_{i,j}\cdot \gamma^{[r]}_j(\xx)},
  \ \ \ \text{for all $\xx\in \mathbb{Z}_{\hpi}^{s_r}$.}
\end{equation}
Again, by the argument we used in the proof of Theorem \ref{tractable-1},
  every non-zero entry of $\DD^{[r]}$ must be a power of $\oo_{\hpi}$.
Therefore, there does exist a function $f^{[r]}$ from $\mathbb{Z}_{\hpi}^{s_r}$
  to $\mathbb{Z}_{\hpi}$ that satisfies (\ref{finaleq1}).
To see $f^{[r]}$ is a quadratic polynomial, by (\ref{finaleq3}), we have
  for every $i\in [s_r]$,
$$
f^{[r]}(\xx+\ee_i)-f^{[r]}(\xx)=\beta_i+\sum_{j\in [h]}\left(\widehat{b}_{i,j}
  \cdot\left(\frac{\hpi}{\pi_j}\right)
  \left(\xx\AA^{[r]}_{*,j}+\widehat{\frak a}^{[r]}_j\right)\right)
,\ \ \ \text{for all $i\in [s_r]$ and $\xx\in \mathbb{
  Z}_{\hpi}^{s_r},$}
$$
which is an affine linear form of $\xx$ with all coefficients from $\mathbb{Z}_{\hpi}$.

By using Lemma \ref{quadraticlemma1} and Lemma \ref{quadraticlemma2},
  we can prove that $f^{[r]}$ is a quadratic polynomial over $\mathbb{Z}_{\hpi}$,
  and this finishes the reduction from $\eval(\FF,\fD)$ to $\eval(\hpi)$.

\input{Decidability}

\section{Acknowledgements}

{We would like to thank
  Miki Ajtai,
  Al Aho,
  Sanjeev Arora,
  Dick Askey,
  Paul Beame,
  Richard Brualdi,
  Andrei Bulatov,
  Xiaotie Deng,
  Alan Frieze,
  Martin Grohe,
  Pavol Hell,
  Lane Hemaspaandra,
  Kazuo Iwama,
  Gabor Kun,
  Dick Lipton,
  Tal Malkin,
  Christos Papadimitriou,
  Mike Paterson,
  Rocco Servedio,
  Endre Szemer\'edi,
  Shang-Hua Teng,
  Joe Traub,
  Osamu Watanabe,
  Avi Wigderson, and
  Mihalis Yannakakis,
  for their interest and many comments.
We thank especially Martin Dyer, Leslie Goldberg, Mark Jerrum, Marc Thurley, Leslie Valiant, and Mingji Xia
  for in-depth discussions.}

\newpage
\begin{flushleft}
\bibliography{Reference}
\end{flushleft}

\end{document}

%% file: Introduction.tex

\section*{Index of Conditions and Problem Definitions\vspace{0.26cm}}
\
\begin{tabular}{lll}
 \ \ \ \ & $(\text{\sl Pinning})$\ \ \ \ \ \ \ \    &     p.\hspace{0.06cm}\pageref{PINNINGCONDITION} \\[0.8ex]
&$(\calU_1)$\hspace{0.05cm}--\hspace{0.05cm}$(\calU_4)$  &  p.\hspace{0.06cm}\pageref{CONDITIONU} \\[0.8ex]
&$(\calU_5)$ & p.\hspace{0.06cm}\pageref{CONDITIONU5} \\[0.8ex]
&$(\calR_1)$\hspace{0.05cm}--\hspace{0.05cm}$(\calR_3)$  &  p.\hspace{0.06cm}\pageref{CONDITIONR} \\[0.8ex]
&$(\calL_1)$\hspace{0.05cm}--\hspace{0.05cm}$(\calL_3)$  &  p.\hspace{0.06cm}\pageref{CONDITIONL} \\[0.8ex]
&$(\calD_1)$\hspace{0.05cm}--\hspace{0.05cm}$(\calD_4)$  &  p.\hspace{0.06cm}\pageref{CONDITIOND} \\[0.8ex]
&$(\calU'_1)$\hspace{0.05cm}--\hspace{0.05cm}$(\calU'_4)$  &  p.\hspace{0.06cm}\pageref{CONDITIONUP} \\[0.8ex]
&$(\calU'_5)$  &  p.\hspace{0.06cm}\pageref{CONDITIONUP5} \\[0.8ex]
&$(\calR_1')$\hspace{0.05cm}--\hspace{0.05cm}$(\calR_3')$  &  p.\hspace{0.06cm}\pageref{CONDITIONRP} \\[0.8ex]
&$(\calL'_1)$\hspace{0.05cm}--\hspace{0.05cm}$(\calL'_2)$  &  p.\hspace{0.06cm}\pageref{CONDITIONLP} \\[0.8ex]
&$(\calD'_1)$\hspace{0.05cm}--\hspace{0.05cm}$(\calD'_2)$ & p.\hspace{0.06cm}\pageref{CONDITIONDP} \\[0.8ex]
&$(\calT_1)$\hspace{0.05cm}--\hspace{0.05cm}$(\calT_3)$  &  p.\hspace{0.06cm}\pageref{CONDITIONT} \\[0.8ex]
&$(\calS_1)$  &  p.\hspace{0.06cm}\pageref{CONDITIONS1} \\[0.8ex]
&$(\calS_2)$\hspace{0.05cm}--\hspace{0.05cm}$(\calS_3)$  &  p.\hspace{0.06cm}\pageref{CONDITIONS23} \\[0.8ex]
&({\sl Shape}$_1$)\hspace{0.05cm}--\hspace{0.05cm}({\sl Shape}$_5$)  &  p.\hspace{0.06cm}\pageref{SHAPECONDITION} \\[0.8ex]
&({\sl Shape}$_6$)  &  p.\hspace{0.06cm}\pageref{SHAPECONDITION6} \\[0.8ex]
&(\GC)  &  p.\hspace{0.06cm}\pageref{CONDITIONGC} \\[0.8ex]
&$(\calF_1)$\hspace{0.05cm}--\hspace{0.05cm}$(\calF_4)$  &  p.\hspace{0.06cm}\pageref{CONDITIONF14} \\[0.8ex]
&$(\calS'_1)$\hspace{0.05cm}--\hspace{0.05cm}$(\calS_2')$  &  p.\hspace{0.06cm}\pageref{CONDITIONSP} \\[0.8ex]
&({\sl Shape}$_1'$)\hspace{0.05cm}--\hspace{0.05cm}({\sl Shape}$_6'$) &  p.\hspace{0.06cm}\pageref{SHAPECONDITIONP} \\[0.8ex]
&$(\calF'_1)$\hspace{0.05cm}--\hspace{0.05cm}$(\calF'_4)$  &  p.\hspace{0.06cm}\pageref{CONDITIONFP} \\[0.8ex] \\

&$Z_\AA(G)$ and $\eval(\AA)$ & p.\hspace{0.06cm}\pageref{Z-def} \\[0.9ex]
&$Z_{\CC,\fD}(G)$ and $\eval(\CC,\fD)$ & p.\hspace{0.06cm}\pageref{ZCD}\\[0.9ex]
&$Z_{\CC,\fD}^{\rightarrow}(G,u)$ & p.\hspace{0.06cm}\pageref{ZCDARROW}\\[0.9ex]
&$Z_{\CC,\fD}^{\leftarrow}(G,u)$ & p.\hspace{0.06cm}\pageref{ZCDARROW}\\[0.9ex]
&$Z_{\AA}(G,w,k)$ and $\evalp(\AA)$ & p.\hspace{0.06cm}\pageref{ZAPINNING} \\[0.9ex]
&$Z_q(f)$ and $\eval(q)$ & p.\hspace{0.06cm}\pageref{ZQF}\\[0.9ex]
&$Z_{\AA}(G,w,S)$ and $\eval(\AA,S)$ & p.\hspace{0.06cm}\pageref{ZAGWS}\\[0.9ex]
&$Z_{\CC,\fD}(G,w,k)$ and $\evalp(\CC,\fD)$ & p.\hspace{0.06cm}\pageref{ZCDI}\\[0.9ex]
&$Z_{\CC,\fD}(G,w,S)$ and $\eval(\CC,\fD,S)$\ \ \ \ \ \ \ \  & p.\hspace{0.06cm}\pageref{ZCDS}\\[0.9ex]
&$\COUNT(\AA)$ & p.\hspace{0.06cm}\pageref{COUNTA}\\
\end{tabular}
\newpage

\section{Introduction}\label{sec1}

Graph homomorphism has been studied intensively over the
years~\cite{lovasz67,HellBook,DyerGreenhill,freedman-l-s,BulatovGrohe,acyclic,GoldbergGJT}.
Given~two graphs $G$~and $H$,
a graph homomorphism from $G$ to $H$ is a map $f$ from
the vertex set $V(G)$ to $V(H)$
such that, whenever $(u,v)$ is an edge in $G$,
$(f(u), f(v))$ is an edge in $H$.
The counting problem for graph homomor\-phism
is to compute the number of homomorphisms from $G$  to $H$.
For a fixed graph $H$, this problem is also known as the \#$H$-coloring problem.
In 1967, Lov\'{a}sz~\cite{lovasz67} proved that $H$ and $H'$ are isomorphic
if and only if for all $G$, the number of homomorphisms from $G$  to $H$ and
from  $G$ to $H'$ are the same.
Graph homomorphism and the associated partition function defined
  below provide an
elegant and general notion of \emph{graph
  properties} \cite{HellBook}.\vspace{0.003cm}

In this paper, all graphs considered are undirected.
We follow standard definitions:
$G$ is allowed to have multiple edges; $H$ can have loops,
  multiple edges, and more generally, edge weights.
(The standard definition of graph homomorphism does not allow
  self-loops for $G$.
However, our result is stronger: We prove polynomial-time tractability
  even for input graphs $G$ with self-loops; and
  at the same time, our hardness results hold for the more
  restricted case of $G$ with no self-loops.)
Formally, we use $\AA$ to denote an $m\times m$ symmetric matrix with entries $(A_{i,j})$,
  $i,j\in [m]=\{1,2,\ldots,m\}$.
Given any undirected graph $G=(V,E)$, we define the graph homomorphism function
\begin{equation}\label{Z-def}
Z_\AA(G)=\sum_{\xi:V\rightarrow [m]}\hspace{0.06cm} \prod_{(u,v)\in E} A_{\xi(u),\xi(v)}.
\end{equation}
This is also called the {\it partition function}
from statistical physics.\vspace{0.006cm}

Graph homomorphism can express many {natural graph properties}.
For example, if we take $H$ to be
  the graph over two vertices $\{0, 1\}$ with an edge $(0,1)$
  and a loop at $1$, then a graph homomorphism from $G$ to $H$
corresponds to a {\sc Vertex Cover} of $G$, and the counting problem
simply counts the number of vertex covers.
As another example, if $H$ is the complete graph over $k$ vertices
(without self-loops), then the problem is exactly the {\sc $k$-Coloring}
problem for $G$.  Many additional graph invariants can be expressed
as $Z_\AA(G)$ for appropriate $\AA$.
Consider the Hadamard matrix
\begin{equation}\label{thehadamard}
{\bf H} =
{\begin{pmatrix} 1 & 1 \\ 1 & -1 \end{pmatrix}},
\end{equation}
where we index the rows and columns by $\{0, 1\}$.
In $Z_{\bf H}(G)$, every product
$$\prod_{(u,v)\in E} H_{\xi(u),\xi(v)} \in
\big\{1, -1\big\},$$
and is $-1$ precisely when the induced subgraph of $G$ on $\xi^{-1}(1)$
has an odd number of edges. Therefore, $$\Big( 2^n - Z_{\bf H}(G)\Big)\Big/2$$
is the number of induced subgraphs of $G$ with an odd number of edges.
Also expressible as $Z_\AA(\cdot)$ are $S$-\-flows where $S$ is a subset
of a finite Abelian group closed under inversion~\cite{freedman-l-s},
and (a scaled version of) the Tutte polynomial $\hat{T}(x,y)$
 where $(x-1)(y-1)$ is a positive integer.
In \cite{freedman-l-s}, Freedman, Lov\'{a}sz and Schrijver characterized  what graph
functions can be expressed as $Z_\AA(\cdot)$.\vspace{0.008cm}

In this paper, we study the complexity of the partition function $Z_\AA(\cdot)$
  where $\AA$ is an \emph{arbitrary fixed symmetric matrix over the algebraic complex numbers}.
Throughout the paper, we let $\mathbb{C}$ denote the set of
  algebraic complex numbers, and refer to them
  simply as complex numbers when it is clear from the context.
More discussion on the model of computation can be found in
  Section \ref{complexmodel}.

The complexity question of $Z_\AA(\cdot)$ has been intensively studied.
Hell and Ne\v{s}et\v{r}il first studied
  the $H$-coloring problem \cite{Hell,HellBook} (that is, given an undirected graph $G$,
  decide whether there exists a graph homomor\-phism from $G$ to $H$)
and proved that for any fixed undirected graph $H$, the problem is either in
  polynomial time or NP-complete.
Results of this type are called \emph{complexity dichotomy theorems}.
They state that
  every member of the class of problems concerned is either tractable (i.e., solvable in P)
  or intractable (i.e., NP-hard or \#P-hard depending on whether
  it is a decision or counting problem).
This includes the well-known
  Schaefer's theorem~\cite{Schaefer78} and more generally the study on
  constraint satisfaction problems (CSP in short)~\cite{CSPbook}.
In particular, the famous dichotomy conjecture by Vardi and Feder~\cite{Feder-Vardi}
  on decision CSP motivated much of subsequent work.\vspace{0.006cm}

In \cite{DyerGreenhill} Dyer and Greenhill studied the counting version of the
  $H$-coloring problem. They proved that for any fixed symmetric $\{0,1\}$-matrix $\AA$,
  computing $Z_\AA(\cdot)$ is either in P or \#P-hard.
Bulatov and Grohe \cite{BulatovGrohe} then gave a sweeping generalization of
  this theorem to all non-negative symmetric matrices $\AA$
  (see Theorem \ref{basicsharp} for the precise statement).
They obtained an elegant dichotomy theorem, which basically says that $Z_\AA(\cdot)$
  is computable in P if each {\it block} of $\AA$ has rank at most one, and is \#P-hard otherwise.
More precisely,
  decompose $\AA$ as a direct sum of $\AA_i$ which
correspond to the connected components $H_i$ of
the undirected graph $H$ defined by the nonzero entries of $\AA$.
Then, $Z_\AA(\cdot)$ is computable in P if every
$Z_{\AA_i}(\cdot)$ is, and \#P-hard otherwise.
For each non-bipartite $H_i$,
the corresponding $Z_{\AA_i}(\cdot)$ is computable in P if
$\AA_i$ has rank at most one, and is \#P-hard otherwise.
For each bipartite $H_i$,
the corresponding $Z_{\AA_i}(\cdot)$ is computable in P if
  $\AA_i$ has the following form:\vspace{-0.06cm}
$$\AA_i = {
\begin{pmatrix} {\bf 0} & \BB_i \\ \BB_i^T & 0 \end{pmatrix}},\vspace{-0.06cm}$$
where $\BB_i$ has rank one, and is \#P-hard otherwise.
\vspace{0.006cm}

The result of Bulatov and Grohe is both sweeping and enormously applicable.
It completely solves the problem for all
  non-negative symmetric matrices.
However, when we are dealing with non-negative matrices,
  there are no cancellations in the exponential sum $Z_\AA(\cdot)$.
These potential cancellations, when $\AA$ is either a real or
  a complex matrix, may in fact be the source of surprisingly efficient algorithms
  for computing $Z_\AA(\cdot)$.
The occurrence of these cancellations, or the mere possibility of
  such occurrence, makes proving any complexity dichotomies more difficult.
Such a proof must identify all polynomial-time algorithms utilizing
  the potential cancellations, such as those found in holographic
  algorithms \cite{AA_FOCS,HA_J,STOC07}, and at the same time carves
  out exactly what is left.
This situation is similar to \emph{monotone} versus \emph{non-monotone}
  circuit complexity.
It turns out that indeed there are more interesting tractable
  cases over the reals 
  and in particular, the $2\times 2$ Hadamard matrix $\HH$ in (\ref{thehadamard})
  turns out to be one of such cases.
This is the starting point of the next great chapter on the
  complexity of $Z_\AA(\cdot)$. \vspace{0.006cm}


In a paper \cite{GoldbergGJT} comprising $73$ pages of beautiful proofs of  both
exceptional depth and conceptual vision,
Goldberg, Jerrum, Grohe, and Thurley proved a complexity dichotomy
theorem for all the real-valued symmetric matrices $\AA$.
Their result is too intricate to give a short
and accurate summary here
but essentially it states that the problem
of computing  $Z_\AA(G)$ for any real $\AA$
is either in P or \#P-hard.  Again, which case it is
depends on the connected components
of $\AA$.
The overall statement remains  that  $Z_\AA(G)$ is
tractable if every
connected component of $\AA$ is, and is \#P-hard  otherwise.
However, the exact description of tractability for  connected $\AA$
is much more technical and involved.
The Hadamard matrix  ${\bf H}$ and
its tensor products ${\bf H} \otimes {\bf H} \otimes \cdots \otimes {\bf H}$
play a major role in the tractable case.
If we index rows and columns of ${\bf H}$
by the finite field $\mathbb{Z}_2$,
then its $(x,y)$ entry is $(-1)^{xy}$.  For the non-bipartite
case, there is another $4 \times 4$ symmetric matrix ${\bf H}_4$,
different from ${\bf H} \otimes {\bf H}$,
where the rows and columns are indexed by $(\mathbb{Z}_2)^2$,
and the entry at $((x_1, x_2), (y_1, y_2))$ is
$$(-1)^{x_1 y_2 + x_2 y_1}.$$
These two matrices, and their arbitrary tensor products,
all correspond to new tractable $Z_\AA(\cdot)$.
In fact, there are some more tractable cases, starting
with what  can be
roughly described as certain rank one modifications
on these tensor products.\vspace{0.006cm}

The proof of \cite{GoldbergGJT} proceeds by establishing a long sequence of
successively more stringent properties that a tractable $\AA$
must satisfy. Ultimately, it arrives at a point where
satisfaction of these properties implies that
$Z_\AA(G)$ can be computed as
\[\sum_{x_1, x_2, \ldots, x_n \in \mathbb{Z}_2}
(-1)^{f_G(x_1, x_2, \ldots, x_n)},\]
where $f_G$ is a quadratic polynomial over $\mathbb{Z}_2$.
This sum is known to be computable in poly\-nomial time in $n$~\cite[Theorem 6.30]{Carlitz,field},
  the number of variables.
In hindsight, the case with the simplest Hadamard matrix
${\bf H}$
which was
an obstacle to the Bulatov-Grohe dichotomy theorem
and was left open
for some  time, could have been directly solved, if
one had adopted the polynomial view point of \cite{GoldbergGJT}.\vspace{0.008cm}

While positive and negative real numbers provide the possibility
  of cancellations, over the complex domain, there is a
  significantly richer variety of possible cancellations.
We independently came to the tractability of $Z_{\bf H}(\cdot)$, with $\HH$
  being the $2\times 2$ Hadamard matrix, from a
slightly different angle. In \cite{holant}, we were studying a
certain type of constraint satisfaction problem. This is motivated
by investigations of a class of counting problems called {\it
Holant} Problems, and it is connected with the technique called
holographic reductions introduced by Valiant~\cite{HA_FOCS,AA_FOCS}.
Let us briefly describe this framework.\vspace{0.008cm}

A {\it signature grid}
$\Omega = (G, {\cal F})$ is a tuple in which $G=(V,E)$ is a graph and
each $v \in V$ is attached a function $F_v \in {\cal F}$. An edge
assignment $\sigma$ for every $e\in E$ gives an evaluation
$$
\prod_{v\in  V} F_v\big(\sigma\hspace{0.03cm}|_{\hspace{0.05cm}E(v)}\big),
$$
where $E(v)$ is the incident edges of $v$.
 The  counting problem on an input instance  $\Omega$
is to compute
$$
{\rm Holant}(\Omega)=\sum_{\text{edge assignments~} \sigma~}
\prod_{v\in V} F_v\big(\sigma\hspace{0.03cm}|_{\hspace{0.05cm}E(v)}\big).
$$
For example, if we take $\sigma: E \rightarrow \{0,1\}$,
and attach the {\sc Exact-One} function at every
vertex $v\in V$, then ${\rm Holant}(\Omega)$ is exactly the number of
perfect matchings. Incidentally, Freedman, Lov\'{a}sz, and Schrijver
  showed in \cite{freedman-l-s} that counting perfect matchings {\it cannot} be expressed
as $Z_\AA(\cdot)$ for any matrix $\AA$ over $\mathbb{R}$.
However, every function $Z_\AA(\cdot)$ (vertex assignment) {\it can} be
simulated by ${\rm Holant}(\cdot)$ (edge assignment) as follows:
$\AA$ defines a function of arity 2 for every edge of $G$.
Consider the bipartite Vertex-Edge incident graph $G' = (V(G), E(G), E')$
of $G$, where $(v,e) \in E'$ iff $e$ is incident to $v$ in $G$.
Attach the {\sc Equality} function at every $v \in V(G)$ and
the function defined by $\AA$ at every $e \in E(G)$.
This defines a signature grid $\Omega$ with the underlying graph $G'$.
Then $Z_\AA(G) = {\rm Holant}(\Omega)$.

Denote a symmetric function on boolean variables $x_1,x_2, \ldots, x_n$
by $[f_0, f_1,
\ldots, f_n]$, where $f_i$ is the
value on inputs of Hamming  weight  $i$.
Thus the {\sc Exact-One} function is
$[0, 1, 0, \ldots, 0]$, and ${\bf H}$ is just
$[1, 1, -1]$.\vspace{0.012cm}
We discovered that the following three families of\vspace{0.03cm}
functions
\begin{eqnarray*}
&&{\cal F}_1=\big\{\hspace{0.08cm} \lambda ([1, 0]^{\otimes k} + i^r[0,
\hspace{0.3cm}1]^{\otimes k} )
 \hspace{0.1cm}\big|\hspace{0.1cm} \lambda  \in \mathbb{C},
 \hspace{0.07cm}  k = 1, 2, \ldots, \mbox{ and }
 r = 0, 1, 2, 3\hspace{0.08cm}\big\};\\[0.5ex]
&&{\cal F}_2=\big\{\hspace{0.08cm} \lambda ([1, 1]^{\otimes k} + i^r[1, -1]^{\otimes k} )
 \hspace{0.1cm}\big|\hspace{0.1cm} \lambda  \in \mathbb{C},
 \hspace{0.07cm}k = 1, 2, \ldots, \mbox{ and }
 r = 0, 1, 2, 3\hspace{0.08cm}\big\};\\[0.5ex]
&&{\cal F}_3=\{\hspace{0.08cm} \lambda ([1,\;i]^{\otimes k} + i^r[1,
  \hspace{0.05cm}-i]^{\otimes k} )
 \hspace{0.1cm}\big|\hspace{0.1cm} \lambda  \in \mathbb{C},
 \hspace{0.07cm} k = 1, 2, \ldots, \mbox{ and }
 r = 0, 1, 2, 3\hspace{0.08cm}\big\}\vspace{0.06cm}
\end{eqnarray*}
give rise to tractable problems:  ${\rm Holant}(\Omega)$
for any  $\Omega = (G, {\cal F}_1 \cup {\cal F}_2 \cup {\cal F}_3)$
is computable in P (here we listed functions in ${\cal F}_i$ in the form
of truth tables on $k$ boolean variables).
  In particular,
 note that by taking $r=1$, $k=2$ and $ \lambda = (1+i)^{-1}$ in ${\cal F}_3$,
we recover the binary function $[1, 1, -1]$
which corresponds to exactly the $2\times 2$ Hadamard matrix ${\bf H}$ in (\ref{thehadamard}).
If we take $r=0$, $\lambda = 1$ in ${\cal F}_1$,
we get the {\sc Equality} function $[1, 0, \ldots, 0, 1]$ on $k$ bits.
This shows that  $Z_{\bf H}(\cdot)$, as a special case,
  can be computed in P.\vspace{0.008cm}

However, more instructive for us is the natural way in which
complex numbers appeared in such counting problems,
especially when applying holographic reductions.
One can say that the presence of powers of $i = \sqrt{-1}$
in ${\cal F}_1 \cup {\cal F}_2 \cup {\cal F}_3$ ``reveals'' the true nature
of ${\bf H}$ (i.e., $[1, 1, -1]$) as belonging to a family of
tractable counting problems, where complex numbers are
the correct language. In fact,\vspace{0.003cm} the tractability of
 ${\rm Holant}(\Omega)$ for   $\Omega = (G, {\cal F}_1 \cup {\cal F}_2 \cup {\cal F}_3)$
all boils down to an exponential sum of the form\vspace{-0.05cm}
\begin{equation}\label{mod4-of-mod2-sum}
\sum_{x_1,\hspace{0.03cm} x_2, \ldots, x_n\hspace{0.03cm} \in\hspace{0.03cm} \{0, 1\}}
i^{L_1 + L_2 +\hspace{0.04cm} \cdots\hspace{0.04cm} + L_s},\vspace{-0.03cm}
\end{equation}
where each $L_j$ is an indicator function of an affine
linear form of $x_1, x_2,\ldots, x_n$ over $\mathbb{Z}_2$
  (and thus, the exponent of $i$ in the equation above
is a mod 4 sum of mod 2 sums of $x_1, x_2,\ldots, x_n$).
From here it is only natural to investigate the
  complexity of $Z_{\AA}(\cdot)$ for symmetric complex matrices,
  since it
  not only is a natural generalization,
  but also can reveal the inner unity and some deeper structural properties.
Interested readers can find more details in \cite{holant}.
Also see the {\bf Remark} at the end of Section~\ref{tractabilitysec}.\vspace{0.006cm}

Our investigation of {complex-valued} graph homomorphisms is also
motivated by the partition fun\-ction in quantum physics. In classical
statistical physics, the partition function is always real-valued.
However, in a generic quantum system for which complex numbers
are the right language, the partition function is in general
complex-valued \cite{feynman}. In particular, if the physics model is
over a discrete graph and is non-orientable, then the edge weights are
  given by a symmetric complex matrix.\vspace{0.01cm}

Our main theorem is the following complexity dichotomy theorem:\vspace{0.06cm}

\begin{theorem}[Main Dichotomy]\label{main-in-intro}
Let $\AA$ be a symmetric matrix with \emph{algebraic complex} entries.\\
Then $Z_{\AA}(\cdot)$ either can be computed in
  polynomial time or is $\#$P-hard.\vspace{0.06cm}
\end{theorem}

Furthermore, we show that, under the model of computation
  described in Section \ref{complexmodel}, the following decision problem
  is solvable in polynomial time: Given any symmetric matrix $\AA$,
  where every entry is an algebraic number,
  decide if $Z_\AA(\cdot)$ is in polynomial time or is \#P-hard.\vspace{0.06cm}

\begin{theorem}[Polynomial-Time Decidability]\label{theo-decidability}
Given a symmetric matrix $\AA$, where all entries
  are\\ algebraic numbers, there is a polynomial-time algorithm that decides
  whether $Z_\AA(\cdot)$ is in polynomial\\ time or is \#P-hard.\vspace{0.02cm}
\end{theorem}

\subsection*{Recent Developments}

Recently Thurley \cite{Hermitian}  announced
a complexity dichotomy theorem for $Z_\AA(\cdot)$,
  where $\AA$ is any complex \emph{Hermitian} matrix.
The polynomial-time tractability result of the present paper (in Section \ref{tractabilitysec})
  is used in \cite{Hermitian}.
In \cite{acyclic}, Dyer, Goldberg, and Paterson proved a dichotomy theorem for
  $Z_H(\cdot)$ with $H$ being a directed \emph{acyclic} graph.
Cai and Chen  proved a dichotomy theorem for $Z_\AA(\cdot)$,
  where $\AA$ is a non-negative matrix \cite{NonnegativeHomo}.
A dichotomy theorem is also proved~\cite{BulatovCSP1,BulatovCSP2}
  for the more general counting constraint
  satisfaction problem, when the constraint functions
  take values in $\{0,1\}$  (with an alternative proof given in \cite{DyerRicherby} which also proves the decidability of
the dichotomy criterion),
  when the functions take non-negative and rational values \cite{RationalCSP},
  and when they are non-negative and algebraic \cite{NonnegativeCSP}.


\subsection*{Organization}

Due to the complexity of the proof of Theorem \ref{main-in-intro}, both in terms of
  its overall structure and in terms of technical difficulty,
  we first give a high level description of the proof
  for the bipartite case in Section \ref{highlevel}.
We then prove the First and Second Pinning Lemmas in Section \ref{pinninglemma}.
A more detailed~out\-line of the proof for the two
  cases (bipartite and non-bipartite) is presented in Section \ref{outlinebipartite} and
  \ref{outlinenonbip}, respectively, with formal definitions and theorems.
We then prove all the lemmas and theorems used in Section \ref{outlinebipartite}
  and \ref{outlinenonbip}, as well as Theorem \ref{theo-decidability},
  in the rest of the paper.

\section{Preliminaries}


In the paper, we let $\mathbb{Q}$ denote the set of rational numbers,
  and let $\mathbb{R}$ and $\mathbb{C}$ denote the set
  of algebraic real and algebraic complex numbers, respectively, for convenience
  (even though many of the supporting lemmas\hspace{0.06cm}/\hspace{0.06cm}theorems actually hold
  for general real or complex numbers, especially when computation
  or polynomial-time reduction is not concerned in the statement).

\subsection{Notation}

For a positive integer $n$, we use $[n]$ to denote the set $\{1,
\ldots, n\}$ (when
  $n=0$, $[0]$ is the empty set).
We also use $[m:n]$, where $m\le n$, to denote the set
$\{m,m+1,\ldots, n\}$.
We use $\11_n$ to denote the all-one vector of dimension $n$.
Sometimes we omit $n$ when the dimension is clear from the context.\vspace{0.005cm}

Let $\xx$ and $\yy$ be two vectors in $\mathbb{C}^n$, then we use
  $\langle \xx,\yy\rangle$ to denote their inner product
$$
\langle\xx,\yy\rangle=\sum_{i=1}^n x_i\cdot \overline{y_i}
$$
and $\xx\circ \yy$ to
  denote their Hadamard product: $\zz=\xx\circ\yy\in \mathbb{C}^n$,
  where $z_i=x_i\cdot y_i$ for all $i\in [n]$.\vspace{0.006cm}

Let $\AA=(A_{i,j})$ be a $k\times \ell$ matrix, and
  $\BB=(B_{i,j})$ be an ${m\times n}$ matrix.
We use $\AA_{i,*}$, $i\in [k]$, to denote the $i${th} row vector,
  and $\AA_{*,j}$, $j\in [\ell]$, to denote the $j${th}
  column vector of $\AA$.
We let $\CC=\AA\otimes \BB$ denote the tensor product of $\AA$ and
$\BB$:
  $\CC$ is a $km\times \ell n$ matrix whose rows and columns are indexed by~$[k]\times [m]$
  and $[\ell]\times [n]$, respectively, such that
  $$C_{(i_1,i_2),(j_1,j_2)}=A_{i_1,j_1}\cdot B_{i_2,j_2},\ \ \ \text{for all
  $i_1\in [k]$, $i_2\in [m]$, $j_1\in [\ell]$ and $j_2\in [n]$.}$$

Let $\AA$ be an $n\times n$ symmetric complex matrix. We build
an undirected graph $G=(V,E)$ from
  $\AA$ as follows: $V=[n]$, and $ij\in E$ iff $A_{i,j}\ne 0$.
We say $\AA$ is \emph{connected} if $G$ is connected;
  and we say $\AA$ has connected components
  $\AA_1,\ldots,\AA_s$, if the connected components of $G$
  are $V_1,\ldots,V_s$ and $\AA_i$ is the $|V_i|\times |V_i|$
  sub-matrix of $\AA$ restricted by $V_i\subseteq [n]$, for all $i\in [s]$.
Moreover, we say $\AA$ is \emph{bipartite}
  if the graph $G$ is bipartite; otherwise, $\AA$ is \emph{non-bipartite}.\vspace{0.006cm}
Let $\Sigma$ and $\Pi$ be two permutations from $[n]$ to itself
  then we use $\AA_{\Sigma,\Pi}$ to denote the $n\times n$ matrix
  whose $(i,j)${\emph{th}} entry, where $i,j\in [n]$, is $A_{\Sigma(i),\Pi(j)}$.\vspace{0.028cm}

We say $\CC$ is the \emph{bipartisation} of a matrix $\FF$ if\vspace{-0.05cm}
$$
\CC=\left(\begin{matrix}\00& \FF\\ \FF^T & \00\end{matrix}\right).\vspace{-0.05cm}
$$
Note that $\CC$  is always a symmetric matrix no matter whether
$\FF$ is or is not.
For a positive integer $N$, we use $\omega_N$ to denote $e^{2\pi
i/N}$, a primitive $N$th root of unity.\vspace{0.008cm}

We say a problem $\cal P$ is tractable if it can be solved in
  polynomial time.
Given two problems $\cal P$ and $\mathcal{Q}$,
  we say $\cal P$ is polynomial-time reducible to $\mathcal{Q}$ (or $\mathcal{P}\le \mathcal{Q}$),
  if there is an polynomial-time algorithm that solves $\cal P$
  using an oracle for $\mathcal{Q}$.

\subsection{Model of Computation}\label{complexmodel}

One technical issue is the \emph{model of computation} with algebraic numbers\hspace{0.02cm}\footnote{For readers who are
  not particularly concerned with details of the model of
  computation with complex numbers, this section can be
  skipped initially.\vspace{0.032cm}}.
We adopt a standard model from \cite{LENSTRA} for computation in
  an algebraic number field.
It has also been used, for example, in \cite{Hermitian, ThurleyThesis}.
We start with some notations.\vspace{0.008cm}

Let $\AA$ be a fixed symmetric matrix where every entry $A_{i,j}$
  is an algebraic number.
We let $\mathscr{A}$ denote the finite set of algebraic numbers consisting of
  entries $A_{i,j}$ of $\AA$.
Then it is easy to see that $Z_\AA(G)$,
  for any undirected graph $G$, is a number in $\mathbb{Q}(\mathscr{A})$,
  the algebraic extension of $\mathbb{Q}$ by $\mathscr{A}$.
By the primitive element theorem \cite{textbook2}, there exists an algebraic number
  $\alpha\in \mathbb{Q}(\mathscr{A})$ such that $\mathbb{Q}(\mathscr{A})=\mathbb{Q}(\alpha)$.
(Essentially, $\mathbb{Q}$ has characterisitc 0, and therefore
the field extension $\mathbb{Q}(\mathscr{A})$ is separable.
We can take the normal closure of $\mathbb{Q}(\mathscr{A})$, which is
a finite dimensional separable and normal extension of $\mathbb{Q}$,
and thus Galois \cite{textbook1}. By Galois correspondence, there are only a finite number
of intermediate fields between $\mathbb{Q}$ and this Galois extension
field and thus {\it a fortiori}
only a finite number
of intermediate fields between $\mathbb{Q}$ and $\mathbb{Q}(\mathscr{A})$.
Then Artin's theorem on primitive elements implies that
$\mathbb{Q}(\mathscr{A})$ is a simple extension $\mathbb{Q}(\alpha)$.)
In the proof of Theorem \ref{main-in-intro}
  when the complexity of a partition function $Z_\AA(\cdot)$ is concerned,
  we are given, as part of the problem description, such a number $\alpha$,
  encoded by a minimal polynomial $F(x)\in \mathbb{Q}[x]$ of $\alpha$.
In addition to $F$, we are also given a sufficiently good rational
  approximation $\hat{\alpha}$ of $\alpha$ which uniquely determines $\alpha$ as
  a root of $F(x)$.\vspace{0.008cm}\hspace{0.03cm}\footnote{This
  is a slight modification to the model of \cite{LENSTRA} and of \cite{Hermitian,ThurleyThesis}.
It will come in handy later in one step of the proof in Section \ref{sec:purification},
  with which we can avoid certain technical subtleties.}

Let $d=\deg(F)$, then every number $c$ in $\mathbb{Q}(\mathscr{A})$, including
  the $A_{i,j}$'s as well as $Z_\AA(G)$ for any $G$, has a unique representation as
  a polynomial of $\alpha$:
$$
c_0+c_1\cdot \alpha+\cdots +c_{d-1}\cdot\alpha^{d-1},\ \ \ \ \text{where
  every $c_i$ is a rational number.}
$$
We will refer to this polynomial as the \emph{standard representation} of $c$.
Given a number $c\in \mathbb{Q}(\mathscr{A})$ in the standard representation,
  its input size is sum of the binary lengths of all the rational coefficients.
It is easy to see that all the field operations over $\mathbb{Q}(\mathscr{A})$
  in this representation can be done efficiently in
  polynomial time in the input size. \vspace{0.008cm}

We emphasize that, when the complexity of $Z_\AA(\cdot)$ is concerned in
  the proof of Theorem \ref{main-in-intro}, all the following are considered as constants
since they are part of the problem description defined by $\AA$
and not part of the input:
  the size of $\AA$, the minimal polynomial $F(x)$ of $\alpha$,
  the approximation $\hat{\alpha}$ of $\alpha$ as well as the
  entries $A_{i,j}$ of $\AA$ encoded in the standard representation.
Given an undirected graph $G$, the problem is then
  to output $Z_\AA(G)\in \mathbb{Q}(\mathscr{A})$ encoded in the standard representation.
We remark that the same model also applies to
  the problem of computing $Z_{\CC,\fD}(\cdot)$, to be defined in Section \ref{evalcd}.\vspace{0.008cm}

However, for the most part of the proof of Theorem \ref{main-in-intro}
  this issue of computation model seems not central, because
  our proof starts with a preprocessing step using the Purification Lemma (see
  Section \ref{highlevel} for a high-level description of the proof,
  and see Section \ref{sec:purification} for the Purification Lemma),
  after which the matrix concerned becomes a \emph{pure} one, meaning that
  every entry is the product of a non-negative integer and a root of unity.
So throughout the proof, we let $\mathbb{C}$ denote the set of
  algebraic numbers and refer to them simply as complex numbers,
  except in the proof of the Purification Lemma in Section \ref{sec:purification},
  where we will be more careful on the model of computation.\vspace{0.008cm}

After the proof of Theorem \ref{main-in-intro}, we consider the decidability
  of the dichotomy theorem, and prove Theorem \ref{theo-decidability}.
The input of the problem is the full description of $\AA$,
  including the minimal polynomial $F(x)$ of $\alpha$,
  the approximation $\hat{\alpha}$ of $\alpha$
  as well as the standard representation of the entries $A_{i,j}$ of $\AA$.
We refer to the binary length of all the components above as the input size of $\AA$.
To prove Theorem \ref{theo-decidability}, we give an algorithm that runs in polynomial
  time in the binary length of $\AA$ and decides whether the problem of
  computing $Z_\AA(\cdot)$ is in polynomial time or \#P-hard.

\subsection{Definitions of $\eval(\AA)$ and $\eval(\CC,\fD)$}\label{evalcd}

Let $\AA\in \mathbb{C}^{m\times m}$ be a symmetric complex matrix
  with entries $(A_{i,j})$.
It defines a graph homomorphism problem $\eval(\AA)$ as follows:
Given
  an undirected graph $G=(V,E)$, compute
$$
Z_\AA(G)=\sum_{\xi:V\rightarrow [m]} \text{wt}_\AA(\xi),\ \ \ \
\text{where\ \ \ \  $\text{wt}_\AA(\xi) =\prod_{(u,v)\in E}
A_{\xi(u),\xi(v)}$}.
$$
We call $\xi$ an \emph{assignment} to the vertices of $G$, and
  $\text{wt}_\AA(\xi)$ the \emph{weight} of $\xi$.\vspace{0.006cm}

To study the complexity of $\eval(\AA)$ and prove Theorem \ref{main-in-intro},
  we introduce a much larger class of $\eval$ problems
  with not only edge weights but also vertex weights.
Moreover, the vertex weights in the problem depend on the degrees
  of vertices of $G$, modulo some integer modulus.
It is a generalization of the edge-vertex weight problems
  introduced in \cite{GoldbergGJT}.
See also \cite{Lovasz2006}.

\begin{definition}
Let $\CC\in \mathbb{C}^{m\times m}$ be a symmetric matrix,
  and $${\frak D}=\{\DD^{[0]},\DD^{[1]}, \ldots,\DD^{[N-1]}\}$$ be
  a sequence of diagonal matrices in $\mathbb{C}^{m\times m}$
  for some $N\ge 1$ \emph{(}we use
  $D^{[r]}_{i}$ to denote the $(i,i)^{th}$ entry of $\DD^{[r]}$\emph{)}.
We define the following problem $\eval(\CC,{\frak D})$: Given an
  undirected graph $G=(V,E)$, compute
\begin{equation*}
Z_{\CC,{\frak D}}(G)=\sum_{\xi:V\rightarrow [m]}\text{\rm
wt}_{\CC,\fD}(\xi),\vspace{-0.22cm}\label{ZCD}
\end{equation*}
where\vspace{-0.05cm}
\begin{equation*}
\text{\rm wt}_{\CC,\fD}(\xi)=\left(\hspace{0.03cm}\prod_{(u,v)\in E}
  C_{\xi(u),\xi(v)}\right)
\left(\hspace{0.05cm} \prod_{v\in V}
  D^{[\text{\rm deg}(v)\bmod\hspace{-0.025cm}N]}_{\xi(v)}\right)
\end{equation*}
and $\deg(v)$ denotes the degree of $v$ in $G$.
\end{definition}

Let $G$ be an undirected graph, and $G_1,\ldots,G_s$ be its
  connected components. Then\vspace{0.02cm}
\begin{lemma}\label{connectedcomponents}
$Z_{\CC,\fD}(G)=Z_{\CC,\fD}(G_1)\times \ldots\times
Z_{\CC,\fD}(G_s)$.\vspace{0.02cm}
\end{lemma}
Lemma \ref{connectedcomponents} implies that, if we want to design
  an efficient algorithm for computing $Z_{\CC,\fD}(\cdot)$, we only need to
  focus on connected graphs.
Furthermore, if we want to construct a reduction from one problem
  $\eval(\CC,\fD)$ to another $\eval(\CC',\fD')$,
  we only need to consider input graphs that are connected.
Also note that, since $\eval(\AA)$ is a special case of
  $\eval(\CC,\fD)$ (in which every $\DD^{[i]}$
  is the identity matrix), Lemma \ref{connectedcomponents} and the remarks above
  also apply to $Z_{\AA}(\cdot)$ and $\eval(\AA)$.\vspace{0.01cm}

Now suppose $\CC$ is the bipartisation of an $m\times n$ matrix
  $\FF$ (so $\CC$ is $(m+n)\times (m+n)$).
For any graph $G$ and vertex $u$ in $G$,
  we define $Z_{\CC,\fD}^{\rightarrow}(G,u)$ and $Z_{\CC,\fD}^{\leftarrow}(G,u)$ as follows.
Let\vspace{0.008cm} $\Xi_1$ denote the set of $\xi: V\rightarrow [m+n]$ with
  $\xi(u)\in [m]$, and $\Xi_2$ denote the set of
  $\xi$ with $\xi(u)\in [m+1:m+n]$, then
$$
Z_{\CC,\fD}^{\rightarrow}(G,u)=\sum_{\xi\in
\Xi_1}\text{wt}_{\CC,\fD}(\xi)\ \ \ \ \text{and} \ \ \ \
Z_{\CC,\fD}^{\leftarrow}(G,u)=\sum_{\xi\in
\Xi_2}\text{wt}_{\CC,\fD}(\xi).\label{ZCDARROW}
$$
It then follows from the definition of $Z_{\CC,\fD}$,
$Z_{\CC,\fD}^\rightarrow$
  and $Z_{\CC,\fD}^\leftarrow$ that\vspace{0.02cm}
\begin{lemma}\label{verytrivial}
For any graph $G$ and vertex $u\in G$, $Z_{\CC,\fD}(G)
  =Z_{\CC,\fD}^\rightarrow(G,u)+Z_{\CC,\fD}^{\leftarrow}(G,u)$.\vspace{0.02cm}
\end{lemma}

The reason we introduce $Z_{\CC,\fD}^\rightarrow$ and
$Z_{\CC,\fD}^\leftarrow$
  is because of the following useful lemma.\vspace{0.02cm}

\begin{lemma}\label{tensorproduct}
For each $i\in \{0,1,2\}$, $\FF^{[i]}$ is an $m_i\times n_i$ complex
matrix for some positive integers $m_i$\\ and $n_i$;
  $\CC^{[i]}$ is the bipartisation of $\FF^{[i]}$; and
$${\frak D}^{[i]}=\{\DD^{[i,0]},\ldots,\DD^{[i,N-1]}\}$$ is
  a sequence of $(m_i+n_i)\times (m_i+n_i)$ diagonal matrices for
  some positive integer $N$, where
$$
\DD^{[i,r]}=\left(\begin{matrix} \PP^{[i,r]} & \\
  & \QQ^{[i,r]}\end{matrix}\right)
$$
and $\PP^{[i,r]}$, $\QQ^{[i,r]}$ are $m_i\times m_i$, $n_i\times n_i$
   diagonal matrices, respectively.
Suppose $m_0=m_1 m_2$, $n_0=n_1 n_2$,
$$\FF^{[0]}=\FF^{[1]}\otimes \FF^{[2]},\ \ \PP^{[0,r]}=
  \PP^{[1,r]}\otimes \PP^{[2,r]},\ \ \text{and}\ \ \
  \QQ^{[0,r]}=\QQ^{[1,r]}\otimes \QQ^{[2,r]},\ \
  \ \text{for all $r\in [0:N-1]$.}$$
Then for any \emph{connected} graph $G$ and any vertex $u^*$ in $G$,
we have
\begin{eqnarray}
&Z_{\CC^{[0]},\fD^{[0]}}^\rightarrow
(G,u^*)=Z_{\CC^{[1]},\fD^{[1]}}^\rightarrow (G,u^*)
  \cdot Z_{\CC^{[2]},\fD^{[2]}}^\rightarrow (G,u^*)&\text{\ \ and}\label{jujublabla}\\[0.8ex]
&Z_{\CC^{[0]},\fD^{[0]}}^\leftarrow
(G,u^*)=Z_{\CC^{[1]},\fD^{[1]}}^\leftarrow (G,u^*)
  \cdot Z_{\CC^{[2]},\fD^{[2]}}^\leftarrow (G,u^*)&\hspace{-0.3cm}\nonumber.
\end{eqnarray}
\end{lemma}
\begin{proof}
We only prove \vspace{-0.04cm}(\ref{jujublabla}) about
$Z^{\rightarrow}$. First note that, if $G$ is not bipartite then
$Z_{\CC^{[i]},\fD^{[i]}}^\rightarrow
  (G,u^*)=0$ for all $i\in \{0,1,2\}$, and (\ref{jujublabla}) holds trivially.\vspace{0.01cm}

Now suppose $G=(U\cup V,E)$ is a bipartite graph, $u^*\in U$,
  and every edge $uv\in E$ has one vertex $u$ from $U$ and one vertex $v$ from $V$.
We let $\Xi_i$, $i\in \{0,1,2\}$, denote the set of assignments
$\xi_i$
  from $U\cup V$ to $[m_i+n_i]$ such that $\xi_i(u)\in [m_i]$ for all $u\in U$
  and $\xi_i(v)\in [m_i+1:m_i+n_i]$ for all $v\in V$.
Since $G$ is \emph{connected}, we have
$$
Z_{\CC^{[i]},\fD^{[i]}}^\rightarrow(G,u^*)=\sum_{\xi_i\in \Xi_i}
  \text{wt}_{\CC^{[i]},\fD^{[i]}}(\xi_i),\ \ \ \text{for $i\in \{0,1,2\}$.}
$$

To prove (\ref{jujublabla}), we define the following map $\rho:
  \Xi_1\times \Xi_2\rightarrow \Xi_0$: $\rho(\xi_1,\xi_2)=\xi_0$, where
  for every $u\in U$, $\xi_0(u)$ is the row index of $\FF^{[0]}$ that corresponds to
  row $\xi_1(u)$ of $\FF^{[1]}$ and row $\xi_2(u)$ of $\FF^{[2]}$ in
  the tensor product $\FF^{[0]}=\FF^{[1]}\otimes \FF^{[2]}$; and similarly,
  for every $v\in V$, $\xi_0(v)-m_0$ is the column index of $\FF^{[0]}$
  that corresponds to column $\xi_1(v)-m_1$ of $\FF^{[1]}$ and
  column $\xi_2(v)-m_2$ of $\FF^{[2]}$ in the tensor product.
One can check that $\rho$ is a bijection, and\vspace{-0.03cm}
$$
\text{wt}_{\CC^{[0]},\fD^{[0]}}(\xi_0)=\text{wt}_{\CC^{[1]},\fD^{[1]}}(\xi_1)
  \cdot \text{wt}_{\CC^{[2]},\fD^{[2]}}(\xi_2),
  \ \ \ \text{if $\rho(\xi_1,\xi_2)=\xi_0$.}\vspace{-0.04cm}
$$
Equation (\ref{jujublabla}) then follows, and the lemma is proven.
\end{proof}

\subsection{Basic \#P-Hardness}

We formally state the complexity dichotomy theorem of Bulatov and Grohe as follows:
\begin{theorem}[\cite{BulatovGrohe}]\label{basicsharp}
Let $\AA$ be a symmetric and
  connected matrix with \emph{non-negative algebraic} entries,\\ then $\eval(\AA)$
is either in \emph{polynomial time} or \emph{\#P-hard}.
Moreover, we have the following cases:
\begin{itemize}
\item If $\AA$ is bipartite, then $\eval(\AA)$ is in
  polynomial time if the rank of $\AA$ is $2$;\\
Otherwise $\eval(\AA)$ is $\#$P-hard.\vspace{-0.1cm}

\item If $\AA$ is not bipartite, then $\eval(\AA)$ is
  in polynomial time if the rank of $\AA$ is at most $1$;
\\ Otherwise $\eval(\AA)$ is $\#$P-hard.
\end{itemize}
\end{theorem}

Theorem \ref{basicsharp} gives us the following useful corollary:

\begin{corollary}\label{usefulhahacoro}
Let $\AA$ be a symmetric and connected matrix with \emph{non-negative algebraic} entries.
If
$$
\begin{pmatrix} A_{i,k} & A_{i,\ell}\\ A_{j,k} & A_{j,\ell}\end{pmatrix}
$$
is a $2\times 2$ sub-matrix of $\AA$
such that all of its four entries are nonzero and $A_{i,k}A_{j,\ell}
  \ne A_{i,\ell}A_{j,k}$, then the problem $\eval(\AA)$ is \emph{\#P-hard}.
\end{corollary}

\section{A High Level Description of the Proof}\label{highlevel}

The first step, in the proof of Theorem~\ref{main-in-intro},
  is to reduce the problem to connected graphs and matrices.
Let $\AA$ be an $m\times m$ symmetric complex matrix.
It is clear that if $G$ has connected components $G_i$ then
\begin{eqnarray*}
&Z_{\AA}(G) = \prod_i\hspace{0.03cm}  Z_{\AA}(G_i);&
\end{eqnarray*}
and if $G$ is connected and $\AA$  has connected components $\AA_j$, then
\begin{eqnarray*}
&Z_{\AA}(G) = \sum_j\hspace{0.03cm} Z_{\AA_j}(G).&
\end{eqnarray*}
Therefore, if every $Z_{\AA_j}(\cdot)$ is computable in polynomial time,
then so is $Z_{\AA}(\cdot)$.\vspace{0.008cm}

The hardness direction is less obvious.
Assume that $Z_{\AA_j}(\cdot)$ is \#P-hard for some $j$,
we want to show that $Z_{\AA}(\cdot)$ is also \#P-hard.
This is done by proving that computing $Z_{\AA_j}(\cdot)$
is reducible to computing $Z_{\AA}(\cdot)$.
Let $G$ be an arbitrary input graph.
To compute $Z_{\AA_j}(G)$, it suffices to compute $Z_{\AA_j}(G_i)$
for all connected components $G_i$ of $G$. Therefore, we may just assume
that $G$ is connected.
Define a {\it pinning} version of the $Z_{\AA}(\cdot)$ function
as follows.
For any chosen vertex $w \in V(G)$, and any $k \in [m]$, let\vspace{-0.01cm}
$$
Z_{\AA}(G,w,k)=\sum_{\xi:V\rightarrow [m],
  \hspace{0.06cm}\xi(w)\hspace{0.04cm}=\hspace{0.04cm}k~}
   \prod_{(u,v)\in E} A_{\xi(u),\xi(v)}.\vspace{-0.01cm}
$$
Then we can prove a {\it Pinning Lemma} (Lemma \ref{pinning1}) which
states that the problem of computing $Z_{\AA}(\cdot)$
is polynomial-time equivalent to computing $Z_{\AA}(\cdot,\cdot,\cdot)$.
Note that if $V_j$ denotes the subset of $[m]$ where $\AA_j$ is the
  sub-matrix of $\AA$ restricted by $V_j$,
then for a connected $G$, we have
$$
Z_{\AA_j} (G) = \sum_{k \in V_j} Z_{\AA}(G,w,k),\label{ZAPINNING}
$$
which gives us a polynomial-time reduction from $Z_{\AA_j}(\cdot)$ to $Z_\AA(\cdot)$.\vspace{0.01cm}

The proof of this Pinning Lemma (Lemma \ref{pinning1})
is a standard adaptation to the complex numbers of the one proved in
\cite{GoldbergGJT}. However, for technical reasons we will need
a total of three Pinning Lemmas (Lemma~\ref{pinning1}, \ref{pinning2}
and \ref{pinning3}), and the proofs of the other two are a bit more involved.
We remark that all three Pinning Lemmas only show the \emph{existence} of
a polynomial-time reduction between $Z_{\AA}(\cdot)$
and $Z_{\AA}(\cdot,\cdot,\cdot)$, but they do not \emph{constructively}
produce such a reduction, given $\AA$.
We also remark that the proof of the Pinning Lemma in \cite{GoldbergGJT}
  used a recent result
by Lov\'{a}sz~\cite{Lovasz2006} for real matrices.
This result is not known for complex matrices.
We give direct proofs of our three lemmas without using \cite{Lovasz2006}.\vspace{0.008cm}

After this preliminary step, we restrict to \emph{connected} and \emph{symmetric} $\AA$.
As indicated, to our work the two most influential predecessor
papers are by Bulatov and Grohe~\cite{BulatovGrohe}
and by Goldberg et al.~\cite{GoldbergGJT}.
In both papers, the polynomial-time algorithms for those tractable
  cases are relatively straightforward and are previously known.
The difficult part of the proof
  is to show that in all other cases the problem is \#P-hard.\vspace{0.006cm}
Our proof follows a similar conceptual framework to that
  of Goldberg et al.~\cite{GoldbergGJT}.
However, over the complex numbers, new difficulties arise in both
  the tractability and the hardness part of the proof.
Therefore, both the overall organization
  and the substantive part of the proof have to be done separately.

First of all, the complex numbers afford a much richer
variety of cancelations, which could lead to surprisingly efficient algorithms
  for computing $Z_\AA(\cdot)$, when the complex matrix $\AA$ satisfies certain nice conditions.
This turns out to be the case, and we obtain additional non-trivial
  tractable cases.
These boil down to the following class of problems:
\begin{flushleft}
\begin{quote}
$Z_q(\cdot)$: Let $q=p^k$ be a fixed prime power for some prime $p$ and
  positive integer $k$.
The input of $Z_q(\cdot)$ is a quadratic polynomial
$$
f(x_1,x_2,\ldots,x_n)=\sum_{i,j\in [n]} a_{i,j} x_ix_j,\ \ \ \
  \text{where $a_{i,j}\in \mathbb{Z}_q$ for all $i,j$;}
$$
and the output is
$$
Z_q(f)=\sum_{x_1,\ldots,x_n\in \mathbb{Z}_q} \omega_q^{f(x_1,\ldots,x_n)}.\label{ZQF}
$$
\end{quote}\end{flushleft}
We show that for any fixed prime power $q$, the problem of computing $Z_q(\cdot)$ is
  in polynomial time.
In the algorithm (see Section \ref{tractabilitysec}), Gauss sums play a crucial role.
The tractability part of our dichotomy theorem is then done by reducing
  $Z_\AA(\cdot)$, assuming $\AA$ satisfies a set of nice structural conditions
  (to be described in the rest of this section)
  imposed by the hardness part, to $Z_q(\cdot)$
  for some appropriate prime power $q$.
While the corresponding sums for finite fields (when $q$ is a prime) are known to be
  computable in polynomial time \cite[Theorem 6.30]{Carlitz,field}, in particular this includes the
  special case of $\mathbb{Z}_2$ which was used in \cite{GoldbergGJT},
  our algorithm over rings $\mathbb{Z}_q$ is new and should be of independent
  interest.\vspace{0.01cm}

Next we briefly describe the proof structure of the hardness part
  of the dichotomy theorem.
Let $\AA$ be a connected and sym\-metric matrix.
The difficulty starts with the most basic proof technique called \emph{gadget constructions}.
With a graph gadget, one can take any input undirected graph $G$ and produce
  a modified graph ${G}^*$ by replacing each edge of $G$ with the gadget.
Moreover, one can define a suitable modified matrix $\AA^*$
  from the fixed matrix  $\AA$ and the gadget
  such that \vspace{-0.03cm}
$$
Z_{ {\AA}^*} (G) = Z_{\AA} ( {G}^*),\ \ \ \ \text{for all undirected graphs $G$.}\vspace{-0.03cm}
$$

A simple\vspace{0.005cm} example of this maneuver is called {\it thickening}
where one replaces each edge in the input $G$ by $t$ parallel edges
to get $ {G}^*$.  Then it is easy to see that if
$ {\AA}^*$ is obtained from $\AA$ by replacing each entry
$A_{i,j}$ by its $t^{th}$ power $(A_{i,j})^t$,
  then the equation above holds and we get a reduction from
  $Z_{\AA^*}(\cdot)$ to $Z_\AA(\cdot)$.
In particular, if $\AA$ is real (as in the case of \cite{GoldbergGJT})
and $t$ is even, this produces a non-negative
matrix $ {\AA}^*$, to which one may apply the Bulatov-Grohe result:
\begin{flushleft}
\begin{enumerate}
\item If $\AA^*$, as a symmetric and non-negative matrix, does not satisfy
  the tractability criteria of Bulatov and Grohe as described in Theorem \ref{basicsharp},
  then both $Z_{\AA^*}(\cdot)$ and $Z_\AA(\cdot)$ are \#P-hard and we are already done;
\item Otherwise, $\AA^*$ satisfies the
  Bulatov-Grohe tractability criteria, from which we know that $\AA$ must
  satisfy certain necessary structural properties since $\AA^*$ is derived from $\AA$.
\end{enumerate}\end{flushleft}

The big picture\hspace{0.06cm}\footnote{The exact proof structure, however, is different
  from this very high-level description, which will become clear through the rest of this section.}
  of the proof of the dichotomy theorem is then to
  design various graph gadgets to show that, assuming $Z_\AA(\cdot)$
  is not \#P-hard, the matrix $\AA$ must satisfy a collection of strong necessary
  conditions over its complex entries $A_{i,j}$.
To finish the proof, we show that for every $\AA$ that satisfies all
  these structural conditions, one can reduce $Z_\AA(\cdot)$ to $Z_q(\cdot)$, for
  some appropriate prime power $q$ (which depends only on $\AA$)
  and thus, $Z_\AA(\cdot)$ is tractable.\vspace{0.008cm}



%
%

For a complex matrice $\AA$, we immediately
encountered the following difficulty.  Any graph gadget
  will only produce a matrix $ {\AA}^*$ whose entries  are
obtained from entries of $\AA$ by arithmetic operations $+$ and $\times$.
While for real numbers any even power guarantees a non-negative
quantity as was done in \cite{GoldbergGJT},
no obvious arithmetic operations on the complex
numbers have this property. 
Pointedly, \emph{conjugation} is not an
arithmetic operation.  However, it is also clear that for roots of unity,
one {\it can} produce conjugation by multiplication.\vspace{0.008cm}

Thus, our proof starts with a process to replace an arbitrary
complex matrix by a {\it purified} complex matrix which
has a special form.  It turns out that we must separate out
the cases where $\AA$ is bipartite or non-bipartite.
A purified bipartite (and symmetric, connected) matrix takes the following form:
$$
\begin{pmatrix} {\bf 0} & \BB \\
                \BB^{T} & {\bf 0}
        \end{pmatrix},$$ where\vspace{-0.15cm} $$
\BB\hspace{0.015cm} = \left(\begin{matrix}
\mu_1\hspace{-0.05cm} \\
& \hspace{-0.05cm}\mu_2\hspace{-0.05cm} \\
& & \hspace{-0.05cm}\ddots\hspace{-0.05cm}\\
& & & \hspace{-0.05cm}\mu_k
\end{matrix}\right)
\left( \begin{matrix}
\zeta_{1,1} & \zeta_{1,2} & \ldots & \zeta_{1,m-k} \\
\zeta_{2,1} & \zeta_{2,2} & \ldots & \zeta_{2,m-k} \\
\vdots      & \vdots      & \ddots & \vdots \\
\zeta_{k,1} & \zeta_{k,2} & \ldots & \zeta_{k,m-k}
\end{matrix}\right)
\left(\begin{matrix}
\mu_{k+1}\hspace{-0.05cm} \\
& \hspace{-0.05cm}\mu_{k+2}\hspace{-0.05cm} \\
& & \hspace{-0.05cm}\ddots\hspace{-0.05cm}\\
& & & \hspace{-0.05cm}\mu_{m}
\end{matrix}\right),\vspace{0.16cm}$$
for some $1\le k<m$, in which every $\mu_i$ is a positive rational number
  and every $\zeta_{i,j}$ is a root of unity.\vspace{0.008cm}

The claim is that, for every
symmetric, connected, and bipartite matrix  $\AA\in \mathbb{C}^{m\times m}$,
either we can already prove the $\#$P-hardness of
computing $Z_\AA(\cdot)$ or there exists a
symmetric, connected and {\it purified} bipartite matrix
$\AA' \in \mathbb{C}^{m\times m}$, such that
computing $Z_{\AA'} (\cdot)$ is polynomial time
equivalent to computing $Z_\AA(\cdot)$ (see Theorem~\ref{bi-step-1}).
For non-bipartite  $\AA$ a corresponding  statement holds
(see Theorem~\ref{t-step-1}).
For convenience, in the discussion below, we only focus on the bipartite case.\vspace{0.008cm}

Continuing now with a purified bipartite matrix
  $\AA'$, the next step is to \emph{further regularize}
its entries.  In particular we need to combine those rows and
columns of the matrix  where they
are essentially the same, apart from a multiple of a root of unity.
This process is called
\emph{Cyclotomic} Reduction. 
In order to carry out this process, we need to use the more general problem
$\eval(\CC,\fD)$ defined earlier in Section \ref{evalcd}.
We also need to introduce the following type of matrices called
\emph{discrete unitary} matrices:\vspace{0.16cm}

\begin{definition}[Discrete Unitary Matrix]
Let $\FF\in \mathbb{C}^{m\times m}$ be a
  matrix with entries $(F_{i,j})$.
We say $\FF$\\ is an \emph{$M$-discrete unitary matrix}, for some positive
  integer $M$, if it satisfies the following conditions:
\begin{enumerate}
\item[1.] Every entry $F_{i,j}$ of $\FF$ is a root of unity,
  and $M = \text{\rm lcm}\hspace{0.07cm}\big\{\text{the order of $F_{i,j}:i,j\in [m]$}\big\}$;\vspace{-0.16cm}
\item[2.] $F_{1,i}=F_{i,1}=1$ for all $i\in [m]$, and
 for all $i\ne j\in [m]$, we have
$$
\langle\FF_{i,*},\FF_{j,*}\rangle=0\ \ \ \text{and}\ \ \
  \langle \FF_{*,i},\FF_{*,j}\rangle=0.
$$
\end{enumerate}
\end{definition}

Some simplest examples of discrete unitary matrices are
as follows:
\[
{\bf H} = \left(\begin{matrix} 1 & 1 \\
             1 & -1 \end{matrix}\right),
~
{\bf H_4} = \left(\begin{matrix} 1 & 1 & 1 & 1\\            1 & 1 & -1 & -1 \\
            1 & -1 & 1 & -1 \\
            1 & -1 & -1 & 1 \end{matrix}\right),
~
\FF_{3}  = \left(\begin{matrix} 1 & 1 & 1 \\
                             1 & \omega & \omega^2 \\
                 1 & \omega^2 & \omega
                  \end{matrix}\right),
~
\FF_{5}  = \left(\begin{matrix} 1 & 1 & 1 & 1 & 1\\
                                 1 & \zeta&\zeta^{-1} & \zeta^2 & \zeta^{-2}\\
                                 1 & \zeta^2&\zeta^{-2} & \zeta^{-1} & \zeta\\
                 1 & \zeta^{-1}& \zeta & \zeta^{-2}&\zeta^2 \\
                                 1 & \zeta^{-2}&\zeta^2 & \zeta&\zeta^{-1}
                  \end{matrix}\right),
\]
where $\omega = e^{2 \pi i /3}$ and $\zeta = e^{2 \pi i /5}$.
Also note that any tensor product of discrete unitary matrices
is also a discrete unitary matrix.
These matrices play a major role in our proof.\vspace{0.008cm}

Now we come back to the proof outline.
We show that $Z_{\AA'} (\cdot)$ is either $\#$P-hard or
polynomial time equivalent to $Z_{\CC,{\frak D}}(\cdot)$
for  some  $\CC \in \mathbb{C}^{2n\times 2n}$ and some ${\frak D}$
of diagonal matrices from $\mathbb{C}^{2n\times 2n}$, where $n\le$ $m$
  and $\CC$ is the bipartisation of a discrete unitary matrix, denoted by $\FF$.
In addition, 
there are further stringent requirements for ${\frak D}$; otherwise
$Z_{\AA'}(\cdot)$ is \#P-hard.
The detailed statements can be found in Theorem~\ref{bi-step-2}
 and \ref{step30}, summarized in properties $(\calU_1)$ to $(\calU_5)$.
Roughly speaking, the first matrix $\DD^{[0]}$ in $\fD$ must be the identity matrix;
and for any matrix $\DD^{[r]}$ in~$\fD$, each entry of $\DD^{[r]}$
  is either zero or a root of unity.
We call these conditions, with some abuse of terminology, the discrete unitary
requirements. The proof of these requirements is demanding
  and among the most difficult in the paper.\vspace{0.006cm}

Next, assume that we have a problem $\eval(\CC,\fD)$ satisfying
the discrete unitary requirements with $\CC$ being
  the bipartisation of $\FF$.\vspace{0.025cm}
\begin{definition}
Let $q>1$ be a prime power,
then the following $q\times q$ matrix $\bcF_q$ is called the \emph{$q$-Fourier matrix}:
The $(x,y)$th entry of $\bcF_q$ is $\omega_q^{xy}$, where $x,y\in [0:q-1]$,
and $\omega_q = e^{2 \pi i /q}$.\vspace{0.025cm}
\end{definition}

We show that, either $Z_{\CC,\fD}(\cdot)$ is $\#$P-hard;
or after a permutation of rows and columns,
$\FF$ becomes the \emph{tensor product} of a collection of suitable Fourier matrices:
$$
\bcF_{q_1}\otimes \bcF_{q_2}\otimes \cdots\otimes \bcF_{q_d},\ \ \ \ \text{where
  $d\ge 1$ and every $q_i$ is a prime power.}
$$
Basically, we show that even with the stringent conditions imposed
  on the pair $(\CC,\fD)$ by the discrete unitary requirements,
  most of $\eval(\CC,\fD)$ are still \#P-hard, unless $\FF$ is the
  tensor product of Fourier matrices.
On the other hand, the tensor product decomposition into Fourier
  matrices finally brings in group theory and Gauss sums.
It gives us a canonical way of writing the entries of $\FF$ in a closed form.
More exactly, we can index the rows and columns of $\FF$ using
$$
\xx=(x_1,\ldots,x_d)\ \text{and}\ \yy=(y_1,\ldots,y_d)\in \mathbb{Z}_{q_1}\times \cdots
  \times \mathbb{Z}_{q_d},
$$
respectively, such that
$$
F_{\xx,\yy}=\prod_{i\in [d]} \omega_{q_i}^{x_iy_i},\ \ \ \ \text{for any $\xx$ and $\yy$.}
$$
Assume $q_1,\ldots,q_d$ are powers of $s\le d$ distinct primes $p_1,\ldots,p_s$.
We can also view the set of indices as
$$
\mathbb{Z}_{q_1}\times \cdots \times \mathbb{Z}_{q_d}=
  G_1\times \cdots \times G_s,
$$
where $G_i$ is the finite Abelian group which is the product of all the $\mathbb{Z}_{q_j}$'s
  with $q_j$ being a power of $p_i$.\vspace{0.006cm}

This canonical tensor product decomposition of $\FF$ also gives
  a natural way to index the rows and columns of $\CC$ and the diagonal
  matrices in $\fD$ using $\xx$.
More exactly, we index the first half of the rows and columns of $\CC$
  and every $\DD^{[r]}$ in $\fD$ using $(0,\xx)$; and index
  the second half of the rows and columns using $(1,\xx)$, where
  $\xx\in \mathbb{Z}_{q_1}\times \cdots\times \mathbb{Z}_{q_d}$.\vspace{0.008cm}

With this canonical expression of $\FF$ and $\CC$, we further inquire
the structure of $\fD$.
Here one more substantial difficulty awaits us.
There are two more properties that we must demand of
  those diagonal matrices in $\fD$.
If $\fD$ does not satisfy these additional properties, then
  $Z_{\CC,\fD}(\cdot)$ is \#P-hard.\vspace{0.008cm}

First, for each $r$,
we define $\Lambda_r$
 and $\Delta_r$
to be the support of $\DD^{[r]}$,
where $\Lambda_r$  refers to the first half of
the entries and  $\Delta_r$  refers to the second half of
the entries: (Here we follow the convention of using $D_i$ to denote
  the $(i,i)$th entry of a diagonal matrix $\DD$)\vspace{0.04cm}
\begin{eqnarray*}
\Lambda_r=\big\{\xx\hspace{0.08cm} \big|\hspace{0.08cm} D^{[r]}_{(0,\xx)}\
\ne 0\big\}\ \ \ \ \text{and}\ \ \ \
\Delta_r=\big\{\xx\hspace{0.08cm} \big|\hspace{0.08cm} D^{[r]}_{(1,\xx)}\ne 0\big\}.\vspace{0.04cm}
\end{eqnarray*}
We let $\cS$ denote the set of subscripts $r$ such that $\Lambda_r\ne \emptyset$
  and $\cT$ denote the set of $r$ such that $\Delta_r\ne \emptyset$.
We can prove that for every
$r\in \cS$, $$\Lambda_r=\prod_{i=1}^s
  \hspace{0.03cm}\Lambda_{r,i}$$ must be a direct product of cosets $\Lambda_{r,i}$ in the Abelian
groups $G_i$, where $i=1,\ldots,s$ correspond
  to the constituent prime powers of the group.
Similarly for every $r\in \cT$, $$\Delta_r=\prod_{i=1}^s\hspace{0.03cm}
  \Delta_{r,i}$$ is also a direct product of cosets in the same
 Abelian groups. Otherwise, $Z_{\CC,\fD}(\cdot)$ is \#P-hard.\vspace{0.006cm}

Second, we show that for each $r \in \cS$ and $r \in \cT$, respectively,
 $\DD^{[r]}$ on its support $\Lambda_{r}$ for the first half of its
entries
and on $\Delta_{r}$ for the second half of its entries, respectively,
  possesses a {\it quadratic} structure; otherwise $Z_{\CC,\fD}(\cdot)$ is \#P-hard.
We can express the quadratic structure
as \emph{a set of exponential difference equations} over
bases which are appropriate roots of unity of orders equal to
various prime powers. The constructions used in this part of the proof
  are among the most demanding ever attempted.\vspace{0.012cm}

After all these necessary conditions, we finally show that
if $\CC$ and $\fD$ satisfy all these requirements,
there is a polynomial-time
algorithm to compute $Z_{\CC,\fD}(\cdot)$ and thus,
  the problem of computing $Z_\AA(\cdot)$ is in polynomial time as well.
To this end, we reduce $Z_{\CC,\fD}(\cdot)$ to $Z_q(\cdot)$
  for some appropriate prime power $q$ (which depends only on
    $\CC$ and $\fD$) and as remarked earlier,
  the tractability of $Z_q(\cdot)$ is new and is of independent interest.

%
%

%% file: Rankone.tex

\newtheorem{Claim}[theorem]{Claim}

\subsection{Step 2.3}\label{D-is-rank-one}

Now we have a pair $(\CC,\fD)$ that satisfies conditions ({\sl Shape}$_1
  $)--({\sl Shape}$_5$) since otherwise, by Lemma \ref{shapecondition},
  $\eval(\CC,\fD)$ is \#P-hard and we are done.

In particular, by using ({\sl Shape}$_5$) we define two diagonal matrices $\KK^{[0]}$
  and $\LL^{[0]}$ as follows.
$\KK^{[0]}$ is an $(s+t)\times (s+t)$ diagonal matrix.
We use $(0,i)$, where $i\in [s]$, to index the first $s$ rows,
  and $(1,j)$, where $j\in [t]$, to index the last $t$ rows of $\KK^{[0]}$.
The diagonal entries of $\KK^{[0]}$ are
$$
K^{[0]}_{(0,i)}=D^{[0]}_{(0,(i,1))}\ \ \ \text{and}\ \ \
K^{[0]}_{(1,j)}=D^{[0]}_{(1,(j,1))},\ \ \ \text{for all $i\in [s]$ and $j\in [t]$.}
$$
The matrix $\LL^{[0]}$ is the $2h\times 2h$ identity matrix.
We use $(0,i)$, where $i\in [h]$, to index the first $h$ rows,
  and $(1,j)$, where $j\in [h]$, to index the last $h$ rows of $\LL^{[0]}$.
By ({\sl Shape}$_5$), we have
\begin{equation}\label{conditionbb}
D^{[0]}_{(0,\xx)}=K^{[0]}_{(0,x_1)}\cdot L^{[0]}_{(0,x_2)}\ \ \ \text{and}\ \ \
D^{[0]}_{(1,\yy)}=K^{[0]}_{(1,y_1)}\cdot L^{[0]}_{(1,y_2)},\ \ \
\text{for all $\xx\in I$ and $\yy\in J$.}
\end{equation}
or equivalently,
\begin{equation}\label{fufufufu}
\DD^{[0]}=\left(\begin{matrix}
\DD^{[0]}_{(0,*)}\\ & \DD^{[0]}_{(1,*)}
\end{matrix}\right)=\left(\begin{matrix}
\KK^{[0]}_{(0,*)}\otimes \LL^{[0]}_{(0,*)}\\
& \KK^{[0]}_{(1,*)}\otimes \LL^{[0]}_{(1,*)}
\end{matrix}\right).
\end{equation}
The main target of this step is to prove a similar
  statement for $\DD^{[r]}$, $r\in [N-1]$.
These equations will allow us to decompose, in Step 2.4,
  the problem $\eval(\CC,\fD)$ into two subproblems.

In the proof of Lemma \ref{shapecondition}, we crucially used the property (from $(\calT_3)$)
  that all the diagonal entries of $\DD^{[0]}$ are positive integers.
However, for $r\ge 1$, $(\calT_3)$ only gives us some very weak properties about $\DD^{[r]}$.
For example, the entries are not guaranteed to be real numbers.
So the proof that we are going to present here is more difficult.
We prove the following lemma:

\begin{lemma}\label{horrible}
Let $(\CC,\fD)$ be a pair that satisfies conditions \emph{({\sl Shape}$_1$)-({\sl Shape}$_5$)},
  then either the problem $\eval(\CC,\fD)$ is \#P-hard, or
  it satisfies the following additional condition:
\begin{enumerate}
\item[]\hspace{-0.65cm}\emph{({\sl Shape}$_6$): There exist diagonal matrices $\KK^{[0]}$
  and $\LL^{[0]}$ such that $\DD^{[0]},\KK^{[0]}$ and $\LL^{[0]}$ satisfy (\ref{fufufufu}).
Every entry of $\KK^{[0]}$ is a positive integer, and $\LL^{[0]}$ is the
  $2h\times 2h$ identity matrix.
For every $r\in [N-1]$, there
  exist two diagonal matrices: $\KK^{[r]}$ and $\LL^{[r]}$. $\KK^{[r]}$
  is an $(s+t)\times (s+t)$ matrix, and $\LL^{[r]}$ is a $2h\times 2h$
  matrix. We index $\KK^{[r]}$ and $\LL^{[r]}$ in the same way we index
  $\KK^{[0]}$ and $\LL^{[0]}$, respectively, and 
$$
\DD^{[r]}=\left(\begin{matrix}
\DD^{[r]}_{(0,*)}\\ & \DD^{[r]}_{(1,*)}
\end{matrix}\right)=\left(\begin{matrix}
\KK^{[r]}_{(0,*)}\otimes \LL^{[r]}_{(0,*)}\\
& \KK^{[r]}_{(1,*)}\otimes \LL^{[r]}_{(1,*)}
\end{matrix}\right).
$$\label{SHAPECONDITION6}
Moreover, the norm of every diagonal entry in $\LL^{[r]}$ is
  either $0$ or $1$, and for any $r\in [N-1]$,}
\begin{eqnarray*}
\KK^{[r]}_{(0,*)}=\00 \ \Longleftrightarrow \ \LL^{[r]}_{(0,*)}=\00\ &\text{and}&\
  \KK^{[r]}_{(1,*)}=\00 \ \Longleftrightarrow \ \LL^{[r]}_{(1,*)}=\00;\\[0.2ex]
\LL^{[r]}_{(0,*)}\ne \00 \ \Longrightarrow\ \exists\hspace{0.05cm}i\in [h],\ L^{[r]}_{(0,
  i)}=1\ &\text{and}&\ \LL^{[r]}_{(1,*)}\ne \00\ \
  \Longrightarrow\ \ \exists\hspace{0.05cm}i\in [h],\ L^{[r]}_{(1,i)}=1.\\[-3.5ex]
\end{eqnarray*}
\end{enumerate}
\end{lemma}

We now present the proof of Lemma \ref{horrible}.
Fix an $r\in [N-1]$ to be any index.
We use the following notation.
Consider the diagonal matrix $\DD^{[r]}$. 
  This matrix has two parts: $$\DD^{[r]}_{(0,*)}\in \mathbb{C}^{
  sh\times sh}\ \ \ \text{and}\ \ \ \DD^{[r]}_{(1,*)}
  \in \mathbb{C}^{th\times th}.$$
The first part has $s$ blocks where each
  block is a diagonal matrix with $h$ entries. We will rearrange
  the entries indexed by $(0,*)$ into another matrix which we will denote as
$\DD$ (just like what we did to $\DD^{[0]}$ in the proof of
  Lemma \ref{shapecondition}), where its $i$-th row $\DD_{i,*}$, for $i\in [s]$,
denotes the values of the $i$-th block and the $j$-th entry of
the $i$-th row $D_{i,j}$, for $j\in [h]$, denotes the
  $j$-th entry of that $i$-th block. 
More exactly,
$$
D_{i,j}=D^{[r]}_{(0,(i,j))},\ \ \ \text{for all $i\in [s]$ and $j\in [h]$.}
$$

We prove the following lemma in Section \ref{proofofrank1}.
A similar statement can be proved for $\DD^{[r]}_{(1,*)}$.

\begin{lemma}\label{rank1}
Either problem $\eval(\CC,\fD)$ is \#P-hard; or
\begin{itemize}
\item[--] \hspace{0.06cm}$\text{\rm rank}(\DD)$ is at most $1$, and for any $i,j,j'\in [h]$, if
  $D_{i,j}\ne 0$ and $D_{i,j'}\ne 0$, then $|D_{i,j}|=|D_{i,j'}|$.
\end{itemize}
\end{lemma}


We now use it to prove the first half of Lemma \ref{horrible}, that is,
  there exist $\KK^{[r]}_{(0,*)}$ and $\LL^{[r]}_{(0,*)}$ such that
\begin{equation}\label{justonce}
\DD^{[r]}_{(0,*)}=\KK^{[r]}_{(0,*)}\otimes \LL^{[r]}_{(0,*)}.\end{equation}
Assume\vspace{-0.01cm} $\DD^{[r]}_{(0,*)}$
  is non-zero; otherwise, the lemma is trivially true by setting $\KK^{[r]}_{(0,*)}$ and
  $\LL^{[r]}_{(0,*)}$\vspace{-0.03cm} to be zero.
Let $a$ be an index in $[s]$ and $b$ be an index in $[h]$ such that $D_{a,b}\ne 0$.
By Lemma \ref{rank1}, we know the rank of $\DD$ is $1$,
  so $\DD_{i,*}=(D_{i,b}/D_{a,b})\cdot \DD_{a,*}$, for any $i\in [s]$.
Then it is clear that, by setting 
$$
K^{[r]}_{(0,i)}= D_{i,b} ,\ \ \ \text{and}\ \ \ \
L^{[r]}_{(0,j)}=\frac{D_{ a,j }}{D_{ a,b }},
$$
we have
$$
D^{[r]}_{(0,(i,j))}=D_{i,j}=K^{[r]}_{(0,i)}\cdot L^{[r]}_{(0,j)},
\ \ \ \text{for all $i\in [s]$ and $j\in [h]$,}$$
and (\ref{justonce}) follows.
The existence of matrices $\KK^{[r]}_{(1,*)}$ and
  $\LL^{[r]}_{(1,*)}$ can be proved similarly.\vspace{-0.1cm}

One can also check that $\KK^{[r]}$ and $\LL^{[r]}$ satisfy
  all the properties stated in ({\sl Shape}$_6$).
This finishes the proof~of Lemma \ref{horrible} (assuming Lemma \ref{rank1}).

\subsubsection{The Vanishing Lemma}

We will use the following Vanishing Lemma in the proof of Lemma \ref{rank1}.

\begin{lemma}[Vanishing Lemma]
Let $k$ be a positive integer and $\{x_{i, n}\}_{n \ge 1}$,
for $1\le i\le k$, be $k$~infinite sequences of non-zero real numbers.
For notational uniformity we also denote by $\{x_{0, n}\}_{n \ge 1}$
the sequence where $x_{0, n} = 1$ for all $n\ge 1$.
Suppose\vspace{-0.15cm}
$$\lim_{n \rightarrow \infty} \frac{x_{i+1, n}}{x_{i, n}} = 0,
  \ \ \ \text{for $0 \le i < k$.}$$
\begin{description}
\item{\rm \bf Part A}\emph{:}
Let $a_i$ and $b_i \in \mathbb{ C}$, for $0 \le i \le k$. Suppose
for some $1 \le \ell \le k$,
$a_i = b_i$ for all $0 \le i < \ell$ and\\ $a_0 = b_0 = 1$.
Also suppose $\text{\emph{Im}}(a_\ell)=\text{\emph{Im}}(b_\ell)$.
If for infinitely many $n$,
\[\left| \sum_{i=0}^{k} a_i x_{i, n} \right|
= \left| \sum_{i=0}^{k} b_i x_{i, n} \right|,\]
then $a_\ell = b_\ell$.
\item{\rm \bf Part B}\emph{:}
Let $a_i \in {\mathbb{C}}$, for $0 \le i \le k$.
Suppose
for infinitely many $n$,
\[\left| \sum_{i=0}^{k} a_i x_{i, n} \right| = 0,\]
then $a_i = 0$ for all $0 \le i \le k$.
\end{description}
\end{lemma}

\begin{proof}
We first prove Part B, which is simpler.
By taking $n \rightarrow \infty$ (Technically we take
a subsequence of $n$ approaching $\infty$ where
the equality holds; same below), we get immediately
$a_0 =0$.
Since $x_{1, n} \not =0$, we can divide out
$|x_{1, n}|$, and get for infinitely many $n$,
\[\left| \sum_{i=1}^{k} a_i x_{i, n}\big/x_{1, n} \right| = 0.\]
Now the result follows by induction.

Next we prove Part A.
Multiplying by its conjugate, we get
\[ \left(\sum_{i=0}^{k} a_i x_{i, n} \right)
\left(\sum_{j=0}^{k} \overline{a_j} x_{j, n} \right)
= \left( \sum_{i=0}^{k} b_i x_{i, n} \right)
\left(\sum_{j=0}^{k} \overline{b_j} x_{j, n} \right).\]
Every term involves a product $x_{i, n} x_{j, n}$.
If $\max\{i, j\} < \ell$, then the terms
$a_i \overline{a_j} x_{i, n} x_{j, n} =
b_i \overline{b_j} x_{i, n} x_{j, n}$ and they cancel (since $a_i=b_i$ and $a_j=b_j$).
If $\max\{i, j\} > \ell$, then both terms
$a_i \overline{a_j} x_{i, n} x_{j, n}$ and
$b_i \overline{b_j} x_{i, n} x_{j, n}$
are $o(|x_{\ell, n}|)$ as $n \rightarrow \infty$.
This is also true if $\max\{i, j\} = \ell$ and $\min\{i, j\}  > 0$.
The only remaining terms correspond to $\max\{i, j\} = \ell$
and $\min\{i, j\}
=0$.  After canceling out identical terms,
we get
\[ ( a_\ell + \overline{a_\ell} ) x_{\ell, n} +
o(|x_{\ell, n}|)
= ( b_\ell + \overline{b_\ell} ) x_{\ell, n} + o(|x_{\ell, n}|),\]
as $n \rightarrow \infty$.
Dividing out $x_{\ell, n}$, and then taking limit
$n \rightarrow \infty$,  we get the real part
\[\text{Re} (a_\ell) = \text{Re}(b_\ell).\]
It follows that $a_\ell = b_\ell$ since $\text{Im}(a_\ell)=\text{Im}(b_\ell)$.
\end{proof}

We remark that Part A of the Vanishing Lemma above cannot
  be extended to arbitrary sequences $\{a_i\}$ and $\{b_i\}$ without the condition
  that $\text{Im}(a_\ell)=\text{Im}(b_{\ell})$,
  as shown by the following example:
Let $$a_1 = 3 + \sqrt{3} i,\ \ a_2 =  3 \left(\frac{1}{2} + \frac{\sqrt{3}}{2} i\right) ,\ \
  \text{and}\ \ b_1 = b_2 = 3.$$
Then the following is an identity for  all real values $x$,
\[\left| 1 + a_1 x + a_2 x^2 \right|
= \left| 1 + b_1 x + b_2 x^2 \right|.
\]
In particular this holds when $x \rightarrow 0$.
We note that $a_1 \not = b_1$.

\subsubsection{Proof of Lemma \ref{rank1}}\label{proofofrank1}

Without loss of generality, we assume $1=\mu_1>\ldots>\mu_s>0$
  and $1=\nu_1>\ldots>\nu_t>0$ (otherwise, we can multiply $\CC$
  with an appropriate scalar so that the new $\CC$ has this property.
  This operation clearly does not affect
  the complexity of $\eval(\CC,\fD)$).
We assume $\eval(\CC,\fD)$ is not \#P-hard.

Again, we let $\fD^*$ denote a sequence of $N$ $m\times m$ diagonal
  matrices in which every matrix is a copy of
  the matrix $\DD^{[0]}$ in $\fD$.
It is clear that $\fD^*$ satisfies condition $(\calT_3)$.

\begin{figure}
\center
\includegraphics[height=6cm]{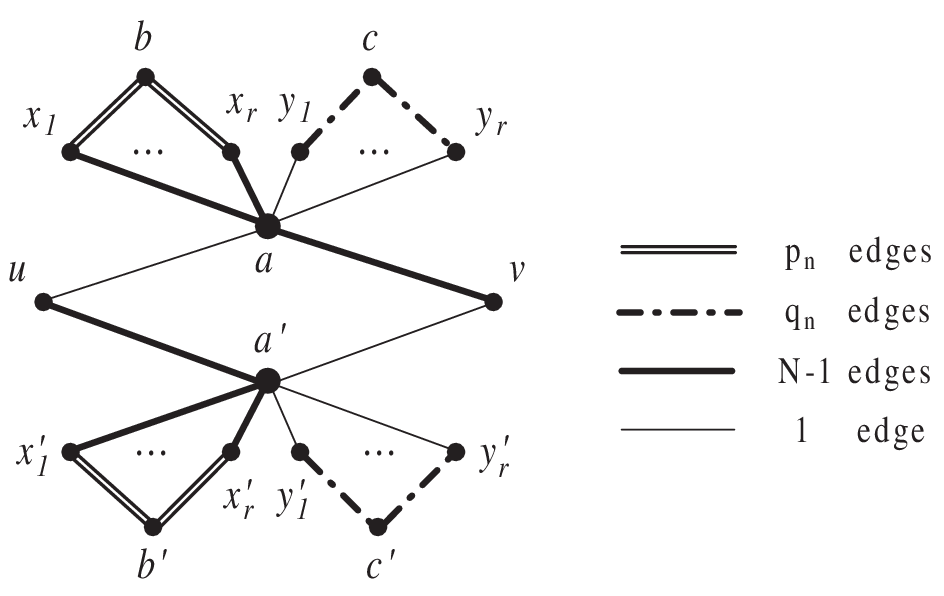}
\caption{Gadget for constructing $G^{[n]}$, $n\ge 1$ (Note that the
  subscript $e$ is suppressed).}\label{figure_4}
\end{figure}

Recall that $r$ is a fixed index in $[N-1]$,
  and the definition of the $s\times h$ matrix $\DD$ from $\DD^{[r]}$.
Let $G=(V,E)$ be an undirected graph. For each $n\ge 1$, we construct
  a new graph $G^{[n]}$ by replacing every edge $uv\in E$
  with a gadget which is shown in Figure \ref{figure_4}.

More exactly, we define $G^{[n]}$ as follows. Let $p_n=n^2N+1$ and
$q_n=nN-1$ (when $n\rightarrow \infty$,
  $q_n$ will be arbitrarily large, and for a given $q_n$,
  $p_n$ will be arbitrarily {\it larger}). Then
$$
V^{[n]}=V\cup
\big\{a_e,x_{e,i},y_{e,i},b_e,c_e,a_e',x_{e,i}',y_{e,i}',b_e',c_e'
  \hspace{0.06cm}\big|\hspace{0.06cm}e\in E,i\in [r]\big\},
$$
and $E^{[n]}$ contains exactly the following edges: For every edge $e=uv\in E$,
\begin{enumerate}
\item One edge between $(u,a_e),(v,a_e'),
  (a_e,y_{e,i})$ and $(a_e',y_{e,i}')$, for all $i\in [r]$;
\item $N-1$ edges between $(v,a_e),(u,a_e'),(a_e,x_{e,i})$ and $
  (a_e',x_{e,i}')$, for all $i\in [r]$;
\item $p_n$ edges between $(b_e,x_{e,i})$ and $(b_e',x_{e,i}')$, for all $i\in [r]$;
\item $q_n$ edges between $(c_e,y_{e,i})$ and $(c_e',y_{e,i}')$, for all $i\in [r]$.
\end{enumerate}
It is easy to check that the degree of every vertex in graph $G^{[n]}$
  is a multiple of $N$ except for $b_e$ and $b_e'$, which have degree $r\bmod N$,
    and $c_e$ and $c_e'$, which have degree $N-r\bmod N$.

Since the gadget is symmetric with respect to vertices $u$ and $v$, the
  construction of $G^{[n]}$ gives us a symmetric $m\times m$ matrix $\RR^{[n]}$
  (recall $m=(s+t)\times h$) such that
$$
Z_{\RR^{[n]},\fD^*}(G)=Z_{\CC,\fD}(G^{[n]}),\ \ \ \text{for all undirected graphs $G$.}
$$
As a result, $\eval(\RR^{[n]},\fD^*)\le \eval(\CC,\fD)$,
  and $\eval(\RR^{[n]},\fD^*)$ is also not \#P-hard.\vspace{0.008cm}

The entries of $\RR^{[n]}$ are as follows: For all $\uu\in I$ and $\vv\in J$,
$$
R^{[n]}_{(0,\uu),(1,\vv)}=R^{[n]}_{(1,\uu),(0,\vv)}=0.
$$
For $\uu,\vv\in J$, we have
\begin{eqnarray*}
R^{[n]}_{(1,\uu),(1,\vv)}=\left(\sum_{\aa,\bb,\cc\in I}
\left(\sum_{\xx\in J} F_{\aa,\xx}^{N-1} F_{\bb,\xx}^{p_n}D^{[0]}_{(1,\xx)}
  \right)^r\left(\sum_{\yy\in J} F_{\aa,\yy}F_{\cc,\yy}^{q_n}
  D^{[0]}_{(1,\yy)}\right)^r F_{\aa,\uu}F_{\aa,\vv}^{N-1}D^{[0]}_{(0,\aa)}D^{[r]}_{(0,\bb)}
  D^{[N-r]}_{(0,\cc)}\right)\\[0.8ex]
\times \left(\sum_{\aa,\bb,\cc\in I}
\left(\sum_{\xx\in J} F_{\aa,\xx}^{N-1} F_{\bb,\xx}^{p_n}D^{[0]}_{(1,\xx)}
  \right)^r\left(\sum_{\yy\in J} F_{\aa,\yy}F_{\cc,\yy}^{q_n}
  D^{[0]}_{(1,\yy)}\right)^r F_{\aa,\uu}^{N-1}F_{\aa,\vv} D^{[0]}_{(0,\aa)}D^{[r]}_{(0,\bb)}
  D^{[N-r]}_{(0,\cc)}\right).
\end{eqnarray*}
Let us simplify the first factor. By using ({\sl Shape}$_2$) and ({\sl Shape}$_5$), we have
\begin{eqnarray}\nonumber
\sum_{\xx\in J} F_{\aa,\xx}^{N-1} F_{\bb,\xx}^{p_n}D^{[0]}_{(1,\xx)}\hspace{-0.15cm}&=&
\hspace{-0.15cm}\mu_{a_1}^{N-1}\mu_{b_1}^{p_n}\hspace{0.09cm}\sum_{\xx\in J}
  (\nu_{x_1})^{N-1+p_n} \overline{H_{a_2,x_2}}H_{b_2,x_2}D^{[0]}_{(1,(x_1,1))}\\[0.8ex]
  \label{check}
  \hspace{-0.15cm}&=&\hspace{-0.15cm}\mu_{a_1}^{N-1}\mu_{b_1}^{p_n}\sum_{x_1\in [t]}
  (\nu_{x_1})^{N-1+p_n}D^{[0]}_{(1,(x_1,1))}\langle \HH_{b_2,*},\HH_{a_2,*}\rangle.
\end{eqnarray}
We use $L$ to denote the following positive number which is independent of
  $\uu,\vv,\aa,\bb$ and $\cc$:
$$
L=h\cdot \sum_{x_1\in [t]} (\nu_{x_1})^{N-1+p_n}D^{[0]}_{(1,(x_1,1))}.
$$
Then by ({\sl Shape}$_4$), (\ref{check}) is equal to $L\cdot \mu_{a_1}^{N-1}\mu_{b_1}^{p_n}$
  if $a_2=b_2$; and $0$ otherwise. Similarly,
$$ \sum_{\yy\in J} F_{\aa,\yy}F_{\cc,\yy}^{q_n}
  D^{[0]}_{(1,\yy)}=L'\cdot \mu_{a_1}\mu_{c_1}^{q_n},\ \ \ \text{if $a_2=c_2$;}$$
  and $0$ otherwise, where $L'$ is a positive number that
  is independent of $\uu,\vv,\aa,\bb$ and $\cc$.

By ({\sl Shape}$_3$), we have $$D^{[N-r]}_{(0,\cc)}=\overline{
  D^{[r]}_{(0,\cc)}}=\overline{D_{c_1,c_2}}.$$
Combining these equations, the first factor of $R^{[n]}_{(1,\uu),(1,\vv)}$ becomes
$$
\nu_{u_1}\nu_{v_1}^{N-1}
  \sum_{\aa\in I,b,c\in [s]}\Big(L\cdot \mu_{a_1}^{N-1}\mu_{b}^{p_n}\Big)^r
  \Big(L'\cdot \mu_{a_1}\mu_{c}^{q_n}\Big)^r
  \mu_{a_1}^N H_{a_2,u_2}\overline{H_{a_2,v_2}}D^{[0]}_{(0,(a_1,1))}
  D_{b,a_2}\overline{D_{c,a_2}}.
$$
Let $Z$ denote the following positive number that is independent of $\uu$ and $\vv$:
$$
Z=\sum_{a_1\in [s]} \Big(L\cdot \mu_{a_1}^{N-1}\Big)^r
  \Big(L'\cdot \mu_{a_1}\Big)^r\mu_{a_1}^ND^{[0]}_{(0,(a_1,1))}.
$$
Let $P_n=rp_n$ and $Q_n=rq_n$, then the first factor becomes
$$
Z\cdot \nu_{u_1}\nu_{v_1}^{N-1}
\sum_{b,c\in [s]}\mu_b^{P_n}\mu_c^{Q_n}\sum_{a\in [h]}D_{b,a}
  \overline{D_{c,a}}H_{a,u_2}\overline{H_{a,v_2}}.
$$
We can also simplify the second factor so that $R^{[n]}_{(1,\uu),(1,\vv)}$ is equal to
$$
Z^2 (\nu_{u_1} \nu_{v_1})^N
  \left(\sum_{b,c\in [s]}\mu_b^{P_n}\mu_c^{Q_n}\sum_{a\in [h]}D_{b,a}
  \overline{D_{c,a}}H_{a,u_2}\overline{H_{a,v_2}}\right) \left(
  \sum_{b',c'\in [s]}\mu_{b'}^{P_n}\mu_{c'}^{Q_n}\sum_{a\in [h]}D_{b',a}
  \overline{D_{c',a}}\overline{H_{a,u_2}}{H_{a,v_2}}\right).
$$
%
%
%
%
Since $\eval(\RR^{[n]},\fD^*)$ is not \#P-hard and $(\RR^{[n]},\fD^*)$ satisfies
  $(\calT)$ for all $n\ge 1$,
  the necessary condition of the Inverse Cyclotomic Reduction Lemma
  (Corollary \ref{inversetwin}) applies to $\RR^{[n]}$.

In the proof below, for notational convenience we suppress the index
  $n\ge 1$ and use $P,Q$ and $\RR$
  to represent sequences $\{P_n\},\{Q_n\}$ and $\{\RR^{[n]}\}$, respectively.
Whenever we state or prove a property about $\RR$, we mean $\RR^{[n]}$
  has this property for any large enough $n$ (sometimes it holds for all $n\ge 1$).
Moreover, since we only use the entries of $\RR^{[n]}$ indexed by $((1,\uu),(1,\vv))$
  with $u_1=v_1=1$, we let
$$
R_{u,v}\equiv R_{(1,(1,u)),(1,(1,v))},\ \ \ \text{for all $u,v\in [h]$}.$$
As a result, we have (note that $\nu_1=1$)
\begin{equation}\label{maineq}
R_{u,v}=Z^2
  \left(\sum_{b,c\in [s]}\mu_b^{P}\mu_c^{Q}\sum_{a\in [h]}D_{b,a}
  \overline{D_{c,a}}H_{a,u}\overline{H_{a,v}}\right) \left(
  \sum_{b',c'\in [s]}\mu_{b'}^{P}\mu_{c'}^{Q}\sum_{a\in [h]}D_{b',a}
  \overline{D_{c',a}}\overline{H_{a,u}}{H_{a,v}}\right).
\end{equation}


We will consider the above expression for $R_{u,v}$
{\it stratified} according to the order of magnitude
of
$$ \mu_b^P \mu_c^Q \mu_{b'}^P \mu_{c'}^Q
  =(\mu_b\mu_{b'})^P(\mu_c\mu_{c'})^Q.$$
Since $P=\Theta(n^2)$ and $Q=\Theta(n)$, when $n\rightarrow \infty$,
  $Q$ is arbitrarily and sufficiently large, and $P$ is
  further arbitrarily and sufficiently
  large compared to $Q$.
Thus, the terms are ordered strictly first by
$\mu_b \mu_{b'}$, and then by $\mu_c \mu_{c'}$.

Inspired by this observation, we define the following
  total order $\le_{\mu}$ over $\calT$, where
$$
\calT=\Big\{ \left(\begin{matrix}
b& c \\ b' & c'
\end{matrix}\right)\hspace{0.06cm}\Big|\hspace{0.06cm}\text{$b,b',c,c'\in [s]$}\hspace{0.06cm}\Big\}.
$$
For $T_1$ and $T_2$ in $\calT$, where\vspace{-0.26cm}
$$
T_1=\tone\ \ \ \text{and}\ \ \ T_2=\ttwo,\vspace{0.06cm}
$$
we have $T_1\lek T_2$\vspace{0.01cm} if either $\mu_{b_1}\mu_{b_1'}<\mu_{b_2}\mu_{b_2'}$;
  or $\mu_{b_1}\mu_{b_1'}=\mu_{b_2}\mu_{b_2'}$
  and $\mu_{c_1}\mu_{c_1'}\le \mu_{c_2}\mu_{c_2'}$.
For\vspace{0.006cm} convenience, whenever
  we denote a $2\times 2$ matrix in $\calT$ by $T_i$ or $T$, we denote
  its entries by
$$
\left(\begin{matrix}b_i & c_i\\ b_i' & c_i'\end{matrix}\right) \ \ \text{or} \ \
\tmatrix,\ \ \text{respectively.}
$$
Using $\lek$, we can divide $\calT$ into classes $\calT_1,\calT_2,\ldots,\calT_d$
  ordered from the largest to the smallest, for
  some positive integer $d$, such that\vspace{0.1cm}
\begin{enumerate}
\item If $T_1,T_2\in \calT_i$, for some $i\in [d]$, then we have $
  \mu_{b_1}\mu_{b_1'}=\mu_{b_2}\mu_{b_2'}$ and
  $\mu_{c_1}\mu_{c_1'}=\mu_{c_2}\mu_{c_2'}$. Note that this is
  an equivalence relation which we denote by $=_\mu$;
\item If $T_1\in \calT_i$, $T_2\in \calT_j$ and $i<j$, then either
  $\mu_{b_1}\mu_{b_1'}>\mu_{b_2}\mu_{b_2'}$; or
  $\mu_{b_1}\mu_{b_1'}=\mu_{b_2}\mu_{b_2'}$ and
  $\mu_{c_1}\mu_{c_1'}>\mu_{c_2}\mu_{c_2'}$.\vspace{0.1cm}
\end{enumerate}
For each $i\in [d]$, we arbitrarily pick a $T\in \calT_i$ and let
  $U_i$ denote $\mu_{b}\mu_{b'}$ and
  $W_i$ denote $\mu_{c}\mu_{c'}$ (note that $U_i$
  and $W_i$ are independent of the choice of $T$).
It is clear that there is exactly one matrix
${1\ 1\choose 1\ 1}$ in $\calT_1.$

Now we can rewrite (\ref{maineq}) as follows
\begin{equation}\label{main2}
R_{u,v}=Z^2\sum_{i\in [d]} U_i^{P}W_i^{Q}
  \sum_{T\in \calT_i} X_{u,v,T},
\end{equation}
where\vspace{-0.3cm}
$$
  X_{u,v,T}=\left(\sum_{a\in [h]}D_{b,a}
  \overline{D_{c,a}}H_{a,u}\overline{H_{a,v}}\right) \left(
  \sum_{a\in [h]}D_{b',a}
  \overline{D_{c',a}}\overline{H_{a,u}}{H_{a,v}}\right),\ \ \ \text{for $T=\tmatrix$.}
$$

Clearly, the term with the maximum possible order in the sum
(\ref{main2}) corresponds to the choice of $T={1\ 1\choose 1\ 1}\in
\calT_1$, since $\mu_1$ is strictly maximum among all $\mu_1,
\ldots, \mu_s$. This\vspace{-0.04cm} is true for every $(u,v)$, and
it will be the actual
  leading term of the sum, provided the coefficient
  of $U_1^PW_1^Q= \mu_1^{2P + 2Q}$ is non-zero.

Consider the diagonal entries where $u=v$:\vspace{-0.035cm}
First, notice that from (\ref{maineq}), we have $R_{u,u}=R_{1,1}$
  for all $u\in [h]$; Second, the coefficient of the leading term $U_1^PW_1^Q$
  is $$X_{u,u,{1\ 1\choose 1\ 1}}=\left(\sum_{a\in [h]} |D_{1,a}|^2\right)^2
  =\|\DD_{1,*}\|^4,$$
which is, again, independent of $u$.
Without loss of generality, we may assume $\DD_{1,*}$ is not identically 0;
otherwise, we can remove all terms involving $\mu_1$
  in Eq.\hspace{0.05cm}(\ref{maineq}) and $\mu_2$ will take its
place, and the proof is completed by induction.  (If all $\DD_{i,*} = \00$,
then the statement that $\DD$ has rank at most one is trivial.)

Assuming that\vspace{-0.03cm} $\DD_{1,*}\ne \00$,
  we have $R_{u,u}=R_{1,1}\ne 0$, for all $u\in [h]$ (and sufficiently large $n$).
This is because, ignoring the positive factor $Z^2$,
  the coefficient $\|\DD_{1,*}\|^4$ of the leading term $U_1^PW_1^Q$ is positive.
By using Corollary \ref{inversetwin},
  we have\vspace{0.03cm}

\begin{property}\label{property1}
For all sufficiently large $n$,
  $|R_{1,1}|>0$ and $|R_{u,v}|\in \big\{0,|R_{1,1}|\big\}$ for all $u,v\in [h]$.
\end{property}

From now on, we focus on $u=1$ and denote by ${\calHH}_{*,v} =
  \HH_{*,1} \circ \overline{\HH_{*,v}}$.
We note that $\{\calHH_{*,v}\}_{v \in [h]}$ forms an orthogonal
basis, with each $\|\calHH_{*,v}\|^2 = h$. We also denote $X_{1,v,T}$ by $X_{v,T}$, so
\begin{equation}\label{factors}
X_{v,T}=\left(\sum_{a\in [h]}D_{b,a}
  \overline{D_{c,a}}\calH_{a,v}\right) \left(
  \sum_{a\in [h]}D_{b',a}
  \overline{D_{c',a}}\overline{\calH_{a,v}} \right)\ \text{for $T=\tmatrix$.}
\end{equation}

We make two more definitions.
Let $K = \{i \in [h] \mid D_{1,i} \not = 0\}.$
By our assumption, $K \not = \emptyset$. Define
$$
A = \{v \in [h]\hspace{0.07cm}\big|\hspace{0.1cm}\forall\hspace{0.05cm}
  i, j \in K, \calH_{i,v}=\calH_{j,v}\}\ \ \
\text{and}\ \ \ B = [h] - A.$$
Note that if $|K| = 1$ then $A=[h]$.
The converse is also true which follows from the fact that
  $\{\calHH_{*,v}\}_{v\in [h]}$ forms an orthogonal basis.
Also since $\calHH_{*,1}$ is the all-one vector, $1\in A$ and
  $A$ is non-empty.
Moreover, if $K = [h]$, then  $A = \{1\}$.
This, again, follows from the fact that
$\{\calHH_{*,v}\}$ forms an orthogonal basis.

Now we\vspace{0.005cm} consider 
  the coefficient $X_{v,T}$ of $U_1^PW_1^Q$ in $R_{1,v}$, where
  $T={1\ 1 \choose 1\ 1}$.
For every $v\in A$,
  it has norm $\|\DD_{1,*}\|^4>0$.
It then follows from Property \ref{property1}
  and Part B of the Vanishing Lemma that

\begin{property}\label{property2}
For any $v\in A$ and sufficiently large $n$, $|R_{1,v}|=|R_{1,1}|$.
\end{property}

If $B \not = \emptyset$, then for any $v \in B$,
the coefficient of $T={1\ 1\choose 1\ 1}$ in $R_{1,v}$ is
$$X_{v,T}=
\left(\sum_{a \in K} |D_{1,a}|^2 \calH_{a,v}\right) \left(\sum_{a
\in K} |D_{1,a}|^2 \overline{\calH_{a,v}}\right) = \left| \sum_{a\in
K} |D_{1,a}|^2 \calH_{a,v} \right|^2\in \mathbb{R}.$$ Since we
assumed $v\in B$,
  $\sum_{a\in K} |D_{1,a}|^2 \calH_{a,v}$
  is a sum of positive terms $|D_{1,a}|^2$ weighted by
  non-constant $\calH_{a,v}$, for $a \in K$, each with complex norm $1$.
Thus its absolute value must be strictly less than $\|\DD_{1,*}\|^2$,
which is only achieved when all $\calH_{a,v}$, for $a \in K$, are
equal to a constant. 
It follows that 
$
X_{v,T}<\|\DD_{1,*}\|^4.
$
Therefore, for $v\in B$ (and $n$ sufficiently large), we have
$|R_{1,v}|<|R_{1,1}|.$
By using Property \ref{property1}
  and Part B of the Vanishing Lemma, we have the following property:

\begin{property}\label{property3}
If $v\in B$, then for all sufficiently large $n$, $R_{1,v}=0$ and thus,
$$
\sum_{T\in \calT_i} X_{v,T}=0,\ \ \ \text{for all $i\in [d]$.}
$$
\end{property}

In particular, by applying Property \ref{property3} to
  $\calT_1=\{{1\ 1\choose 1\ 1}\}$, we have
$$\sum_{a\in K} |D_{1,a}|^2 \calH_{a,v}=\sum_{a\in K} |D_{1,a}|^2
  \overline{\calH_{a,v}}=\langle |\DD_{1,*}|^2,\calHH_{*,v}\rangle =0,
  \ \ \ \text{for every $v\in B$,}$$
since $|D_{1,a}|$ is real.
Furthermore, because $\{\calHH_{*,v}\}$ forms an orthogonal basis,
$|\DD_{1,*}|^2$ must be expressible as
 a linear combination of $\{\calHH_{*,v}\hspace{0.06cm}|\hspace{0.06cm}
   v \in A\}$, over ${\mathbb{C}}$.
From such an expression, we have $|D_{1,i}|^2 = |D_{1,j}|^2$
  for all $i, j \in K$, by the definition of $K$.
Since $\DD_{1,*}$ is only non-zero on $K$,  $|D_{1,i}|$ is a constant on $K$,
  and $D_{1,i} = 0$ for any $i \in [h] - K$.
(The above proof does not actually assume $B \not = \emptyset$;
if $B = \emptyset$,
then $A = [h]$ and by $\{\calHH_{*,v}\}$ being an orthogonal basis,
$|K| = 1$. Then the above  statement about $\DD_{1,*}$   is still valid, namely
$\DD_{1,*}$ has a unique non-zero entry and zero elsewhere.)\vspace{0.01cm}

We summarize the above as follows:

\begin{claim}\label{claim-D-firstrow}
$|\DD_{1,*}|^2 \perp \calHH_{*,v}$ for all $v \in B$,
and $|\DD_{1,*}|^2$ is a constant on $K$ and 0 elsewhere.
In particular the vector $\chi_K$, which is 1 on $K$ and 0 elsewhere,
is in the span of $\{\calHH_{*,v}\hspace{0.07cm}|\hspace{0.07cm} v \in A\}$, and is
  orthogonal to all $\{\calHH_{*,v}\hspace{0.07cm}|\hspace{0.07cm} v \in B\}$.
\end{claim}

Our next goal is to show that on set $K$,
$\DD_{2,*}$ is a constant multiple of $\DD_{1,*}$.
Clearly if $B = \emptyset$, then $|K| = 1$ as noted above and thus,
it is trivially true that $\DD_{2,*}$ is a constant multiple of $\DD_{1,*}$
on $K$.
So we assume $B  \not = \emptyset$.
We now consider\vspace{-0.15cm}
$$T_1 =  \begin{pmatrix} 2 & 1 \\ 1 & 2\end{pmatrix}\ \ \ \text{and}
\ \ \ T_2 =  \begin{pmatrix} 1 & 2 \\ 2 & 1\end{pmatrix}.$$
$T_1$ and $T_2$ belong to the same $\calT_g$, for some $g\in [d]$.
By Property \ref{property3}, we have $\sum_{T\in \calT_g} X_{v,T}=0$
  for every $v\in B$.
So we focus on terms $X_{v,T}$, where $T\in \calT_g$ (i.e., $T\eqk T_1$).
Suppose $T\eqk T_1$, then by definition, we have $\mu_{b}\mu_{b'}=\mu_1\mu_2$
  and $\mu_c\mu_{c'}=\mu_1\mu_2$.
Thus, $\{b,b'\}=\{c,c'\}=\{1,2\}$.
As a result,
$$
\calT_g=\left\{T_1,T_2,T_3=\begin{pmatrix}1&1\\2&2\end{pmatrix},
  T_4 = \begin{pmatrix} 2 & 2 \\ 1 & 1 \end{pmatrix}
\right\}.
$$
However, due to the presence of a row $(1\  1)$,
  the sum $\sum_{a =1}^h |D_{1, a}|^2 \calH_{a,v}
= \sum_{a =1}^h |D_{1, a}|^2 \overline{\calH_{a,v}} = 0$
for any $v \in B$ as shown above. Therefore,
the coefficients $X_{v,T_3}$, $X_{v,T_4}$ corresponding to $T_3$ and $T_4$ are
both 0.\vspace{0.01cm}

We make one more observation:
\begin{enumerate}
\item[]{\bf Observation}:
We say a matrix $T\in \calT$ is of a {\emph{Conjugate-Pair}} form if it
is of the form $$T=\begin{pmatrix} b & c\\ c & b\end{pmatrix}.$$
For a matrix $T$ in Conjugate-Pair form,
the corresponding coefficient $X_{v,T}$ 
is of the form
\[
X_{v,T}=\left|
\sum_{a=1}^{h} D_{b,a}
        \overline{D_{c,a}} \calH_{a,v} \right|^2,\]
which is always non-negative.
\end{enumerate}

Now the remaining two matrices $T_1$ and $T_2$ in $\calT_g$
  both have this form, so both $X_{v,T_1}$ and
  $X_{v,T_2}$ are non-negative.
Since $X_{v,T_1}+X_{v,T_2}=0$,
  both $X_{v,T_1}$ and $X_{v,T_2}$ must be zero.
This gives us
\[\sum_{a\in [h]} \overline{D_{1,a}} D_{2,a} \overline{\calH_{a,v}} = 0,
\ \ \ \text{for all $v \in B$.}
\]
Hence the vector $\overline{\DD_{1,*}} \circ \DD_{2,*} \perp \calHH_{*,v}$
for all $v \in B$.
It follows that the vector $\overline{\DD_{1,*}} \circ \DD_{2,*}$
  is expressible as a linear combination of $\calHH_{*,v}$
over $v \in A$. By the definition of $A$,
this expression has a constant value
on entries indexed by $a \in K$, where $|D_{1,a}|$ is a positive constant.
Therefore, over $K$, $\DD_{2,*}$ is
  a constant multiple of $\DD_{1,*}$.
This accomplished our goal stated above, which we summarize as

\begin{claim}\label{claim-2-D-second-row-multi-1row}
There exists some complex number $\lambda$,
such that $D_{2,a} = \lambda D_{1,a}$, for all $a \in K$.
\end{claim}

Let ${K}_2 = \{i \in [h]\hspace{0.08cm}|\hspace{0.06cm} D_{2,i} \not = 0\}$.
Note that the $\lambda$ above could be $0$ so it is possible that $K\not\subset K_2$.
Our next goal is to show that for every $v \in A$,  $\calHH_{*,v}$
takes a constant value on ${K}_2$.
This means that for all $v \in A$,
$\calH_{i,v} = \calH_{j,v}$, for all $i, j \in {K}_2$.
Without loss of generality, we assume $\DD_{2,*} \not = \00$ since otherwise
 ${K}_2 = \emptyset$ and everything below regarding $\DD_{2,*}$
 and regarding $\calHH_{*,v}$ on $K_2$ are trivially true.\vspace{0.01cm}

Toward this end, we will consider the class
$$\calT_g=\left\{T_1=\begin{pmatrix} 2 & 1 \\ 1 & 2 \end{pmatrix},
T_2=\begin{pmatrix} 1 & 2 \\ 2 & 1 \end{pmatrix},
T_3 =  \begin{pmatrix} 1 & 1 \\ 2 & 2\end{pmatrix},
T_4 =  \begin{pmatrix} 2 & 2 \\ 1 & 1\end{pmatrix}\right\}$$
and their corresponding coefficients $X_{v,T_i}$ for any $v \in A$.
We will apply the more delicate
  Part A of the Vanishing Lemma 
  on $R_{1,v}$ and $R_{1,1}$, for an arbitrary $v\in A$.
Our target is to show that
\begin{equation}\label{target1}
\sum_{T\in \calT_g} X_{v,T}=\sum_{T\in \calT_g} X_{1,T},\ \ \ \text{for any $v\in A$}.
\end{equation}
By Property \ref{property2}, we already know that $|R_{1,v}|=|R_{1,1}|$ for any
  sufficiently large $n$.
So in order to apply the Vanishing Lemma, we need first to show that
  terms which have a higher order of magnitude satisfy
\begin{equation}\label{higherorder}
\sum_{T\in \calT_{g'}} X_{v,T}=\sum_{T\in \calT_{g'}} X_{1,T},\ \ \
  \text{for all $1\le g'<g$ and $v\in A$.}
\end{equation}
We also need to show that\vspace{-0.1cm}
\begin{equation}\label{imagepart}\text{Im}\left(\sum_{T\in \calT_g} X_{v,T }\right)=
  \text{Im}\left(\sum_{T\in \calT_g} X_{1,T }\right).\vspace{0.05cm}\end{equation}

By definition, any $T\ge_{\mu} T_1$ must satisfy
  $\mu_b\mu_{b'}\ge \mu_1\mu_2$.
Thus the first column of $T$ is\vspace{-0.05cm}
$$\text{either $\begin{pmatrix} 1 \\ 1 \end{pmatrix}$,
$\begin{pmatrix} 1 \\ 2 \end{pmatrix}$ or
$\begin{pmatrix} 2 \\ 1 \end{pmatrix}$.}$$

Firstly, consider those matrices $T\ge_{\mu} T_1$
where each row of  $T$ has at least one 1's.
For every $v \in A$,
 the two inner product factors in (\ref{factors}),
namely,  $\sum_{a=1}^{h} D_{b,a}
        \overline{D_{c,a}} \calH_{a,v}$,
and $\sum_{a=1}^{h} D_{b',a}
        \overline{D_{c',a}}\hspace{0.06cm}\overline{\calH_{a,v}}$,
must be actually a sum over $a \in K$, since $\DD_{1,*}$ is zero elsewhere.
But for $a \in K$,  $\calH_{a,v}$ is just a constant $\alpha_v$ of norm 1 (a root of
unity),
independent of $a \in K$.  Thus
$$\sum_{a=1}^{h} D_{b,a}
        \overline{D_{c,a}} \calH_{a,v} = \alpha_v \sum_{a \in K} D_{b,a}
\overline{D_{c,a}}\ \ \ \text{and}\ \ \
\sum_{a=1}^{h} D_{b',a}
        \overline{D_{c',a}} ~ \overline{\calH_{a,v}} = \overline{\alpha_v}
\sum_{a \in K} D_{b',a} \overline{D_{c',a}}.$$
Since $\alpha_v \overline{\alpha_v} = |\alpha_v|^2 = 1$,
it follows that their product is
\[ \left(\sum_{a=1}^{h} D_{b,a} \overline{D_{c,a}} \calH_{a,v} \right)
 \left( \sum_{a=1}^{h} D_{b',a} \overline{D_{c',a}} \overline{\calH_{a,v}} \right)
= \left(\sum_{a \in K} D_{b,a} \overline{D_{c,a}} \right)
\left( \sum_{a \in K} D_{b',a} \overline{D_{c',a}} \right),\]
which is the same as the coefficient $X_{1,T}$ corresponding to
  $T$ for $v_0 =1 \in A$.
Thus for all such $T$, their respective contributions
  to $R_{1, v}$ and to $R_{1, 1}$ are the same, for any $v \in A$.

Such matrices $T\ge_{\mu} T_1$ with at least one 1's in each row
  include any matrix of the form
$$\text{$\begin{pmatrix} 1 & c \\ 1 & c' \end{pmatrix}$,\ \
$\begin{pmatrix} 1 & 1 \\ 2 & 1 \end{pmatrix}$\ \
or\ \
$\begin{pmatrix} 2 & 1 \\ 1 & 1 \end{pmatrix}$.}$$
These exhaust all $T>_{\mu} T_1$, and (\ref{higherorder}) follows.

Such matrices $T\ge_\mu T_1$  also include $T_1$ and $T_2$ in $\calT_g$.
  So $X_{v,T_1}=X_{1,T_1}$
  and $X_{v,T_2}=X_{1,T_2}$, for any $v\in A$.
Now we deal with matrices $T_3$ and $T_4$. 
We note that the {\it sum} of $X_{v,T_3}$ and $X_{v,T_4}$, at any $v$, is
\begin{equation}\label{contribution-sum-of-two}
\left( \sum_{a \in K} |D_{1,a}|^2 \calH_{a,v} \right)
   \left( \sum_{a =1}^h |D_{2,a}|^2 \overline{\calH_{a,v}} \right)
+
\left( \sum_{a =1}^h |D_{2,a}|^2 \calH_{a,v} \right)
\left( \sum_{a \in K} |D_{1,a}|^2 \overline{\calH_{a,v}} \right),
\end{equation}
which is a real number. (\ref{imagepart}) then follows.

Now we can apply Part A of the Vanishing Lemma 
  which gives us (\ref{target1}).
Because $X_{v,T_1}=X_{1,T_1}$ and $X_{v,T_2}=X_{1,T_2}$,
we have
$$
X_{v,T_3}+X_{v,T_4}=X_{1,T_3}+X_{1,T_4}= 2\cdot \|\DD_{1,*}\|^2\|\DD_{2,*}\|^2.$$
However this is clearly the maximum possible value of
(\ref{contribution-sum-of-two})
(By our assumption, $\|\DD_{1,*}\|^2 \|\DD_{2,*}\|^2 >0$).
The only way the sum in (\ref{contribution-sum-of-two})
also achieves this maximum at $v \in A$
is for $\calH_{a,v}$ to take a constant value $\beta_v$
for all $a \in {K}_2$, and  $\calH_{a,v}$
to take a constant value $\alpha_v$
for all $a \in K$, for some two complex numbers $\alpha_v$
and $\beta_v$ of norm $1$.
Moreover, by (\ref{contribution-sum-of-two}), we have
\[\alpha_v \overline{\beta_v} + \overline{\alpha_v} \beta_v = 2.\]
It follows that $\alpha_v = \beta_v$.
Thus, $\calH_{a,v}$ is a constant on $a\in K \cup {K}_2$
for each $v \in A$.

We summarize it as follows:
\begin{claim}\label{claim-3-h-is-const-onKhat}
For every $v \in A$, there exists a complex number $\alpha_v$
of norm 1, such that $\calH_{a,v} = \alpha_v$ for all $a$ in $K \cup {K}_2$.
\end{claim}

We eventually want to prove ${K}_2 = K$.
Our next goal is to prove that $|\DD_{2,*}|^2 \perp \calHH_{*,v}$,
for all $v \in B$.
Of course if $B = \emptyset$ then this is vacously true.
We assume $B \not = \emptyset$.

For this purpose we will examine
$$T^* = \begin{pmatrix} 2 & 2 \\ 2 & 2\end{pmatrix},$$
and the class $\calT_g$ it belongs to.
By Property \ref{property3}, we have
$$
\sum_{T\in \calT_g}X_{v,T}=0,\ \ \ \text{for any $v\in B$.}
$$
Thus we will examine $T\in \calT_g$, namely,
  $\mu_{b}\mu_{b'}=\mu_{c}\mu_{c'}=\mu_2^2$.

Now there might be some other pair $(b,b') \not = (2,2)$ such that
$\mu_b \mu_{b'} = \mu_2\mu_2$.
If such a pair exists,
  it is essentially unique,
and is of the form $(1,s)$ or $(s,1)$, where $s > 2$.
Then $\calT_g$ consists of precisely the following matrices, namely
each column must be either
\begin{equation}\label{qusiba}\begin{pmatrix} 2 \\ 2\end{pmatrix}\ \
\text{or}\ \  \begin{pmatrix} s \\ 1 \end{pmatrix}\ \ \text{or}\ \
\begin{pmatrix} 1 \\ s \end{pmatrix}.\end{equation}

Let's\vspace{-0.225cm} examine such a matrix $T =
\begin{pmatrix} b & c \\ b' & c' \end{pmatrix}$ in more detail.
Suppose $T\in \calT_g $ has a row that is either $(1 \ 1)$
or $( 1 \  2)$
or $( 2 \ 1 )$.
Then, 
$$X_{v,T}=\left( \sum_{a=1}^{h} D_{b,a}
        \overline{D_{c,a}} \calH_{a,v} \right)
\left( \sum_{a=1}^{h} D_{b',a}
        \overline{D_{c',a}}\hspace{0.05cm}\overline{\calH_{a,v}}
        \right) =0,\ \ \ \text{for any $v\in B$.}$$
This is because the following: The presence of $\DD_{1,*}$
  restricts the sum to $a \in K$.
By Claim~\ref{claim-D-firstrow}  we know
that for every $v \in B$, $|\DD_{1,*}|^2 \perp \calHH_{*,v}$.
Moreover, on set $K$, we know
from Claim~\ref{claim-2-D-second-row-multi-1row},
that both vectors
  $\overline{\DD_{1,*}}\circ \DD_{2,*}$ and ${\DD_{1,*}}\circ \overline{\DD_{2,*}}$
can be replaced by a constant multiple of the vector
  $|\DD_{1,*}|^2$ (the constant could be $0$),
  thus also perpendicular to $\calHH_{*,v}$ (and to $\overline{\calHH_{*,v}}$).\vspace{0.01cm}

Now suppose $T$ is a matrix in $\calT_g$,
and yet it does not have
a row which is either $(1 \ 1)$
or $(1 \  2)$ or $(2 \ 1)$. By (\ref{qusiba}),
it is easy to check that the only cases are
$$\text{$T^*=\begin{pmatrix} 2 & 2 \\ 2 & 2 \end{pmatrix}$,\ \ \ \
$T_1 = \begin{pmatrix} 1 & s \\ s & 1 \end{pmatrix}$\ \ \ \text{and}\ \ \
$T_2 = \begin{pmatrix} s & 1 \\ 1 & s \end{pmatrix}$.}$$
Thus $X_{v,T^*}+X_{v,T_1}+X_{v,T_2}=0$ for all $v\in B$.
However, as noted above, all three matrices $T^*,T_1$ and $T_2$
have the Conjugate-Pair form, so their contributions
\[\left|
\sum_{a=1}^{h} D_{2,a}
        \overline{D_{2,a}} \calH_{a,v} \right|^2, ~~~
\left|
\sum_{a=1}^{h} D_{1,a}
        \overline{D_{s,a}} \calH_{a,v} \right|^2~~~ \mbox{and} ~~~
\left|
\sum_{a=1}^{h} D_{s,a}
        \overline{D_{1,a}} \calH_{a,v} \right|^2\]
are all non-negative.
It follows that all three  sums are simultaneously zero.
In particular, from $X_{v,T^*}$, we get $|\DD_{2,*}|^2 \perp \calHH_{*,v}$
for all $v \in B$.

It follows that the vector $|\DD_{2,*}|^2$ is
in the span of $\{\calHH_{*,v}\hspace{0.07cm}|\hspace{0.07cm}v \in A\}$.
This linear combination produces a constant value
at any entry $|D_{2,a}|^2$, for $a \in K \cup {K}_2$.
This is because each vector $\calHH_{*,v}$ for $v \in A$ has this property
by Claim~\ref{claim-3-h-is-const-onKhat}.

As we assumed $\DD_{2,*} \not = 0$, and $\DD_{2,*}$ is 0
outside of ${K}_2$ (by the definition of ${K}_2$),
this constant value produced at each entry $| D_{2,a}|^2$
for $a \in K \cup {K}_2$ must be non-zero.
In particular, $D_{2,a} \not = 0$ at $a \in K$.
It follows that $K \subseteq {K}_2$.
It also implies that the vector, which is $1$ on $K\cup K_2=K_2$
  and $0$ elsewhere, is in the span of
  $\{\calHH_{*,v}\hspace{0.07cm}|\hspace{0.07cm} v \in A\}$.

Next we prove that $K = {K}_2$, by showing that $|K| = |{K}_2|$ (since we already
  know $K\subseteq {K}_2$).
Let $\chi_K$ denote the $h$-dimensional characteristic vector for $K$,
which is 1 for any index $a \in K$ and 0 elsewhere.
Similarly denote by $\chi_{{K}_2}$ the characteristic vector
 for ${K}_2$.
We know that both vectors $\chi_K$  and $\chi_{{K}_2}$ are
in the linear span of $\{\calHH_{*,v}\hspace{0.07cm}|\hspace{0.07cm} v \in A\}$.
Write $\chi_K = \sum_{v \in A} x_v \calHH_{*,v}$,
where $x_v \in {\mathbb{C}}$,
then $$x_v \|\calHH_{*,v}\|^2 = \langle \chi_K,\calHH_{*,v}\rangle
  =\sum_{a=1}^h \chi_K(a) \overline{\calH_{a,v}}
=  \sum_{a \in K} \overline{\calH_{a,v}}
= |K| \overline{\alpha_v},$$ by Claim~\ref{claim-3-h-is-const-onKhat}.
It follows that $|x_v|h = |K|$ for each $v \in A$.
Thus
\[ |K|=\|\chi_K\|^2 = \sum_{v \in A} |x_v|^2\cdot \|\calHH_{*,v}\|^2
= |A| \left(\frac{|K|}{h}\right)^2 h = \frac{|A| |K|^2}{h},\]
and it follows that $|K| = h/|A|$.
Exactly the same argument gives also $|{K}_2| = h/|A|$.
Hence $|K| = |{K}_2|$, and $K = {K}_2$.
At this point the statement in Claim~\ref{claim-2-D-second-row-multi-1row}
can be strengthened to

\begin{claim}\label{claim-4-D2-linear-dep-on-D1}
There exists some complex number $\lambda$,
such that $\DD_{2,*} = \lambda \DD_{1,*}.$
\end{claim}

Our final goal is to generalize this proof to all $\DD_{\ell,*}$,
for $\ell = 1, 2, \ldots, s$. We prove this by induction.
\begin{enumerate}
\item[]
{\bf Inductive Hypothesis}:
  For some $\ell \ge 2$,
all rows $\DD_{1,*}, \ldots, \DD_{\ell-1,*}$ are
linearly dependent:
\[ \DD_{i,*} = \lambda_i \cdot \DD_{1,*},\ \ \ \ \text{
for some $\lambda_i$, and $1 \le i < \ell$.}\]
\end{enumerate}
The proof below will mainly follow the proof for the case
$\ell =2$ above, except for  one crucial argument at the end.
We presented the special case $\ell=2$ alone for ease of understanding.

We now prove that $\DD_{\ell,*}=\lambda_\ell\cdot \DD_{1,*}$
  for some $\lambda_\ell$.
Clearly we may assume $\DD_{\ell,*} \not = \00$, for otherwise
the inductive step is trivial.
To start with, we consider the following two matrices
$$T_1 =  \begin{pmatrix} \ell & 1 \\ 1 & \ell\end{pmatrix}\ \ \
\text{and}\ \ \ T_2 =  \begin{pmatrix} 1 & \ell \\ \ell & 1\end{pmatrix},$$
and the corresponding class $\calT_g$ they belong to.
By Property \ref{property3}, we have for every $v\in B$,
$$
\sum_{T\in \calT_g}X_{v,T}=0.
$$
We only need to examine\vspace{-0.02cm}
  each $T\in \calT_g$ with exactly the same order as that of $T_1$, $T_2$:
  $\mu_{b}\mu_{b'}=\mu_c\mu_{c'}=\mu_1\mu_\ell$.
To satisfy this condition, both columns
$ b \choose b' $
and $ c \choose c' $ of $T$
must have entries $\{1, \ell\}$ or have both entries $< \ell$.
Clearly, no entry in $\{b, b', c, c'\}$ can be $> \ell$.
There are two cases now:  Case 1: There is a row
$(b \ c)$
or $(b' \ c')$ (or both)
which has both entries $< \ell$; Case 2: Both rows
have an entry $= \ell$.

In Case 1, at least one of the inner product sums in
  the following product
\[X_{v,T}=\left( \sum_{a=1}^{h} D_{b,a}
        \overline{D_{c,a}} \calH_{a,v} \right)
\left( \sum_{a=1}^{h} D_{b',a}
        \overline{D_{c',a}}\hspace{0.06cm}\overline{\calH_{a,v}} \right)\]
actually takes place over $a \in K$. This follows
from the Inductive Hypothesis.
In fact that  inner product is a constant
multiple of $\sum_{a \in K} |D_{1, a}|^2 \calH_{a,v}$
or its conjugate $\sum_{a \in K} |D_{1, a}|^2  \overline{\calH_{a,v}}$
which are 0 according to Claim~\ref{claim-D-firstrow},
for all $v \in B$.

In Case 2, it is easy to verify that to have the same order
$\mu_1\mu_l$, $T$
must be equal to either $T_1$ or $T_2$.
Now observe that both $T_1$ and $T_2$ have the Conjugate-Pair
form. Therefore, their contributions $X_{v,T_1}$ and $X_{v,T_2}$
  are both non-negative. Since $X_{v,T_1}+X_{v,T_2}=0$,
both of them have to vanish:
$$\sum_{a\in [h]} \overline{D_{1,a}} D_{\ell,a} \overline{\calH_{a,v}} = 0,
~~~\text{and}\ \ \
\sum_{a\in [h]} {D_{1,a}} \overline{D_{\ell,a}}\hspace{0.06cm}\overline{\calH_{a,v}} = 0,
\ \ \ \text{for all $v \in B$.}$$
Hence the vector $\overline{\DD_{1,*}} \circ \DD_{\ell,*} \perp \calHH_{*,v}$,
for all $v \in B$.
It follows that the vector $\overline{\DD_{1,*}} \circ \DD_{\ell,*}$
belongs to the linear span of $\{\calHH_{*,v} \hspace{0.07cm}|\hspace{0.07cm}
  v \in A \}$.
By the definition of $A$,
this expression has a constant value
on entries indexed by $a \in K$. Therefore, on $K$, $\DD_{\ell,*}$ is
a constant multiple of  $\DD_{1,*}$.  We summarize this as follows

\begin{claim}\label{claim-5-D-ell}
There exists some complex number $\lambda_\ell$,
such that $D_{\ell,a} = \lambda_\ell\cdot  D_{1,a},$ for all $a \in K$.
\end{claim}

Let ${K}_\ell = \{i \in [r] \mid D_{\ell,i} \not = 0\}$.
Next, we prove that  for every $v \in A$,  $\calHH_{*,v}$
takes a constant value on ${K}_\ell$, i.e.,
$\calH_{i,v} = \calH_{j,v}$, for all indices $i, j \in {K}_\ell$.
We had assumed $\DD_{\ell,*} \not = 0$, since otherwise
the induction is completed for $\ell$.
Then ${K}_\ell \not = \emptyset$.

To show that  $\calHH_{*,v}$ is a constant on ${K}_\ell$,\vspace{-0.1cm}
we consider $$T_3 =  \begin{pmatrix} \ell & \ell \\ 1 & 1\end{pmatrix}\ \ \
\text{and}\ \ \ T_4 =  \begin{pmatrix} 1 & 1 \\ \ell & \ell\end{pmatrix},$$
and the class $\calT_g$ they belong to.
We want to apply Part A of the Vanishing Lemma to show that
\begin{equation}\label{hahahahahahaha}
\sum_{T\in\calT_g}X_{v,T}=\sum_{T\in \calT_g}X_{1,T},
  \ \ \ \text{for any $v\in A$.}
\end{equation}  
For this purpose, we need to compare the respective terms of the sum
  (\ref{main2}), for an arbitrary $v \in A$ and
  for the particular $v_0 =1 \in A$. More exactly,
  we will show that
\begin{equation}\label{checkcheckhy}
\sum_{T\in \calT_{g'}}X_{v,T}=\sum_{T\in \calT_{g'}}X_{1,T},\ \ \ \ \text{and}\ \ \ \
\text{Im}\left(\sum_{T\in \calT_{g}}X_{v,T}\right)=\text{Im}\left(
  \sum_{T\in \calT_{g}}X_{1,T}\right),
\end{equation}
for all $v\in A$ and $g'<g$. Then (\ref{hahahahahahaha}) follows from
  Part A of the Vanishing Lemma.

To this end, we first consider any matrix $T$ which has
  an order of magnitude strictly larger than that of $T_3$ and $T_4$.\vspace{-0.05cm}
We have $$\text{either}\ \
 \mu_b \mu_{b'} > \mu_1 \mu_\ell,\ \ \text{or}\ \ \big[\hspace{0.05cm}\mu_b \mu_{b'}
  = \mu_1 \mu_\ell\ \ \text{and}\ \ \mu_c \mu_{c'} > \mu_1 \mu_\ell\hspace{0.05cm}\big].\vspace{-0.05cm}$$
The first alternative implies that both $b$ and $ b' < \ell$.  The second alternative
implies that $c$ and $ c' < \ell$.

In both cases, each row of $T$ has\vspace{0.012cm}
at least one entry $< \ell$.
By the Inductive Hypothesis, both inner products
in (\ref{factors}), namely,  $\sum_{a=1}^{h} D_{b,a}
        \overline{D_{c,a}} \calH_{a,v}$
and $\sum_{a=1}^{h} D_{b',a}
        \overline{D_{c',a}}\hspace{0.06cm}\overline{\calH_{a,v}}$,\vspace{0.012cm}
must be actually a sum over $K$ since $\DD_{1,*}$ is zero elsewhere.
However for any $a \in K$,
$\calH_{a,v}$ is a constant $\alpha_v$ of norm 1 (a root of
unity),
independent of $a \in K$. Thus
$$\sum_{a\in [h]} D_{b,a}
        \overline{D_{c,a}} \calH_{a,v} = \alpha_v \sum_{a \in K} D_{b,a}
\overline{D_{c,a}}\ \ \ \ \text{and}\ \ \ \ \sum_{a\in [h]} D_{b',a}
        \overline{D_{c',a}}\hspace{0.06cm}\overline{\calH_{a,v}} = \overline{\alpha_v}
\sum_{a \in K} D_{b',a} \overline{D_{c',a}}.$$
Since $\alpha_v \overline{\alpha_v} = |\alpha_v|^2 = 1$,
it follows that their product
\[ X_{v,T} 
= \left(\sum_{a \in K} D_{b,a} \overline{D_{c,a}} \right)
\left( \sum_{a \in K} D_{b',a} \overline{D_{c',a}} \right),\]
which is exactly the same as the coefficient $X_{1,T}$ for $v_0 =1 \in A$.
Thus for any $T$, where each row has
  at least one entry $< \ell$, $X_{v,T}=X_{1,T}$, for any $v\in A$.
This includes {\it all} matrices  $T>_\mu T_3$
(as well as {\it some}  matrices  $T=_\mu T_3\in \calT_g$), 
and the first part of (\ref{checkcheckhy}) follows.\vspace{0.005cm}

Now we consider any matrix $T\in \calT_g$.
If each row of $T$ has at least one entry $<\ell$, then by
  the proof above, we know $X_{v,T}=X_{1,T}$ for any $v\in A$.
Suppose $T\in \calT_g$ does not have this property.
\vspace{-0.02cm}Then each column of such a matrix must consist of $\{1, \ell\}$.
We have four such matrices: $T_1,T_2,T_3$ and $T_4$.
But the former two matrices
  already belong to the case covered above. 
So we have $$
\sum_{T\in \calT_g}X_{v,T}-\sum_{T\in \calT_g} X_{1,T}= X_{v,T_3}+X_{v,T_4}-
  \left(X_{1,T_3}+X_{1,T_4}\right),\ \ \ \text{for any $v\in A$.}
$$

Now to the matrices $T_3,T_4$ themselves.
We note that the {\it sum} of their coefficients $X_{v,T_3}+X_{v,T_4}$ is
\begin{equation}\label{contribution-sum-of-two-at-ell}
\left( \sum_{a \in K} |D_{1,a}|^2 \calH_{a,v} \right)
   \left( \sum_{a =1}^h |D_{\ell,a}|^2 \overline{\calH_{a,v}} \right)
+
\left( \sum_{a =1}^h |D_{\ell,a}|^2 \calH_{a,v} \right)
\left( \sum_{a \in K} |D_{1,a}|^2 \overline{\calH_{a,v}} \right),\ \
  \text{at any $v\in A$.}
\end{equation}
This is a real number, and the second part of (\ref{checkcheckhy})
follows.

Now we can apply Part A of the Vanishing Lemma to conclude that
$$X_{v,T_3}+X_{v,T_4}=X_{1,T_3}+X_{1,T_4}=2 \cdot\|\DD_{1,*}\|^2 \|\DD_{\ell,*}\|^2,
\ \ \ \text{for any $v\in A$}.$$
This is the maximum possible value of
(\ref{contribution-sum-of-two-at-ell}).
By our assumption $\|\DD_{1,*}\|^2 \|\DD_{\ell,*}\|^2 >0$.
The only way the sum in (\ref{contribution-sum-of-two-at-ell})\vspace{-0.01cm}
also achieves this maximum at $v \in A$
is for  $\calH_{a,v}$ to take a constant value $\gamma_v$
for all $a \in {K}_\ell$, (and we already know
that $\calH_{a,v}$
takes a constant value $\alpha_v$
for all $a \in K$),
where $\alpha_v$
and $\gamma_v$ are of norm 1.
Moreover, by (\ref{contribution-sum-of-two-at-ell}), we have
\[\alpha_v \overline{\gamma_v} + \overline{\alpha_v} \gamma_v = 2.\]
It follows that $\alpha_v = \gamma_v$.
Thus $\calHH_{*,v}$ is a constant on $ K \cup {K}_\ell$
  for each $v \in A$. We summarize it as

\begin{claim}\label{claim-6-h-is-const-onwidehatK}
For every $v \in A$, there exists a complex number $\alpha_v$
of norm 1, such that $\calH_{v,a} = \alpha_v$ for all $a \in K \cup {K}_\ell$.
\end{claim}

Our next goal is to show that $|\DD_{\ell,*}|^2 \perp \calHH_{*,v}$
for all $v \in B$.
Of course if $B = \emptyset$ then this is vacously true.
We assume $B \not = \emptyset$.
For this purpose, we examine
$$T^* = \begin{pmatrix} \ell & \ell \\  \ell & \ell \end{pmatrix},$$
and the class $\calT_g$ it belongs to.
By Property \ref{property3}, we have $\sum_{T\in \calT_g} X_{v,T}=0$
  for any $v\in B$, and our target is to show that $X_{v,T^*}=0$.
To prove this,
  we need to examine terms $X_{v,T}$ for all $T=_\mu T^* \in \calT_g$.


It is now possible to have a number of pairs,
$(a_1, b_1), (a_2, b_2), \ldots, (a_k, b_k)$,
for some $k \ge 0$, such that $\mu_{a_i} \mu_{b_i} = \mu_{\ell}^2$,
for $1 \le i \le k$. (When $\ell=2$, such a pair, if it exists,
  is essentially unique, but for $\ell>2$ there could be
  many such pairs. This is a complication\vspace{-0.02cm} for $\ell>2$).
For every matrix $T\in \calT_g$, it must have each column
chosen from either $  \ell \choose  \ell$
or one of the
pairs $ a_i \choose b_i $
or $ b_i \choose a_i $.
Note that if such pairs do not exist,
i.e., $k=0$, then $\calT_g=\{T^*\}$ and we have
\[X_{v,T^*}=\left( \sum_{a=1}^{h} |D_{\ell, a}|^2 \calH_{a,v} \right)
\left( \sum_{a=1}^{h}  |D_{\ell, a}|^2 \overline{\calH_{a,v}} \right) =0,\ \ \
\text{at any $v\in B$.}\]
 The following proof is to show that
even when  such  pairs exist ($k \ge 1$), we still have $X_{v,T^*}=0$.
For\vspace{-0.02cm} this purpose, we show that $\sum_{T\in \calT_g,T\ne T^*}X_{v,T}\ge 0$.

Suppose $k \ge 1$.
We may assume $a_i < \ell < b_i$, for all $i\in [k]$.  Let's examine all the
matrices $T\in \calT_g$ other than $T^*$. 
If $T$ has at least one row, say $(b \ \hspace{0.05cm}c)$,
with $\max\{b, c\} \le \ell$ and $\min\{b, c\} < \ell$,
then by the Inductive Hypothesis and Claim~\ref{claim-5-D-ell},
the corresponding inner product actually takes place
over $K$.  In fact, the inner product
is a constant multiple of the projection of $|\DD_{1,*}|^2$
on either $\calHH_{*,v}$ or $\overline{\calHH_{*,v}}$.
But we already know that this projection is zero for all $v \in B$.

For the remaining $T$\vspace{-0.01cm}
where both rows satisfy
[\hspace{0.04cm}$\max\{b, c\} > \ell$ or $\min\{b, c\} \ge \ell$\hspace{0.04cm}],
if $T\ne T^*$ then one of its two columns\vspace{-0.01cm} $\ne {\ell \choose\ell}$, and
  one entry of this column is $a_i<\ell$, for some $i\in [k]$.
It then follows that
  the other entry in the same row as $a_i$ must be $b_j>\ell$, for some $j\in [k]$.
As a result, the only matrices remaining are of the form
$$\begin{pmatrix} a_i & b_j \\ b_i & a_j \end{pmatrix}\ \ \
\text{or}\ \ \ \begin{pmatrix} b_i & a_j \\ a_i & b_j \end{pmatrix},
\ \ \ \text{for some $1 \le i, j \le k.$}$$

We consider the first type
$\begin{pmatrix} a_i & b_j \\ b_i & a_j \end{pmatrix}$.
The total contribution of these matrices is\vspace{0.1cm}
\begin{eqnarray*}
&& \sum_{i, j =1}^k \left( \sum_{a=1}^{h} D_{a_i, a}
 \overline{D_{b_j, a}} \calH_{a,v} \right)
\left( \sum_{a'=1}^{h} D_{b_i, a'} \overline{D_{a_j, a'}}\hspace{0.08cm}
\overline{\calH_{a',v}}\right)\\[0.3ex]
&=&  \sum_{i, j =1}^k \left( \sum_{a=1}^{h} \lambda_{a_i}
D_{1,a} \overline{D_{b_j, a}} \calH_{a,v} \right)
\left( \sum_{a'=1}^{h} D_{b_i, a'} \overline{\lambda_{a_j}}\hspace{0.08cm}
\overline{D_{1, a'}}\hspace{0.08cm} \overline{\calH_{a',v}} \right)\\[0.3ex]
&=& \sum_{i, j =1}^k \sum_{a, a'=1}^{h}
\overline{\lambda_{a_j}} D_{1,a} \overline{D_{b_j, a}} \calH_{a,v}
\cdot
\lambda_{a_i}  D_{b_i, a'} \overline{D_{1, a'}}\hspace{0.08cm}\overline{\calH_{a',v}}\\[0.3ex]
&=&
\left[ \sum_{a=1}^{h} D_{1,a} \calH_{a,v} \left( \sum_{j=1}^k
             \overline{\lambda_{a_j}}\hspace{0.08cm}\overline{D_{b_j, a}} \right) \right]
\cdot
\left[ \sum_{a'=1}^{h} \overline{D_{1, a'}}\hspace{0.08cm}\overline{\calH_{a',v}}
 \left( \sum_{i=1}^k \lambda_{a_i} D_{b_i, a'}  \right) \right]\\[0.6ex]
&=& \left| \sum_{a=1}^{h} D_{1,a} \calH_{a,v} \left( \sum_{j=1}^k
             \overline{\lambda_{a_j}}\hspace{0.08cm} \overline{D_{b_j, a}} \right) \right|^2
 \ge 0.
\end{eqnarray*}
Here in the first equality we used the Inductive Hypothesis
for $a_i, a_j < \ell$.

The argument for the second type of matrices
is symmetric.

Note also that the matrix $T^*$ 
has the Conjugate-Pair form, and therefore
its contribution $X_{v,T^*}$ at any $v \in B$ is also non-negative.
It follows from $\sum_{T\in \calT_g}X_{v,T}=0$ (Property \ref{property3})
  that $X_{v,T^*}=0$ and\vspace{-0.15cm}
$$\left|
\sum_{a=1}^{h} |D_{\ell,a}|^2
         \overline{\calH_{a,v}} \right|^2 = 0,\ \ \ \text{for all $v \in B$.}$$
This means that $|\DD_{\ell,*}|^2 \perp \calHH_{*,v}$
for all $v \in B$ and thus, $|\DD_{\ell,*}|^2$ is in the linear span of
  $\{\calHH_{*,v}\hspace{0.07cm}|\hspace{0.07cm} v \in A\}$.

Now by exactly the same argument as for $\ell =2$
we obtain $K = {K}_\ell$. We summarize as follows

\begin{claim}\label{claim-7-Dell--linear-dep-on-D1}
There exists some complex number $\lambda_\ell$,
such that $\DD_{\ell,*} = \lambda_\ell \cdot\DD_{1,*}.$
\end{claim}

This completes the proof by induction that $\DD$ has rank at most one.


\subsection{Step 2.4}
\def\MM{\mathbf{M}}

After Step 2.3, we get a pair $(\CC,\fD)$ that satisfies conditions
  ({\sl Shape}$_1$)-({\sl Shape}$_6$).
By ({\sl Shape}$_2$), we have
$$
\CC=\left(\begin{matrix}\hspace{-0.15cm}\00& \FF\\ \FF^T&\00\end{matrix}\right)=
\left(\begin{matrix}\hspace{-0.15cm}\00&\MM\otimes
  \HH\\ (\MM\otimes \HH)^T&\00\end{matrix}\right),
$$
where $\MM$ is an $s\times t$ matrix of rank $1$: $M_{i,j}=\mu_i\nu_j$,
  and $\HH$ is the $h\times h$ matrix defined in ({\sl Shape}$_2$).
By ({\sl Shape}$_5$) and ({\sl Shape}$_6$), we have
$$
\DD^{[r]}=\left(\begin{matrix}
\DD^{[r]}_{(0,*)} &  \\ & \DD^{[r]}_{(1,*)}
\end{matrix}\right)=\left(\begin{matrix}
\KK^{[r]}_{(0,*)}\otimes \LL^{[r]}_{(0,*)}& \\
 & \KK^{[r]}_{(1,*)}\otimes \LL^{[r]}_{(1,*)}
\end{matrix}\right), \ \ \ \text{for every $r\in [0:N-1]$.}
$$
Moreover, every diagonal entry in $\LL^{[r]}$ either is $0$ or has norm $1$
  and $\LL^{[0]}$ is the $2h\times 2h$ identity matrix.

Using\vspace{0.005cm} these matrices, we define two new pairs $(\CC',{\frak K})$ and
  $(\CC'',{\frak L})$, which give rise to two problems
  $\eval(\CC',{\frak K})$ and $\eval(\CC'',{\frak L})$:
First,\vspace{0.005cm} $\CC'$ is the bipartisation of $\MM$, so it is $(s+t)\times (s+t)$;
  and $\frak K$ is a sequence of $N$ diagonal matrices of the
  same size: $\{\KK^{[0]},\ldots,\KK^{[N-1]}\}$.
Second, $\CC''$ is the bipartisation of $\HH$, so it is $2h\times 2h$;
  and $\frak L$ is a sequence of $N$ diagonal matrices: $\{\LL^{[0]},\ldots,
  \LL^{[N-1]}\}$.
The following lemma shows that
  $\eval(\CC,\fD)$ has the same complexity as $\eval(\CC'',{\frak L})$.

\begin{lemma}\label{jajajaja}
$\eval(\CC,\fD)\equiv \eval(\CC'',{\frak L})$.
\end{lemma}
\begin{proof}
Let $G$ be a connected undirected graph and $u^*$ be one of its vertices,
  then by Lemma \ref{verytrivial} and Lemma \ref{tensorproduct}, we have
$
Z_{\CC,\fD}(G)=Z_{\CC,\fD}^{\rightarrow}(G,u^*)+Z_{\CC,\fD}^{\leftarrow}(G,u^*)$,
$$Z_{\CC,\fD}^{\rightarrow}(G,u^*)=
  Z_{\CC',{\frak K}}^{\rightarrow}(G,u^*)\cdot Z_{\CC'',
  {\frak L}}^{\rightarrow}(G,u^*),\ \ \ \text{and}\ \ \
Z_{\CC ,\fD }^{\leftarrow}(G,u^*)=
  Z_{\CC',{\frak K}}^{\leftarrow}(G,u^*)\cdot
  Z_{\CC'',{\frak L}}^{\leftarrow}(G,u^*).
$$
Because $\MM$ is of rank $1$, both $Z_{\CC',{\frak K}}^{\rightarrow}$
  and $Z_{\CC',{\frak K}}^{\leftarrow}$ can be computed in polynomial time.
We only prove for $Z_{\CC',{\frak K}}^{\rightarrow}$ here:
If $G$ is not bipartite, then $Z_{\CC',{\frak K}}^{\rightarrow}(G,u^*)$
  is trivially $0$;
Otherwise let $U\cup V$ be the vertex set of $G$, $u^*\in U$, and every edge $uv\in E$ has
  one vertex $u$ from $U$ and one vertex $v$ from $V$.
We use $\Xi$ to denote the set of assignments $\xi$ which maps $U$ to $[s]$ and
  $V$ to $[t]$. Then we have (note that we use
  $\KK^{[r]}$ to denote $\KK^{[r\bmod N]}$, for any $r\ge N$)
\begin{eqnarray*}
Z_{\CC',{\frak K}}^{\rightarrow}(G,u^*)\hspace{-0.15cm}&=&\hspace{-0.1cm}\sum_{\xi\in \Xi}
  \left(\prod_{uv\in E} \mu_{\xi(u)}\cdot \nu_{\xi(v)}\right) \left(\prod_{u\in U}
  K^{[\text{deg}(u) ]}_{(0,\xi(u))}\right)\left(
  \prod_{v\in V} K^{[\text{deg}(v) ]}_{(1,\xi(v))}\right)\\[0.35ex]
  \hspace{-0.15cm}&=&\hspace{-0.15cm}
  \prod_{u\in U} \left(\sum_{i\in [s]} (\mu_i)^{\text{deg}(u)}
  \cdot K^{[\text{deg}(u) ]}_{(0,i)}\right)
  \times \prod_{v\in V} \left(\sum_{j\in [t]} (\nu_j)^{\text{deg}(v)}
  \cdot K^{[\text{deg}(v) ]}_{(1,j)}\right) ,
\end{eqnarray*}
which can be computed in polynomial time.

Moreover, because pair $(\CC'',{\frak L})$ satisfies {(\sl Pinning}),
  by the Second Pinning Lemma (Lemma \ref{pinning2}) the problem of computing
  $Z_{\CC'',{\frak L}}^{\rightarrow}$ and $Z_{\CC'',{\frak L}}^{\leftarrow}$
  is reducible to $\eval(\CC'',{\frak L})$.
It then follows that\vspace{-0.08cm}
$$
\eval(\CC,\fD)\le \eval(\CC'',{\frak L}).\vspace{-0.08cm}
$$

We next prove the reverse direction. First note that,
  by the Third Pinning Lemma (Corollary \ref{pinning3}),
  computing $Z_{\CC,\fD}^{\rightarrow}$ and $Z_{\CC,\fD}^{\leftarrow}$
  is reducible to $\eval(\CC,\fD)$.
However, this does not finish the proof because $Z_{\CC',{\frak K}}^{\rightarrow}$
  (or $Z_{\CC',{\frak K}}^{\leftarrow}$) could be $0$ at $(G,u^*)$.
To deal with this case, we prove the following claim:\vspace{0.06cm}

\begin{claim}\label{hahaclaim}
Given any connected bipartite graph $G=(U\cup V,E)$ and $u^*\in U$,
  either we can construct\\
  a new connected bipartite graph $G'=(U'\cup V',E')$ in polynomial time
  such that $u^*\in U\subset U'$,
\begin{equation}\label{gjgi}
Z^{\rightarrow}_{\CC'',{\frak L}}(G',u^*)=h^{|U\cup V|}\cdot Z^{
  \rightarrow}_{\CC'',{\frak L}}(G,u^*),
\end{equation}
and $Z^{\rightarrow}_{\CC',{\frak K}}(G',u^*)\ne 0$;
  or we can show that $Z_{\CC'',{\frak L}}^{\rightarrow}(G, u^*)=0$.\vspace{0.06cm}
\end{claim}
Claim \ref{hahaclaim}
  gives us a polynomial-time reduction from $Z^{\rightarrow}_{
  \CC'',{\frak L}}$ to $Z^{\rightarrow}_{\CC,\fD}$.
A similar claim can be proved for $Z^{\leftarrow}$, and Lemma \ref{jajajaja} follows.
We now prove Claim \ref{hahaclaim}.

For every $u\in U$ (and $v\in V$), we let $r_u$ (and $r_v$) denote its degree in graph $G$.
To construct $G'$, we need an integer $\ell_u\in [s]$ for every $u\in U$,
  and an integer $\ell_v\in [t]$ for every $v\in V$, such that
\begin{equation}\label{opop}
\sum_{i\in [s]} \mu_i^{\ell_uN+r_u}\cdot  K^{[r_u ]}_{(0
  ,i)}\ne 0,\ \ \ \text{and}\ \ \
\sum_{i\in [t]} \nu_i^{\ell_vN+r_v}\cdot  K^{[r_v ]}_{(1,i)}\ne 0.
\end{equation}
Assume there exists a $u\in U$ such that no $\ell_u\in [s]$ satisfies (\ref{opop}).
In this case, note that the $s$ equations for $\ell_u=1,$ $\ldots,s$ form
  a Vandermonde system since $\mu_1>\ldots>\mu_s>0$.
As a result,\vspace{-0.145cm} we have $$\KK^{[r_u ]}_{(0,
  *)}=\00\ \ \Longrightarrow\ \ \LL^{[r_u]}_{(0,*)}=\00,$$
by ({\sl Shape}$_6$).
It follows that $Z_{\CC'',{\frak L}}^{\rightarrow}(G,u^*)=0$, and
  we are done.
Similarly, we have $Z_{\CC'',{\frak L}}^{\rightarrow}(G,u^*)=0$
  if there exists a $v\in V$ such that no $\ell_v\in [t]$ satisfies (\ref{opop}).

Otherwise,\vspace{0.01cm} suppose there exist an $\ell_u\in [s]$
  for every $u\in U$, and an $\ell_v\in [t]$
  for every $v\in V$, which satisfy (\ref{opop}).
We construct a bipartite graph $G'=(U'\cup V',E')$ as follows: First,
$$
U'=U\cup \widehat{V},\ \ \ \text{and}\ \ \ V'=V\cup \widehat{U},\ \ \ \text{where\ \
  $\widehat{V}=\big\{\widehat{v}\hspace{0.07cm}|\hspace{0.07cm}v\in V\big\}$\ \  and\ \
  $\widehat{U}=\big\{\widehat{u}\hspace{0.07cm}|\hspace{0.07cm}u\in U\big\}$.}
$$
Edge set $E'$ contains $E$ over $U\cup V$, and the following edges:
$\ell_uN$ parallel edges between $u$ and $\widehat{u}$, for every $u\in U$;
  and $\ell_vN$ parallel edges between $v$ and $\widehat{v}$, for every $v\in V$.

It is clear that $G'$ is a connected and bipartite graph. The degree of
  $u\in U$ (or $v\in V$) is $r_u+\ell_uN$ (or $r_v+\ell_vN$), and the degree of
  $\widehat{u}$ (or $\widehat{v}$) is $\ell_uN$ (or $\ell_vN$).
We now use $G'$ to prove Claim \ref{hahaclaim}.\vspace{0.01cm}

First, $Z_{\CC',{\frak K}}^{\rightarrow}(G',u^*)$ is equal to
  (the summation is over all $\xi$ that maps $U'$ to $[s]$ and $V'$ to $[t]$)\vspace{-0.05cm}
\begin{eqnarray*}
&&\hspace{-0.6cm}\sum_{\xi}
\left(\prod_{uv\in E} M_{\xi(u),\xi(v)}
\prod_{u\in U} M_{\xi(u),\xi(\widehat{u})}^{\ell_u N}\prod_{v\in V}
  M_{\xi(\widehat{v}),\xi({v})}^{\ell_vN} \right) \left(
\prod_{u\in U} K^{[r_u ]}_{(0,\xi(u))}
  K^{[0]}_{(1,\xi(\widehat{u}))}\right)\left(
\prod_{v\in V} K^{[r_v ]}_{(1,\xi(v))}
  K^{[0]}_{(0,\xi(\widehat{v}))}\right)\\[0.75ex]
&& \hspace{-0.7cm}=\hspace{-0.0cm}
\prod_{u\in U}\left(\sum_{i\in [s]} \mu_i^{\ell_uN+r_u}\cdot
  K^{[r_u ]}_{(0,i)} \right)
  \prod_{v\in V}\left(\sum_{i\in [t]} \nu_i^{\ell_vN+r_v}\cdot
  K^{[r_v ]}_{(1,i)}\right)
\prod_{\widehat{u}\in \widehat{U}}\left(\sum_{i\in [t]} \nu_i^{\ell_uN} \cdot K^{[0]}_{(1,i)}
\right) \prod_{\widehat{v}\in \widehat{V}}\left(\sum_{i\in [s]}
  \mu_i^{\ell_vN} \cdot K^{[0]}_{(0,i)}\right).
\\[-0.95ex]
\end{eqnarray*}
It is non-zero: the first two factors are non-zero because of
  the way we picked $\ell_u$ and $\ell_v$;
the latter two factors are non-zero because $\mu_i,\nu_i>0$,
  and by ({\sl Shape}$_6$), every entry of $\KK^{[0]}$ is a positive integer.\newpage

The only thing left is to prove (\ref{gjgi}).
We let $\eta$ be any assignment over $U\cup V$, which maps
  $U$ to $[s]$ and $V$ to $[t]$.
Given $\eta$, we let $\Xi$ denote the set of assignments $\xi$ over $U'\cup V'$
  which map $U'$ to $[s]$, $V'$ to $[t]$, and satisfies $\xi(u)=\eta(u)$,
  $\xi(v)=\eta(v)$ for all $u\in U$ and $v\in V$.
We have
\begin{eqnarray*}
\sum_{\xi\in \Xi}\text{wt}_{\CC'',{\frak L}}(\xi)\hspace{-0.1cm}&=&
  \hspace{-0.1cm}  \sum_{\xi\in \Xi}\hspace{0.1cm}\left(\prod_{uv\in E}H_{\eta(u),\eta(v)}
  \prod_{u\in U} (H_{\eta(u),\xi(\widehat{u})})^{\ell_uN}
  \prod_{v\in V} (H_{\xi(\widehat{v}),\eta(v)})^{\ell_vN}\right)\\[0.3ex] &&\hspace{0.7cm}\times
  \left(\prod_{u\in U} L^{[r_u ]}_{(0,\eta(u))}
  L^{[0]}_{(1,\xi(\widehat{u}))}\right)
  \left(\prod_{v\in V} L^{[r_v ]}_{(1,\eta(v))}
  L^{[0]}_{(0,\xi(\widehat{v}))}\right)\\[0.2ex]
&=&\hspace{-0.1cm}\sum_{\xi\in \Xi}\text{wt}_{\CC'',{\frak L}}(\eta)
  \hspace{0.06cm}=\hspace{0.06cm}h^{|\widehat{U}\cup
  \widehat{V}|}\cdot \text{wt}_{\CC'',{\frak L}}(\eta).
\end{eqnarray*}
The second equation uses the fact that every entry of $\HH$ is a power of
  $\oo_N$ (thus $(H_{i,j})^N=1$) and $\LL^{[0]}$ is the identity matrix.
(\ref{gjgi}) then follows.
\end{proof}

\subsection{Step 2.5}\label{step25}

We are almost done with Step 2.
The only conditions $(\calU_i)$'s that are possibly violated by $(\CC'',{\frak L})$
  are $(\calU_1)$ ($N$ might be odd), and $(\calU_2)$ ($H_{i,1}$ and $H_{1,j}$
  might not be $1$). We deal with $(\calU_2)$ first.

What we will do below is to normalize $\HH$ (in $\CC''$) so that
  it becomes a discrete unitary matrix for some positive integer $M$ that divides $N$,
while not changing the complexity of $\eval(\CC'',{\frak L})$.

First, without loss of generality, we may assume $\HH$ satisfies
$H_{1,1}=1$ since otherwise, we can divide $\HH$ with $H_{1,1}$, which does
  not affect the complexity of $\eval(\CC'',{\frak L})$.
Second, we construct the following pair $(\XX,{\frak Y})$:
  $\XX$ is the bipartisation of an $h\times h$ matrix over $\mathbb{C}$,
  whose $(i,j)^{th}$ entry is
$$
H_{i,j}\cdot \overline{H_{1,j}H_{i,1}},\ \ \ \text{for all $i,j\in [h]$;}
$$
and ${\frak Y}=\{\YY^{[0]},...,\YY^{[N-1]}\}$
  is a sequence of $2h\times 2h$ diagonal matrices:
$\YY^{[0]}$ is the identity matrix;
Let 
$$
\calS=\{r\in [0:N-1]\hspace{0.07cm}\big|\hspace{0.07cm}\LL^{[r]}_{(0,*)}\ne \00\}\ \ \
\text{and}\ \ \
\calT=\{r\in [0:N-1]\hspace{0.07cm}\big|\hspace{0.07cm}\LL^{[r]}_{(1,*)}\ne \00\},
$$
then
$$\YY^{[r]}_{(0,*)}=\00,\ \ \text{for all $r\notin \calS$;\ \ \ and}\ \ \
  \YY^{[r]}_{(1,*)}=\00,\ \ \text{for all $r\notin \calT$.}\vspace{0.04cm}$$
For every $r\in \calS$ (and $r\in \calT$), by
  ({\sl Shape}$_6$), there must exist an
  $a_r\in [h]$ (and $b_r\in [h]$, resp.) such that
$$L^{[r]}_{(0,a_r)}=1\ \ \ \ \Big(\text{and $L^{[r]}_{(1,b_r)}=1$, resp.}\Big).$$
Set\vspace{-0.15cm}
$$
Y^{[r]}_{(0,i)}=L^{[r]}_{(0,i)}\cdot \left(\frac{H_{i,1}}{H_{a_r,1}}\right)^r,\ \  \text{for
  all $i\in [h]$;} \ \ \
Y^{[r]}_{(1,j)}=L^{[r]}_{(1,j)}\cdot \left(\frac{H_{1,j}}{ H_{1,b_r}}\right)^r,\ \
  \text{for all $j\in [h]$.}\vspace{0.1cm}
$$
We show that $\eval(\CC'',{\frak L})\equiv \eval(\XX,{\frak Y})$.\vspace{0.01cm}

First, we prove that $\eval(\XX,{\frak Y})\le \eval(\CC'',{\frak L})$.
Let $G=(U\cup V, E)$ be a connected undirected graph and $u^*$ be a vertex in $U$.
For every $r \in \calS$ (and $r\in \calT$), we use $U_r\subseteq U$
  (and $V_r\subseteq V$, resp.) to denote the subset of
  vertices with degree $r\bmod N$.
It is clear that if $U_r\ne \emptyset$ for some $r\notin \calS$
  or if $V_r\ne$  $\emptyset$ for some $r\notin \calT$,
  both $Z_{\CC'',{\frak L}}^\rightarrow(G,u^*)$ and
  $Z_{\XX,{\frak Y}}^\rightarrow(G,u^*)$ are trivially zero.
Otherwise, we have
\begin{eqnarray}\label{xaxaabove}
Z^{\rightarrow}_{\CC'',{\frak L}}(G,u^*)=\left(\prod_{r\in \calS }
(H_{a_r,1})^{r|U_r|}\right)\left(\prod_{r\in \calT }
(H_{1,b_r})^{r|V_r|} \right)\cdot Z^{\rightarrow}_{\XX,{\frak Y}}(G,u^*).
\end{eqnarray}
So the problem of computing $Z_{\XX,{\frak Y}}^\rightarrow$
  is reducible to computing $Z_{\CC'',{\frak L}}^\rightarrow$.
By combining it with the Second Pinning Lemma (Lemma \ref{pinning2}),
  we know that computing $Z^{\rightarrow}_{\XX,{\frak Y}}$ is reducible
  to $\eval(\CC'',{\frak L})$.
A similar statement can be proved for $Z_{\XX,{\frak Y}}^\leftarrow$,
  and it follows that $$\eval(\XX,{\frak Y})\le \eval(\CC'',{\frak L}).$$
The other direction, $\eval(\CC'',{\frak L})\le \eval(\XX,{\frak Y})$,
  can be proved similarly.\vspace{0.005cm}


One can check that $(\XX,{\frak Y})$ satisfies $(\calU_1)$-$(\calU_4)$
  except that $N$ might be odd.
In particular the upper-right $h\times h$ block of $\XX$ is an $M$-discrete
  unitary matrix for some positive integer $M\hspace{0.06cm}|\hspace{0.06cm}N$;
and ${\frak Y}$ satisfies both $(\calU_3)$ and $(\calU_4)$ (which follow from the
  fact that every entry of $\HH$ is a power of $\oo_N$).

If $N$ is even then we are done with Step 2; otherwise we extend
  ${\frak Y}$ to be $${\frak Y}'=\{\YY^{[0]},
  \ldots,\YY^{[N-1]},\YY^{[N]},\ldots,\YY^{[2N-1]}\}, $$
  where $\YY^{[r]}=\YY^{[r-N]}$, for all $r\in [N:2N-1]$.
We have $\eval(\XX,{\frak Y})\equiv \eval(\XX,{\frak Y}')$, since
$$
Z_{\XX,{\frak Y}}(G)=Z_{\XX,{\frak Y}'}(G),\ \ \ \text{for all undirected graphs $G$,}
$$
and the new tuple $((M,2N),\XX,{\frak Y}')$ now satisfies conditions $(\calU_1)$--$(\calU_4)$.

%% file: Tractability.tex

Now we turn to the proof of Theorem \ref{pinyan}:
  $\eval(q)$ is tractable for any prime power $q$.

Actually, there is a well-known polynomial-time
  algorithm for $\eval(q)$ when $q$ is a prime (see \cite[Theorem 6.30]{Carlitz,field}:
  The algorithm works for any finite field).
In this section, we present a polynomial-time algorithm that works for any prime power $q$.
We start with the easier case when $q$ is odd.


\begin{lemma}\label{tractable_odd}
Let $p$ be an odd prime, and $q=p^k$ for some positive integer $k$.
Let $f \in \mathbb{Z}_q [x_1,\ldots, x_n]$~be a quadratic polynomial over $n$ variables
$x_1, \ldots, x_n $. Then the following sum\vspace{-0.03cm}
$$Z_{q}(f)=\sum_{x_1, \ldots, x_n \in \mathbb{Z}_q }
\omega_{q}^{f(x_1, \ldots, x_n\vspace{-0.03cm}
)}$$
can be evaluated in polynomial time \emph{(}in $n$\emph{)}.
Here by a quadratic polynomial over $n$ variables
we mean a polynomial where every monomial term has degree
at most 2.
\end{lemma}


\begin{proof}
In the proof, we assume $f(x_1,x_2,\ldots,x_n)$ has the following form:\vspace{-0.04cm}
\begin{equation}\label{quadraticform}
f(x_1,\ldots,x_n)=\sum_{i\le j\in [n]} c_{i,j}x_ix_j+\sum_{i\in [n]} c_ix_i+c_0.\vspace{-0.04cm}
\end{equation}
where all the $c_{i,j}$ and $c_i$ are elements in $\mathbb{Z}_q$.

First, as a warm up, we give an algorithm
  and prove its correctness for the case $k=1$. In this
  case $q=p$ is an odd prime. Note that if $f$ is an affine
linear function, then the evaluation can be trivially done in
polynomial time. In fact the sum simply decouples
into a product of $n$ sums\vspace{-0.1cm}
\[\sum_{x_1,x_2,\ldots, x_n \in \mathbb{Z}_q }
\omega_{q}^{f(x_1,x_2,\ldots, x_n)}
= \sum_{x_1,x_2,\ldots, x_n \in \mathbb{Z}_q }
\omega_{q}^{\sum_{i=1}^n c_i x_i+c_0}
= \oo_q^{c_0}\times
\prod_{i=1}^n\hspace{0.06cm} \sum_{x_i \in \mathbb{Z}_q } \omega_{q}^{c_i x_i}.
\]
This sum is equal to $0$ if any $c_i\in \mathbb{Z}_q$ is non-zero, and
  is equal to $q^n\oo^{c_0}_q$ otherwise.

Now assume $f(x_1,\ldots, x_n)$ is
not affine linear.
Then in each round (which we will describe below),
  the algorithm will decrease the number of
  variables by at least one, in polynomial time.

Assume $f$ contains some
quadratic terms. There are two cases: $f$ has at least one square term;
  or $f$ does not have any square term.
In the first case, without loss of generality,
  we assume that $c_{1,1} \in \mathbb{Z}_q$ is non-zero.
Then there exist an affine linear
function $g \in \mathbb{Z}_q[x_2,x_3,\ldots,x_n]$,
and a quadratic polynomial $f' \in \mathbb{Z}_q[x_2,x_3,\ldots,x_n]$,
both over $n-1$
variables $x_2,x_3,\ldots,x_n$, such that\vspace{-0.04cm}
\[f(x_1,x_2,\ldots, x_n
)=c_{1,1}\big(x_1+g(x_2,x_3,\ldots,x_n)\big)^2+f'(x_2,x_3,\ldots,x_n).\vspace{-0.03cm}\]
Here we used the fact that both $2$ and $c_{1,1} \in \mathbb{Z}_q$
are invertible in the field  $\mathbb{Z}_q$ (Recall we assumed that
$q =p$ is an odd prime).  Thus we can factor out a coefficient
$2c_{1,1}$ from the cross term $x_1 x_i$, for every $i >1$, and from
  the linear term $x_1$,
to get the expression $c_{1,1}(x_1+g(x_2,\ldots,x_n))^2$.\vspace{0.008cm}

For any fixed $x_2,\ldots,x_n \in \mathbb{Z}_q$, when $x_1$ goes over
$\mathbb{Z}_q$, $x_1+g(x_2,\ldots,x_n)$ also goes over
$\mathbb{Z}_q$. Thus,
\begin{eqnarray*}
\sum_{x_1,x_2,\ldots, x_n \in \mathbb{Z}_q }
\omega_{q}^{f(x_1,x_2,\ldots, x_n
)}
=
\sum_{x_2,\ldots,x_n \in \mathbb{Z}_q }
\omega_{q}^{f'(x_2,\ldots, x_n )}
\sum_{x_1 \in \mathbb{Z}_q} \omega_{q}^{c_{1,1} (x_1+g(x_2,\ldots,x_n))^2}
=
 \sum_{x \in \mathbb{Z}_q}
\omega_{q}^{c_{1,1}x^2}\cdot Z_q(f').
\end{eqnarray*}
The first factor can be
evaluated in constant time (which is independent of $n$)
  and the computation of $Z_{q}(f)$ is reduced
  to the computation of $Z_q(f')$ in which $f'$ has at most $n-1$ variables.
\begin{quote}
{\bf Remark}: The claim of $\sum_{x}
\omega_{q}^{c x^2}$ being ``computable in constant time''
is a trivial statement, since we consider $q=p$ to be a fixed constant.
However, for a general prime $p$, we remark that the sum
is the famous Gauss quadratic sum, and has the closed
formula\vspace{-0.1cm}
$$\text{$\sum_{x \in \mathbb{Z}_p} \omega_{p}^{c x^2} = p$, \ if $c =0$,\  and
it is\   $\left( \frac{c}{p} \right) \cdot G$,\ if $c \not = 0$,\
where\ $G= \sum_{x \in \mathbb{Z}_p}\left( \frac{x}{p} \right) \omega^{x}$.}\vspace{-0.3cm}
$$
Here \vspace{-0.07cm}$\left( \frac{c}{p} \right)$ is
the Legendre symbol, which can be computed in polynomial time
in the binary
length of $c$ and $p$, and $G$ has the closed form
$G = + \sqrt{p}$ if $p  \equiv 1 \bmod 4$
and $G = + i \sqrt{p}$ if $p  \equiv 3 \bmod 4$
\footnote{It had been known to Gauss since 1801 that
$G^2 = \left( \frac{-1}{p} \right) p$.  Thus \vspace{-0.066cm}
$G = \pm \sqrt{p}$ if $p  \equiv 1 \hspace{-0.07cm}\pmod 4$
and $G = \pm i \sqrt{p}$ if $p  \equiv 3 \pmod 4$. The fact that $G$
always takes the sign $+$ was conjectured by Gauss in his diary
in May 1801.
Four years later, on Sept 3, 1805, he wrote,
... Seldom had a week passed for four years that he had not tried
in vein to prove this very elegant theorem mentioned in 1801\hspace{0.06cm}...
``Wie der Blitz einschl\"{a}gt, hat sich das R\"{a}thsel gel\"{o}st ..."
(``as lightning strikes was
the puzzle solved ...").}.\end{quote}

The second case is that all the quadratic terms in $f$ are cross terms
  (in particular this implies that $n\ge 2$). In
this case we assume, without loss
of generality, that $c_{1,2} $ is non-zero.
We apply the following transformation: $x_1=x'_1+x'_2$ and $x_2=x'_1- x'_2$. As $2$ is
invertible in $\mathbb{Z}_q$, when $x'_1$ and $x'_2$ go over
$\mathbb{Z}^2_q$, $x_1$ and $x_2$ also go over $\mathbb{Z}^2_q$. Therefore,
we have
\[\sum_{x_1,x_2, \ldots, x_n \in \mathbb{Z}_q }
\omega_{q}^{f(x_1,x_2, \ldots, x_n
)}=\sum_{x'_1,x'_2, ,\ldots, x_n \in \mathbb{Z}_q}
\omega_{q}^{f(x'_1+x'_2,x'_1-x'_2, \ldots, x_n )} .\] If we view
$f(x'_1+x'_2,x'_1-x'_2 ,\ldots, x_n)$ as a new quadratic polynomial $f'$ of
$x'_1,x'_2,\ldots, x_n$, its coefficient of $x_1'^2$ is
  exactly $c_{1,2}\ne 0$,
so $f'$ contains at least one square term.
This reduces our problem back to
the first case, and we can use the method above to
  reduce the number of variables.

By repeating this process,
  we get a polynomial-time algorithm for computing $Z_q(f)$
  when $q=p$ is an odd prime.
Now we consider the case when $q=p^k$.

For any non-zero $a \in \mathbb{Z}_q$, we can
write it as $a=p^t a'$, where $t$ is a unique non-negative
 integer, such that $p\nmid a'$. We
call $t$ the order of $a$  (with respect to $p$). Again, if $f$ is an affine
linear function, $Z_q(f)$ is easy to compute,
as the sum factors into $n$ sums as before.
Now we assume $f$ has non-zero quadratic
terms. Let $t_0$ be the smallest order of all the non-zero quadratic
coefficients $c_{i,j}$ of $f$. We consider the following two cases:
  there exists at least one square term with
  coefficient of order $t_0$ or not.

In the first case, without loss of generality,
  we assume $c_{1,1}=p^{t_0}c$ and $p\nmid c$ (so $c$ is
  invertible in $\mathbb{Z}_q$).
Then by the minimality of $t_0$, every non-zero coefficient of a quadratic term has
  a factor  $p^{t_0}$.
Now we factor out $c_{1,1}$ from every quadratic term
  involving $x_1$, namely from $x_1^2, x_1 x_2, \ldots, x_1 x_n$
(clearly it does not matter if the coefficient of a term
$x_1 x_i$, $i\ne 1$, is 0). We can write
\[f(x_1,x_2,\ldots, x_n) = c_{1,1} \big(x_1+g(x_2,\ldots,x_n)\big)^2
+ c_1 x_1 +\hspace{0.04cm} \mbox{\emph{a quadratic polynomial in} } (x_2,\ldots,x_n),\]
where $g$ is a linear form over $x_2,\ldots,x_n$.
By adding and then subtracting $c_1 g(x_2,\ldots,x_n)$, we get
\[f(x_1,x_2,\ldots, x_n
)=c_{1,1}
\big(x_1+g(x_2,\ldots,x_n)\big)^2+c_1\big(x_1+g(x_2,\ldots,x_n)\big)
+f'(x_2,\ldots,x_n),\] where $f'(x_2,\ldots, x_n) \in
\mathbb{Z}_q[x_2,\ldots,x_n]$ is a quadratic polynomial over
$x_2,\ldots x_n$.

For any fixed $x_2, \ldots,x_n \in \mathbb{Z}_q$, when $x_1$ goes over
$\mathbb{Z}_q$, $x_1+g(x_2, \ldots,x_n)$ also goes over
$\mathbb{Z}_q$. Thus,
\[\sum_{x_1, \ldots, x_n \in \mathbb{Z}_q }
\omega_{q}^{f(x_1, \ldots, x_n
)}=\left(\sum_{x \in \mathbb{Z}_q} \omega_{q}^{c_{1,1}
  x^2+c_1x}\right)\left(\sum_{x_2,\ldots,x_n \in \mathbb{Z}_q }
\omega_{q}^{f'(x_2,\ldots, x_n )}\right)= \sum_{x \in \mathbb{Z}_q} \omega_{q}^{c_{1,1}
  x^2+c_1x} \cdot Z_{q}(f').\]
The first term can be
evaluated in constant time and the problem is reduced to $Z_q(f')$
  in which $f'$ has at most $n-1$ variables.

In the second case, all the square terms of $f$ are either $0$ or
  have orders larger than
$t_0$. Then we assume, without loss of generality, that $c_{1,2}=p^{t_0}c$ and
  $p\nmid c$.
We apply the following transformation:
  $x_1=x'_1+x'_2$ and $x_2=x'_1- x'_2$. Since $2$ is invertible in
$\mathbb{Z}_q$, when $x'_1$ and $x'_2$ go over $\mathbb{Z}^2_q$, $x_1$ and $x_2$
also go over $\mathbb{Z}^2_q$.
After the transformation, we get a new quadratic polynomial over
  $x'_1,x'_2,x_3,\ldots, x_n$ such that $Z_q(f')=Z_q(f)$.
It is easy to check that $t_0$ is
  still the smallest order of all the quadratic terms of $f'$:
The terms $x_1^2$ and $x_2^2$ (in $f$) produce terms with coefficients
divisible by $p^{t_0+1}$, the term $x_1 x_2$ (in $f$) produces
terms ${x'_1}^2$ and ${x'_2}^2$ with coefficients of order exactly $t_0$,
and terms $x_1 x_i$ or $x_2 x_i$, for $i \not =1, 2$,
produce terms $x'_1 x_i$ and $x'_2 x_i$ with coefficients
divisible by $p^{t_0}$.
In particular, the coefficient of $(x'_1)^2$ in $f'$ has order
  exactly $t_0$, so we can reduce the problem
to the first case.

To sum up, we have a polynomial-time algorithm for every $q=p^k$, when $p\ne 2$.
\end{proof}

Now we deal with the more difficult case when $q=2^k$ is a power of $2$,
for some $k \ge 1$.
We note that the property of an element $c\in \mathbb{Z}_{2^k}$
  being even or odd is well-defined.
We will use the following simple but important
observation,  the proof of which is straightforward:
\begin{lemma}\label{simple_observation}
For any integer $x$ and integer $k>1$, $(x+2^{k-1})^2\equiv x^2 \pmod{2^k}.$
\end{lemma}

\begin{lemma}\label{tractable_even_q}
Let $q=2^k$ for some positive integer $k$.
Let $f \in \mathbb{Z}_q[x_1,\ldots, x_n]$
be a quadratic polynomial over $n$ variables
$x_1, \ldots, x_n$. Then $Z_q(f)$ can
be evaluated in polynomial time \emph{(}in $n$\emph{)}.
\end{lemma}

\begin{proof}
If $k=1$, $Z_q(f)$ is computable in polynomial time according
to \cite[Theorem 6.30]{Carlitz,field} so we assume $k>1$. We also assume $f$ has the form
as in (\ref{quadraticform}).

The algorithm goes as follows: For each round, we can, in polynomial time, either
\begin{enumerate}
\item output the correct value of $Z_q(f)$; or
\item construct a new quadratic polynomial $g\in \mathbb{Z}_{q/2}[x_1,\ldots,x_n]$
  and reduce the computation\\ of $Z_q(f)$ to the computation of $Z_{q/2}(g)$; or\vspace{-0.035cm}
\item construct a new quadratic polynomial $g\in \mathbb{Z}_q[x_1,\ldots,x_{n-1}]$,
  and reduce the computation\\ of $Z_q(f)$ to the computation of $Z_q(g)$.
\end{enumerate}
This gives us a polynomial-time algorithm for $\eval(q)$ since we know how to solve
  the two base cases when
  $k=1$ or $n=1$ efficiently.

Suppose we have a quadratic polynomial $f\in \mathbb{Z}_q[x_1,\ldots,x_n]$.
Our first step is to transform $f$ so that
  all the coefficients of its cross terms ($c_{i,j}$, where $i\ne j$) and linear terms
  ($c_i$) are divisible by $2$.
Assume $f$
does not yet have this property.
We let $t$ be the smallest index in $[n]$ such
  that one of $\{c_t, c_{t,j}\hspace{-0.05cm}:\hspace{-0.05cm} j>t\}$ is not divisible by $2$.
By separating out the terms involving $x_t$, we rewrite $f$ as follows
\begin{equation}\label{expre-f-as-f1-f2}
f=c_{t,t}\cdot x^2_t + x_t\cdot f_1 (x_1, \ldots, \widehat{x_t}, \ldots, x_n)
+ f_2 (x_1, \ldots, \widehat{x_t}, \ldots, x_n),
\end{equation}
where $f_1$ is an affine linear function and $f_2$ is a quadratic polynomial.
Both $f_1$ and $f_2$ are over variables $\{x_1, \ldots, x_n\} - \{x_t\}.$
Here the notation $\widehat{x_t}$ means that $x_t$ does not appear in the polynomial.
Moreover,
\begin{equation}\label{checkcheck}f_1(x_1, \ldots, \widehat{x_t}, \ldots, x_n)
= \sum_{i<t} c_{i,t} x_i + \sum_{j>t} c_{t,j} x_j + c_t.
\end{equation}
By the minimality of $t$, $c_{i,t}$ is even for all $i<t$,
  and at least one of $\{c_{t,j},c_t:j>t\}$ is odd.

We claim that
\begin{equation}\label{quadraticimp}
Z_q(f)=\sum_{x_1, \ldots, x_n \in \mathbb{Z}_q }
\omega_{q}^{f(x_1, \ldots, x_n)}
=\sum_{\substack{x_1, \ldots,x_n\in \mathbb{Z}_q \\
f_1(x_1,\ldots,\widehat{x}_t,\ldots,x_n)\hspace{0.03cm}\equiv\hspace{0.03cm} 0 \bmod 2}}
  \omega_{q}^{f(x_1, \ldots, x_n)}.
\end{equation}
This is because\vspace{-0.3cm}
\begin{eqnarray*}
\sum_{\substack{x_1,\ldots, x_n \in \mathbb{Z}_q\\
  f_1\hspace{0.03cm}\equiv\hspace{0.03cm} 1 \bmod 2}}
\omega_{q}^{f(x_1,\ldots, x_n)}
= \sum_{\substack{x_1, \ldots, \widehat{x}_t, \ldots, x_n \in \mathbb{Z}_q\\
  f_1\hspace{0.03cm}\equiv\hspace{0.03cm} 1 \bmod 2}}\hspace{0.1cm}
 \sum_{x_t\in \mathbb{Z}_q} \omega_{2^k}^{c_{t,t} x^2_t + x_t f_1 + f_2}.
\end{eqnarray*}
However, for any fixed $x_1,\ldots,\widehat{x_t},\ldots,x_n$,
  $\sum_{x_t\in \mathbb{Z}_q} \omega_{2^k}^{c_{t,t} x^2_t + x_t f_1 + f_2}$
  is equal to $\oo_{2^k}^{f_2}$ times\vspace{0.1cm}
\begin{eqnarray*}
 \sum_{x_t\in [0:2^{k-1}-1]}
 \omega_{2^k}^{c_{t,t} x^2_t + x_t f_1}
 +\omega_{2^k}^{c_{t,t} (x_t+2^{k-1})^2 + (x_t+2^{k-1})f_1}
= \left(1+(-1)^{f_1}\right)
 \sum_{x_t\in [0:2^{k-1}-1]}\omega_{2^k}^{c_{t,t} x^2_t + x_t f_1}=0,\vspace{0.06cm}
\end{eqnarray*}
since $f_1\equiv 1\bmod 2$, and $1+(-1)^{f_1}=0$. Note that we used Lemma
\ref{simple_observation} in the first equation.

Recall $f_1$ (see (\ref{checkcheck}))
  is an affine linear form of $\{x_1,\ldots,\widehat{x}_t,\ldots,x_n\}$.
Also note that $c_{i,t}$ is even for all $i<t$, and one of
  $\{c_{t,j},c_t:j>t\}$ is odd.
We consider the following two cases.

In the first case, $c_{t,j}$ is even for all $j>t$ and $c_t$ is odd.
Then for any assignment $(x_1,\ldots,
  \widehat{x}_t,\ldots,x_n)$ in $\mathbb{Z}_q^{n-1}$, $f_1$ is odd.
As a result, by (\ref{quadraticimp}), $Z_q(f)$ is trivially zero.

In the second case, there exists at least one $j>t$ such that
  $c_{t,j}$ is odd.
Let $\ell > t$ be the smallest of such $j's$.
Then we substitute the variable $x_\ell$ in $f$
  with a new variable $x_{\ell}'$ over $\mathbb{Z}_q$, where (since
  $c_{t,\ell}$ is odd, $c_{t,\ell}$ is invertible in $\mathbb{Z}_q$)\vspace{-0.35cm}
\begin{equation}\label{replace}
x_{\ell} = c_{t,\ell}^{-1} \left( 2 x'_{\ell} - \left(\sum_{i<t}c_{i,t}x_i+
  \sum_{j>t,j\ne \ell} c_{t,j}x_j+c_t\right) \right).
\end{equation}
and let $f'$ denote the new quadratic polynomial
  in $\mathbb{Z}_q[x_1,\ldots,x_{\ell}',\ldots,x_n]$.

We claim that
$$
Z_{q}(f')=
2\cdot Z_{q}(f)=2\cdot \sum_{\substack{x_1,\ldots,x_n\in \mathbb{Z}_{q}\\
   f_1\hspace{0.03cm}\equiv\hspace{0.03cm} 0\bmod 2}} \oo_q^{f(x_1,\ldots,x_n)}.\vspace{0.1cm}
$$
To see this, we define the following map from $\mathbb{Z}_q^n$
  to $\mathbb{Z}_q^n$:\vspace{-0.03cm}
$$(x_1,\ldots,x_\ell',\ldots,x_n)\mapsto
  (x_1,\ldots,x_\ell,\ldots,x_n),\vspace{-0.03cm}$$
where $x_\ell$ satisfies (\ref{replace}).
It is easy to show that the range of the map is the
  set of $(x_1,\ldots,x_\ell,\ldots ,x_n)$ in $\mathbb{Z}_q^n$ such that $f_1$ is even.
Moreover, for every such tuple $(x_1,\ldots,x_\ell,\ldots ,x_n)$
  the number of its preimages in $\mathbb{Z}_q^n$ is exactly two.
The claim then follows.

So to compute $Z_q(f)$, we only need to compute $Z_q(f')$.
The advantage of $f'\in \mathbb{Z}_q[x_1,\ldots,x_\ell',\ldots,x_n]$
  over $f$ is the following property that we are going to prove:\vspace{0.05cm}
\begin{enumerate}
\item[]\hspace{-0.3cm}({\sl Even}): For every cross term and linear term that involves
  $x_1,\ldots,x_t$, its coefficient in $f'$ is even.\vspace{0.05cm}
\end{enumerate}
To prove this, we divide the terms of $f'$
  (that we are interested in) into three groups:
Cross and linear terms that involve $x_t$;
linear terms $x_s$, $s<t$;
and cross terms of the form $x_sx_{s'}$, where $s<s',s<t$.

Firstly, we consider the expression (\ref{expre-f-as-f1-f2}) of $f$
  after the substitution.
The first term $c_{t,t}x_t^2$ remains the same;
The second term $x_tf_1$ becomes $2x_tx_\ell'$ by (\ref{replace});
and $x_t$ does not appear in the third term, even after the substitution.
Therefore, condition ({\sl Even}) holds for $x_t$.

Secondly, we consider the coefficient $c_s'$ of the linear term $x_s$ in $f'$,
  where $s<t$.
Only the following terms in $f$ can possibly contribute to $c_s'$:\vspace{-0.05cm}
$$
c_sx_s,\ c_{\ell,\ell}x_\ell^2,\ c_{s,\ell}x_s
  x_{\ell},\ \text{and}\ \ c_\ell x_{\ell}.\vspace{-0.05cm}
$$
By the minimality of $t$, both $c_s$ and $c_{s,\ell}$ are even.
For $c_{\ell,\ell}x_\ell^2$ and $c_\ell x_\ell$,
  although we do not know whether $c_{\ell,\ell}$ and $c_\ell$ are even or odd,
  we know that the coefficient $-c_{t,\ell}^{-1}c_{s,t}$ of $x_s$ in (\ref{replace})
  is even since $c_{s,t}$ is even.
As a result, for every term in the list above,
  its contribution to $c_s'$ is even and thus, $c_s'$ is even.

Finally, we consider the coefficient $c_{s,s'}'$ of the term $x_sx_{s'}$ in $f'$,
  where $s<s'$ and $s<t$.
Similarly, only the following terms in $f$ can possibly contribute to $c_{s,s'}'$
  (Here we consider the general case when $s'\ne \ell$.
  The special case when $s'=\ell$ is easier)\vspace{-0.05cm}
\begin{eqnarray*}
c_{s,s'}x_{s}x_{s'},\ c_{\ell,\ell}x_\ell^2,\ c_{s,\ell}x_sx_{\ell},\ \text{and}\
  \ c_{\ell,s'}x_\ell x_{s'}\ \text{(or\ \hspace{0.06cm}$c_{s',
  \ell}x_{s'}x_{\ell}$)}.\vspace{-0.05cm}
\end{eqnarray*}
Again, by the minimality of $t$, $c_{s,s'}$ and $c_{s,\ell}$ are even.
Moreover, the coefficient $-c_{t,\ell}^{-1}c_{s,t}$ of $x_s$ in (\ref{replace})
  is even.
As a result, for every term listed above, its contribution to $c_{s,s'}'$
  is even and thus, $c_{s,s'}'$ is even.

To summarize, after substituting $x_\ell$ with $x_\ell'$ using (\ref{replace}),
  we get a new quadratic polynomial $f'$ such that $Z_q(f')=2\cdot Z_q(f)$,
  and for every cross term and linear term that involves $x_1,\ldots,x_t$,
  its coefficient in $f'$ is even.
We can repeat this substitution procedure on $f'$:
Either we show that $Z_q(f')$ is trivially $0$,
  or we get a quadratic polynomial $f''$ such that $Z_q(f'')=2\cdot Z_q(f')$
  and the parameter $t$ increases by at least one.
As a result, given any quadratic polynomial $f$,
  we can, in polynomial time, either show that $Z_q(f)$ is zero,
  or construct a new quadratic polynomial $g\in \mathbb{Z}_q[x_1,\ldots,x_n]$
  such that $Z_q(f)=2^{k'}\cdot Z_q(g)$, for some known integer $k'\in [0:n]$,
  and every cross term and linear term has an even coefficient in $g$.





Now we only need to compute $Z_q(g)$.
We will show that, given such a polynomial $g$ in $n$ variables,
  we can reduce it to either $\eval(2^{k-1})=\eval(q/2)$,
  or to the computation of $Z_q(g')$, in which $g'$ is a
  quadratic polynomial in $n-1$ variables.

Let
$$
g=\sum_{i\le j\in [n]}a_{i,j}x_ix_j+\sum_{i\in [n]}a_ix_i+a,
$$
then we consider the following two cases:
$a_{i,i}$ is even for all $i\in [n]$; or at least one of the
  $a_{i,i}$'s is odd.

In the first case, we know $a_{i,j}$ and $a_i$ are even
  for all $i\le j\in [n]$. \vspace{-0.04cm}We let $a_{i,j}'$ and $a_i'$ denote integers
  in $[0:2^{k-1}-1]$ such that $a_{i,j}\equiv 2a_{i,j}'\hspace{-0.06cm}\pmod{q}$
  and $a_i\equiv 2a_i'\hspace{-0.05cm}\pmod{q}$, respectively. Then,$$
Z_q(g)=\oo_q^a\cdot \sum_{x_1,\ldots, x_n \in \mathbb{Z}_{q}}
\omega_{q}^{2\big(\sum_{i\le j\in [n]}a_{i,j}'x_ix_j+\sum_{i\in [n]}a_i'x_i\big)}
=2^n\cdot \oo_q^a\cdot Z_{2^{k-1}}(g'),$$
where\vspace{-0.2cm} $$g'=\sum_{i\le j\in [n]}a_{i,j}'x_ix_j+\sum_{i\in [n]}a_i'x_i
  \vspace{0.08cm}$$ is
  a quadratic polynomial over $\mathbb{Z}_{q/2}=\mathbb{Z}_{2^{k-1}}$.
This reduces the computation of $Z_q(g)$ to $Z_{q/2}(g')$.

In the second case, without loss of generality, we assume $a_{1,1}$
  is odd. Then we have
$$f=a_{1,1}(x_1^2+2x_1g_1)+g_2=a_{1,1}(x_1+g_1)^2+g',$$
where $g_1$ is an affine linear form, and $g_2,g'$ are quadratic polynomials,
  all of which are over $x_2,\ldots,x_n$.
We are able to do this because $a_{1,j}$ and $a_1$, for all $j\ge 2$, are even.
Now we have\vspace{0.05cm}
\begin{eqnarray*}
Z_q(g)=\sum_{x_1,\ldots,x_n\in \mathbb{Z}_q}
  \oo_q^{a_{1,1}(x_1+g_1)^2+g'}
=\sum_{x_2,\ldots,x_n\in \mathbb{Z}_q}
  \oo_q^{g'}\cdot \sum_{x_1\in \mathbb{Z}_q}
  \oo_q^{a_{1,1}(x_1+g_1)^2}
=\left(\sum_{x\in \mathbb{Z}_q}\oo_q^{a_{1,1}x^2}\right)
\cdot Z_q(g').\\[-1.2ex]
\end{eqnarray*}
The last equation is because the sum over $x_1\in \mathbb{Z}_q$
  is independent of the value of $g_1$.
This reduces the computation of $Z_q(g)$ to $Z_q(g')$ in
  which $g'$ is a quadratic polynomial in $n-1$ variables.


To sum up, given any quadratic polynomial $f$, we can, in polynomial time,
  either output the correct value of $Z_q(f)$; or reduce one of the two
  parameters, $k$ or $n$, by at lease one.
This gives us a polynomial time algorithm to evaluate $Z_q(f)$.
\end{proof}

\noindent
{\bf Remark:}
We remark that back in Section~\ref{sec1} Introduction
we mentioned that ${\rm Holant}(\Omega)$ 
for   $\Omega = (G, {\cal F}_1 \cup {\cal F}_2 \cup {\cal F}_3)$
 are all tractable, and the tractability 
boils down to the exponential sum in (\ref{mod4-of-mod2-sum})
is computable in polynomial time.
This can also be  derived from Theorem~\ref{pinyan}.

First, each mod 2 sum $L_j$ in (\ref{mod4-of-mod2-sum})
can be replaced by its square $(L_j)^2$. We note that
$L_j  = 0, 1 \pmod 2$ if and only if $(L_j)^2 = 0, 1 \pmod 4$,
respectively.  Hence 
\[\sum_{x_1, x_2, \ldots, x_n \in \{0, 1\}}
i^{L_1 + L_2 +\hspace{0.04cm} \cdots\hspace{0.04cm} + L_s},\]
can be expressed as a sum of the form $i^{Q(x_1, \ldots, x_n)}$,
where $Q$ is an (ordinary)
 sum of squares of affine linear forms with integer coefficients,
in particular
a quadratic polynomial with integer coefficients.
For a sum of squares of affine linear forms $Q$,
if we evaluate each $x_i \in \{0, 1, 2, 3\}$,
we may take $x_i \bmod 2$, and therefore
\[\sum_{x_1, x_2, \ldots, x_n \in \mathbb{Z}_4} 
i^{Q(x_1, \ldots, x_n)} 
= 2^n \sum_{x_1, x_2, \ldots, x_n \in \{0, 1\}}
i^{Q(x_1, \ldots, x_n)}.\]

It can also be seen easily that in the sum 
$\sum_{x_1, x_2, \ldots, x_n \in \{0, 1\}} i^{Q(x_1, \ldots, x_n)}$,
a quadratic polynomial $Q$ with integer coefficients
can be expressed as a sum of squares of affine linear forms
iff all cross terms $x_i x_j$, where $i \not = j$,
have even coefficients.  Thus, this is exactly the same class
of sums considered in (\ref{mod4-of-mod2-sum}).

%% file: Decidability.tex





\section{Decidability in Polynomial Time: Proof of Theorem \ref{theo-decidability}}\label{sec:dec}

Finally, we prove Theorem \ref{theo-decidability}, i.e.,
  the following decision problem is computable in polynomial
  time: Given a symmetric matrix $\AA \in {\mathbb C}^{m
\times m}$ where all the entries $A_{i,j}$ in $\AA$ are algebraic
  numbers, decide if $\eval(\AA)$ is tractable or is \#P-hard.
We follow the model of computation discussed in Section \ref{complexmodel}.
Let
$$
\mathscr{A}=\{A_{i,j}:i,j\in [m]\}=\{a_j:j\in [n]\},
$$
and let $\alpha$ be a primitive element of $\mathbb{Q}(\mathscr{A})$ and thus,\vspace{0.08cm}
  $\mathbb{Q}(\mathscr{A})=\mathbb{Q}(\alpha)$.
The input of the problem consists of the following three parts:
\begin{enumerate}
\item A minimal polynomial $F(x)\in \mathbb{Q}[x]$ of $\alpha$;\vspace{-0.08cm}
\item A rational approximation $\hat{\alpha}$ of $\alpha$, which
  uniquely determines $\alpha$ as a root of $F(x)$; and \vspace{-0.08cm}
\item The standard representation, with respect to $\alpha$ and $F(x)$, of $A_{i,j}$,
  for all $i,j\in [m]$.\vspace{0.08cm}
\end{enumerate}
The input size is then the length of the binary string needed to describe all
  these three parts.
\vspace{0.008cm}


Given $\AA$, we follow the proof of Theorem \ref{main-in-intro}.
First by Lemma \ref{connected}, we can assume without loss of
  generality that $\AA$ is connected.
Then we follow the proof sketch described in
  Section \ref{outlinebipartite} and Section \ref{outlinenonbip},
  depending on whether the matrix $\AA$ is bipartite or non-bipartite.
We assume that $\AA$ is connected and bipartite.
The proof for the non-bipartite case is similar.

\subsection{Step 1}

In Step 1, we either conclude that $\eval(\AA)$
  is \#P-hard or construct a purified matrix $\AA'$
  such that $$\eval(\AA)\equiv \eval(\AA')$$
and then pass $\AA'$ down to Step 2.
We follow the proof of Theorem \ref{bi-step-1}.
First, we show that given $\mathscr{A}$ $=\{a_j:j\in [n]\}$,
  a generating set $\mathscr{G}=\{g_1,\ldots,g_d\}\subset \mathbb{Q}(\mathscr{A})$ of $\mathscr{A}$
  can be computed in polynomial time.\vspace{0.008cm}
Recall the definition of a generating set from Section \ref{generating-sec}.
We denote the input size as $\widehat{m}$. Thus $\widehat{m}\ge m$.\vspace{0.008cm}

\begin{theorem}\label{lattice}
Given a finite set of non-zero algebraic numbers $\mathscr{A}$
  \emph{(under the model of computation described in Section \ref{complexmodel})},
  one can in polynomial time \emph{(}in
  $\widehat{m}$\emph{)} find a generating set $\{g_1,\ldots,g_d\}$
  of $\mathscr{A}$.
Moreover, for each $a\in \mathscr{A}$, one can in polynomial time find the unique
  tuple $(k_1,\ldots,k_d)\in \mathbb{Z}^d$ such that
$$
\frac{a}{g_1^{k_1}\cdots g_d^{k_d}}\ \ \ \ \text{is a root of unity.}
$$
\end{theorem}

We start with the following Lemma.

\begin{lemma}\label{lattice2}
Let
\[L = \Big\{\big(x_1, \ldots, x_n\big) \in {\mathbb Z}^n
\hspace{0.1cm} \Big| \hspace{0.1cm} a_1^{x_1} \cdots a_n^{x_n} = 1 \Big\}.\]
Let $S$ be the ${\mathbb Q}$-span of  $L$,
and let $L' = {\mathbb Z}^n \cap S$, then
\begin{equation}\label{root-unity-eqn}
L' = \Big\{\big(x_1, \ldots, x_n\big) \in {\mathbb Z}^n
\hspace{0.1cm} \Big| \hspace{0.1cm} a_1^{x_1} \cdots a_n^{x_n} =
  \mbox{a root of unity}\hspace{0.08cm} \Big\}.
\end{equation}
\end{lemma}

\begin{proof}
Clearly $L$ is a lattice, being a discrete subgroup of ${\mathbb Z}^n$.
Also $L'$ is a lattice, and $L \subseteq L'$.

Suppose $(x_1, \ldots, x_n) \in {\mathbb Z}^n$
is in the lattice in (\ref{root-unity-eqn}).
Then there exists some non-zero integer $\ell$ such that
$(a_1^{x_1} \cdots a_n^{x_n})^\ell =1$.
As a result, $\ell (x_1, \ldots, x_n) \in L$ and thus,
  $(x_1, \ldots, x_n) \in S$, the ${\mathbb Q}$-span of  $L$.\vspace{0.012cm}

Conversely, if $\dim(L) = 0$, then clearly
$L = \{(0, \ldots, 0)\} = S= L'$.
Suppose $\dim(L) >0$, and we let ${\bf b}_1, \ldots, {\bf b}_t$
be a basis for $L$, where $1 \le t \le n$.
Let $(x_1, \ldots, x_n) \in \mathbb{Z}^n \cap S$,
then there exists some rational numbers $r_1, \ldots, r_t$
such that $(x_1, \ldots, x_n)  = \sum_{i=1}^t r_i {\bf b}_i$.
Then
\[ a_1^{x_1} \cdots a_n^{x_n} =
\prod_{j=1}^n a_j^{\sum_{i=1}^t r_i b_{i,j}}.\]
Let $N$ be a positive integer such that $N r_i$ are all
integers, for $ 1 \le i \le t$.
Then
\[\big(a_1^{x_1} \cdots a_n^{x_n}\big)^N
= \prod_{i=1}^t \left( \prod_{j=1}^n a_j^{ b_{i,j}} \right)^{N r_i}
=1.\]
Thus $a_1^{x_1} \cdots a_n^{x_n}$ is a root of unity and
  $(x_1,\ldots,x_n)$ is in the lattice in (\ref{root-unity-eqn}).
\end{proof}

To prove Theorem \ref{lattice}, we will also need the following theorem by Ge \cite{Ge1,Ge2}:

\begin{theorem}[\hspace{0.03cm}\cite{Ge1,Ge2}\hspace{0.03cm}]\label{ge}
Given a finite set of non-zero algebraic numbers $\mathscr{A}=\{a_1,\ldots,a_n\}$
  \emph{(under the model of computation described in Section \ref{complexmodel})},
  one can in polynomial time find a lattice basis for the lattice $L$ given by
\[L = \Big\{ \xx=\big(x_1,\ldots,x_n\big) \in
{\mathbb Z}^n \hspace{0.1cm} \Big| \hspace{0.1cm} a_1^{x_1}\cdots a_n^{x_n}=1\Big\}.\]
\end{theorem}


\begin{proof}[Proof of Theorem \ref{lattice}]
We prove Theorem \ref{lattice}.
Conceptually this is what we will do:
We first use Ge's algorithm to compute a basis for $L$. Then we show how to
  compute a basis for $L'$ efficiently. Finally, we compute a basis for ${\mathbb Z}^n/L'$.
This basis for ${\mathbb Z}^n/L'$ will define our generating set
  for $\mathscr{A}$.\vspace{0.006cm}

More precisely, given the set $\mathscr{A} = \{a_1, \ldots, a_n\}$,
  we use ${\mathbb \kappa} =\{ {\bf k}_1, \ldots, {\bf k}_t \}$ to
  denote the lattice basis for $L$
  found by Ge's algorithm \cite{Ge1,Ge2} where $0\le t \le n$.
This basis has polynomially many bits in each integer entry $k_{i,j}$.
The following two cases are easy to deal with:\vspace{0.1cm}
\begin{flushleft}\begin{enumerate}
\item If $t=0$, then we can take $g_i = a_i$ as the generators,
$1 \le i \le n$. There is no non-trivial
relation $$a_1^{k_1} \cdots a_n^{k_n} =\hspace{0.08cm} \text{a root of unity},$$
for any $(k_1, \ldots, k_n) \in {\mathbb Z}^n$ other than ${\bf 0}$,
otherwise a suitable non-zero integer power gives
a non-trivial lattice point in $L$.\vspace{-0.06cm}

\item If $t=n$, then $S = {\mathbb Q}^n$ and $L' =  {\mathbb Z}^n$,
hence every $a_i$ is a root of unity.  In this case, the empty set
$\emptyset$ is a generating set for $\mathscr{A}$.\vspace{0.1cm}
\end{enumerate}\end{flushleft}

Now we suppose $0 < t < n$.
We will compute from the basis ${\mathbb \kappa}$
a basis ${\mathbb \beta}$ for $L' = {\mathbb Z}^n \cap S$
where  $S$ is the ${\mathbb Q}$-span of  $L$.
Then we compute a basis ${\mathbb \gamma}$ for the quotient lattice
${\mathbb Z}^n/L'$.
Both lattice bases  ${\mathbb \gamma}$ and ${\mathbb \beta}$
will have polynomially many bits in each integer entry.\vspace{0.016cm}

Before showing how to compute $\beta$ and $\gamma$,
  it is clear that
$$
\dim L' = \dim L = t\ \ \ \text{and}\ \ \ \dim \big({\mathbb Z}^n/L'\big)=n-t.\vspace{-0.1cm}
$$
Let \vspace{-0.15cm}
$$
{\mathbb \gamma} = \big\{{\bf x}_1, \ldots, {\bf x}_{n-t} \big\}\ \ \ \text{and}\ \ \
  {\mathbb \beta}  = \big\{{\bf y}_1, \ldots, {\bf y}_t \big\}.\vspace{0.05cm}
$$
We define the following set $\{g_1, \ldots, g_{n-t}\}$ from $\gamma$ as follows:\vspace{0.05cm}
$$
g_j = a_1^{x_{j,1}} a_2^{x_{j,2}} \cdots a_n^{x_{j,n}},\ \ \ \
  \text{where ${\bf x}_{j} = (x_{j,1}, x_{j,2}, \ldots, x_{j,n})$.}
$$
We check that $\{g_1,\ldots,g_{n-t}\}$ is a generating set of $\mathscr{A}$.
Clearly, being exponentials, all $g_j \not = 0$.
Suppose for some $(c_1, \ldots, c_{n-t}) \in {\mathbb Z}^{n-t}$,
$g_1^{c_1} \cdots g_{n-t}^{c_{n-t}}$ is a root of
unity.  Since
\[g_1^{c_1} g_2^{c_2} \cdots g_{n-t}^{c_{n-t}}
= a_1^{\sum_{j=1}^{n-t} c_j x_{j,1}} a_2^{\sum_{j=1}^{n-t} c_j x_{j,2}}
\cdots a_n^{\sum_{j=1}^{n-t} c_j x_{j,n}},\]
we have
\[\left(\sum_{j=1}^{n-t} c_j x_{j,1}, \sum_{j=1}^{n-t} c_j x_{j,2},
\ldots, \sum_{j=1}^{n-t} c_j x_{j,n}\right) = \sum_{j=1}^{n-t} c_j {\bf x}_j
 \in L'.\vspace{0.16cm}\]
It follows that $c_j=0$, for all $1 \le j \le n-t$.
On the other hand, by the definition of $\mathbb{Z}^n/L'$,
  notice that for every $(k_1, \ldots, k_n) \in {\mathbb Z}^n$,
  there exists a unique sequence of integers $c_1, \ldots, c_{n-t} \in {\mathbb Z}$
such that $$(k_1, \ldots, k_n) - \sum_{j=1}^{n-t}
c_j {\bf x}_j \in L'.$$
In particular for ${\bf e}_i = (0, \ldots, 1, \ldots, 0)$,
where there is a single $1$ in the $i$th position,
there exist integers $c_{i,j}$, $1 \le i \le n$ and $1 \le j \le n-t$,
such that $${\bf e}_i - \sum_{j=1}^{n-t} c_{i,j} {\bf x}_j \in L'.$$
As a result, we have
\[\frac{a_i}
{a_1^{\sum_{j=1}^{n-t} c_{i,j} x_{j,1}} a_2^{\sum_{j=1}^{n-t} c_{i,j} x_{j,2}}
\cdots a_n^{\sum_{j=1}^{n-t} c_{i,j} x_{j,n}}}
=
\frac{a_i}
{g_1^{c_{i,1}} \cdots g_{n-t}^{c_{i,n-t}}},\]
is a root of unity.

This completes the construction of the generating set $\{g_1,\ldots,g_{n-t}\}$ for $\mathscr{A}
  =\{a_1,\ldots,a_n\}$.
In the following we describe how we compute the bases
$\gamma$ and $\beta$ in polynomial time, given $\kappa$.

Firstly, we may change the first vector ${\bf k}_1
= (k_{1,1}, \ldots, k_{1,n})$ in $\kappa$ to be a \emph{primitive} vector,
meaning that ${\rm gcd}(k_{1,1}, \ldots, k_{1,n}) =1$,
by factoring out the gcd.
If the gcd is greater than $1$ then this changes the la\-ttice $L$,
but it does not change the ${\mathbb Q}$-span $S$, and thus
no change to $L'$.

In addition, there exists a unimodular matrix $\MM_1$ such that
$$\big(k_{1,1}, \ldots, k_{1,n}\big)\hspace{0.06cm}
  \MM_1 = \big(1, 0, \ldots, 0\big)\in \mathbb{Z}^n.$$
This is just the extended Euclidean algorithm.
(An integer matrix $\MM_1$ is \emph{unimodular} if and only if its determinant is $\pm 1$,
or equivalently it has an integral inverse matrix.)
Now consider the $t \times n$ matrix
\[\begin{pmatrix} u_{1,1} & \ldots &  u_{1,n}\\
\vdots & \ddots & \vdots\\
u_{t,1} & \ldots &  u_{t,n}
\end{pmatrix} =
\begin{pmatrix} k_{1,1} & \ldots &  k_{1,n}\\
\vdots & \ddots & \vdots\\
k_{t,1} & \ldots &  k_{t,n}
\end{pmatrix} \MM_1.\]
This is also an integral matrix since $\MM_1$ is integral.
Moreover its first row is $(1, 0, \ldots, 0)$.
Now we may perform row transformations to make $u_{2,1} =0$,
\ldots, $u_{t,1} =0$. Performing the same transformations on
  the RHS replaces the basis $\kappa$ by another basis for the same lattice,
and thus  $L'$ is unchanged. We still use $\kappa=\{{\bf k}_1,\ldots,{\bf k}_t\}$ to denote this
new basis.

Next, we consider the entries $u_{2,2}, \ldots,  u_{2,n}$.
If ${\rm gcd}(u_{2,2}, \ldots,  u_{2,n}) >1$,
  we may divide out this gcd. Since the second row satisfies
$$\big(k_{2,1}, k_{2,2}, \ldots, k_{2,n}\big) =
  \big(0, u_{2,2}, \ldots, u_{2,n}\big)\hspace{0.06cm} \MM_1^{-1},$$
this gcd must also divide $k_{2,1}, k_{2,2}, \ldots, k_{2,n}$.
(In fact, this is also the gcd of $(k_{2,1}, k_{2,2}, \ldots, k_{2,n})$.)
This division updates the basis $\kappa$ by another basis,
which changes the lattice $L$, but still it does not change the
${\mathbb Q}$-span $S$, and thus the lattice $L'$ remains unchanged.
We continue to use the same $\kappa$ to denote this
updated basis.

For the same reason, there exists an $(n-1) \times (n-1)$ unimodular matrix $\MM'$
such that $$\big(u_{2,2}, \ldots, u_{2,n}\big)\hspace{0.06cm} \MM'
=\big(1, 0, \ldots, 0\big) \in {\mathbb Z}^{n-1}.$$
Append a 1 at the $(1,1)$ position, this defines
a second $n \times n$ unimodular matrix $\MM_2$
such that we may update the matrix equation as follows
\[\begin{pmatrix} 1 & 0 & 0 & \ldots & 0\\
0 & 1 & 0 & \ldots & 0\\
0 & u_{3,2} & u_{3,3} & \ldots & u_{3,n}\\
\vdots & \vdots & \vdots & \ddots & \vdots\\
0 & u_{t,2} & u_{t,3} & \ldots & u_{t,n}
\end{pmatrix} =
\begin{pmatrix} k_{1,1} & \ldots &  k_{1,n}\\
\vdots & \ddots & \vdots\\
k_{t,1} & \ldots &  k_{t,n}
\end{pmatrix} \MM_1 \MM_2.\]
Now we may kill off the entries $u_{3,2}, \ldots, u_{t,2}$,
accomplished by row transformations which do not change $L$ nor $L'$.
It follows that we can finally find a unimodular matrix $\MM^*$
such that the updated $\kappa$ satisfies
\begin{equation}\label{unimodular-final}
\begin{pmatrix} k_{1,1} & \ldots &  k_{1,n}\\
\vdots & \ddots & \vdots\\
k_{t,1} & \ldots &  k_{t,n}
\end{pmatrix} \MM^*
=
\begin{pmatrix} 1 & 0 & \ldots & 0 & 0 & \ldots & 0\\
0 & 1 & \ldots & 0 & 0 & \ldots & 0\\
\vdots & \vdots &  \ddots & \vdots & \vdots & \ddots & \vdots\\
0 & 0 & \ldots & 1 & 0 & \ldots & 0
\end{pmatrix},
\end{equation}
where the RHS is the  $t \times t$ identity matrix ${\bf I}_t$ appended
  by an all zero $t \times (n-t)$ matrix.
The updated $\kappa$ here is a lattice basis for a lattice $\widehat{L}$
  which has the same ${\mathbb Q}$-span $S$ as $L$.
It is also a full dimensional sublattice of (the unchanged) $L'$.

We claim this updated $\kappa=\{{\bf k}_1,\ldots,{\bf k}_t\}$ is
  actually a lattice basis for $L'$ and thus, $\widehat{L} = L'$.
Assume for some rational numbers $r_1, \ldots, r_t$,
the vector $\sum_{i=1}^{t} r_i {\bf k}_i \in {\mathbb Z}^n$,
then  multiplying $(r_1, \ldots, r_t)$ to the left in
  (\ref{unimodular-final}) impies that all $r_1, \ldots, r_t$ are integers.
This completes the computation of a basis for $L'$.
Since the only operations we perform are Gaussian eliminations
and gcd computations, this is in polynomial time, and the number of
bits in every entry is always polynomially bounded.

Finally we describe the computation of a basis for
the  quotient lattice
${\mathbb Z}^n/L'$.
We start with a basis $\kappa$ for $L'$ as computed above, and extend
it to a basis for ${\mathbb Z}^n$.  The extended part
will then be a basis for ${\mathbb Z}^n/L'$.
Suppose that we are given the basis $\kappa$ for $L'$
together with a unimodular matrix $\MM^*$ satisfying
(\ref{unimodular-final}).
Then consider the $n \times n$ matrix $(\MM^*)^{-1}$.
As $$(\MM^*)^{-1} = {\bf I}_n (\MM^*)^{-1},$$
the first $t$ rows of $(\MM^*)^{-1}$ is precisely the $\kappa$ matrix.
We define the basis for ${\mathbb Z}^n/L'$
to be the last $n-t$ row vectors of $(\MM^*)^{-1}$.
It can be easily verified that this is a lattice basis for ${\mathbb Z}^n/L'$.
\end{proof}

With Theorem \ref{lattice}, we can now follow the proof of Theorem \ref{bi-step-1}.
First, by using the generating set, we construct
  the matrix $\BB$ as in Section \ref{generating-sec}.
Every entry of $\BB$ is the product of a non-negative integer and a root of unity,
  and it satisfies $\eval(\AA)\equiv \eval(\BB).$

We then check whether $\BB'$, where $B_{i,j}'=|B_{i,j}|$ for all $i$ and $j$,
  satisfies the conditions imposed by the dichotomy theorem of Bulatov and Grohe.
(Note that every entry of $\BB'$ is a non-negative integer.)
If $\BB'$ does not satisfy, then $\eval(\BB')$ is \#P-hard, and so is $\eval(\AA)$
  by Lemma \ref{absolutevalue}.
Otherwise, $\BB$ must be a purified matrix and we pass it down to the next step.

\subsection{Step 2}


In Step 2, we follow closely the proof of Theorem \ref{bi-step-2}.
After rearranging the rows and columns of the purified matrix $\BB$, we check the orthogonality
  condition as imposed by Lemma \ref{hahajaja}.
If $\BB$ satisfies the orthogonality condition, we can use the cyclotomic reduction to
  construct efficiently a pair $(\CC,\fD)$ from $\BB$,
  which satisfies the conditions ({\sl Shape}$_1$),
  ({\sl Shape}$_2$), ({\sl Shape}$_3$) and
$$
\eval(\BB)\equiv \eval(\CC,\fD).
$$
Next, we check whether the pair $(\CC,\fD)$ satisfies ({\sl Shape}$_4$) and ({\sl Shape}$_5$).
If any of these two conditions is not satisfied, we know that $\eval(\CC,\fD)$ is
  \#P-hard and so is $\eval(\BB)$.
Finally, we check the rank-$1$ condition (which implies ({\sl Shape}$_6$)) as imposed by
  Lemma \ref{rank1} on $(\CC,\fD)$.
With
  ({\sl Shape}$_1$)\hspace{0.05cm}--\hspace{0.05cm}({\sl Shape}$_6$),
  we can finally follow Section \ref{step25} to construct a tuple $((M,2N),\XX,{\frak Y}')$ that
  satisfies $(\calU_1)$\hspace{0.05cm}--\hspace{0.05cm}$(\calU_4)$ and
$$
\eval(\CC,\fD)\equiv \eval(\XX,{\frak Y}').
$$
We then pass the tuple $((M,2N),\XX,{\frak Y}')$ down to Step 3.

\subsection{Step 3}

In Step 3, we follow Theorem \ref{step30}, \ref{bi-step-3},
   \ref{bi-step-4}, and \ref{bi-step-5}.
First it is clear that the condition $(\calU_5)$ in Theorem \ref{step30}
  can be verified efficiently.
Then in Theorem \ref{bi-step-3} we need to check whether
  the matrix $\FF$ has a Fourier decomposition
  after an appropriate permutation of its rows and columns.
This decomposition, if $\FF$ has one, can be computed efficiently
  by first checking the group condition in Lemma \ref{groupcondition1}
  and then following the proof of both Lemma \ref{decomp1} and Lemma \ref{decomp2}.
Finally it is easy to see that all the conditions imposed by
  Theorem \ref{bi-step-4} and Theorem \ref{bi-step-5}
  can be checked in polynomial time.

If $\AA$ and other matrices\hspace{0.03cm}/\hspace{0.03cm}pairs\hspace{0.03cm}/\hspace{0.03cm}tuples
  derived from $\AA$ satisfy
  all the conditions in these three steps, then by
  the tractability part of the dichotomy theorem,
  we immediately know that $\eval(\AA)$ is
  solvable in polynomial time.
From this, we obtain the polynomial-time decidability of the complexity dichotomy
  and Theorem \ref{theo-decidability} is proven.